\documentclass[fleqn,usenatbib]{mnras}

\usepackage{newtxtext,newtxmath}
\usepackage{arydshln}
\usepackage{amsmath}
\usepackage[T1]{fontenc}

\DeclareRobustCommand{\VAN}[3]{#2}
\let\VANthebibliography\thebibliography
\def\thebibliography{\DeclareRobustCommand{\VAN}[3]{##3}\VANthebibliography}

\usepackage{graphicx}	% Including figure files
\usepackage{amsmath}	% Advanced maths commands
\defcitealias{Quintana2025}{Q25}
\defcitealias{PantaleoniGonzalez2025}{PG25}
\defcitealias{Zari2021}{Z21}
\defcitealias{Poggio2021}{P21}
\defcitealias{Chen2019}{C19}
\defcitealias{GaiaDR3GoldenSample}{G23}
\defcitealias{Khalatyan2024}{K24}
\defcitealias{Maltsev2025}{M25}
\defcitealias{HuntReffert2024}{HR24}
\defcitealias{Ekstrom}{E12}

\usepackage{ulem}
\usepackage{soul}

\title[Unveiling the Milky Way with OB-type stars]{Unveiling the Milky Way with a Gaia DR3 census of OB-type stars within 2 kpc. I. Tracing local Galactic structure, massive star-forming regions and core-collapse supernova progenitors}

\author[A. L. Quintana, et al. ]{Alexis L. Quintana$^{1}$\thanks{E-mail: alexis.quintana@obspm.fr},  Kiril Maltsev$^{2}$,   Eloisa Poggio$^{3}$,  Emily L. Hunt$^{4}$, Nicholas J. Wright$^{5}$\thanks{E-mail: n.j.wright@keele.ac.uk}, \newauthor Sara R. Berlanas$^{6,7}$, Laia Casamiquela$^{1}$, Abel de Burgos$^{8}$, Hanna Parul$^{1}$,  Misha Haywood$^{1}$, \newauthor Paola Di Matteo$^{1}$,  Chervin Laporte$^{1,9,10}$ and Juan Mart\'inez Garc\'ia$^{5}$ \\
$^{1}$LIRA, Observatoire de Paris, Université PSL, Sorbonne Université, Université Paris Cité, CY Cergy Paris Université, CNRS, 92190 Meudon, France\\
$^{2}$Max-Planck-Institut f\"ur Astronomie, K\"onigstuhl 17, 69117 Heidelberg, Germany \\
$^{3}$INAF – Osservatorio Astrofisico di Torino, via Osservatorio 20, 10025 Pino Torinese (TO), Italy \\
$^{4}$Department of Astrophysics, University of Vienna, Türkenschanzstrasse 17, 1180 Wien, Austria\\
$^{5}$Astrophysics Group, Keele University, Keele ST5 5BG, UK\\
$^{6}$Instituto de Astrofísica de Canarias, E-38 200 La Laguna, Tenerife, Spain \\
$^{7}$Departamento de Astrofísica, Universidad de La Laguna, E-38205 La Laguna, Tenerife, Spain \\
$^{8}$European Southern Observatory, Alonso de Córdova 3107, Vitacura, Santiago, Chile \\
$^{9}$Institut de Ciències del Cosmos (ICCUB), Universitat de Barcelona, Martí i Franquès 1, E-08028 Barcelona, Spain \\
$^{10}$Kavli IPMU (WPI), UTIAS, The University of Tokyo, Kashiwa, Chiba 277-8583, Japan \\}

\date{Accepted 2026 November 3. Received 2026 October 15; in original form 2026 July 6}

\pubyear{\the\year{}}

\begin{document}
\label{firstpage}
\pagerange{\pageref{firstpage}--\pageref{lastpage}}
\maketitle

\begin{abstract}
O- and B-type stars are young and hot, thereby serving as vital tracers of the star formation and spiral arm structure of the Milky Way. At the dusk of the \textit{Gaia} DR3 era, a high-confidence and accurate catalogue appears timely. We have characterized 105,971 OB-type stars (T$_{\rm eff} >$ 10,000 K) within 2 kpc from the Sun, using an astro-photometric Bayesian inference tool. Our map unveils a complex view of the young stellar populations across the thin disk, with prominent large-scale features such as the Cepheus Spur, the Giant Oval Cavity, and a segment the Sagittarius-Carina arm all visible. Their inhomogeneous spatial distribution implies that massive star formation has taken place clustered across a few highly concentrated regions. We find a correlation between the overdensities of OB-type stars and young open clusters, and attribute departure from it to extinction effects. We identify 4200 stars as model-dependent core-collapse supernova (ccSN) or direct-collapse black hole (BH) progenitor candidates, including 6 ccSN progenitor candidates within 125 pc -- the maximum radius within which a ccSN explosion could be harmful to Earth's biosphere -- but our analysis suggests none of them exploding within the next 5 Myrs. Finally, we anticipate more evolved OB-type stars to collapse to form BHs within the next 1 Myr than ccSN progenitors will explode, despite the former's much scarcer number. A possible interpretation is a recent star formation burst in the local Milky Way, consistent with overdensities of ccSN progenitor candidates in Cyg OB2 and the Sagittarius-Carina arm.
\end{abstract}

%%%%%%%%%%%%%%%%%%%%%%%%%%%%%%%%%%%%%%%%%%%%%%%%%%

%%%%%%%%%%%%%%%%% BODY OF PAPER %%%%%%%%%%%%%%%%%%

\section{Introduction}
\label{introduction}

Our Milky Way is a barred-spiral galaxy, whose structure and properties can be analysed in more details than its distant/extragalactic counterparts (e.g. \citealt{BlandHawthornGerhard2016,Brown2021}), and whose study is critical to better understand how disk galaxies form and evolve (e.g. \citealt{vanderKruit2011,GaiaDR3_AsymmetricalDisk}). Within its main components, the spiral arms are the drivers of the chemical, kinematic and energetic evolution of galaxies (e.g. \citealt{Romanelli2025}). Nevertheless, from our internal perspective, we are only able to map them out indirectly, through the exploitation of tracers. Given that star formation in the Milky Way preferentially occurs within its spiral arms (e.g. \citealt{Elmegreen2011}), OB stars (e.g. \citealt{Chen2019,Poggio2021,Zari2021}) are such tracers, alongside giant molecular clouds (e.g. \citealt{HouHan2014}), masers (e.g. \citealt{Reid2019}), interstellar dust (e.g. \citealt{Rezeai2018}) and young star clusters (e.g. \citealt{CastroGinard2021}).

As the hottest and most massive members of the Harvard spectral classification \citep{Payne1925}, OB stars have a short lifetime (a few Myrs to a few hundreds Myrs, e.g.; see \citealt{Crowther2012}) that make them powerful tracers of the Milky Way spiral arms (e.g. \citealt{Chen2019}). They act as regulators of star formation, since their feedback processes (protostellar outflows, stellar winds, supernova explosions ...) repel and excite the molecular gas from which stars arise (e.g. \citealt{Krumholz2019}), a physical phenomenon known as residual gas expulsion (e.g. \citealt{BaumgardatKroupa2007}). On the other hand, core-collapse supernova (ccSN) explosions produced by dying OB stars  eject heavy chemical elements into the interstellar medium (see e.g. \citealt{Jones2013}) which are used to create the subsequent generations of stars and planets (e.g. \citealt{DeRossi2010}). Furthermore, young OB stars are the most prominent members of OB associations (e.g. \citealt{McKeeWilliams1997}), which are stellar groups characterized by their low-density ($<$ 0.1 M$_{\odot}$ pc$^{-3}$); and because these OB associations are gravitationally unbound \citep{Ambartsumian1947}, they will expand and dissolve into the Galactic field after a few tens of Myrs. They thus serve as short-lived tracers of star formation and Milky Way structure \citep{Wright2020}, and should be characterized and mapped out jointly with the general population of OB stars.

ESA's \textit{Gaia} satellite \citep{Gaia} has revolutionized our view of the Milky Way over the past decade. This is particularly true in the solar neighbourhood \citep{Zucker2023}, with the identification of important structures such as the Radcliffe Wave \citep{Alves2020}, the Split \citep{Lallement2019}, the Cepheus \citep{PantaleoniGonzalez2021} and Sagittarius \citep{Kuhn2021} spurs, superseding historical features such as the Gould Belt \citep{PantaleoniGonzalez2026}. Amongst other breakthroughs, \textit{Gaia} data has enabled the compilation of all-sky catalogues of OB stars larger by several orders of magnitudes compared with their historical counterparts (e.g. \citealt{Poggio2021,Zari2021,PantaleoniGonzalez2025}), including from the \textit{Gaia} consortium themselves as the golden sample of OBA stars from \citet{GaiaDR3GoldenSample}.

In \citeauthor{Quintana2025} (\citeyear{Quintana2025}; hereafter, \citetalias{Quintana2025}) we compiled a census of $\sim$25,000 O- and B-type stars within 1 kpc to map out their spatial distribution and various characteristics derived from photometry. The objective was to produce a catalogue as accurate and complete as possible, valuable for the scientific community. Most notably, this catalogue provides targets for upcoming ground-based spectroscopic follow-up such as the SCIP survey of the William Herschel Telescope Enhanced Area Velocity Explorer (WEAVE, \citealt{WEAVE}) and the 4-metre Multi-Object Spectroscopic Telescope (4MOST, \citealt{4MOST}).

This catalogue improved our understanding of star formation in the local Milky Way ($<$1 kpc), by providing new values of the core-collapse supernova (ccSN) rate of 15--30 per Myr, star formation rate (SFR) of $2896^{+417}_{-1}$ M$_{\odot}$ Myr$^{-1}$, and surface density star formation rate ($\Sigma_{\rm SFR}$) of $922^{+133}_{-1}$ M$_{\odot}$ Myr$^{-1}$ kpc$^{-2}$. The latter value is compatible with the clustered model of star formation from \citet{LadaLada2003} once it was compared with its equivalent from open clusters ($\Sigma_{\rm SFR,OC}$), taking advantage of the catalogue from \citet{HuntReffert2023,HuntReffert2024} (see \citealt{QuintanaHuntParul2025}). From the sample in \citetalias{Quintana2025}, the SFR and ccSN rates were extrapolated to the entire Milky Way (respectively $0.67^{+0.08}_{-0.01}$ M$_{\odot}$ yr$^{-1}$ and 0.4--0.5 per century), lower values compared to past estimates thanks to improvements in the stellar census and evolutionary models. Finally, we also exploited this catalogue to identify 56 OB associations, doubling the historical census \citep{Wright2020}, and providing an updated list that enables the study of the star formation history of the local Milky Way (see \citealt{Quintana2026} for details).

In this work, we expand it up to a distance of 2 kpc. This corresponds to an increase of the surface area covered of a factor of four, which will allow us to start intercepting the Perseus and Sagittarius-Carina arms (e.g. \citealt{Reid2019,Zari2021}) and therefore to extract more accurate properties of the entire Milky Way. Mapping out stellar populations down to late B-type stars enables us to disentangle both the scale of OB associations as well as larger features (spurs, superbubbles and spiral arms). The resulting catalogue can then be compared with other Milky Way components identified and catalogued with unprecedented accuracy with \textit{Gaia} data, such as star clusters (e.g. \citealt{HuntReffert2023}) and interstellar dust \& gas (e.g. \citealt{Edenhofer2024,Kormann2026,McCallum2026}). Before the upcoming release of \textit{Gaia} DR4, which will notably benefit from astrometry $\sim$2.25 times more precise than \textit{Gaia} DR3 \citep{Brown2019}, this study aims at improving our understanding of the structure,  star formation and ccSN explosion landscape within the local Milky Way, by compiling an accurate and complete census of OB stars within 2 kpc and combining it with other catalogues.

This paper is divided as follows. In Section \ref{identification}, we outline the astro-photometric fitting method used to identify OB stars, and compile a catalogue of their population in the local field. In Section \ref{catalogue_validation}, we test the validity of our catalogue by comparing the estimated physical parameters with external spectroscopic and Gaia-based catalogues, and by crossmatching it with other lists of OB(A) stars. In Section \ref{galactic_structure}, we visualize the spatial distribution of OB stars from our sample and study the structures they trace, particularly focused on the spiral arms. In Section \ref{open_clusters}, we compare OB-star density distribution with the star cluster catalogue from \citet{HuntReffert2023,HuntReffert2024} in order to analyse the model of clustered star formation across the Milky Way's thin disk. In Section \ref{observed_ccsn}, we identify and characterize OB-type star ccSN progenitor candidates and estimate the waiting time between consecutive ccSN events in the local Milky Way. Our findings are summarized in Section \ref{conclusions}.

\section{Identifying OB stars}
\label{identification}

%In this section we describe the methodology adopted to identify OB stars, first by selecting candidate OB stars across the dereddened \textit{Gaia} DR3 CMD (outlined in Section \ref{data}), then by applying an astro-photometric fitting process to estimate their physical properties (summarized in Section \ref{SEDfitter}), with the primary results summarised in Section \ref{gen_results}. 

 %\textcolor{red}{Moreover, the lifetime of a late B-type star (e.g. \citealt{Ekstrom,Zari2025}) roughly corresponds to the value of an orbital period of the Sun around the centre of the Milky Way (e.g. \citealt{Nitschai2020}): therefore, the defined sample enables us to analyse the properties of the young stellar populations of the thin disk up to one Galactic year.}

\subsection{Selection of candidate OB stars}
\label{data}

Our selection process of candidate OB stars within 2 kpc is based on astrometry and photometry from \textit{Gaia} DR3 \citep{GaiaDR3}, which contains astrometric information for $\sim$1.5 billion sources. Given the increasing uncertainties of the \textit{Gaia} DR3 parallaxes with the distance (e.g. \citealt{Quintana2023}), as well as the incompleteness arising from growing interstellar extinction towards the Galactic Centre (e.g. \citealt{Nogueras2019}), we decided to limit the sample to stars within 2 kpc. Furthermore, we stress that, just like in \citetalias{Quintana2025}, we will not follow the sometimes-used definition of an OB star, of stars of spectral type B2 and earlier for main-sequence (dwarf) stars, B5 and earlier for giants and any O- and B-type stars for supergiants \citep{Morgan1951,PantaleoniGonzalez2025}. Instead, we focus on stars with an effective temperature greater than 10,000 K, effectively including every main-sequence O- and B-type star alongside blue supergiants (i.e., every star classified as "O" or "B" according to the Harvard spectral classification, \citealt{Mamajek}). %Including late B-type stars notably allows us to better estimate general stellar properties and fit the initial mass function (the latter of which is beyond the scope of the present work and will be left for a follow-up study)}.

To that end, we followed a similar approach as in \citetalias{Quintana2025} where we performed a cylindrical cut in order to reduce edge effects. We have relied on the \texttt{astroquery} Python package \citep{Astroquery} to select \textit{Gaia} DR3 sources that fell within a distance of 2.2 kpc\footnote{We have chosen an upper distance threshold akin to \citetalias{Quintana2025}, 10\% above the distance limit to account for \textit{Gaia} parallax uncertainties.}; this is,  with $\sqrt{X^2+Y^2} < 2.2$ kpc, where we used the heliocentric Galactic Cartesian coordinates $X = d \, \cos(l) \, \cos(b)$ and $Y = d \, \sin(l) \, \cos(b)$, with $d$ being the geometric line-of-sight distance from \citet{BailerEDR3}. Then we corrected the \textit{Gaia} DR3 photometry and astrometry based on the approach from \citet{Maiz2021}, \citet{Maiz2022} and \citet{MaizApellanizWeiler2025}. In doing so, we discarded sources with $RUWE > 8$, that we complemented by a cut based on parallax uncertainties, i.e. we only kept sources with $\frac{\varpi}{\sigma_{\varpi}} > 2$, where $\varpi$ and $\sigma_{\varpi}$ are the corrected \textit{Gaia} DR3 parallax and the corrected \textit{Gaia} DR3 parallax uncertainty, respectively.

We have also only kept sources with valid G-band photometry, a condition that was met in all cases except whenever $\sigma_G > 2 \, \sigma_{G_{\rm BP/RP}}$, given that there are partially unresolved binaries whose dispersion was thus too large for a \textit{Gaia} detection to be reliable \citep{Maiz2018,Maiz2023}. Furthermore, we have also discarded the photometry from the $G_{\rm BP}$ $G_{\rm RP}$ bands with $\sigma_{G_{\rm BP/RP}} > 2 \, \sigma_G)$, and with $|C^*| < 3 \, \sigma_{C*}$, where $C^*$ is the corrected excess flux factor in the $G_{\rm BP}$ and $G_{\rm RP}$ bands, whilst $\sigma_{C*}$ stands for the power-law on the $G$ band with a chosen 3$\sigma$ level \citep{Riello}. Still based on \citet{Riello}, we have also corrected the three \textit{Gaia} DR3 photometric bands for saturation. 

Since we focus on hot stars ($T_{\rm eff} >$ 10,000 K), candidate OB stars sit in the upper-left part of the absolute \textit{Gaia} DR3 CMD, after correcting for distance and extinction. To do so, we adopt the geometric distance from \citet{BailerEDR3}, and derive $A_V$ as a function of said distance for each source through an interpolation from the \texttt{Edenhofer2024} 3D extinction map \citep{Edenhofer2024}\footnote{We use the main map for line-of-sight distances below 1.25 kpc and the reconstructed map between 1.25 and 2 kpc. For sources further than 2 kpc, we adopt their extinction at 2 kpc because they typically sit at higher Galactic latitudes where their extinction tends to be low.}. The selection on magnitude and colour writes as:

\begin{equation}
\label{mgsec}
M_G = G - 5 \, \log(d) + 5 - 1.25 \, A_G < 1.5 \, \rm mag
\end{equation}

\begin{equation}
\label{bprpsec}
(BP-RP)_0 = (BP-RP) - (A_{BP} - A_{RP}) < 0.5 \, \rm mag
\end{equation}

\noindent where, in both cases, we chose a more liberal threshold than the values of $M_G$ = 1 mag and $BP-RP = -0.037$ mag for an A0V star in \citet{Mamajek}, to account for uncertainties. Especially, to convert from $A_V$ to $A_G$, $A_{BP}$ and  $A_{RP}$, we have adopted the factors from \citet{WangChen2019}, even though there is a temperature dependence on this conversion (see \citealt{Fouesneau2023}). We also included sources with invalid $G_{\rm BP}$ or $G_{\rm RP}$ photometry, on which we only applied the cut from Eq. \eqref{mgsec}.

We crossmatched the resulting sample with several photometric surveys\footnote{We do not impose that the stars in our catalogue are in all these surveys, but we adopt their photometric measurements when available}. First we used \texttt{astroquery} to find 2MASS counterparts \citep{2MASS}\footnote{2 Micron All Sky survey} using the \texttt{gaiadr3.tmass\_psc\_xsc\_best\_neighbour} parameter, that associates every \textit{Gaia} DR3 source with their best neighbour in 2MASS \citep{Marrese2017,Marrese2019}. We then also crossmatched this catalogue with the IGAPS DR1\footnote{the INT Galactic Plane Survey} \citep{Drew,Mongui} and VPHAS+ DR2\footnote{The VST Photometric H$\alpha$ Survey of the Southern Galactic Plane and Bulge} \citep{VPHAS} surveys. For 2MASS, we required that each individual photometric band has a `good' photometric quality flag (i.e., `A', `B', `C' or `D'). For IGAPS, we only included sources identified as "star" or "probable star", that are unsaturated, and with an ellipticity lower than 0.3 and an errBits parameter (used as processing warning and error quality flag) smaller than 10, which removes sources with unreliable photometry (e.g., with saturated or bad pixels, see \citealt{Mongui} for more details). For VPHAS+, we only used the sources classified as `clean', that is, with a good point spread function (PSF) fit and a significant detection \citep{VPHAS}.

This selection process leads to a preliminary catalogue of 1,049,399 candidate OB stars within 2.2 kpc. This is the sample on which we will apply our SED fitter to identify the subset of highly-reliable OB stars.

\subsection{SED fitting}
\label{SEDfitter}

To estimate physical parameters of our candidate OB stars, we apply a spectral energy distribution (SED) fitting process, wherein we compute a forward-modelled SED, based on the combination of stellar atmosphere and evolutionary models with a 3D extinction map, to an observed SED. We perform the fitting with the \texttt{emcee} Python package \citep{Emcee} using a maximum-likelihood, and free parameters of the initial mass (which, in our stellar evolutionary models, corresponds to the stellar mass at the zero-age main-sequence, hereafter ZAMS), $\log(M_{\rm ZAMS}/M_{\odot})$, fractional age $Fr(\tau_f)$ and distance $d$, alongside the additional uncertainty $\ln (\rm f)$ that effectively serves as an indicator of the quality of the fit. A full description of the SED fitter is provided in \citetalias{Quintana2025}, with the main changes summarized below and the methodological details (including a discussion on the quality of the SED fitting process) in Appendix \ref{quality_SED_fitting}.

The observed SED combines \textit{Gaia} DR3 parallax with photometry from \textit{Gaia} DR3, 2MASS, IGAPS DR1 and VPHAS+ DR2 (see Section \ref{data}). Aside from the $g$, $r$ and $i$ bands from IGAPS and VPHAS+ that we already incorporated in \citetalias{Quintana2025}, we also include their U-band. While the U-band from VPHAS+ does not require any specific calibration, the U-band from IGAPS does. The correction used in this work is based on the method from \citet{DeBurgos2025}, which consists of a slit cut around the point of largest density in the $U-g$ vs $g-r$ colour-colour diagram, and was tailored to select target OB stars for the WEAVE-SCIP-LR survey \citep{DeBurgos2025}. Here we apply this correction to every IGAPS field (consists of 5x5 deg$^2$ areas from $l$= [$30\degr$,$215\degr$] and $b$= [$-4\degr$,$4\degr$]), and therefore leverages all the IGAPS data within a given field, obtaining  a list of corrections for the U-band. This correction improves slightly our SED fitter as the peak emission from OB stars is located in the near-UV domain, though the U-band is generally satured for O-type stars within 2 kpc.

The model SED was constructed with the same stellar atmosphere models as in \citetalias{Quintana2025} at solar metallicity and $\log g = 4$, with the BT-NextGen models from \citet{Allard2012} for $T_{\rm eff} = 3000-5000$ K, from the Kurucz models \citep{Coelho} for $T_{\rm eff} = 6000-20,000$ K and from the Tubingen Non Local Thermodynamical Equilibrium (NLTE) Model Atmosphere Package \citep{Werner1999,Rauch,Werner} for $T_{\rm eff} = 21,000-50,000$ K. Likewise, we still interpolate $\log(T_{\rm eff})$ and $\log(L/L_{\odot})$ from the single stellar evolutionary models from \citeauthor{Ekstrom} (\citeyear{Ekstrom}; hereafter, \citetalias{Ekstrom}). With this technique, we therefore model every source as a single star even though O- and B-type stars are more likely to belong to multiple systems (e.g. \citealt{sana2012,DucheneKraus2013,Offner2023}). We implicitly assume that OB star primary outshines a fainter binary companion (if present) that cannot be resolved using broad-band photometry information alone. Here and throughout this paper, with the term ``OB star'' we refer to the OB star primary, but point at the caveat that a fraction of OB-type sources we identify could be false positives due to unresolved binarity.

Finally, we constrain the visual extinction $A_V$ as a function of distance with the 3D extinction map from \citet{Edenhofer2024}, interpolating from the main map for sources closer than 1.25 kpc, and from the reconstructed map for sources located between 1.25 and 2 kpc\footnote{The data from both versions of the dust map have been queried through the \texttt{dustmaps} Python package \citep{Dustmaps}. We have specifically used the version v1.0 and the files \texttt{mean\_and\_std\_lbd.fits} and \texttt{validation\_with\_less\_data\_but\_2kpc\_mean\_and\_std\_lbd.fits}, which are available on Zenodo (\url{https://doi.org/10.5281/zenodo.8187943}).}. Further details about the implementation of visual extinction and the validation of the reconstructed map are provided in Appendix \ref{extinction_effects}.

\subsection{Final catalogue}
\label{gen_results}

We applied the SED fitter to all the OB star candidates, deriving stellar parameters for all of them. As in \citetalias{Quintana2025}, early A-type stars and late B-type stars dominate the sample, with a median $\log(M_{\rm ZAMS}/M_{\odot})$ of 0.31 dex (and a 1$\sigma$ value of 0.14 dex), a reflection of our selection choices in the dereddened \textit{Gaia} DR3 CMD in Section \ref{data}.

We adopted $\log(T_{\rm eff}) > 4$ as threshold for selecting OB stars and added the missing sources from the Bright Stars Catalogue \citep{Schlesinger1930,Hoffleit1991} (hereafter BSC) without or with bad \textit{Gaia} DR3 data (see \citetalias{Quintana2025} for details). This selection resulted in a subset of 147,639 O- and B-type stars (hereafter, OB stars). Then we only kept the stars within $D_{\rm XY} = \sqrt{X^2+Y^2} <$ 2 kpc, where $X$ and $Y$ are defined as in Section \ref{data} but with our median SED-fitted distances $d$ rather than the geometric distances from \citet{BailerEDR3}\footnote{In practice, both these distance estimates are very close within the area covered in this work, as illustrated in Fig. \ref{Comp_Dist}.}, yielding a catalogue of 105,971 SED-fitted OB stars within 2 kpc. 

\begin{figure*}
    \centering
    \includegraphics[scale = 0.35]{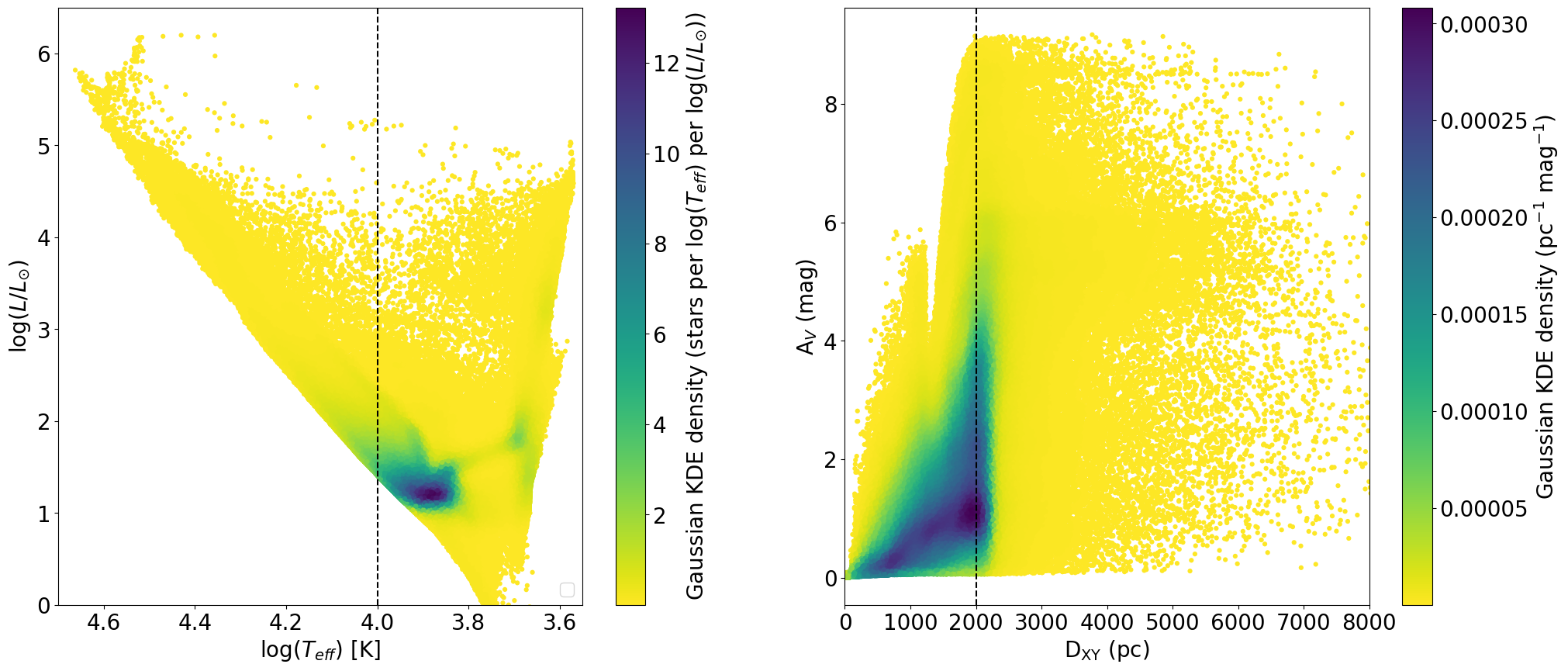}
    \caption{Left panel: HR diagram of the 1,049,399 candidate OB stars from their median SED-fitted effective temperatures and luminosities. Right panel: median $D_{\rm XY} = \sqrt{X^2+Y^2}$ values as a function of the median A$_V$ for these same stars. Each point has been colour-coded by gaussian KDE density. The vertical dashed lines correspond to the threshold of $\log(T_{\rm eff}) =$ 4 and $D_{\rm XY} =$ 2000 pc adopted to create our final catalogue of 105,971 SED-fitted OB stars within 2 kpc.}
    \label{Parameters_Candidate_OBstars}
\end{figure*}

Fig. \ref{Parameters_Candidate_OBstars} shows the Gaussian KDE density of the 1,049,399 candidate OB stars, where the above-mentioned cuts are indicated: the left panel displays the HR diagram derived from the median SED-fitted effective temperatures and luminosity, and the right panel shows the median A$_V$ from \citet{Edenhofer2024} as a function of $D_{\rm XY}$. 

The left panel again reflects our liberal selection choices from the \textit{Gaia} DR3 CMD, with the presence of main-sequence early A-type stars and cool giants. The right panel, on the other hand, shows that a minority of our candidate OB stars have been fitted beyond $D_{\rm XY} >$ 2200 pc, illustrating a small deviation between our SED-fitted distances and the geometric distances from \citet{BailerEDR3}. Visible in the right panel from Fig. \ref{Parameters_Candidate_OBstars} is also a small discontinuity in the relation between distance and reddening around 1.25 kpc, inherent to the adopted 3D extinction map (as the high-resolution map from \citet{Edenhofer2024} only extends up to that distance, beyond which we have used the reconstructed version of the map).

\section{Catalogue validation}
\label{catalogue_validation}

\subsection{Completeness}
\label{incompleteness}

\subsubsection{Sources without G-band photometry}

In Section \ref{data}, we excluded de facto all \textit{Gaia} DR3 sources with missing G-band photometry. They correspond to $\sim$0.3 \% of the entire \textit{Gaia} DR3 catalogue, shared among stellar and extragalactic objects \citep{GaiaDR3}. 

A notable cause of missing G-band photometry is saturation, as it has been estimated that $\sim$20 \% of stars with $G < 3$ mag are missing from \textit{Gaia} DR3    \citep{Fabricius2021}, an issue we have compensated for by the inclusion of the missing OB stars from the BSC, as mentioned in Section \ref{gen_results}. Therefore, we argue that the impact of missing G-band photometry is negligible, as this catalogue includes all stars with $V$ < 6 mag.

\subsubsection{Interstellar extinction}
\label{missing_extinction}

Interstellar extinction can significantly dim the visual magnitude of OB stars, becoming too faint to be detected by \textit{Gaia}. Our SED-fitted OB stars have a median and 1$\sigma$ $G$-band photometry of 11.5 mag and 2.0 mag, respectively, far brighter than the magnitude limit of $G \approx 21$ mag \citep{Riello}. Most of the reddening across our volume is relatively low (the $A_V$ derived from \citealt{Edenhofer2024} have median and 1$\sigma$ values of 1.17 mag and 1.20 mag, respectively), allowing us to map out the vast majority of the OB stars.

Our catalogue actually contains only $\sim$100 OB stars with $A_V >$ 8 mag, and they are all are located within the Cyg OB2 region, which is famous for its high extinction ($A_V =$ [5--10] mag for its population of massive stars, see, e.g. \citealt{Wright2015}). Consequently, aside from a few potential stars obscured due to the high-reddening from Cyg OB2 (knowing that an overdensity of A-type stars has also been found in the Cygnus region, \citealt{Ardevol2023}), we argue that our catalogue of OB stars within 2 kpc is expected to be nearly complete to the faint end. 

\subsubsection{Uncertainties on extinction and distance}

The initial selection boundaries defined in Eqs. \eqref{mgsec} and \eqref{bprpsec} rely on external estimations of distances and extinction, which are mutually-dependent due to the usage of a 3D dust map. On the \textit{Gaia} DR3 CMD, our targets of interest will be characterized by the smaller values of $M_G$ and $BP-RP$. This means that under-estimating either the stellar distance or reddening (e.g., if a line-of-sight is not properly resolved by the dust map) could potentially cause genuine OB stars to be discarded.

In principle, the high-resolution from \texttt{Edenhofer2024} prevents us from introducing a significant amount of false negatives, but the exploitation of the reconstructed version between 1.25 and 2 kpc implies a lower resolution, on top of the intrinsic systematics of the dust map itself (we discuss these caveats in Appendix \ref{extinction_effects}).

The distance estimates from \citet{BailerEDR3}, on the other hand, are based on a Bayesian approach and include priors based on a general Galactic stellar population. \citet{PantaleoniGonzalez2025} compared their distant estimates with priors tailored for a young and massive stellar population, and found that the disrepancies between their values and those from \citet{BailerEDR3} were generally lower than 2.5 \% within 1 kpc. Nevertheless, they also found that \citet{BailerEDR3} typically underestimated distance estimates by a factor 2 for massive stars (as they did not apply corrections to parallax uncertainties). While our SED-fitted distances apply a similar correction for parallax uncertainties than \citet{PantaleoniGonzalez2025} (see Section \ref{data}), Eq. \eqref{mgsec} could be affected by that, although our sample of candidate OB stars is dominated by late B-type and early A-type stars excluded from \citet{PantaleoniGonzalez2025}. Consequently, both the choices of distance estimates and extinction map could slightly affect the completeness of our sample.

\subsubsection{Bad or missing astrometric data}

We have estimated the level of loss of potential OB stars due to bad or missing \textit{Gaia} DR3 astrometry similar to the approach from \citetalias{Quintana2025}, this is, by calculating the fraction of sources discarded as a function of $G$-band magnitude, for each step of the astrometric cuts performed in Section \ref{data}. In this work, the three astrometric cuts are sources with high RUWE values ($> 8$), large parallaxes over parallax uncertainties ($\frac{\varpi}{\sigma_{\varpi}} > 2$), and 2-parameter solutions in \textit{Gaia} DR3. For the latter, we again used the fractions displayed in Table 6 from \citet{Lindegren2021} for sources with $G \geq$ 9 mag, and directly downloaded all the $\sim$177,000 sources in \textit{Gaia} DR3 brighter than $G = 9$ mag in order to estimate which fractions of sources have 2-parameter solutions as a function of $G$-band magnitude. 

\begin{figure}
    \centering
    \includegraphics[scale=0.32]{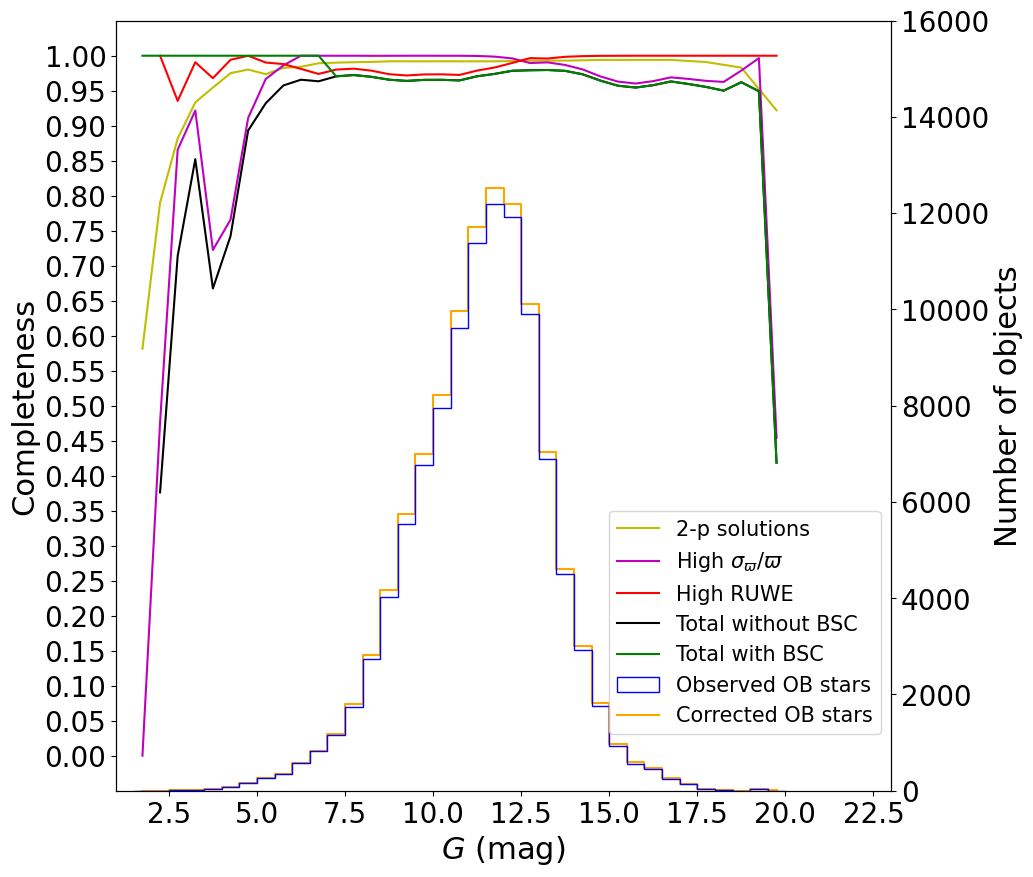}
    \caption{Observed (in blue) and completeness-corrected (in orange) numbers of SED-fitted OB stars within 2 kpc (right-hand Y axis). Each top line represents the fraction of sources that passed the different cuts (based on astrometric data quality) from Section \ref{data} and discussed in Section  \ref{incompleteness}, with the total obtained through the convolution of each curve (left-hand Y axis).} 
    \label{IncompletenessSample}
\end{figure}

Fig. \ref{IncompletenessSample} illustrates the incompleteness of our catalogue of SED-fitted OB stars within 2 kpc as a function of $G$-band magnitude, for the different astrometric cuts applied in Section \ref{data}. The observed population, as well as the completeness-corrected one, have also been displayed. 

It is clear that our sample benefits from a high-level of completeness, with a completeness over 95 \% across all magnitudes. Most of the incompleteness arises either from the saturation at the brightest magnitudes (compensated by the inclusion of the BSC) or at the magnitude limit of \textit{Gaia} DR3. As discussed in Section \ref{missing_extinction}, the majority of the sources concerned by this issue are located towards the highly-obscured Cyg OB2, so again we argue that our catalogue is mostly complete aside from some the visually faintest late B-type stars within (or behind) Cyg OB2.

\subsection{Comparison with previous surveys}
We compare two critical aspects of our sample with previous surveys of OB stars, namely comparing the SED-fitted $\log T_\mathrm{eff}$-estimates with spectroscopic catalogues in Section \ref{comp_spectro} and the SED-fitted initial stellar masses with the open clusters catalogue from \citeauthor{HuntReffert2024} (\citeyear{HuntReffert2024}; hereafter, \citetalias{HuntReffert2024}) in Section \ref{comp_mass}. Furthermore, in Appendix \ref{comparison_obstars} we compare our sample with external catalogues of OB(A) stars.

\subsubsection{Spectroscopically inferred $\log T_\mathrm{eff}$}
\label{comp_spectro}

\label{spectro}

Our SED-fitted effective temperatures were estimated based on astro-photometric data. To assess their quality, we compare them with effective temperatures derived from spectroscopic measurements, for all our 773,077 candidate OB stars within 2 kpc. 

To that end, we have selected the following samples:

\begin{itemize}
    \item The \textit{Gaia} consortium estimated stellar physical parameters as part of the \textit{Gaia} DR3 release \citep{Creevey2023}. This includes the Extended Stellar Parametrizer for hot stars (ESP-HS) module, based on the combination of the BP/RP spectra with the Radial Velocity Spectrometer (RVS) spectra (whenever available), with $T_{\rm eff} >$ 7500 K and $G \leq$ 17.65 mag. A crossmatch of our candidate OB stars with this catalogue gives a subset of 398,363 stars.
    \item The Apache Point Observatory Galactic Evolution Experiment (APOGEE) DR17 spectroscopic sample \citep{Garcia,Abdu} has estimated stellar physical parameters, including $T_{\rm eff}$ up to 20,000 K. From this catalogue, we have selected sources with a measured value of $T_{\rm eff}$ and without a warning on their temperature measurement, obtaining a comparison subset of 13,897 stars.
    \item \citet{Xiang2022} estimated stellar physical parameters for $\sim$330,000 OBA stars ($T_{\rm eff} >$ 7500 K) based on the LAMOST DR6 pipeline, the luminous OBA stars from \citet{Zari2021} and the OB stars catalogue from \citet{Liu2019}. Through crossmatching we obtain a comparison subset of 37,132 stars, selecting those with S/N $>$ 30 from \citet{Xiang2022}.
    \item \citet{Khalatyan2024} exploited \textit{Gaia} DR3 data, including its XP spectral coefficients, together with 2MASS and AllWISE data, from which they estimated stellar physical parameters based on a machine learning regression technique (based on the supervised training of spectroscopic data). This allowed them to produce a catalogue named SHBoost 2024, from which we select those with an input flag corresponding to all data used (\textit{Gaia} DR3 XP spectral coefficients, astrometry and broad-band photometry, as well as 2MASS and AllWISE photometry) and a good quality flag on effective temperature, which gives a comparison subset of 463,866 stars. For this comparison, we use the columns of "logTeffmean" and "s\_logTeffmean" from \citet{Khalatyan2024} for the values and uncertainties, respectively.
    
\end{itemize}

\begin{figure*}
    \centering
     \includegraphics[scale=0.20]{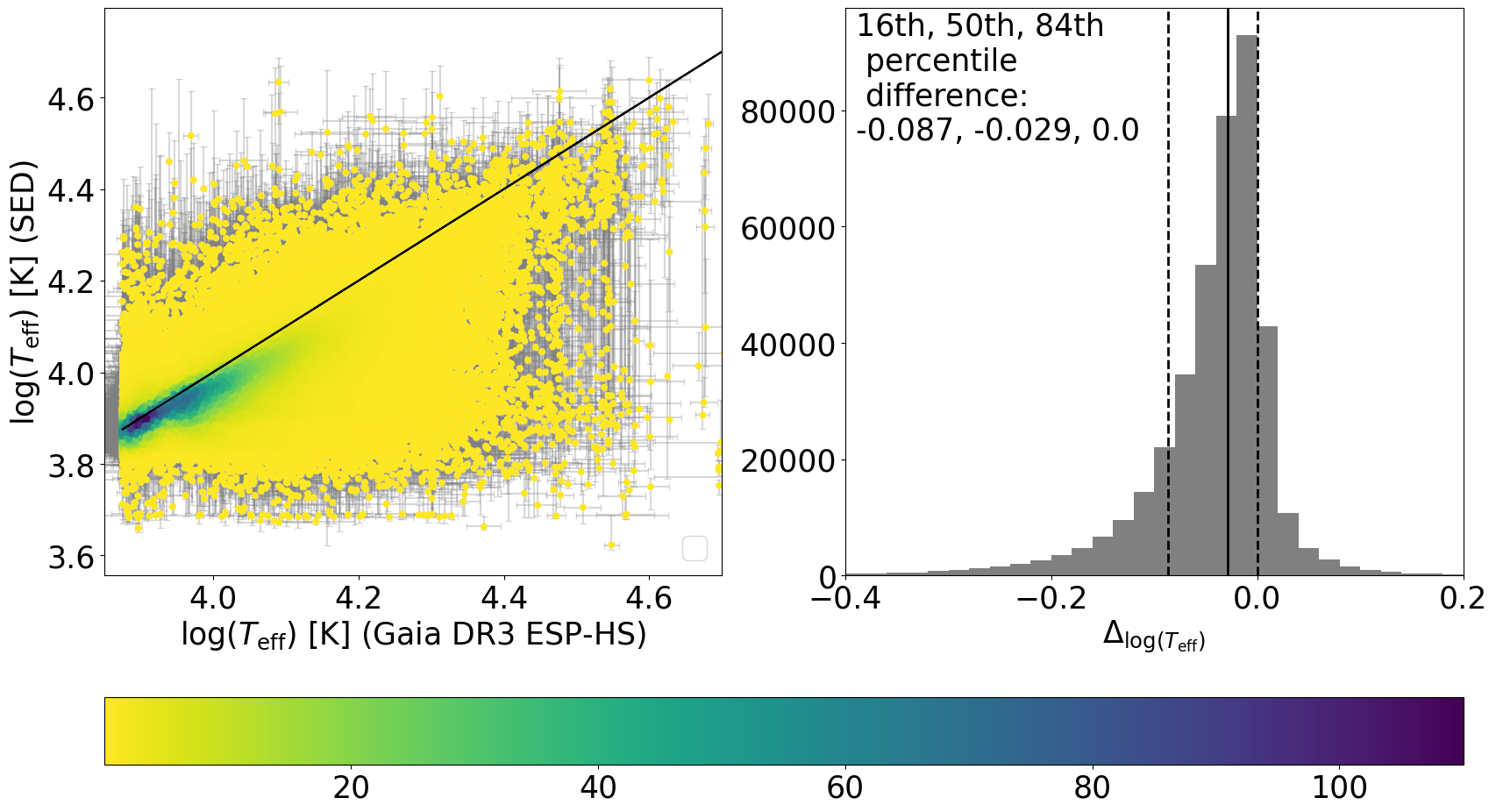}
     \includegraphics[scale=0.20]{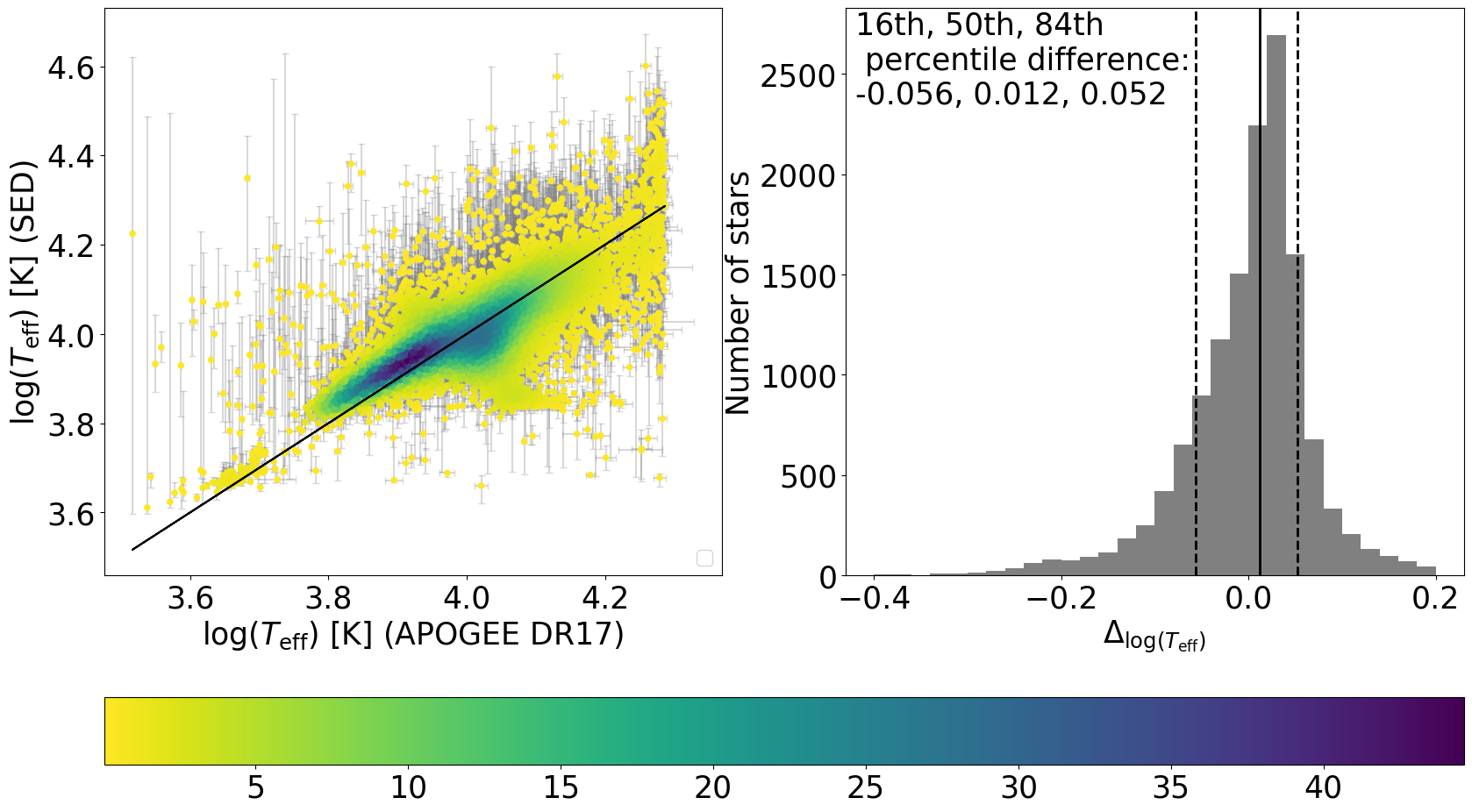}
     \includegraphics[scale=0.20]{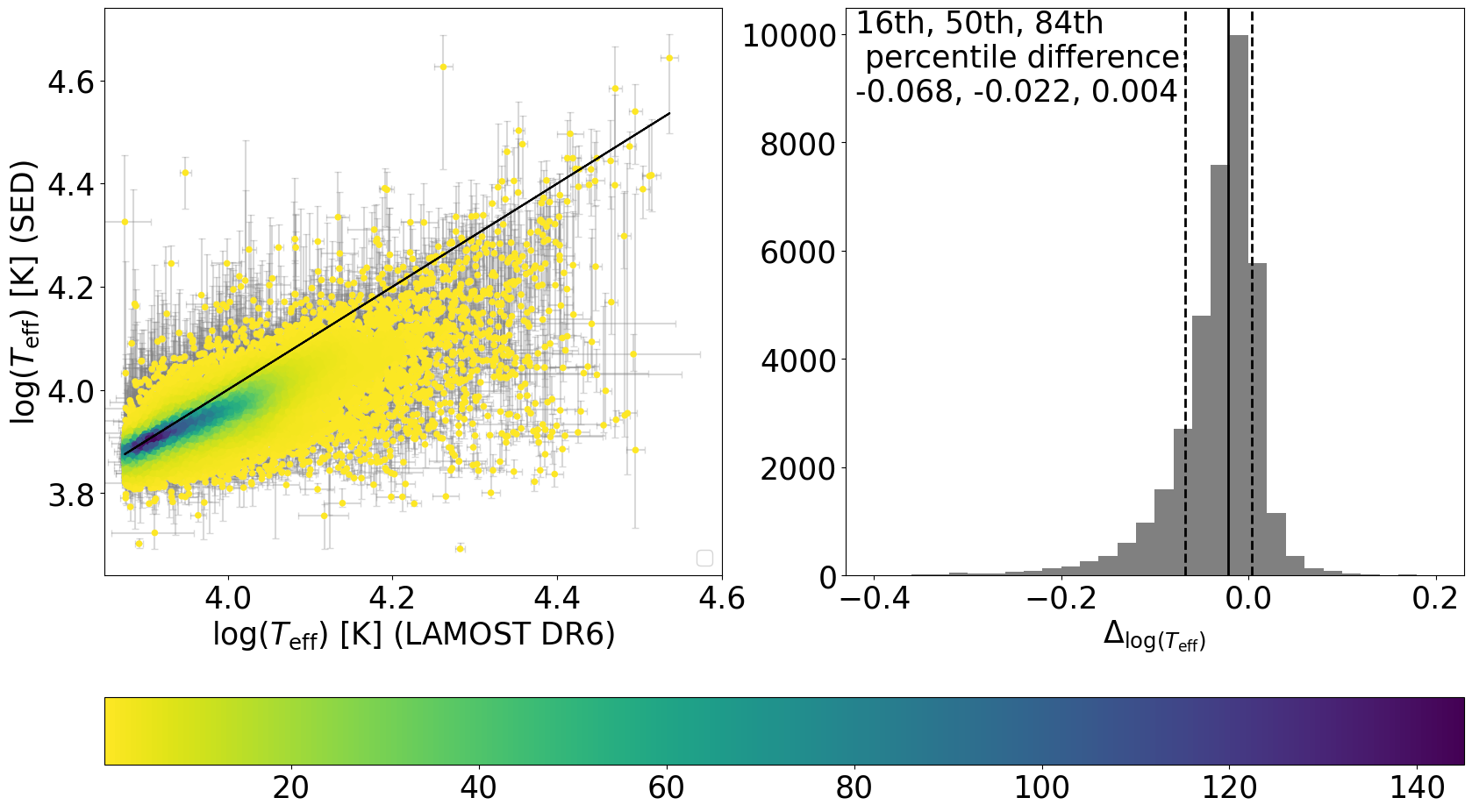} 
     \includegraphics[scale=0.20]{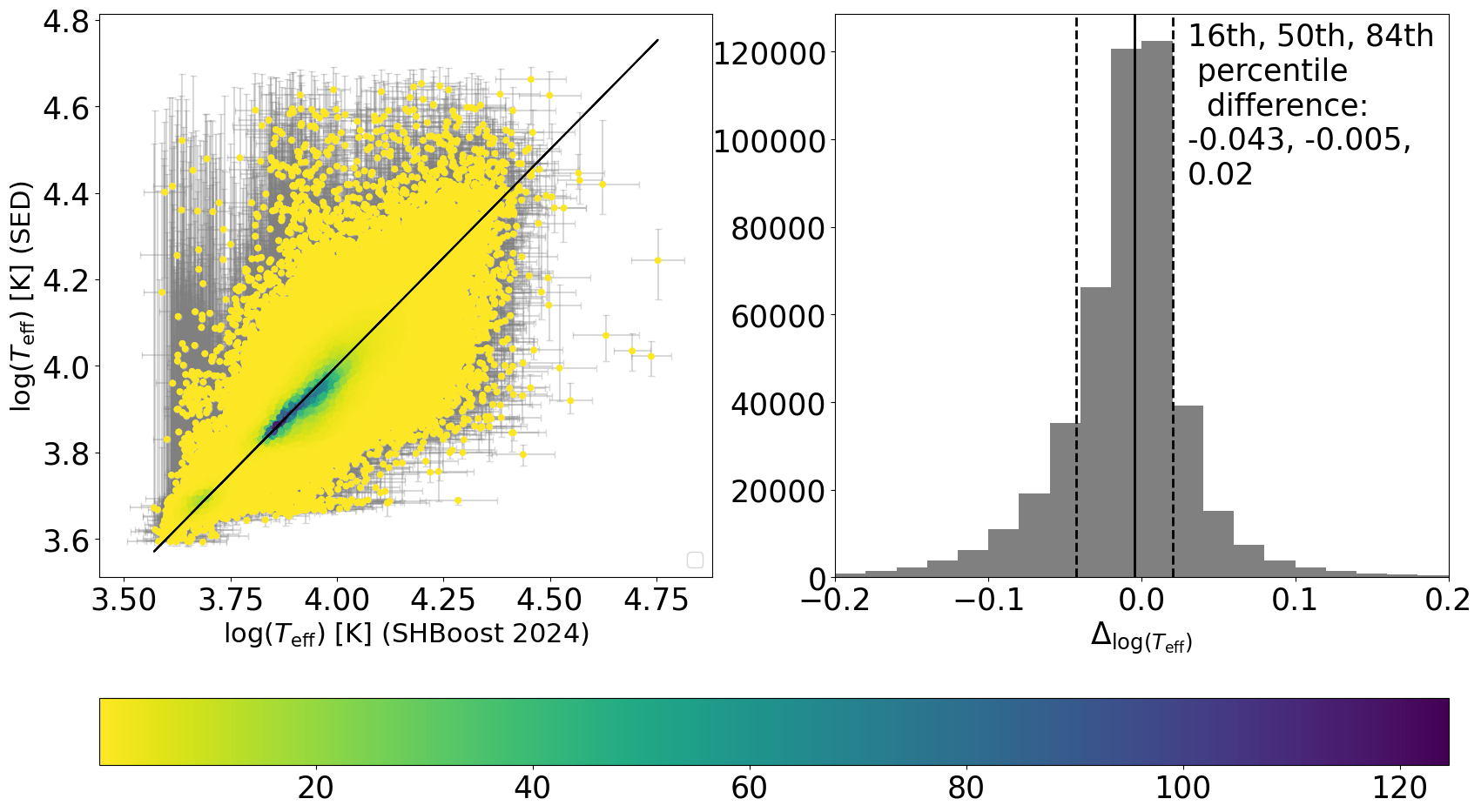}
    \caption{Far left and middle right panels: Comparison of our SED-fitted effective temperatures (ordinate) with the spectroscopic temperatures (abscissa) from the \textit{Gaia} DR3 ESP-HS (top left), APOGEE DR17 (top right), LAMOST DR6 (bottom left) and SHBoost 2024 (bottom right) catalogues, with each star colour-coded by its Gaussian KDE density (in units of stars per $\log(T_{\rm eff})^2$), and including a black 1:1 line. Middle left and far right panels:  histograms of the difference (written as $\Delta_{\log(T_{\rm eff})}$) between our median $\log(T_{\rm eff})$ and those of the corresponding spectroscopic sample, where we have indicated the 16th, 50th and 84th percentile differences on the top left.}

   % \kiril{Write into the caption what the colors represent? The logTeff vs logTeff plots look like there is a major discrepancy between the Teff inferred, with the greatest departures of order -0.6 and +0.4 dex, assuming that perfect agreement is when all points align on the diagonal. These seem more major than those on the histogram plot?}
    \label{CompTeff}
\end{figure*}

Fig. \ref{CompTeff} shows the comparison between our SED-fitted effective temperatures and the spectroscopic temperatures from these different surveys. We report similar trends as in \citetalias{Quintana2025}, where we find good agreements with literature values, as our median differences (or biases) are 0.01--0.03 dex (200--700 K at 10,000 K), with standard deviations around 0.05 dex (around 1200 K at 10,000 K). Furthermore, we notice a tendency of slightly underestimating effective temperatures compared with those from spectroscopic surveys, except for APOGEE where the opposite is observed. 

As observed in \citetalias{Quintana2025}, the upper temperature limit of 20,000 K for APOGEE is visible in Fig. \ref{CompTeff}, and suggests that some of the stars close to that limit could have underestimated temperatures: this is because the grid of models used in APOGEE reaches a maximum of 20,000 K \citep{Jonsson2020}. Similarly, while not as prominent as for our comparison within 1 kpc, there are `wings' extending around $\log(T_{\rm eff})$ = 4, that we attributed to intrinsic biases in APOGEE as they were also visible when comparing APOGEE effective temperatures with those from the \textit{Gaia} DR3 Apsis modules in \citetalias{Quintana2025}. Likewise, a significant level of contamination by cooler stars has been identified in the hot stars from ESP-HS \citep{Fremat2024,Maiz2025}, and could thus explain some of the differences observed here. %As for the comparison with \citet{Khalatyan2024}, in their Appendix A they have only compared their effective temperature for cooler stars than their B-type catalogue, although their Section 5.2 discussed the contrast with other maps of OB(A) stars, for which our dedicated discussion is in Section \ref{comparison_obstars}.

\subsubsection{Photometry-inferred $M_\mathrm{ZAMS}$}
\label{comp_mass}

Another parameter estimated through our SED fitter is the initial stellar mass. This value has also been derived for star cluster members in \citet{HuntReffert2023,HuntReffert2024}, which we know has a strong overlap with our OB stars \citepalias{Quintana2025} and OB associations within 1 kpc \citep{Quintana2026}. Both our SED fitter and the method from \citet{HuntReffert2024} assume that stars are single, making this comparison suitable. 

To aid this comparison, we have selected the star clusters from \citet{HuntReffert2023,HuntReffert2024} with low-reddening (median A$_V$ < 1 mag) and high-quality (cluster significance test, CST $>$ 10), such that the results from their isochrone fitting are more reliable. In doing so we restricted their list to a subset of 919 clusters shared between 225,372 members, 10,550 of which successfully crossmatch with our catalogue of candidate OB stars.

\begin{figure*}
    \centering
    \includegraphics[scale = 0.3]{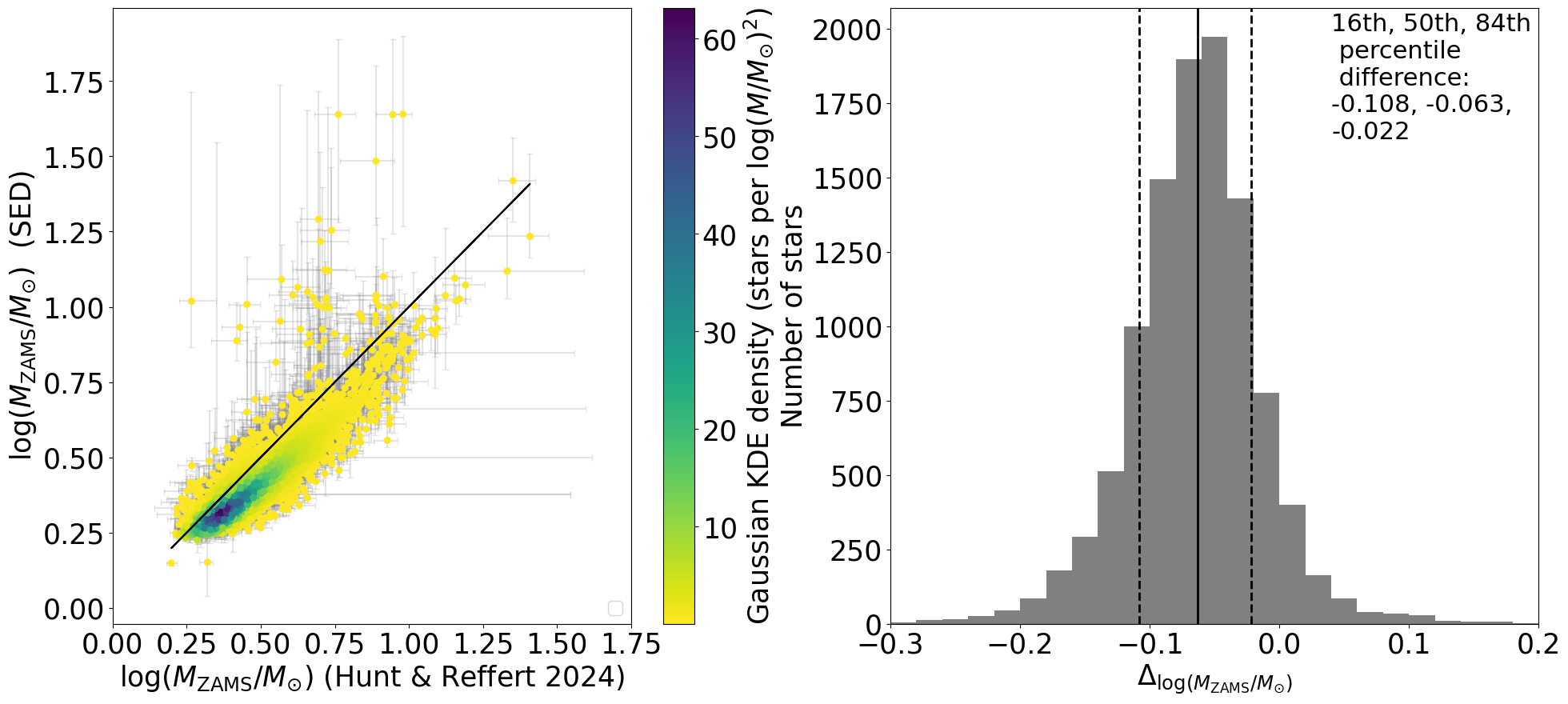}
    \caption{Same as Fig. \ref{CompTeff} but displaying the comparison for the initial stellar masses instead. The catalogue of comparison are the members of the high-quality (CST $>$ 10) and low-reddening (median A$_V$ < 1 mag) star clusters from \citet{HuntReffert2023,HuntReffert2024}.}
    \label{Comp_Mass_HR24}
\end{figure*}

Fig. \ref{Comp_Mass_HR24} shows a comparison between the initial stellar masses of both our catalogues. Immediately apparent in this plot is a vertical offset, as we underestimate $\log(M_{\rm ZAMS}/M_{\odot})$ by a median value of 0.06 dex. This trend cannot be attributed to interstellar extinction (as we have selected only low-reddening clusters from \citetalias{HuntReffert2024}). Instead, given its uniformity across the HR diagram, we attribute this offset to genuine methodological differences. \citet{HuntReffert2023} based their calculations on the SPISEA code from \citet{Hosek2020}, that uses the PARSEC isochrones \citep{Marigo2017}, as well as the ATLAS9 \citep{CastelliKurucz} and PHOENIX \citep{Husser2013} stellar atmosphere models, that are different from the ones we have used for our SED fitter (Section \ref{SEDfitter}).

\section{Spatial distribution of OB stars and local Galactic structure}
\label{galactic_structure}

\subsection{Our population of OB stars}
\label{population}

\begin{figure*}
    \centering
    \includegraphics[scale =0.17]{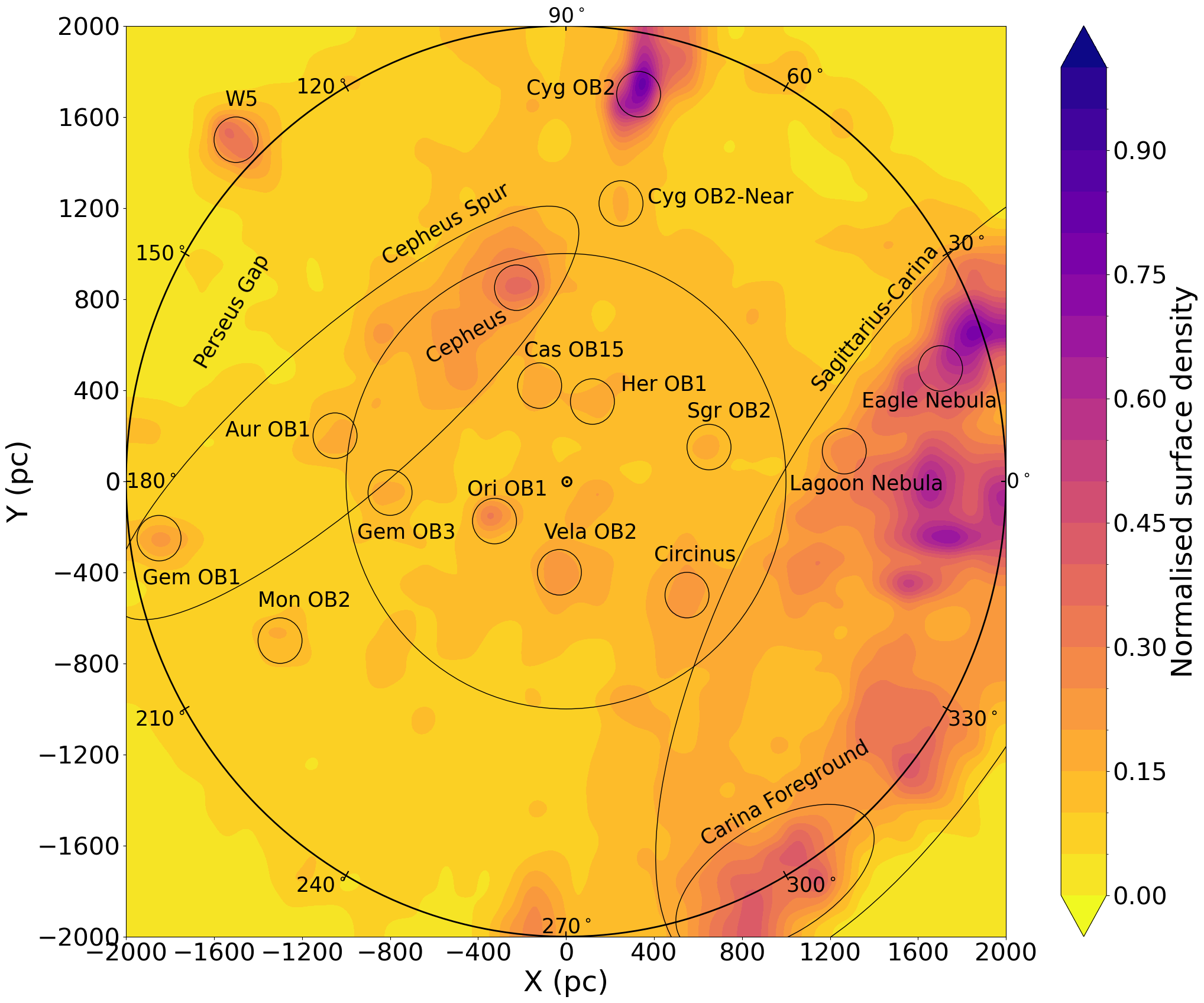}
    \includegraphics[scale=0.17]{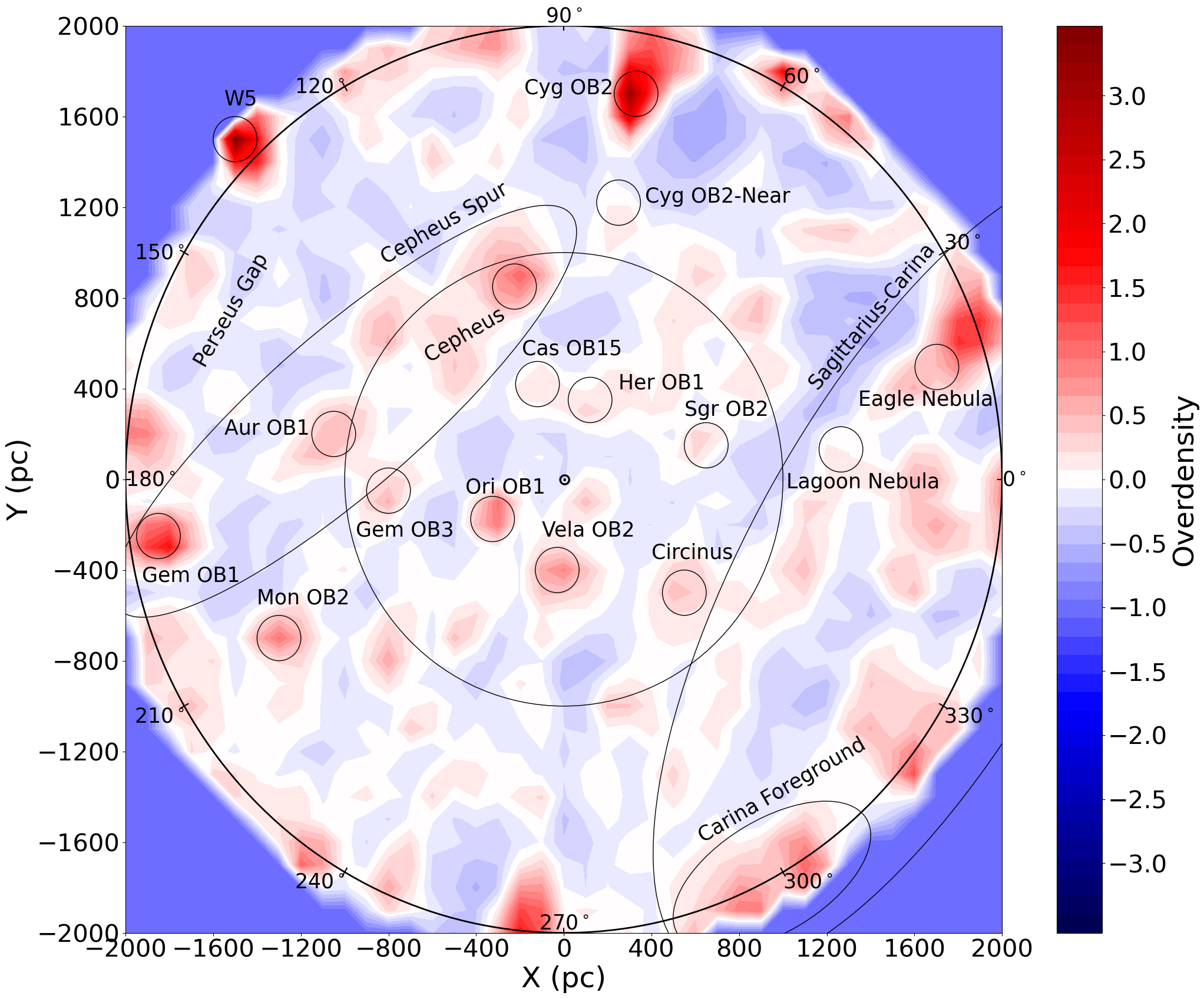}
    \caption{Left panel: normalised surface density of the 147,639 OB stars across the X-Y plane (X is positive towards the direction of Galactic Centre and Y towards the direction of Galactic rotation). The small annotations correspond to OB associations and/or massive star-forming regions, whilst the larger ones correspond to broader features, as described in Section \ref{population}. The inner circle encompasses a radius of $\sqrt{X^2+Y^2} =$ 1 kpc (i.e. the limit of the catalogue from \citetalias{Quintana2025}) whereas the outer circle represents our newer coverage with a radius of $\sqrt{X^2+Y^2} =$ 2 kpc containing the 105,971 OB stars of our census. Right panel; same as the left panel but displaying the overdensities of OB stars instead (as defined in Appendix \ref{technique_overdensities}), adopting a local density bandwidth of 0.1 kpc and mean density bandwitdh of 0.5 kpc. For both panels, we have also included ticks of Galactic longitude every 30 degrees.}
    \label{OBmap}
\end{figure*}

Fig. \ref{OBmap} shows the spatial distribution in Galactic Cartesian coordinates of the 147,639 SED-fitted OB sample stars\footnote{Given that the separation of binaries is much smaller than the typical separation between stars, the spatial distribution of OB primaries traces the general distribution of OB stars.}. To best highlight the overdensities, we have chosen two approaches. For the first one, we have proceeded as in \citet{Zari2021} and \citetalias{Quintana2025}, displaying the normalised surface density and increasing the bin sizes from 50 pc to 100 pc compared with \citetalias{Quintana2025} to account for the growing \textit{Gaia} DR3 parallax uncertainties at larger distances.

For the second approach, we have followed the method from \citet{Poggio2021} to better contrast the overdensities from the underdensities, using the overdensity parameter $\Delta_{\Sigma}$. The details of this technique are provided in Appendix \ref{technique_overdensities}, and the resulting distribution of OB stars colour-coded by $\Delta_{\Sigma}$ is shown on the right panel in Fig. \ref{OBmap}. Compared with the left panel, where large-scale features are better emphasized, the right panel better highlights OB associations, and thereby offers a complementary picture of the spatial distribution of our population of OB stars.

We have annotated overdensities in both panels from Fig. \ref{OBmap} that correspond to well-known OB associations and/or massive star-forming regions. For the inner circle (encompassing the OB stars within 1 kpc from \citetalias{Quintana2025}), we have notably encircled OB associations newly-identified in \citet{Quintana2026}, such as Sgr OB2 and Gem OB3. As for the outer circle, it includes Gem OB1 (which also stands out in \citet{Khalatyan2024}, though it was mistakingly referred to as "Gem OB2"), while "Cyg OB2-Near" corresponds to the foreground population of Cyg OB2 identified in \citet{Berlanas2019}. 
Larger features are also shown in Fig. \ref{OBmap}. By far the most prominent one is the Sagittarius-Carina spiral arm, for which we focus a dedicated discussion in Section \ref{sagitarrius_arm}. We have also included the location of the Cepheus Spur (an inter-arm overdensity extending between the Local and Perseus arm, \citealt{Morgan1953,PantaleoniGonzalez2021,PantaleoniGonzalez2025}), together with the Perseus Gap, a low-density area previously noted in the distribution of young Galactic tracers \citep[e.g.,][]{HouHan2014,Reid2019,Skowron2019,Zari2021,Chen2025}.

\begin{figure*}
    \centering
    \includegraphics[scale =0.18]{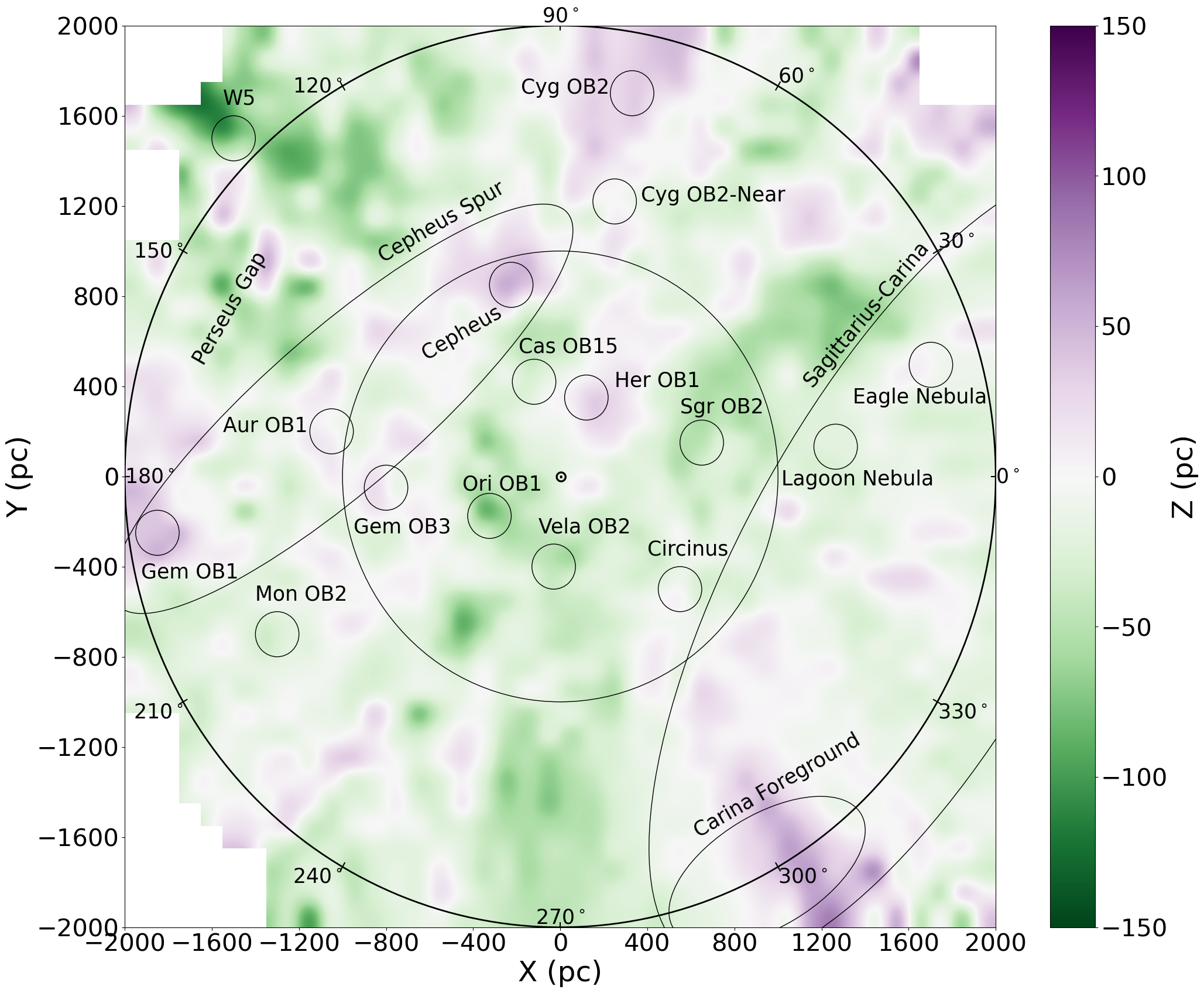}
    \includegraphics[scale=0.34]{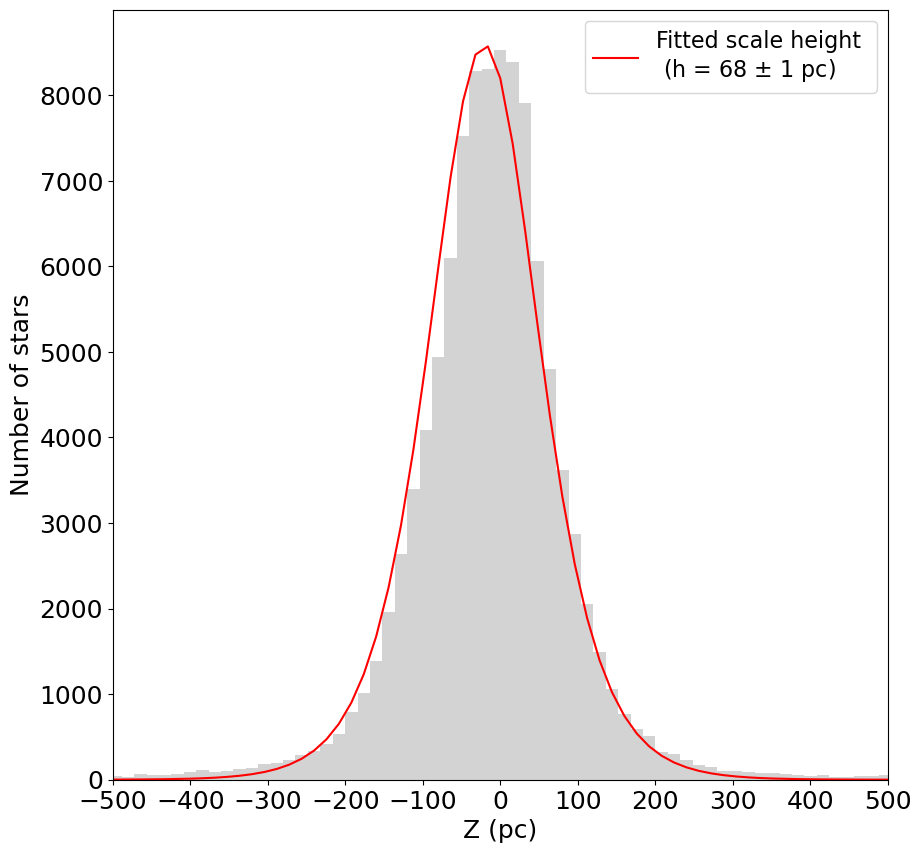}
    \caption{Left panel: Gaussian KDE (adopting a bandwidth of 100 pc) of the median $Z$-height for the 130,441 SED-fitted OB stars situated in the mid-Galactic plane ($|Z| < 150$ pc), across the X-Y plane. Right panel: Sech$^2$ fit of the distribution of $Z$ values of the 105,971 SED-fitted OB stars within 2 kpc.} 
    % \kiril{By eye, it looks like the Sech2 fit is not optimal: overpredicting the number of stars in the negative Z-direction and underpredicting in the positive. It seems if you shifted the fit by 20 pc to the right, it would be centered around zero (now it does not seem to be) and }
    \label{OBmap_Zvalues}
\end{figure*}

We extended our analysis of the distribution of OB stars to the third dimension, as illustrated in Fig. \ref{OBmap_Zvalues}. To that end, we have followed the approach from \citet{PantaleoniGonzalez2021} and \citeauthor{PantaleoniGonzalez2025} (\citeyear{PantaleoniGonzalez2025}; hereafter, \citetalias{PantaleoniGonzalez2025}) and selected the 130,441 SED-fitted OB stars within the mid-Galactic plane ($|Z| < 150$ pc), including 93,650 within 2 kpc, in order to visualize the vertical distribution of the OB stars. This is shown in the left panel of Fig. \ref{OBmap_Zvalues}. Unveiled is a complementary picture to the ALS catalogue \citep{PantaleoniGonzalez2021,PantaleoniGonzalez2025}, with Cyg OB2 and Ori OB1 noticeably located above and below the mid-Galactic plane, respectively. We also observe some differences inherent to the different definitions between our catalogues: the Cepheus Spur is not as prominent in our map as it is in theirs, except in the Cepheus region and in Gem OB1, and we therefore conclude that the Cepheus Spur is better traced by the youngest and most massive OB stars.

We have also calculated the scale height of the 105,971 OB stars within 2 kpc, using the \texttt{Sech$^2$} fit method from \citet{BobylevBajkova2016} as in \citetalias{Quintana2025}. In doing so we have derived a scale height of $68 \pm 1$ pc, illustrated on the right panel from Fig. \ref{OBmap_Zvalues}. As expected, most of the OB stars sit closer to the Galactic mid-Plane, and our value is slightly smaller than the value of $76 \pm 1$ pc from \citetalias{Quintana2025}. We notably notice, on the right panel from Fig. \ref{OBmap_Zvalues}, a slight offset between the fit and the peak of the distribution of our OB stars, despite applying the same approach than \citetalias{Quintana2025} with $Z_{\odot}$ = 20.8 pc from \citet{BennettBovy2019}.

\begin{figure*}
    \centering
    \includegraphics[scale =0.17]{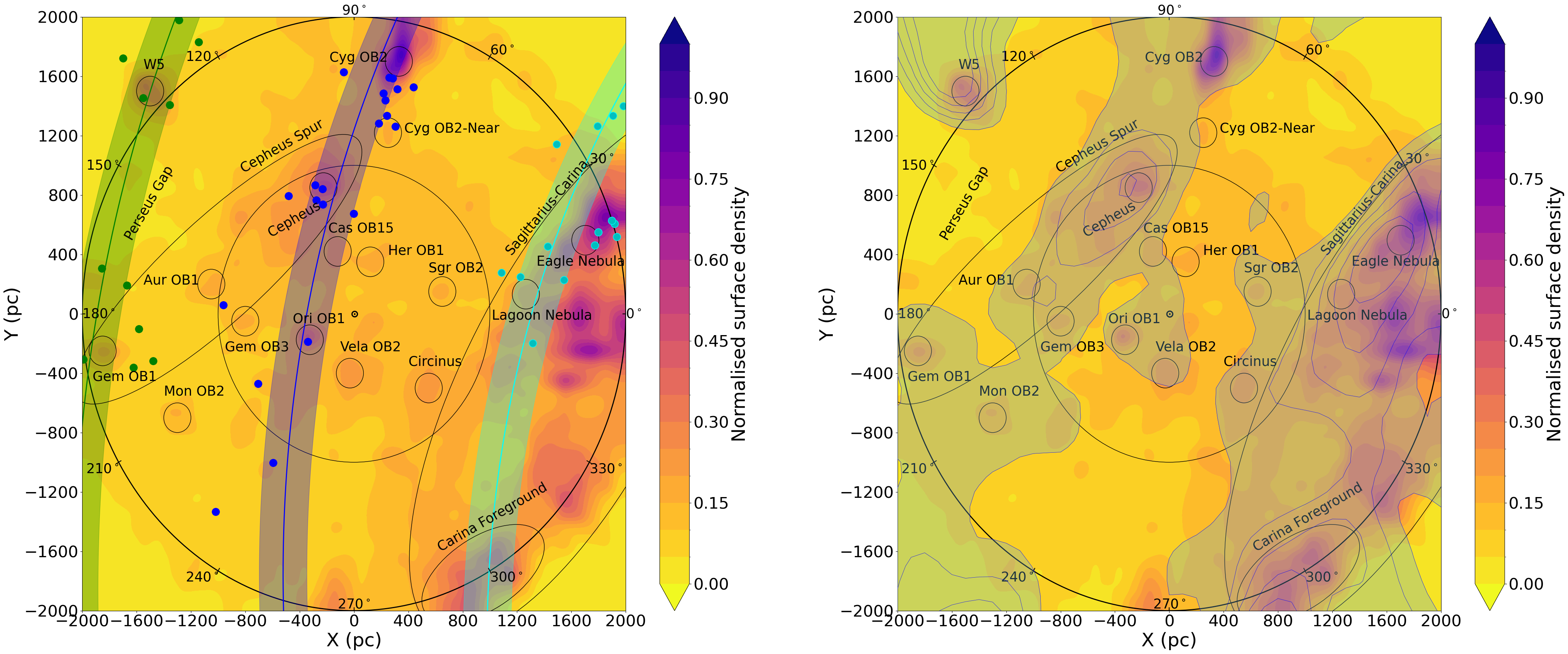}
    \caption{Left panel: Same as the left panel from Fig. \ref{OBmap} but with the fits of the Galactic spiral arms from \citet{Reid2019}, showing (from left to right) the Perseus, Orion-Cygnus and Sagittarius-Carina arms, assuming a thickness of 300 pc. Similarly to \citet{Zari2021}, we have also colour-coded with dots the positions of the masers used for the fit, by inverting their parallax inferred from the VLBI measurements in \citet{Reid2019} and propagating their 2D position into Epoch 2016 as in \textit{Gaia} DR3 using their proper motions. Right panel: same as the left panel from Fig. \ref{OBmap} but including the contours traced by the \textit{Gaia} OB stars from \citet{GaiaDR3_AsymmetricalDisk}.}
    \label{OBmap_SpiralArms}
\end{figure*}

In addition, we have overplotted the contours from the \textit{Gaia} OB stars used in \citet{GaiaDR3_AsymmetricalDisk} in the right panel from Fig. \ref{OBmap_SpiralArms}. We observe a significant overlap between their contours and our overdensities, as they encompass the largest features such as the Cepheus Spur and the Sagittarius-Carina arm, but also newly-identified OB associations from \citet{Quintana2026} such as Gem OB3 and Sgr OB2. 

Star formation in the Milky Way preferentially occurs within its spiral arms \citep{Sparke2000}, and because they are short-lived, OB stars are known to trace their location \citep[e.g.,][]{Russeil2003,Chen2019,Zari2021}. Nevertheless, spiral arms in our Galaxy are usually identified on a much larger scale than those considered in this paper \citep[e.g.][]{TaylorCordes1993, Levine2006, Reid2019, Drimmel2025, Poggio2026}. We have still overplotted the location of the Perseus, Orion-Cygnus (Local) and Sagittarius-Carina arms from the maser fit of \citet{Reid2019} on the normalised surface density map of the X-Y plane in the left panel from Fig. \ref{OBmap_SpiralArms}. In the next subsections, we will discuss the contrast between the overdensities of O- and B-type stars and each individual arm.

\subsection{The Perseus arm}
\label{perseus_arm}

The Perseus arm is one of the main Milky Way spiral arms \citep[see, e.g.,][]{Churchwell2009}. As seen in Fig. \ref{OBmap_SpiralArms}, within 2 kpc we only start to intercept it (one of its tracers, W5, on the upper left that corresponds to the Cassiopeia region, is mostly located beyond our limit). This arm has been noticeably hard to map out in the region known as the "Perseus Gap" that extends between the Auriga and Cassiopeia regions (i.e., $l$ = [140$\degr$-170$\degr$]), where few stellar and molecular tracers have been identified \citep[e.g.,][]{MarcoNegueruela2016}. This gap, also referred to as the "Giant Oval Cavity" and also traced by dust and molecular clouds \citep{Vergely2022,Wang2025}, has been identified as the largest known superbubble of the Milky Way, supported by the energy and momentum of the supernova explosions it underwent \citep{Chen2025}.

Tracers of the Perseus arm at lower Galactic longitudes include the Cassiopeia OB associations such as Cas OB5 aged 10--20 Myr \citep[e.g.,][]{QuintanaNegueruelaBerlanas2025}, and Aur OB2 aged $<$ 10 Myr \citep[e.g.,][]{Quintana2023}, both mostly located beyond the edges of our map. 

One of the most prominent overdensities is  Gem OB1, located on the third quadrant of Fig. \ref{OBmap_SpiralArms}. It sits right in the middle of the Perseus spiral arm fit from \citet{Reid2019}. Given that the position of  Gem OB1 also matches with the water masers at around $l = 190 \degr$ from \citet{Choi2014}, it is highly probable that  Gem OB1 is associated with a young massive star-forming region, and thereby constitutes a tracer of this spiral arm. Nevertheless, the contours shown on the right panel from Fig. \ref{OBmap_SpiralArms} would suggest that  Gem OB1 rather belongs to the Orion-Cygnus arm. Further into the third Galactic quadrant, tracers of the Perseus arm  appear again \citep[e.g.,][]{Vazquez2008}, but are situated beyond the map limits.

However, we should stress that the spiral arm fit from \citet{Reid2019} has been questioned, particularly when it comes to the Perseus Arm. Figure 14 from \citet{GaiaDR3_AsymmetricalDisk} also displays the alternative model of this arm from \citet{Levine2006} characterized by another pitch angle, wherein the Perseus Gap becomes an inter-arm region (and so do all the tracers located in the third quadrant, such as  Gem OB1 and Aur OB2). We refer to Appendix \ref{spiral_arm_models} where we compare the spatial distribution of OB stars with alternative spiral arm models.

\subsection{The Orion-Cygnus arm}
\label{local_arm}

The Orion-Cygnus (also called Local) spiral arm, the closest to the Sun, has been notoriously harder to trace using the position of OB stars in spite of attempts \citep[e.g.,][]{Morgan1953,Ge2024}. On the left panel from Fig. \ref{OBmap_SpiralArms}, the fit from \citet{Reid2019} is intercepted by several overdensities of OB stars such as Ori OB1, Cepheus and Cyg OB2, all well-known massive star-forming complexes ($\lesssim$ 10 Myr; e.g. \citealt{Uyaniker2001,Wright2015,Kerr2023,SanchezSanjuan2024,Quintana2026}). The location of Cas OB15 is also consistent with its young age (10--15 Myr, e.g. \citealt{Kerr2023,Quintana2026}), while Her OB1 and Vela OB2 being slightly older (20--30 Myr, e.g. \citealt{CantatGaudin2019,Kerr2023,Quintana2026}) means they could have formed within the Orion-Cygnus arm before moving away from it.

Similarly to the Perseus arm (Section \ref{perseus_arm}), the geometry and orientation of the arm is different depending on the considered model/map. For instances, the contours of the Orion-Cygnus arm as mapped by \citet{GaiaDR3_AsymmetricalDisk} have a more open geometry compared with the model from \citet{Reid2019}. While being similar on the upper part of the region (Cygnus and Cepheus), their orientation is different in the lower parts.

\subsection{The Sagittarius-Carina arm}
\label{sagitarrius_arm}

The Sagittarius-Carina arm is by far the most prominent feature of our map, that we have highlighted by an ellipse. In particular, at $l=
[-30\degr-30\degr]$ sit several concentrations that likely correspond to the Sagittarius, Scutum and Serpens OB associations and harbor the
famous Lagoon and Eagle Nebulae that are related to those associations \citep{Wright2019,Wright2020,HuntReffert2021,Stoop2023}.Another famous tracer of the Sagittarius-Carina arm is the Carina Nebula complex that includes Car OB1 at $l =286.5\degr$ \citep{MaizApellaniz2020,MaizApellaniz2022,Berlanas2023,Berlanas2025}. Because most of this complex is located beyond the edges of our map (at distances of $\sim$2.3--2.4 kpc, see e.g. \citealt{SmithBrooks2008,GopplPreibisch2025}), we refer to the overdensity in that region as "Carina Foreground" in our map. Our fourth quadrant also includes other overdensities of OB stars at the fourth quadrant of Fig. \ref{OBmap_SpiralArms}, that likely correspond to the Crux and Centaurus OB associations catalogued in \citet{Wright2020}.

The fit of this arm by \citet{Reid2019} reveals a slight offset compared with our overdensities of OB stars. This was previously reported by \citet{Zari2021}, who concluded that the distribution of young stars was more complex than simply following the spiral arms. However, the luminous OBA stars from their catalogue could live up to a few hundreds of Myrs, and this is also the case for the late B-type stars in our catalogue, which could have arisen from this arm before moving away from it. Another cause of the offset between the fit from \citet{Reid2019} and our overdensities of OB stars could stem from the uncertainties in the fit from \citet{Reid2019}: in Fig. \ref{OBmap_SpiralArms}, we note a clear lack of masers in the fourth quadrant, where the offset is the most pronounced.

\section{Clustered star formation in the local Milky Way}
\label{open_clusters}

\subsection{OB star overdensities and open clusters}
It is instructive to contrast the distribution and overdensities of OB stars with those of young open clusters (hereafter OCs), as the two are complementary tracers of recent star formation in the Milky Way. To that end we used the catalogue of \citet{HuntReffert2023}, the largest, most homogeneous and de-duplicated list of star clusters produced to date.

\begin{figure*}
 \includegraphics[scale =0.3]{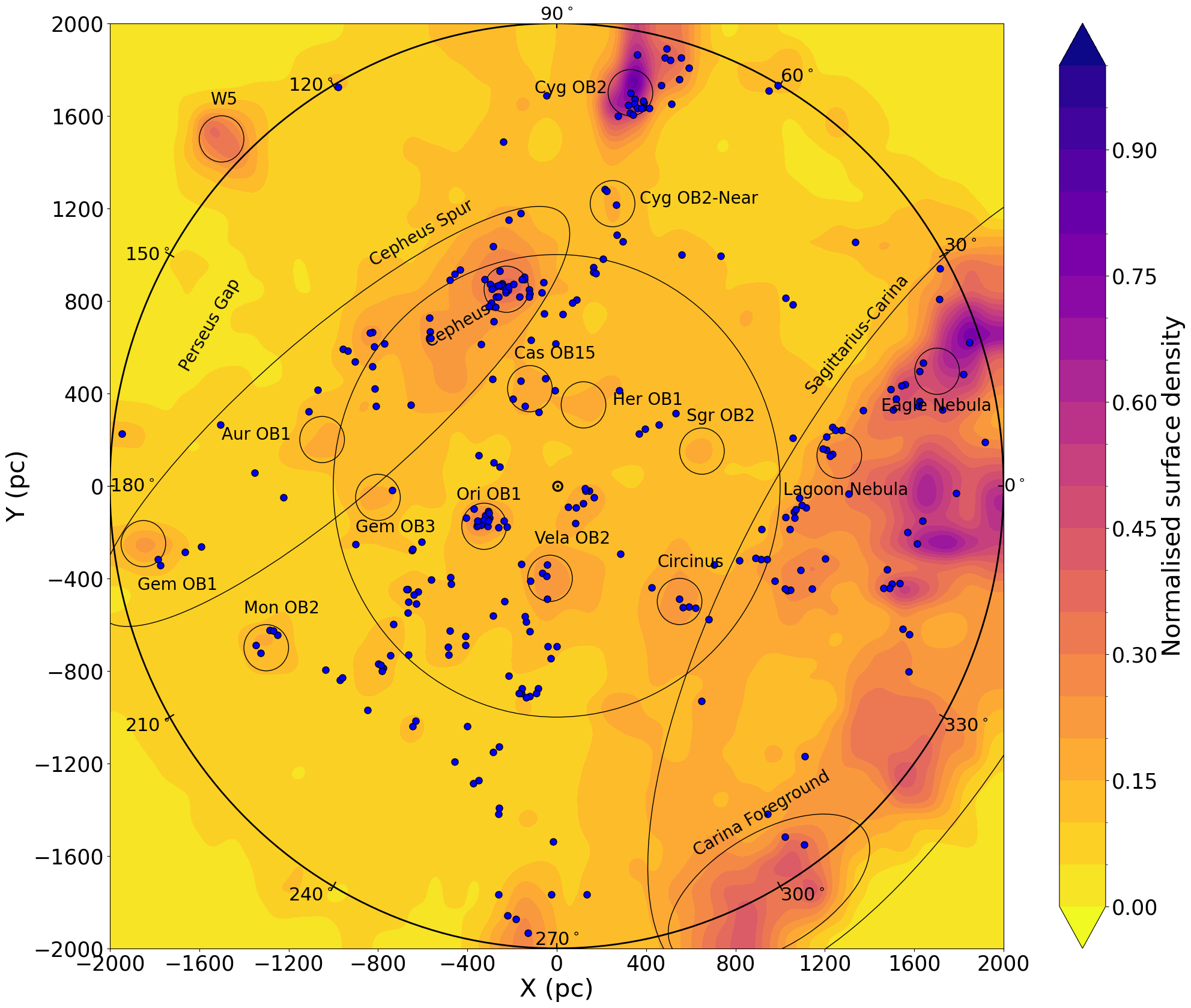}
 \caption{Same as the left panel from Fig. \ref{OBmap} but including the subset of 295 young ($<$ 20 Myr), high-quality, compact OCs within 2 kpc from \citetalias{HuntReffert2024} displayed as blue dots.}
 \label{OBmap_OCs}
\end{figure*}

We selected a subset of the catalogue from \citetalias{HuntReffert2024} as follows. We restricted ourselves to the young ($<$ 20 Myr) compact (classified as type "o" in \citetalias{HuntReffert2024}), high-quality OCs within $\sqrt{X²+Y²} <$ 2 kpc, with $X$ and $Y$ derived from the Galactic coordinates and the median line-of-sight distances of each cluster. The high-quality subset was defined using the two criteria. The first one was that OCs are required to have an astrometric signal-to-noise ratio (cluster significance test, hereafter CST) greater than 5$\sigma$, following the recommendations from \citet{HuntReffert2021}. We have, however, only applied the second condition (a median CMD class, $Q_{\rm CMD}$, greater than 0.5) to the OCs older than 10 Myr. This was done because, as mentioned in \citet{QuintanaHuntParul2025}, many of the youngest clusters have their CMD broadened by extinction. Applying the following cuts, we have obtained a subset of 295 OCs within 2 kpc. Fig. \ref{OBmap_OCs} shows their distribution on top of the normalized surface density of the OB stars in our catalogue.

Fig. \ref{OBmap_OCs} shows a positive correlation between the overdensities of OB stars and those of young OCs, as expected. This is particularly visible in Ori OB1 and Cepheus, both famous for being some of the youngest massive star-forming regions in the local Milky Way (e.g. \citealt{Szilagyi2023,Quintana2026,SanchezSanjuan2026}).

We also observe a shift in some OB associations between the centre of the OB star overdensities with those of OCs, notably in Cyg OB2 and Ori OB1. One explanation of these offsets could be that these regions are too obscured for their OCs to be visible. As illustrated in Fig. 9 from \citet{Hunt2026}, Cyg OB2 corresponds to one of the sightlines from the Galactic disk where the detection probability of OCs is the lowest. This situation is mitigated for the bright OB stars, with the exception of a few extreme objects, such as 2MASS J20395358+4222505, whose high luminosities and extinction (A$_V$ = 9.7 mag, see \citealt{Herrero2022}), prevented our SED fitter to correctly fit it as blue supergiant and include it in our catalogue. The same is also true, albeit to a lesser extent, for the Sagittarius and Serpens OB associations around l = 0$^{\degr}$, where such offset is also visible, and where the visual extinction A$_V$ can be superior to 6 mag.

%\kiril{As an alternative explanation: migration of OB assoc. away from their "home" OCs?}

Noticeable in Fig. \ref{OBmap_OCs} is also the complete lack of young OCs across the Perseus Gap, consistent with the low-density of OB stars present there. This is confirmed by the dedicated census in the Perseus direction from \citet{CantatGaudin2019_Perseus}, wherein they identified 46 OCs, 41 of which are also found in the catalogue of \citetalias{HuntReffert2024}. However, all but one are older than 20 Myr (median age $\sim$150 Myr). The only exception is COIN-Gaia\_16, which is aged $\sim$12 Myr and related to the Auriga OB associations \citep{Quintana2023}. It is thus not part of the Perseus Gap, confirming the absence of young features within this superbubble. 

To take this comparison further, we have proceeded as follows. We crossmatched our 147,639 OB stars with all the star cluster members from \citetalias{HuntReffert2024} and found 12,352 stars in common (including 9612 stars within $\sqrt{X^2+Y^2} <$ 2 kpc). This represents a fraction of $\sim$10 \% of OB stars that are found to be within a clustered environment. To contrast, in \citet{Quintana2026} we found a similar fraction of OB stars within 1 kpc to be members of an OB association $\sim$38 \% of these OB association members are also star cluster members from \citetalias{HuntReffert2024}. Extrapolated within 2 kpc, this would correspond to only $\sim$16 \% of the OB stars to be either member of an OC or an OB association. For such statistics to be compatible with the clustered model from \citet{LadaLada2003}, this would imply that most late B-stars initially born clustered eventually became field stars, given that their lifetime can reach up to about 400 Myrs, higher than the typical dissolution time of star clusters $\sim$93 Myr for an OC of initial mass of 130 M$_{\odot}$ (see e.g. \citealt{Lamers2005,Anders2021}), as discussed in \citet{QuintanaHuntParul2025}\footnote{A fraction of these OB stars can have also been ejected from their natal cluster or association (e.g. \citealt{DorigoJones2020,MartinezGarcia2026}.}. Another explanation, as mentioned above, is that \textit{Gaia} DR3 sensitivity is better tailored to identify OB star populations with a higher completeness than OCs, allowing us to detect the population of OB stars in high-extinction regions (such as Cyg OB2). This warrants an extension of the census of OB associations within 1 kpc from \citet{Quintana2026} up to a distance of 2 kpc.

\subsection{Spatial distribution of the densest star-forming regions}

In this census, we cover a surface area four times higher than in \citet{Quintana2026}, and therefore a more significant portion of the Milky Way's thin disk. This enables us to study the variation of star formation as a function of spatial distribution. To that end, we have binned the 12,352 clustered OB stars (of size 50 pc) across the X-Y plane. We have also divided this sample of clustered OB stars into different thresholds (i.e. $\log(T_{\rm eff}) > 4, 4.1, 4.2$ and 4.3) to contrast their spatial distribution as a function of increasing effective temperature.

\begin{figure*}
    \centering
    \includegraphics[scale=0.18]{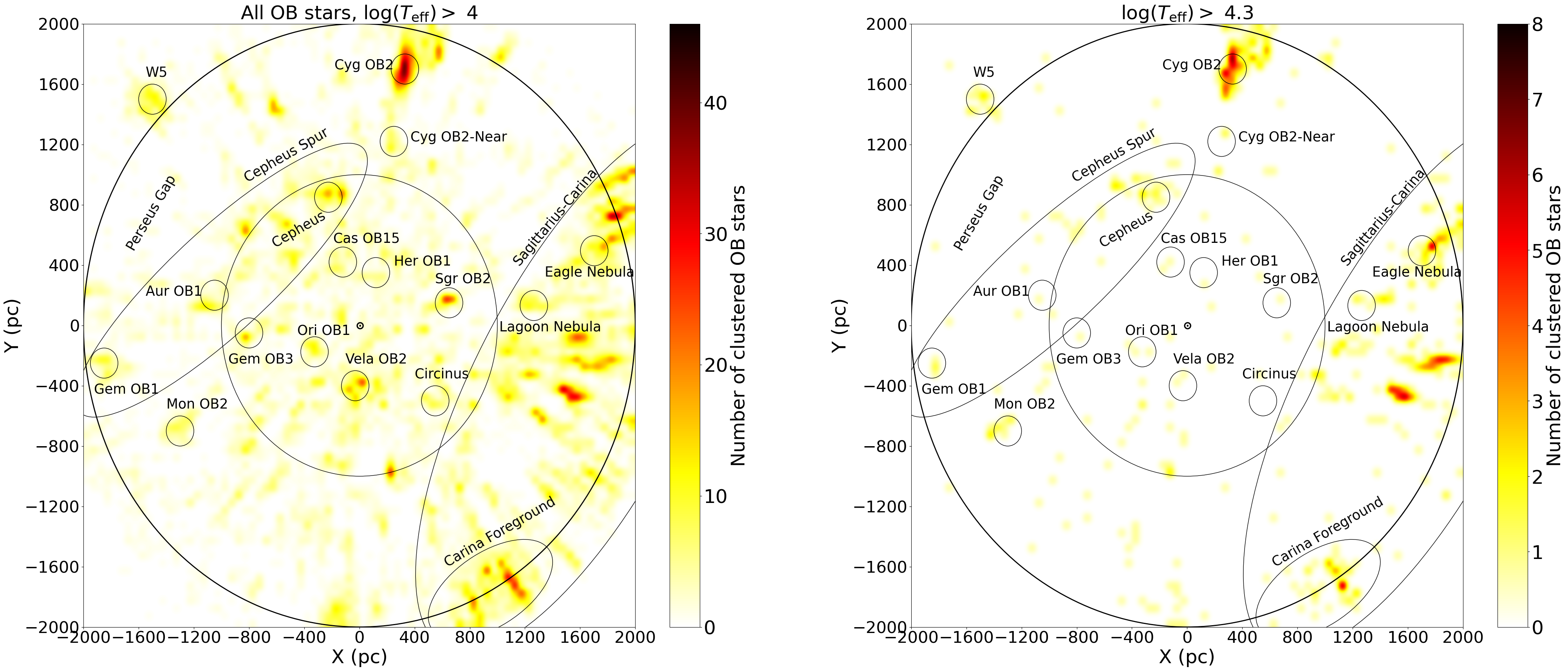}
    \caption{
    Top-down view (across the X-Y plane) of our clustered population of stars, colour-coded by the number of OB stars found as member of \citetalias{HuntReffert2024} across bins of 50 pc size. On the left panel are displayed all 12,352 clustered OB stars, and on the right panel the 611 clustered OB stars with $\log(T_{\rm eff}) > 4.3$.}
    \label{XY_ClusteredOBstars}
\end{figure*}

The resulting top-down view is displayed in Fig. \ref{XY_ClusteredOBstars}, with two trends highlighted. Firstly, there is a higher concentration of clustered OB stars towards the Galactic Centre directions (left edge of the panel) compared with the Galactic Anticentre (right edge of the panel). Secondly, as we restrict our sample to hotter stars, they tend to sit in more concentrated regions, particularly in the Sagittarius-Carina arms and in Cyg OB2.

To quantify these trends, we have calculated the Galactocentric radius R$_ {\rm GC}$ of each clustered OB star, again adopting R$_0$ = 8.23 kpc from \citet{Leung2023} as the distance from the Sun to the Milky Way centre, and show a histogram of the number of clustered OB stars as a function of R$_ {\rm GC}$ (corresponding to the number of clustered OB stars divided by the total number of OB stars), adopting bin sizes of 100 pc.

\begin{figure*}
    \centering
    \includegraphics[scale=0.22]{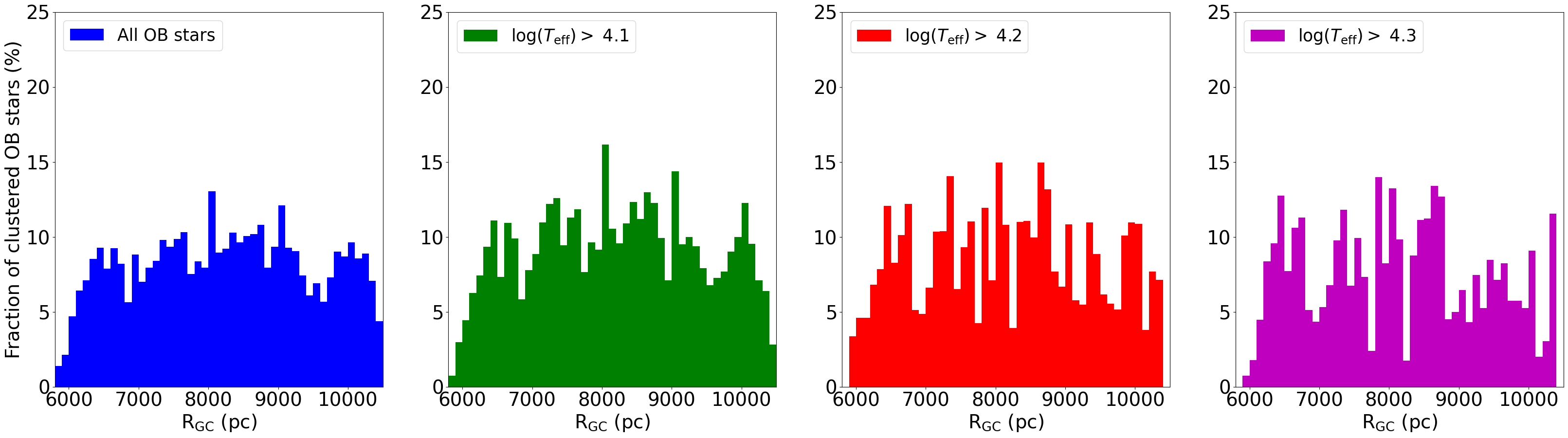}
    \caption{Fraction of clustered OB stars (i.e., that are members of the OCs from \citetalias{HuntReffert2024}) in our catalogue for each bin of 100 pc size of Galactocentric radius R$_{\rm GC}$, where we have adopted R$_0$ = 8.23 kpc from \citet{Leung2023} for the distance of the Sun to the Galactic Centre. Every subset is defined by the effective temperature threshold as written on the different panels.}
    \label{Fraction_Clustered}
\end{figure*}

By restricting our catalogue to hotter stars, we observe a sharper contrast between the bins of small and large fractions as evidenced by the more highlighted peaks on the lower panels. These peaks reach values of only 10-15 \% as we sweep through various regions of the X-Y plane, including both high and small densities of clustered OB stars within a same bin of R$_{\rm GC}$ (these large fractions are therefore sometimes driven by small-number statistics).

Furthermore, the peaks in Fig. \ref{Fraction_Clustered} are scattered throughout the whole region, with little dependence over the value of R$_{\rm GC}$, contrary to the trend we initially observed in Fig. \ref{XY_ClusteredOBstars}. This makes it challenging to draw any conclusion about clustered star formation in the Milky Way as a function of Galactocentric radius. %This would imply that star-forming environnements across the nearest few kpc are quite similar to each other, and to identify a physical trend, this would require us to extend the census to higher distances.}

\subsection{Surface density star formation rate estimation}

In \citet{QuintanaHuntParul2025}, we derived a value for the surface density star formation rate from compact clusters ($\Sigma_{\rm SFR,OC}$) in the local Milky Way between $736^{+159}_{-176}$ and $875^{+34}_{-35}$ M$_{\odot}$ Myr$^{-1}$ kpc$^{-2}$, that we compared with the $\Sigma_{\rm SFR}$ value of $922^{+133}_{-1}$ M$_{\odot}$ Myr$^{-1}$ kpc$^{-2}$ estimated in \citetalias{Quintana2025} for all stars from a general population of O- and B-type stars within 1 kpc ($\Sigma_{\rm SFR}$). We concluded that most stars arise from compact clusters in the local Milky Way. Our extended census of OB stars now offers an opportunity to revisit the clustered model of star formation within a larger volume.

The value of $\Sigma_{\rm SFR,OC} = 791^{+162}_{-191}$ M$_{\odot}$ Myr$^{-1}$ estimated by fitting a power-law to the mass function in \citet{QuintanaHuntParul2025} was described as "valid across a broad local region of the Milky Way". This is because the applied correction for incompleteness depended on volume, due to lower-mass clusters being harder to detect at larger distances compared with their more massive counterparts. In this work, we go one step further and re-estimate this value by splitting the list of young and compact clusters from \citetalias{HuntReffert2024} ($<$10 Myr) into a subset towards the Galactic Anticentre (labelled "GA") and another subset towards the Galactic Centre (labelled "GC")\footnote{Here we will not be able to apply the alternative method for estimating $\Sigma_{\rm SFR,OC}$ used in \citet{QuintanaHuntParul2025}. Indeed, this second approach was based on summing the masses of the youngest compact clusters within  $\sqrt{X^2+Y^2} < 1$ kpc: contrary to the first method, we could not apply a completeness correction, which was mitigated by the fact that the sample of OCs was complete down to 100 M$_{\odot}$ at 1 kpc, closer to the lower limit of 40--60 M$_{\odot}$ (e.g. \citealt{Almeida2025}). Would we extend this calculation to 2 kpc, this would result in a significantly underestimated estimation of $\Sigma_{\rm SFR,OC}$, as this all-sky census of OCs is only compete down to $\sim$230  M$_{\odot}$ at 1.8 kpc \citepalias{HuntReffert2024}.}. For both directions, we calculate $\Sigma_{\rm SFR,OC}$ using both the power-law method with a threshold on $X$ ($>$ and $<$ 0 pc, respectively) and $R_{\rm GC }$ ($<$ and $>$ 8230 pc, respectively), multiplying the resulting values by two to account for slicing the area in half.

\begin{table}
	\centering
	\caption{Surface density star formation rates in open clusters ($\Sigma_{\rm SFR,OC}$) derived from fitting the completeness-corrected mass function of compact OCs from \citetalias{HuntReffert2024} with a power-law, respectively for the Galactic Centre and Anticentre directions. \label{SFR_OC_values}}
	\renewcommand{\arraystretch}{1.5} 
	\begin{tabular}{lcccr} 
		\hline
		Parameter & Value & Units  \\
		\hline
        $\Sigma_{\rm SFR,OC-GC}$ (X $>$ 0 pc) & $836^{+304}_{-354}$ & M$_{\odot}$ Myr$^{-1}$ kpc$^{-2}$ \\
        %$\Sigma_{\rm SFR,OC} /\Sigma_{\rm SFR}$ &  \% & \\
         $\Sigma_{\rm SFR,OC-GC}$ (R$_{\rm GC} <$ 8230 pc) & $921^{+264}_{-303}$ & M$_{\odot}$ Myr$^{-1}$ kpc$^{-2}$ \\
       % $\Sigma_{\rm SFR,OC-GC} /\Sigma_{\rm SFR}$ &  \% & \\
        \hline
         $\Sigma_{\rm SFR,OC-GA}$ (X $<$ 0 pc) & $534^{+196}_{-281}$ & M$_{\odot}$ Myr$^{-1}$ kpc$^{-2}$ \\
        %$\Sigma_{\rm SFR,OC} /\Sigma_{\rm SFR}$ &  \% & \\
         $\Sigma_{\rm SFR,OC-GA}$ (R$_{\rm GC} >$ 8230 pc) & $624^{+174}_{-224}$ & M$_{\odot}$ Myr$^{-1}$ kpc$^{-2}$ \\
        %$\Sigma_{\rm SFR,OC-GA} /\Sigma_{\rm SFR}$ &  \% & \\
		\hline
	\end{tabular}
\end{table}

Table \ref{SFR_OC_values} includes a summary of these results. As expected, both values of $\Sigma_{\rm SFR,OC-GC}$ are larger than those of $\Sigma_{\rm SFR,OC-GA}$, although, due to the large size of the error bars (resulting from applying the power-law method with a smaller subset of OCs than in \citealt{QuintanaHuntParul2025}), they overlap within 1$\sigma$. 

Several studies show that the star formation rate in disk galaxies increases with the gas density towards their centre, meaning that our results are consistent with this scenario (e.g. \citealt{Boissier2003,KennicuttEvans2012,Oemler2017}). This is also consistent with past studies finding a higher density of OCs towards the Galactic Centre \citep{Bonatto2006}. Nevertheless, the large error bars obtained for our $\Sigma_{\rm SFR,OC}$ values in Table \ref{SFR_OC_values}, prevent us from drawing a robust physical conclusion from our current census of OB stars, warranting an extension to greater distances in the future. 

\section{Future ccSN explosion and black hole formation events in the local Milky Way}
\label{observed_ccsn}

Using the SED-fitted zero-age main-sequence (ZAMS) mass distributions and the ccSN model introduced in this section, we identify OB-type stars within 2 kpc that are ccSN candidates, to terminate their lives in a ccSN explosion, BH formation or both. 

\subsection{The final fates of massive OB-type star primaries}
\label{sec:ff}

\begin{table}
	\centering
	\caption{The final fate model for massive single-stars of solar metallicity from \citetalias{Maltsev2025}, reformulated as a function of $M_\mathrm{ZAMS}$ and extended to estimate the least ZAMS mass for a star to end its life in a ccSN explosion.    
    \label{MassiveStars_Fates}}
	\renewcommand{\arraystretch}{1.3} 
	\begin{tabular}{llcr} 
		\hline
		M$_{\rm ZAMS}$-range: & Final fate: \\
		\hline
        \hline
        $M_\mathrm{ZAMS}/M_\odot < 8.55$ & No SN, WD remnant \\
        \hline
        $8.55 \leq M_\mathrm{ZAMS}/M_\odot < 8.85$ & ECSN, NS remnant \\
        $8.85 \leq M_\mathrm{ZAMS}/M_\odot < 22.20$ & FeCCSN, NS remnant \\
        \hline
        $22.20 \leq M_\mathrm{ZAMS}/M_\odot < 23.32$ & Failed SN, BH remnant \\
        \hline
        $23.32 \leq M_\mathrm{ZAMS}/M_\odot <34.05$ & FeCCSN, NS or BH remnant \\ 
        \hline
        $M_\mathrm{ZAMS}/M_\odot \geq  34.05$ & Failed SN, BH remnant \\
		\hline
	\end{tabular}
\end{table}

\subsubsection{Choice of ccSN model}
We distinguish between electron-capture SNe (ECSNe) and iron-core collapse SNe (FeCCSNe), which are initiated after the infall of the oxygen-neon core and of the iron-core, respectively, and both of which are neutrino-driven (e.g. \citealt{Radice2017,WangBurrows2024})
We use the statistical ccSN model introduced in \citeauthor{Maltsev2025} (\citeyear{Maltsev2025}; hereafter, \citetalias{Maltsev2025}) to map out the final fates of massive OB star primaries in our sample.
%wherein the SN explosion itself is driven by the same neutrino energy transport mechanism as in the more massive stars undergoing Fe core-collapse \citepalias{Maltsev2025}. 
Not all FeCCSNe succeed to explode, and if they do not, we assume a BH forms \citep{Burrows1988}. In those cases, for simplicity, we refer to failed ccSNe as ``direct-collapse'' BH formation events, even though the BH initially forms from the collapse of an intermediary proto-neutron star before the outer layers follow. 
%before the outer layers are accreted 
%it is not direct per say, as there is an intermediate protoneutron star that precede the secondary collapse into a BH 
In this work, we are primarily interested in the discrimination between successful and failed ccSNe, and therefore do not deepen modeling explosive BH formation, which may occur via multiple channels \citep{Burrows2025,Eggenberger2026}. %into the type of compact remnant (NS or BH) left behind a successful ccSN.}%-BH formation channel via a weak ccSNe \citep{Colgate1971} that is also part of the \citetalias{Maltsev2025} model. 

\citetalias{Maltsev2025} predict the outcome of iron-core collapse (successful versus failed ccSN explosion) in a massive star, given its carbon-oxygen core mass, $M_\mathrm{CO}$, and metallicity, while distinguishing between hydrogen-rich envelope retaining and envelope-stripped ccSN progenitors. The resulting final fate landscape is characterized by the following bimodal structure. Stars are predicted to successfully explode, unless their $M_\mathrm{CO}$ is within a narrow band correlating with failed ccSNe, and unless their $M_\mathrm{CO}$ is above a high-mass threshold towards a plateau of failed ccSN outcomes extending up to the pair-instability mass gap \citep{woosley2021}. 
Recent gravitational wave observations of binary BH mergers favour \citetalias{Maltsev2025} over several other competing SN models \citep{Willcox2026}. 
To apply \citetalias{Maltsev2025}, we convert the three $M_\mathrm{CO}$-thresholds for SN-BH transitions for massive solar-metallicity single-stars using the (almost linear) $M_\mathrm{CO}$-to-$M_\mathrm{ZAMS}$ relation inherent in the stellar evolution models from \citet{Schneider2021} that underlay the construction of \citetalias{Maltsev2025}. The result is summarized in Table \ref{MassiveStars_Fates}.

We stress that the association of ZAMS masses of massive solar-metallicity single-stars with explodability outcomes is uncertain for the following two main reasons: 
\begin{itemize}
    \item Stellar evolution from the ZAMS up to the evolutionary endpoint is model-dependent, and therefore conditional on the adopted physics e.g. on core mass growth, rotation, stellar wind mass loss and nuclear reaction rates. A variation of the adopted physics induces changes in stellar observables throughout their lifetimes and final stellar structure properties at the evolutionary endpoints.
    \item For the same ccSN progenitor characterized by final stellar structure properties, different research groups modeling ccSNe using detailed multidimensional simulations do not always agree on its final fate outcome and ccSN explosion properties (see e.g. the review from \citealt{Janka2025}).
\end{itemize}

\subsubsection{Determination of the minimal $M_\mathrm{ZAMS}$ for a ccSN explosion}
The least massive stars to explode are super-AGB stars which undergo off-center carbon-flash burning and exceed the effective Chandrasekhar mass \citep{Timmes1996} 
for collapse of the degeneracy-supported oxygen-neon (ONe) core. The corresponding core mass limit is set to about 1.37 M$_{\odot}$ \citep{Nomoto1987,Posdialowski2004,Takahashi2013}. Super-AGB stars of lower ONe core mass do not exceed the effective Chandrasekhar mass and thus remain degeneracy-supported to end up as ONe white dwarves (WDs). Stars with ONe cores exceeding this limit but remaining cool enough not to ignite core-neon burning end their lives with explosive neutron star (NS) formation in SNe initiated by the electron-capture (EC) onto magnesium and neon in the core \citep{Nomoto1980}.
Those which exceed an ONe core mass of about 1.43 M$_{\odot}$ \citep{Tauris2015} are hot enough to ignite neon and evolve through the subsequent burning phases up to the development and collapse of an iron-core. We interpolate the stellar evolution models from \citet{Temaj2024}--  which extend those from \citet{Schneider2021} to the lower $M_\mathrm{ZAMS}$-range -- to determine the $M_\mathrm{ZAMS}$-mass threshold for the transition from non-explosive WD formation and explosive NS formation. From these core-mass conditions and our set of stellar evolution models, we derive a minimum M$_{\rm ZAMS}$ threshold of 8.55 M$_{\odot}$ for a star to explode as an ECSN, and of 8.85 M$_{\odot}$ to explode as a FeCCSN (see Table \ref{MassiveStars_Fates}).

\subsection{Identification of OB-type ccSN progenitor candidate stars}
\label{ccsn_prog_obs}

\begin{figure*}
    \centering
    \includegraphics[scale=0.22]{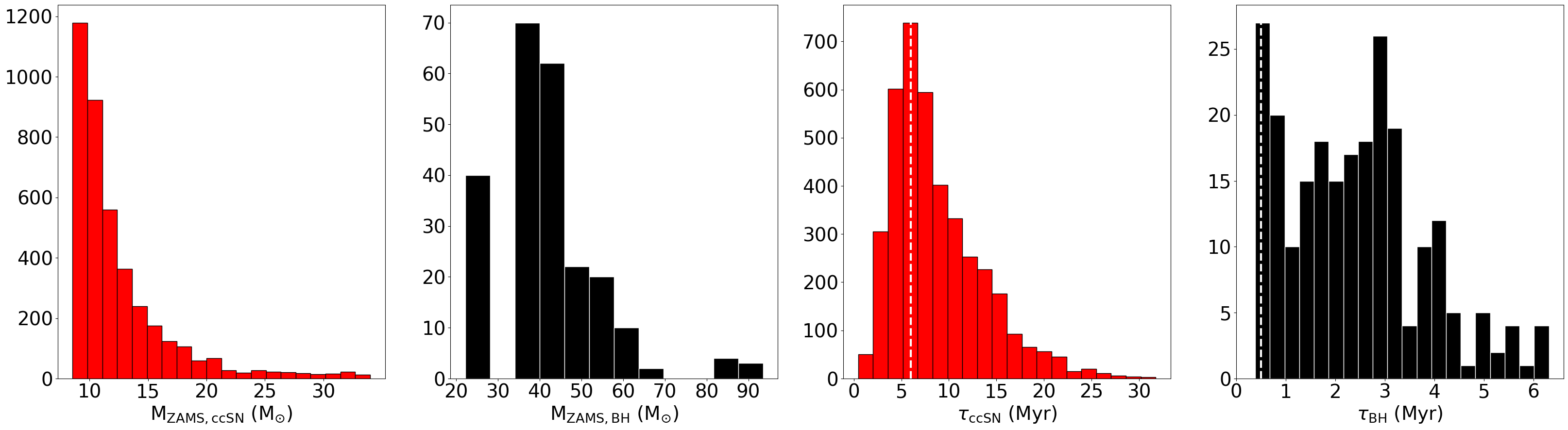}  
    \caption{Far left panel: observed, median SED-fitted ZAMS mass distribution of the ccSN explosion progenitor candidates (M$_\mathrm{ZAMS,ccSN}$, in red) from our catalogue of OB-type stars. Middle right panel: distribution of waiting times ($\tau_\mathrm{ccSN}$) before the next ccSN explosion for this subset, estimated from the solar-metallicity, rotating single-star evolutionary models from \citetalias{Ekstrom}, as outlined in Section \ref{waiting_times}. Middle left and far right panels: same variables (M$_\mathrm{ZAMS,BH}$ and $\tau_\mathrm{BH}$) but for the direct-collapse BH formation progenitor candidates. For the middle right and far right panels, the white dashed line shows the peak time.}
    \label{SN_Progenitors} 
\end{figure*}

Out of the 105,971 OB-type stars within 2 kpc in our sample, we identify a subset of successful (3998, of which we predict 294 to explode as ECSNe) and failed (233) ccSN progenitor candidates, based on their median SED-fitted ZAMS mass. We stress that these are model-dependent candidate counts defined by posterior median masses, rather than secure physical classifications\footnote{Especially, the narrow interval of failed SN within $22.20 \leq M_\mathrm{ZAMS}/M_\odot < 23.32$ in Table \ref{MassiveStars_Fates} is much narrower than the typical uncertainty in $M_\mathrm{ZAMS}$ of our SED fitter for massive stars.}. These numbers represent a significant increase over the $816^{+84}_{-93}$ OB-star ccSN explosion and $13^{+4}_{-2}$ direct-collapse BH progenitors identified within 3 kpc in \citet{Schmidt2014}, although this is a pre-Gaia census with a low completeness at high distances (see their Table B2). 

\begin{figure*}
\includegraphics[scale=0.3]{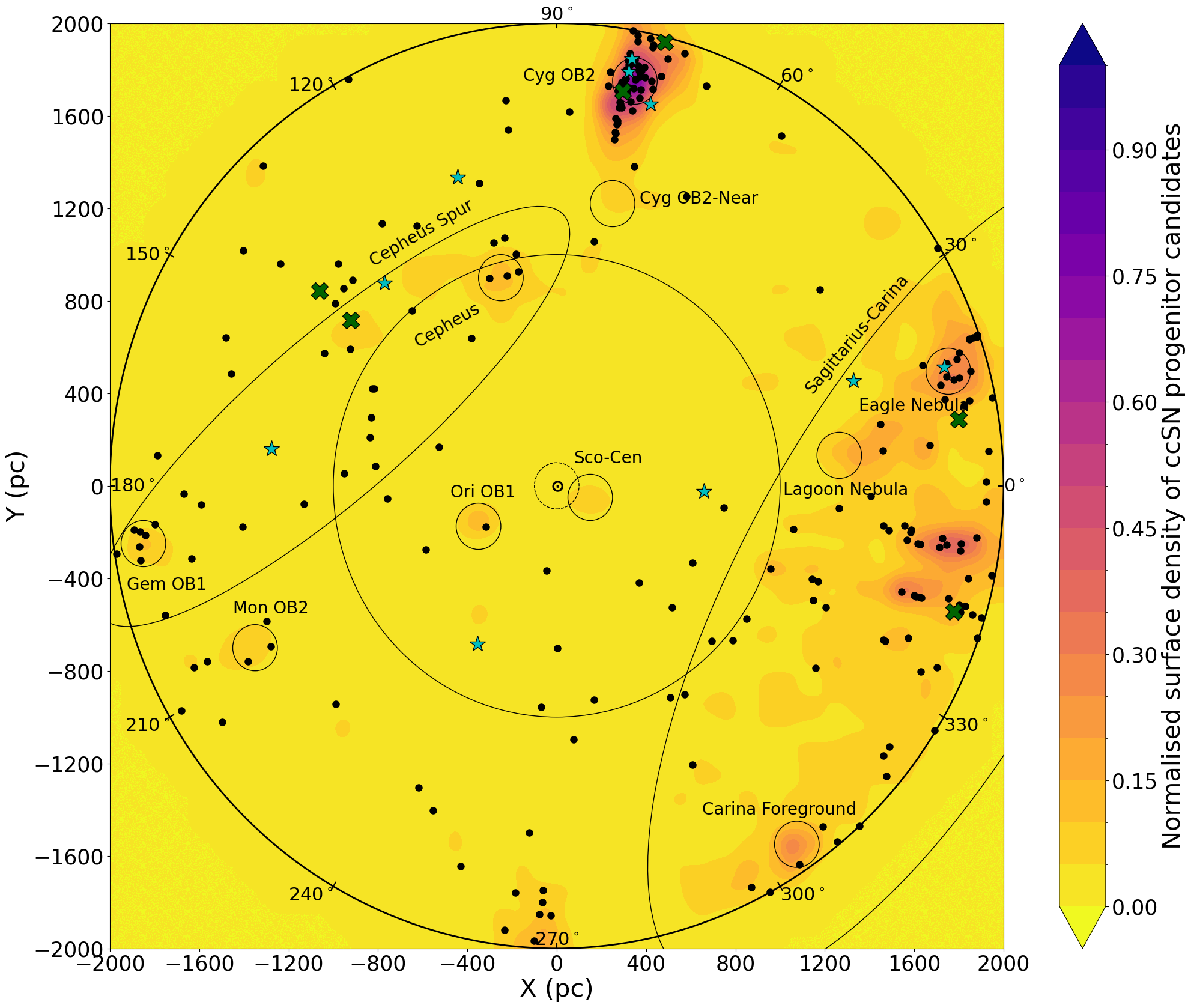}
\caption{Same as Fig. \ref{OBmap} but only displaying the density distribution (color-coded) of the subset of OB-type stars which we predict to end their lives with ccSN explosions, while the rarer failed ccSN BH progenitor candidates are plotted as points (black-dotted). %Both selected based on the thresholds defined in Table \ref{MassiveStars_Fates}. 
In addition, we plot the 10 ccSN progenitor candidates that could explode within less than 1 Myr (as blue stars) and the failed ccSN progenitor candidates that could form black holes within the next 0.5 Myr (as green crosses). The dashed circle surrounding the Sun corresponds to a radius of 100 pc from the Sun, marking the maximal distance at which a nearby ccSN explosion could be harmful to Earth. The electromagnetic transient accompanying a SN going off at this radius would become observable on Earth only after a delay time of 326 years (compared with 3260 and 6520 years at the 1 kpc and 2 kpc radii, respectively).} 
\label{XY_Progenitors} 
\end{figure*}

Fig. \ref{SN_Progenitors} shows the histograms of the median-ZAMS masses of these observed ccSN explosion and direct-collapse BH progenitor candidates, whereas Fig. \ref{XY_Progenitors} shows a top-down view of their spatial distribution, characterized by a greater density contrast compared with Fig. \ref{OBmap}. Our findings suggest that the ccSN statistic is completely dominated by lower-mass stars with ZAMS masses up to about $15-20 \, M_\odot$, as was to be expected due to the initial mass function (IMF) \citep{Kroupa2001}. The relatively high abundance of high-mass ($> 34.0.5 \, M_\odot$) BH progenitor candidates compared to their $\sim 22.2 \, M_\odot$ counterparts -- adverse to the expectation from the IMF -- could be an indication of mistaking unresolved massive binaries as single-star sources, especially so since the majority of BH progenitor candidates in our sample are further than 1 kpc away.

Given that our catalogue of OB stars is limited to 2 kpc, and characterized by an inhomogeneous spatial distribution (see Fig. \ref{OBmap}), these results are dominated by individual, massive star-forming regions. 
Prominent in Fig. \ref{XY_Progenitors} are the sites of massive star formation, notably Cyg OB2, Cepheus, Gem OB1 and the Sagittarius-Carina arm. Sco-Cen, Ori OB1, Mon OB2 and the Carina Foreground also harbour a significant number of candidate ccSN progenitors.

\subsection{Distribution of OB-type star ccSN delay times}
\label{waiting_times}

To further characterize the explosion landscape in the local Milky Way, we have estimated the distribution of delay times (hereafter, $\tau_{cc}$) until the end of OB-type stars' lives. To do so, we have converted the fractional ages of all successful and failed ccSN progenitors from the SED fitter into %actual
present-day ages, $\tau$, and predicted their maximum age, $\tau_\mathrm{f}$, from their median ZAMS mass based on solar-metallicity rotating single-star evolutionary models from \citetalias{Ekstrom}. Then, $\tau_\mathrm{cc} = \tau_\mathrm{f}-\tau$
%difference between their maximum age and their actual age is labeled $\tau_{cc}$, 
and their uncertainties %have been 
are estimated from those of the SED-fitted fractional ages\footnote{In our stellar evolution models, the evolutionary cut-off is at core-carbon depletion, but the remaining evolutionary sequence up to core-collapse typically} happens within less than 100 years \citep{Kippenhahn1990} and is thus negligible.
Both the candidate successful and failed ccSN event progenitor subsets are then sorted by increasing $\tau_\mathrm{cc}$ to obtain the delay time distribution of upcoming ccSN explosion ($\{ \tau_\mathrm{ccSN} \}$-set) and direct-collapse BH formation ($\{ \tau_\mathrm{BH} \}$-set) candidate events, respectively. 

Fig. \ref{SN_Progenitors} shows the highly inhomogeneous $\{ \tau_\mathrm{ccSN} \}$- and $\{ \tau_\mathrm{BH} \}$-distributions, respectively. The upcoming ccSN explosion event candidate events we infer spread out over a timescale of the upcoming $\sim$32 Myr while the BH formation candidate events are dated to within the next $\sim$6 Myr. The right-skewed, unimodal $\tau_\mathrm{ccSN}$-distribution has a pronounced peak at $\sim$6 Myr from now, a time during which the expected number of ccSN explosions per 1 Myr interval is more than a factor of 10 compared to their number within the upcoming 1 Myr from now. Since the lifetime of the least massive and therefore most abundant ccSN progenitors is $\sim$13 Myr, this could be indicative of a massive star formation burst having taken place in the local Milky Way about $7 = 13-6$ Myr ago.

In contrast, $\tau_\mathrm{BH}$-distribution is multimodal, with peak times within the next 500,000 years and in $\sim$ 3 Myr from now. We stress that the inference made here and in the following is conditional on the assumption that the single-star SED fitting of the OB-star BH progenitor candidates is appropriate for characterizing the properties of these OB-type sources.

According to the stellar evolution model we use, the lifetime of the most massive ($>$ 80 \, M$_\odot$) stars is $\tau_f \sim 500,000 \, \mathrm{yr}$, while that of the most massive single-star ccSN explosion progenitor ($M_\mathrm{ZAMS}/M_\odot<34.05$)  is $\sim 3.6$ Myr. Within the framework of our modeling based on single-stars only, these define the least delay time until -- due to ongoing star formation in the local Milky Way -- a newly formed OB-star ccSN explosion and BH formation progenitor, respectively, adds to the event counts. Therefore, the properties of the $\tau_\mathrm{ccSN}$ and $\tau_\mathrm{BH}$ distributions are continuously modified with time progressing after the aforementioned delay times.

%\subsubsection{Near-future ccSN explosion and BH formation rates}
Despite all the aforementioned shortcomings and caveats inherent in our modeling and inference, we dare estimates of the near-future ccSN explosion ($\overline{\tau}_\mathrm{ccSN,1Myr}$) and direct-BH formation ($\overline{\tau}_\mathrm{BH,500kyr}$) event rates on the timescales of the next 1 Myr and 500,000 yr, respectively. On these respective timescales, our analysis suggests 10 ccSN explosion and 6 direct-collapse BH formation events. These translate into the mean waiting times of
\begin{equation}
    \overline{\tau}_\mathrm{ccSN,1Myr}/\mathrm{[yr]} = 100,000 \, \mathrm{yr}
\end{equation}
and
\begin{equation}
    \overline{\tau}_\mathrm{BH,500kyr}/\mathrm{[yr]} = 83,000 \, \mathrm{yr}
\end{equation}
between successive events of each event class. 

%\subsubsection{Average ccSN explosion and BH formation rates}
From the $\tau_\mathrm{ccSN}$- and $\tau_\mathrm{BH}$-  distributions, we also estimate a mean 
\begin{equation}
    \overline{\tau}_\mathrm{ccSN}/[\mathrm{yr}]\simeq7800
\end{equation}
and a mean
\begin{equation}
    \overline{\tau}_\mathrm{BH}/[\mathrm{yr}]\simeq25,000
\end{equation}
%$$ of $\sim$7800 years and 25,000 years 
between successive ccSN explosion and direct-collapse BH formation events, respectively, on the 2 kpc scale. 
We take these as average measures of the ccSN explosion and direct-collapse BH formation event incidences over the upcoming $\sim$32 Myr and $6 \, \mathrm{Myr}$ ahead, respectively. The contrast of the $\overline{\tau}_\mathrm{ccSN,1Myr}$-to-$\overline{\tau}_\mathrm{ccSN}$ and of the $\overline{\tau}_\mathrm{BH,500kyr}$-to-$\overline{\tau}_\mathrm{BH}$ ratios highlights the temporal variability of ccSN explosion and BH formation event incidences in our local Milky Way.
Upon this background, we also stress the ``Galactic snapshot'' character of our candidate successful and failed ccSN progenitor map, which fluctuates due to the death of massive stars and, on even shorter timescales (see Sect. \ref{open_clusters}) due to ongoing massive star formation that eventually produces new ccSN progenitors. 

%Our observationally inferred mean waiting time until subsequent ccSN explosion events of OB-type stellar progenitors, \sout{valid over the timescale of at least} the next 3 Myr} an average measure over the next $\sim$32 Myr ahead}, is much shorter than $15,150_{-4375}^{+1639}$ years inferred from the local ccSN rate we found in \citetalias{Quintana2025} (which is valid on a timescale of roughly 400 Myr), when we extrapolate it from the 1 kpc to the 2 kpc scale \kiril{we have to remember what we did here to compare the SN rates on the 1 kpc scale from Q25 to those inferred here on the 2 kpc scale. This is non-trivial. If we don't arrive at clarity, I'd suggest to cut this entire paragraph out for the time being. We should come back to this when re-discussing it in paper II, so you could copy-paste it directly there. However, the text following after this ("This departure ..." could then be moved elsewhere, where we bring the argument of the recent SF burst based on massive BH formation progenitors still alive or the peak in the SN times}. This departure} stresses that ccSN explosions in the local Milky Way are inhomogeneously distributed in both time and space, and that we currently could be living through} an era characterized by an overdensity of OB stars \sout{nearby} within 2 kpc} compared to \sout{the} their} temporal average on the larger 400 Myr timescale.

\subsubsection{Age constraints on Cyg-OB2}

Apparent from Fig. \ref{SN_Progenitors} is the contrast in number, value range and distribution (unimodal vs. multimodal) of $\{\tau_\mathrm{BH}\}$ vs. $\{\tau_\mathrm{ccSN}\}$. One of the possible interpretations of the inferred greater number of imminent BH progenitors despite the scarcer total number of BH progenitors is that another relatively recent massive star formation burst has taken place within 2 kpc within less than 3 Myr. This interpretation is consistent with the results from \citet{Zari2023} who found a burst of star formation over the Milky Way's recent past, although these were based on a much larger portion of the Galactic disk (6 kpc $\times$ 6 kpc area) and within the last 5--10 Myrs of its history.

If our interpretation is correct, then the region where the most concentrated massive star formation burst would have taken place is Cyg OB2. We find that it dominates the present formation of massive stars in the local Milky Way, and therefore the formation of stellar-mass BHs. The high number of high-mass direct-collapse BH progenitors "still alive" on the right panel from Fig. \ref{SN_Progenitors} would favour an age $<$3 Myr, at the lower end of its age estimate of 1--7 Myr from \citet{Wright2015}.

\subsubsection{The case of Sco-Cen}
\label{scocen}
\begin{figure}
    \centering
    \includegraphics[scale = 0.35]{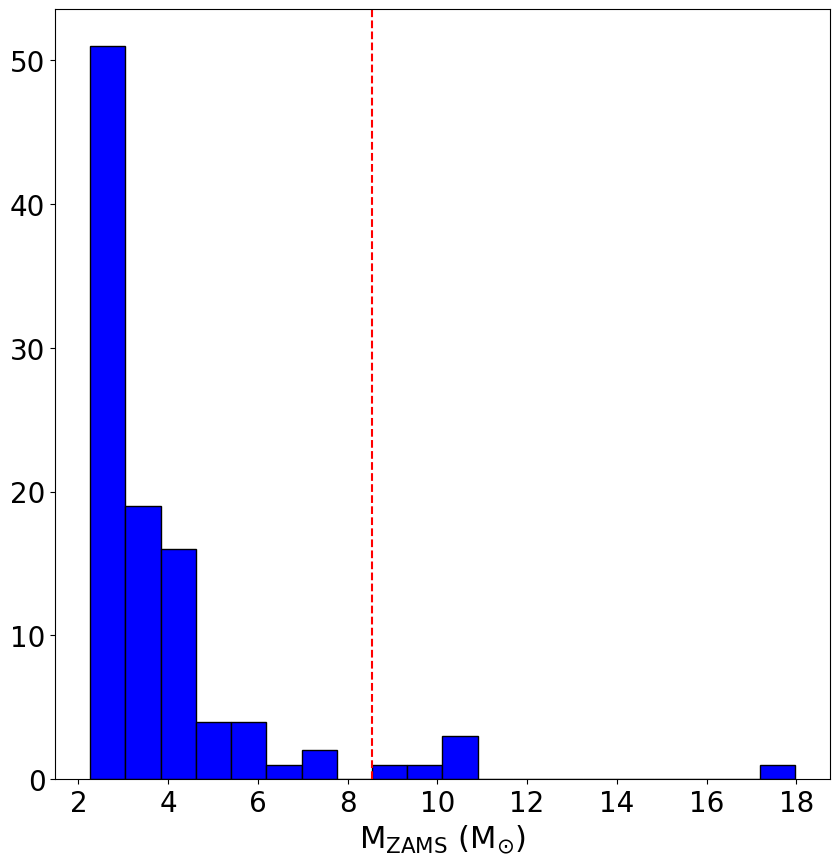}
    \caption{Histogram of the median SED-fitted ZAMS masses of the Sco-Cen members identified in \citet{Quintana2026}, with the vertical dashed line representing the 8.55 M$_{\odot}$ threshold to define a ccSN progenitor adopted from the \citetalias{Maltsev2025} model in Table \ref{MassiveStars_Fates}.}
    \label{ScoCen_Masses}
\end{figure}

Fig. \ref{XY_Progenitors} reveals that the nearest BH progenitor according to our models is located in Ori OB1, whereas the nearest massive star-forming region, would be devoid of such objects. The most massive star we identify in Sco-Cen has a median ZAMS mass of $\sim$18 M$_{\odot}$ whereas the second most massive has a median ZAMS mass of $\sim$11 M$_{\odot}$ (see Fig. \ref{ScoCen_Masses}). To contrast, \citetalias{Maltsev2025} predicts that the minimum ZAMS mass for a star to collapse as a BH is 22.20 M$_{\odot}$ (see Table \ref{MassiveStars_Fates}). Recent studies have unveiled the presence of tens of subgroups in Sco-Cen characterized a significant age gradient (3--19 Myr) and a bursty star formation history that peaked $\sim$15 Myr ago \citep{Ratzenbock2023,Posch2025}. Our findings of few ccSN progenitors remaining are therefore consistent with the inferred star formation history of this complex.

%According to our models, stellar populations in Sco-Cen either didn not form too massive stars (so are limited in their population size) or must be older than the greatest delay time until BH formation, which is around $\sim$6 Myr.}

\subsection{Individual OB-type star ccSN events in the local Milky Way}

\subsubsection{The closest OB-type stars that near the end of their lives}
\label{sec:shortestSNtau}

\begin{table*}
	\centering
	\caption{List of successful and failed ccSN progenitor candidates in our catalogue (separated by a dashed line) that we predict to explode/collapse within the next 1 or 0.5 Myr, respectively, as outlined in Section \ref{waiting_times}. These are ordered by increasing median SED-fitted line-of-sight distance. The ZAMS masses, distances, effective temperatures and luminosities stem from the SED fitting, whereas both the name and the spectral type have been taken from the SIMBAD database. 
    \label{Next_explosions}}
	\renewcommand{\arraystretch}{1.3} 
	\begin{tabular}{ccccccccc} 
    \hline
    Name & Spectral type & l (deg) & b (deg) & M$_{\rm ZAMS}$ (M$_{\odot}$) & d (pc) & $\log(T_{\rm eff})$ [K] & $\log(L/L_{\odot})$ & $\tau_\mathrm{ccSN/BH}$ [Myr] \\
    \hline
    * chi Oph & B2Vne & 357.9 & 20.7 & 23.47$^{+29.68}_{-14.38}$ & 705$^{+1975}_{-351}$ & 4.02$^{+0.58}_{-0.12}$ & 5.21$^{+0.66}_{-1.15}$ & 0.81$^{+4.38}_{-0.46}$ \\
    * eta CMa & B5Ia & 242.6 & -6.5 & 26.50$^{+36.11}_{-10.29}$ & 775$^{+504}_{-276}$ & 4.35$^{+0.27}_{-0.37}$ & 5.31$^{+0.61=}_{-0.39}$ & 0.83$^{+2.58}_{-0.60}$  \\
    * phi Per & B1.5V:e-shell & 131.3 & -11.3 & 29.07$^{+36.87}_{-15.22}$ & 1192$^{+1900}_{-684}$ & 4.37$^{+0.24}_{-0.34}$ & 5.39$^{+0.60}_{-0.92}$ & 1.00$^{+3.27}_{-0.59}$  \\
    *  19 Aur & A5II+ & 172.8 & -1.8 & 21.93$^{+39.49}_{-13.69}$ & 1287$^{+1227}_{-326}$ & 4.03$^{+0.60}_{-0.13}$ & 5.18$^{+0.70}_{-1.14}$ & 0.83$^{+3.85}_{-0.58}$  \\
    HD 172594 & F2Ib & 18.9 & -4.5 & 34.01$^{+30.95}_{-25.22}$ & 1408$^{+371}_{-197}$ & 4.53$^{+0.12}_{-0.68}$ & 5.46$^{+0.46}_{-1.32}$ & 0.95$^{+2.66}_{-0.69}$  \\
    *  73 Dra & A9:VpCrSrEu & 108.4 & 20.0 & 27.40$^{+45.84}_{-8.85}$ & 1495$^{+1800}_{-452}$ & 4.36$^{+0.20}_{-0.09}$ & 5.36$^{+0.69}_{-0.37}$ & 0.77$^{+2.15}_{-0.50}$  \\
    HD 229059 & B2Iabe & 75.7 & 0.4 & 26.46$^{+31.88}_{-17.97}$ & 1704$^{+143}_{-183}$ & 4.46$^{+0.15}_{-0.65}$ & 5.28$^{+0.61}_{-1.18}$ & 0.46$^{+2.49}_{-0.23}$ \\
    HD 169438 & A0Iab & 16.5 & -1.2 & 32.11$^{+30.43}_{-24.75}$ & 1809$^{+269}_{-169}$ & 4.49$^{+0.11}_{-0.72}$ & 5.47$^{+0.48}_{-1.67}$ & 0.60$^{+1.20}_{-0.37}$  \\
    {[CPR2002]} A12 & B1Ib & 79.8 & 0.4 & 17.85$^{+38.77}_{-9.34}$ & 1820$^{+176}_{-118}$ & 4.12$^{+0.46}_{-0.30}$ & 4.99$^{+0.93}_{-0.89}$ & 0.70$^{+1.09}_{-0.37}$  \\
    EM* AS  422 & WN7o/CE+O7V(f) &  79.7 & 0.7 & 32.68$^{+18.62}_{-13.77}$ & 1875$^{+61}_{-86}$ & 4.49$^{+0.11}_{-0.25}$ & 5.52$^{+0.29}_{-0.45}$ & 0.68$^{+1.23}_{-0.50}$  \\
    \hdashline
    HD 21389 & A0Ia & 142.2 & 2.1 & 90.64$^{+7.22}_{-30.65}$ & 1168$^{+103}_{-12}$ & 4.43$^{+0.09}_{-0.10}$ & 6.2$^{+0.05}_{-0.13}$ & 0.42$^{+0.20}_{-0.12}$ \\
    HD 21291 & B9Ia & 141.5 & 2.9 & 34.45$^{+46.40}_{-12.22}$ & 1356$^{+252}_{-148}$ & 4.13$^{+0.22}_{-0.11}$ & 5.63$^{+0.53}_{-0.40}$ & 0.43$^{+0.19}_{-0.17}$ \\
    BD +40 4220 & O7Iafep & 80.1 & 0.9 & 88.86$^{+9.38}_{-13.42}$ & 1732$^{+19}_{-19}$ & 4.47$^{+0.05}_{-0.04}$ & 6.19$^{+0.07}_{-0.07}$ & 0.44$^{+0.21}_{-0.12}$ \\
    HD 165784 & A2Iab & 9.1 & -0.7 & 55.45$^{+28.69}_{-37.83}$ & 1822$^{+231}_{-195}$ & 4.51$^{+0.04}_{-0.47}$ & 5.92$^{+0.24}_{-0.90}$ & 0.47$^{+0.61}_{-0.35}$ \\
    * zet01 Sco & B1.5Ia+ & 343.0 & 0.9 & 93.40$^{+5.32}_{-42.46}$ & 1859$^{+41}_{-119}$ & 4.36$^{+0.06}_{-0.14}$ & 6.20$^{+0.05}_{-0.23}$ & 0.39$^{+0.06}_{-0.11}$ \\
    * P Cyg & B1-2Ia-0ep & 75.8 & 1.3 & 23.10$^{+49.15}_{-5.63}$ & 1979$^{+624}_{-318}$ & 4.01$^{+0.50}_{-0.06}$ & 5.27$^{+0.82}_{-0.25}$ & 0.48$^{+0.44}_{-0.27}$ \\
    \hline
	\end{tabular}
\end{table*}

In Table \ref{Next_explosions} are displayed the 10 ccSN progenitor and the 6 failed ccSN progenitor candidates that have the least median-estimated waiting time until SN explosion and direct-collapse BH formation, respectively (see also Fig. \ref{HR_Diagram_Fates}). We stress that this listing serves primarily to sketch the method to pre-determine the next ccSN explosion and direct-collapse BH formation events from OB-type progenitor stars in the local Milky Way. 
In addition to the aforementioned caveats associated with relating ZAMS masses and explodability outcomes, a faithful inference is further impeded by observational constraints.
The parallax method and SED-fitting yield bolometric luminosities, effective temperatures and distance estimates that are not constraining enough to obtain accurate ZAMS masses and ages needed to obtain accurate $\tau_\mathrm{cc}$-estimates for individual sources (see also Fig. \ref{HR_Diagram_Fates}). Table \ref{Next_explosions} notably reveals some inconsistencies between the SIMBAD spectral classifications and some of our median SED-fitted temperatures. With that in mind, we present these objects as candidates requiring follow-up analysis, in particular with:
\begin{itemize}
    \item \textit{Gaia} DR4 photometry (which would yield improved distance and bolometric luminosity estimates) and
    \item spectroscopy (which would yield improved effective temperature estimates and help identify possible binary companions).
\end{itemize}

Our preliminary analysis suggests that none of the OB-type star  ccSN progenitor candidates will explode within the next 1 Myr within the ``Impact Zone'' of 100 pc -- a scenario that could be harmful for Earth's climate and biosphere \citep{Thomas2016,Melott2017}. Further out, we identify only two OB-type star ccSN progenitor candidates at a distance $d < 1 \, \mathrm{kpc}$ that we predict might explode within the next 1 Myr. However, by construction (see Section \ref{data}), our catalogue of OB-type stars does not include any cool supergiants, and therefore an important contribution to the population of evolved ccSN progenitors. A well-known example of a such is Betelgeuse, located at a distance of 222$^{+48}_{-34}$ pc \citep{Harper2017}.
Its ZAMS mass is estimated to be $\sim$18--19 M$_\odot$ though the waiting time until its SN explosion is debated (e.g. \citealt{Joyce2020, Saio2023, Molnar2023, Goldberg2024}), as it is tied to the model-dependent characterization of Betelgeuse's observed pulsations. \citet{Joyce2020} concluded such pulsations would take place during the early core-helium burning phase while \citet{Saio2023} dates them to the end of the core-carbon burning phase. The remaining lifetimes would approximately be $<1 \, \mathrm{Myr}$ vs $<$1000 \, $\mathrm{yr}$, respectively. (e.g. \citealt{Saio2023}). Another example is Antares A, a red supergiant in Sco-Cen (e.g. \citealt{Ratzenbock2023,Grobschedl2026}), predicted to explode within the next 1.0--1.4 Myr \citep{Neuhauser2022} from isochrone fitting.

\subsubsection{The nearest OB-type ccSN progenitor candidate stars}
\label{nearest}

\begin{table*}
	\centering
	\caption{The nearest ($d <$ 125 pc) ccSN progenitor candidates  from our catalogue, as described in Section \ref{nearest}. Just like Table \ref{Next_explosions}, we have indicated their main SIMBAD name and spectral type, whereas the ZAMS masses, distances, effective temperatures, luminosities and waiting time before ccSNe come from our SED fitting.}
    \label{Nearest_ccSN_progenitors}
	\renewcommand{\arraystretch}{1.3} 
	\begin{tabular}{ccccccccc} 
    \hline
    Name & Spectral type & l (deg) & b (deg) & M$_{\rm ZAMS}$ (M$_{\odot}$) & d (pc) & $\log(T_{\rm eff})$ [K] & $\log(L/L_{\odot})$ & $\tau_\mathrm{ccSN}$ [Myr] \\
    \hline
  * alf Vir & B1V & 316.1 & 50.9 & $8.94^{+1.97}_{-1.82}$ & $77^{+10}_{-9}$ & $4.32 \pm 0.08$  &  $3.84^{+0.16}_{-0.25}$ & $9.59^{+12.63}_{-3.85}$  \\
  * bet Cru & B1IV & 302.5 & 3.2 & $
  9.34^{+1.92}_{-1.59}$ & $86^{+12}_{-8}$ & $4.34 \pm 0.07$ & $3.89^{+0.15}_{-0.19}$ & $9.56^{+12.17}_{-4.10}$  \\
  * alf01 Cru & B0.5IV & 300.1 & -0.4 & $10.71^{+2.19}_{-1.52}$ & $100^{+8}_{-7}$ & $
  4.38 \pm 0.07$ & $4.05^{+0.12}_{-0.10}$ & $8.6^{+10.77}_{-4.36}$ \\
  * alf02 Cru & B1V & 300.1 & -0.4 & $10.28^{+1.78}_{-1.88}$ & $100^{+12}_{-9}$ & $4.38^{+0.05}_{-0.08}$ & $3.93^{+0.15}_{-0.16}$ & $12.76^{+9.68}_{-7.19}$  \\
  * bet Cen & B1III & 311.8 & 1.3 & $17.97^{+4.79}_{-3.86}$ & $123^{+12}_{-8}$ & $4.49^{+0.06}_{-0.09}$ &$4.65^{+0.17}_{-0.13}$ & $5.88^{+4.16}_{-3.49}$  \\
  * eps CMa & B1.5II & 239.8 & -11.3 & $11.18^{+2.32}_{-1.40}$ & $124^{+9}_{-8}$ & $4.38^{+0.07}_{-0.06}$ & $4.14^{+0.10}_{-0.09}$ & $7.17^{+9.69}_{-3.49}$  \\

    \hline
	\end{tabular}
\end{table*}

To complement the discussion from Section \ref{sec:shortestSNtau}, we have also listed in Table \ref{Nearest_ccSN_progenitors} the 6 nearest ($d <$ 125 pc) OB-type ccSN progenitor candidate stars. All of these stars are known ccSN progenitors in the literature, which gives confidence to our method, and are also part of multiple systems, with both components of Acrux resolved by Hipparcos, though the primary star (*alf01 Cru) is expected to undergo a red supergiant phase \citep{Kaler2002} whereas there is still debate whether the secondary star (*alf02 Cru) will explode as a ESCN \citep{WoosleyHeger2015} or become a white dwarf \citep{Kaler2002}.

In particular, we notice 2 OB-type ccSN progenitors below 100 pc, hence within the "Impact-Zone" defined in Section \ref{sec:shortestSNtau}. The closest one is the primary component of the Spica system, which is not expected to explode before a few Myrs \citep{Firestone2014}, consistent with our $\tau_\mathrm{ccSN}$ estimated in Table \ref{Nearest_ccSN_progenitors}. Likewise, \citet{Cotton2022} recently confirmed the ccSN progenitor status of * bet Cru A by fitting an asteroseismic mass of $14.5 \pm 0.1$ M$_{\odot}$ and an age of 9.7--12.8 Myr, implying we underestimated the ZAMS mass of its primary star by fitting the SED formed by the light arising from the entire system. Still our model does not predict any of these systems to undergo a ccSNe within the next 5--10 Myr.

 %whose spatial distribution is displayed in Fig. \ref{} in order to place into context of the nearest OB associations and the Local Bubble (see the dedicated discussion in Appendix \ref{Nearest_200pc}).

 \begin{figure*}
    \centering
    \includegraphics[scale=0.5]{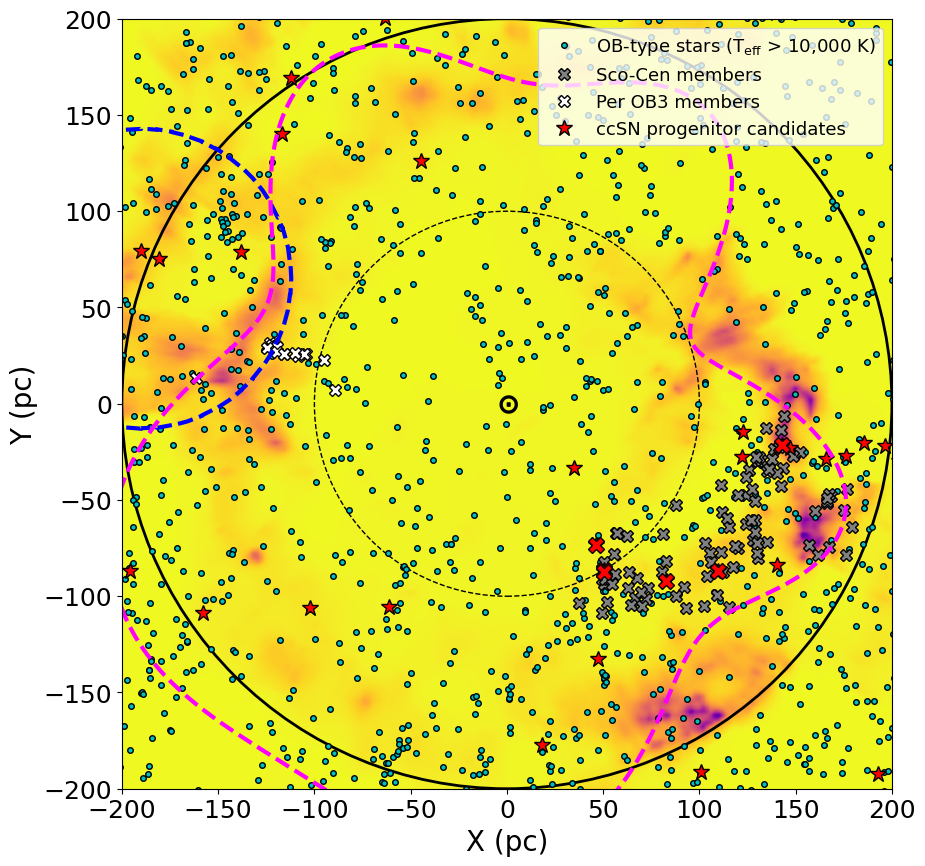}
    \caption{Spatial distribution of the 971 SED-fitted OB-type stars as well as the 24 OB-type ccSN progenitor candidates within 200 pc from the Sun, with the "Impact-Zone" (100 pc radius) shown as a dashed circle. The Local Bubble is both represented with blue dashed contours (taken from \citealt{Pelgrim}) and with the dust density map from \citet{ONeill2024} (itself created with the \texttt{Edenhofer2024} dust map). We have also included the Sco-Cen and Per OB3 members identified in \citet{Quintana2026} and the Per-Tau shell contours (with dashed blue lines) from \citet{Bialy2021}.}
    \label{XY_200pc}
\end{figure*}

To study this view in further details, we have displayed in Fig. \ref{XY_200pc} all the SED-fitted OB-type stars together with the cSN progenitor candidates within 200 pc from the Sun, which we contrast with the identified OB members of the nearest OB associations (i.e., Per OB3 and Sco-Cen) and the Local Bubble, a well-known cavity of low-gas density with a radius of 165 $\pm$ 6 pc (e.g. \citealt{Zucker2022}), which explains the dearth of ccSN progenitor candidates in the local Milky Way, particularly within 100 pc from the Sun.

We do not identify any ccSN progenitor candidate in Per OB3, whose location coincides with the edges of the Per--Tau shell also displayed in Fig. \ref{XY_200pc}, and which contains the Perseus and Taurus molecular clouds. The Per--Tau shell is a superbubble that was likely created by the feedback from past supernova 6--22 Myr ago \citep{Bialy2021}. On the other edge of the Local Bubble lies Sco-Cen which is the main candidate for having generated this structure via past SNe 10--20 Myr ago \citep{MaizApellaniz2001,Zucker2022}. It is therefore not surprising that out of the 24 ccSN progenitor candidates within 200 pc, 6 of them belong to Sco-Cen, as illustrated in Fig. \ref{ScoCen_Masses}, including * bet Cru, *alf01 Cru and *alf02 Cru, 3 out of the 6 ccSN progenitor candidates within 125 pc from the Sun (see Table \ref{Nearest_ccSN_progenitors}). As stressed in Section \ref{scocen}, the lack of very massive stars (and the gap between $\sim$11 and 18 M$_{\odot}$) could be reminiscent of the recent sequence of ccSNe (over the last 15 Myrs).

\subsubsection{The final fate of *P Cyg}

Our method is afflicted with systematic uncertainties also due to the neglected binary interaction effects and to the specifics of the adopted physics of the single-star evolutionary models we use for parameter inference (e.g. the choice of a fixed rotation rate), both of which influence the life expectancy and explodability of massive stars. 
%purposes, as an astro-photometric method is too limited to accurately constrain the ages of massive stars and their ultimate fate, particularly as 
%for all stars and neglects binary interaction, both of which can strongly influence the life expectancy of a massive star.
We consider an example to elucidate on how these act in practice. Table \ref{Next_explosions} includes *P Cyg, an hypergiant luminous blue variable, whose median ZAMS mass implies a direct-collapse BH formation endpoint, but whose error bars are much larger than the interval of 1.2 M$_{\odot}$ for failed ccSNe outcomes according to our model (see Table \ref{MassiveStars_Fates}). Its upper error bar is so extended that it is consistent its ZAMS mass surpassing the threshold to the plateau of direct-collapse BH formation outcomes. Luminous supergiants have large error bars on parameter estimates with the SED fitting method, especially when they are highly extinguished and variable. In previous work, *P Cyg has been predicted to be a candidate Type IIb supernova progenitor (e.g. \citealt{Groh2013}). With M$_\mathrm{ZAMS}=$ 37 M$_{\odot}$ inferred in \citet{Rivet2020} (who used a  set of single-star evolution and SN models different from ours), according to the single-star SN-BH thresholds of \citetalias{Maltsev2025}, it would undergo direct-collapse to a BH. However, since *P Cyg is likely a (Case B) donor transferring envelope mass to a binary companion \citep{Mahy2022}, its \citetalias{Maltsev2025}-explodability is appropriately classified using the stripped-star -- not the single-star -- prescription, for which the threshold to a plateau of failed ccSN outcomes is $M_\mathrm{ZAMS} \simeq 70 \, \mathrm{M_\odot}$, for solar metallicity progenitors (see their Fig. A.5). According to this scheme, *P Cyg therefore will more likely explode at the end of its life.

The upshot from the analysis of the *P Cyg example is that -- to conclude on an OB-type star's final fate and remaining lifetime, it is crucial to find out whether it currently is or eventually will be part of an interacting binary system, which is likely \citep{sana2012, Moe2017}. The combination of astrometric and spectroscopic follow-up analyses could shed light on this. With improved distance estimates (e.g. with \textit{Gaia} DR4) and after having gained insight on the binarity status, we encourage the community to analyze our list of the next and/or nearest OB-type ccSN explosion and direct-collapse BH formation progenitor candidate star observations with detailed single- and binary evolution models to draw more faithful conclusions on their fates and associated ccSN waiting times. 
Even in the absence of evidence for a binary companion to an OB star primary, the percentage of OB stars that are stellar merger products is expected to be $\sim8\%$ \citep{deMink2014}.

For these reasons, in the present work, for the time being, we trust our estimates of statistical measures characterizing population properties (Sect. \ref{waiting_times}) more than our predictions of individual ccSN events.

\section{Conclusions}
\label{conclusions}

In this study, we have identified a population of 105,971 O- and B-type stars within 2 kpc of the Sun, applying an astro-photometric (SED fitting) Bayesian inference technique to estimate their stellar physical parameters. As the \textit{Gaia} DR3 era comes to an end, this catalogue therefore provides high-confidence targets for spectroscopic follow-up.

Comparing our census with various spectroscopic surveys, we have a tendency of slightly underestimating effective temperatures (except for APOGEE DR17 where we observe the reverse trend). Furthermore, for the sample of OB stars that also is in \citetalias{HuntReffert2024}, we have lower $\log(M_{\rm ZAMS}/M_{\odot})$ by a median value of 0.06
dex, an offset we attribute to differences in adopted stellar evolution and atmosphere models.

This catalogue benefits from a high-level of completeness, both at the bright end thanks to the inclusion of the Bright Stars Catalogue to compensate for the saturation in \textit{Gaia} DR3, and at the faint end except for the latest B-type stars in regions of extreme reddening such as Cyg OB2, validating the reconstructed (i.e., between 1.25 and 2 kpc) version of the \texttt{Edenhofer2024} extinction map.

Our map of OB stars unveils a complex view of the young stellar populations across the Galactic thin disk, with a sharp contrast between the Galactic Centre and Anticentre directions, and with most of the OB stars located close to the Galactic mid-plane ($h_Z = 68 \pm 1$ pc). This inhomogeneous distribution is shaped by local overdensities such as Cyg OB2, as well as other OB associations and massive star-forming regions such as Ori OB1, Mon OB1,  Gem OB1 and Cepheus. Prominent large-scale overdensities include a significant segment of the Sagittarius-Carina spiral arm, whilst the Perseus Gap (or Giant Oval Cavity) also stands out as a notable dearth of OB stars. The spatial distribution of OB stars is consistent with \citetalias{Zari2021} on the spiral arm location and width, but further work needs to be carried out in that direction, particularly in line of the different spiral arm models using different tracers, with currently no clear consensus.

Our catalogue serves as a tracer of star formation in the local Milky Way complementary to that of the young open clusters from \citet{HuntReffert2023,HuntReffert2024}. OB associations are hosted in OCs at a relative probability of  $\sim$10 \%, and our bright OB stars allow us to penetrate deeper into the high-extinction regions. Thee derived surface density star formation rates in open clusters ($\Sigma_{\rm SFR,OC}$) are larger in the Galactic Centre directions (between $836^{+304}_{-354}$ and $921^{+264}_{-303}$ M$_{\odot}$ Myr$^{-1}$ kpc$^{-2}$) compared with the Galactic Anticentre directions (between $534^{+196}_{-281}$ 
and $624^{+174}_{-224}$ M$_{\odot}$ Myr$^{-1}$ kpc$^{-2}$). This decrease of star formation towards the Outer Galaxy is consistent with star formation model of disk galaxies that attributes it to exponential drop in gas density (e.g. \citealt{KennicuttEvans2012,Oemler2017}), but the large error bars on $\Sigma_{\rm SFR,OC}$ prevent us from concluding it with confidence. More accurate censuses of the population of OB stars and open clusters will allow us to solve these questions. Aside from the improved astrometry that will be delivered by the two last data releases from the \textit{Gaia} mission, its proposed successor, GaiaNIR, could help us to pierce through the high-extinction regions towards the Galactic Centre (beyond 2--3 kpc, e.g. \citealt{Nogueras2019}). Indeed, GaiaNIR is set to observe $\sim$50 billion stars in the near-infrared, and will therefore resolve currently-hidden populations of OB stars and open clusters in that direction \citep{HobbsHog2018,Hobbs2024}, helping us to better understand star formation in the Milky Way.

Beyond tracing the structure of the local Milky Way, our OB star catalogue can serve as a valuable tracer of (massive) star formation and explodability. Leveraging a recently introduced supernova model, we have identified 3998 core-collapse supernova (ccSN) explosion and 233 direct-collapse black hole (BH) progenitor candidate OB stars within 2 kpc, providing a first-order picture of a potential landscape of upcoming ccSN events in the local Milky Way. 
We have inferred a highly inhomogeneous temporal distribution of ccSN explosion and direct-collapse BH formation events in the local Milky Way, implying that we currently live through an era characterized by a dearth in ccSN explosion events but a peak of direct-collapse BH formation events. This is reflected in the mean waiting times between successive ccSN explosion and direct-collapse BH formation events from OB star progenitors on the timescale of the next 1 Myr and 500,000 yr, respectively, on the 2 kpc scale, which we infer to be 
$\overline{\tau}_\mathrm{ccSN,1Myr}/\mathrm{yr}=100,000$ 
and 
$\overline{\tau}_\mathrm{BH,500kyr}/\mathrm{yr}=83,000$.
We caveat that the inferred great abundance of the most massive OB star BH progenitors could be due to mistaking unresolved massive binaries at > 1 kpc distances as overly massive single-star sources.

Furthermore, we have estimated 
$\overline{\tau}_\mathrm{ccSN}/\mathrm{yr}\simeq7,800$ and $\overline{\tau}_\mathrm{BH}/\mathrm{yr} \simeq 25,000$, respectively, 
as the average delay times between successive ccSN explosion and direct-collapse BH formation events from OB-star sources, valid on a timescale of $\sim$32 Myr and $\sim$6 Myr, respectively. These event rates are derived from OB star primaries and therefore are lower bounds on the integral ccSN event rates, which are further contributed by non-OB star ccSN progenitors and binary companions to OB star primaries, some of which may also be ccSN progenitors while outshined by the OB star primaries.
We infer that the peak time for ccSN explosion incidence in the local Milky Way, more than a factor of 10 higher than within the next 1 Myr, will be in 6 Myr from now.  Dating back the birth time of the corresponding OB star ccSN progenitor candidates, we suggest that a massive star formation episode has taken place in the local Milky Way about 7 Myr ago.

Our analysis of direct-collapse BH formation times further suggests another, more
 recent episode of nearby massive star formation having taken place, less than 3 Myr ago, concentrated in Cyg OB2 and along the Sagittarius-Carina spiral arm where we find the greatest OB-type star BH progenitor overdensities. 
%Our interpretation receives even more support by the finding that -- under the premise of faithfulness of our stellar parameter inference and underlying modeling assumptions -- we expect more BH progenitors to collapse within the next 1 Myr than ccSN progenitors to explode. The most massive stars have the least delay times until the end of their lives, and our finding that an abundant number of them still ``alive'', greater than the number of ccSN explosion progenitors which fill a continuum, speaks in favour of a recent and spatially extended massive star formation episode. 
Grossly viewed, our interpretation is in line with the results from \citet{Zari2023}, although they considered a much wider Galactic field and temporal scale than we do in the present work. Our findings corroborate the spatiotemporal variability of incidence of ccSN events in the local Milky Way.

Furthermore, we give a first-order assessment of future explosion events near-Earth. We re-discover 6 OB-type star ccSN progenitor candidates within 125 pc (the liberal upper distance limit at which long-term cosmic ray exposure triggered by a nearby ccSN explosion could be harmful to Earth's biosphere) but anticipate that none of them will explode within the next 5 Myr. We encourage dedicated follow-up analyses with \textit{Gaia} DR4 photometry and spectroscopy to clarify the nature of 10 other OB-type ccSN progenitor candidates we identify further out but within 2 kpc, and the use of binary evolution models to re-estimate their remaining lifetimes.

In Paper II, we will instead look into the past, and re-estimate the star formation, core-collapse supernova explosion and black hole formation rates during the last $\sim 410$ Myr.

\section*{Acknowledgements}

ALQ acknowledges a Paris Sciences et Lettres (PSL) fellowship granted by the Scientific Council from the Paris Observatory. EP was supported in part by the Italian Space Agency (ASI) through contract ASI-INAF 2025-10-HH.0 to the National Institute for Astrophysics (INAF).

We thank the anonymous referee for their detailed comments that helped us improving the manuscript. The authors also would like to thank Gordian Edenhofer for his suggestions on how best to exploit his 3D extinction map, Ronald Drimmel and Shourya Khanna for suggestions on the spiral arm models, Adam Burrows for sharing insight on black hole formation channels and recent developments in core-collapse supernova explosion modeling, Catherine Zucker for providing the dataset and discussion on the Local Bubble, Kevin Jardine and Friedrich Anders for clarifications on the labelling of overdensities of OB stars, as well as David Katz, Eleonora Zari, Hans-Walter Rix, Coryn Bailer-Jones and Pasquale Panuzzo for other insightful scientific discussions.

This paper exploited the data processed by the Gaia Data Processing and Analysis Consortium (DPAC, https://www.cosmos.esa.int/web/gaia/dpac/consortium) and obtained by the Gaia mission from the European Space Agency (ESA) (https://www.cosmos.esa.int/gaia), as well as the INT Galactic Plane Survey (IGAPS) from the Isaac Newton Telescope (INT) operated in the Spanish Observatorio del Roque de los Muchachos, the observations made with ESO Telescopes at the La Silla Paranal Observatory under programme ID 177.D-3023, as part of the VST Photometric H$\alpha$ Survey of the Southern Galactic Plane and Bulge (VPHAS+, www.vphas.eu), alongside the Two Micron All Star Survey, which is a combined mission of the Infrared Processing and Analysis Center/California Institute of Technology and the University of Massachusetts.

Finally, this work benefited from the use of \textit{TOPCAT} \citep{Topcat}, Astropy \citep{Astropy} and the Vizier and SIMBAD database, both operated at CDS, Strasbourg, France.

%%%%%%%%%%%%%%%%%%%%%%%%%%%%%%%%%%%%%%%%%%%%%%%%%%
\section*{Data Availability}

The catalogue of OB stars within 2 kpc will be uploaded to Vizier.



\bibliographystyle{mnras}
\bibliography{bibliography} % if your bibtex file is called example.bib

@ARTICLE{GaiaDR3,
       author = {{Gaia Collaboration} and {Vallenari}, A. and {Brown}, A.~G.~A. and {Prusti}, T. and {de Bruijne}, J.~H.~J. and {Arenou}, F. and {Babusiaux}, C. and {Biermann}, M. and {Creevey}, O.~L. and {Ducourant}, C. and {Evans}, D.~W. and {Eyer}, L. and {Guerra}, R. and {Hutton}, A. and {Jordi}, C. and {Klioner}, S.~A. and {Lammers}, U.~L. and {Lindegren}, L. and {Luri}, X. and {Mignard}, F. and {Panem}, C. and {Pourbaix}, D. and {Randich}, S. and {Sartoretti}, P. and {Soubiran}, C. and {Tanga}, P. and {Walton}, N.~A. and {Bailer-Jones}, C.~A.~L. and {Bastian}, U. and {Drimmel}, R. and {Jansen}, F. and {Katz}, D. and {Lattanzi}, M.~G. and {van Leeuwen}, F. and {Bakker}, J. and {Cacciari}, C. and {Casta{\~n}eda}, J. and {De Angeli}, F. and {Fabricius}, C. and {Fouesneau}, M. and {Fr{\'e}mat}, Y. and {Galluccio}, L. and {Guerrier}, A. and {Heiter}, U. and {Masana}, E. and {Messineo}, R. and {Mowlavi}, N. and {Nicolas}, C. and {Nienartowicz}, K. and {Pailler}, F. and {Panuzzo}, P. and {Riclet}, F. and {Roux}, W. and {Seabroke}, G.~M. and {Sordo}, R. and {Th{\'e}venin}, F. and {Gracia-Abril}, G. and {Portell}, J. and {Teyssier}, D. and {Altmann}, M. and {Andrae}, R. and {Audard}, M. and {Bellas-Velidis}, I. and {Benson}, K. and {Berthier}, J. and {Blomme}, R. and {Burgess}, P.~W. and {Busonero}, D. and {Busso}, G. and {C{\'a}novas}, H. and {Carry}, B. and {Cellino}, A. and {Cheek}, N. and {Clementini}, G. and {Damerdji}, Y. and {Davidson}, M. and {de Teodoro}, P. and {Nu{\~n}ez Campos}, M. and {Delchambre}, L. and {Dell'Oro}, A. and {Esquej}, P. and {Fern{\'a}ndez-Hern{\'a}ndez}, J. and {Fraile}, E. and {Garabato}, D. and {Garc{\'\i}a-Lario}, P. and {Gosset}, E. and {Haigron}, R. and {Halbwachs}, J. -L. and {Hambly}, N.~C. and {Harrison}, D.~L. and {Hern{\'a}ndez}, J. and {Hestroffer}, D. and {Hodgkin}, S.~T. and {Holl}, B. and {Jan{\ss}en}, K. and {Jevardat de Fombelle}, G. and {Jordan}, S. and {Krone-Martins}, A. and {Lanzafame}, A.~C. and {L{\"o}ffler}, W. and {Marchal}, O. and {Marrese}, P.~M. and {Moitinho}, A. and {Muinonen}, K. and {Osborne}, P. and {Pancino}, E. and {Pauwels}, T. and {Recio-Blanco}, A. and {Reyl{\'e}}, C. and {Riello}, M. and {Rimoldini}, L. and {Roegiers}, T. and {Rybizki}, J. and {Sarro}, L.~M. and {Siopis}, C. and {Smith}, M. and {Sozzetti}, A. and {Utrilla}, E. and {van Leeuwen}, M. and {Abbas}, U. and {{\'A}brah{\'a}m}, P. and {Abreu Aramburu}, A. and {Aerts}, C. and {Aguado}, J.~J. and {Ajaj}, M. and {Aldea-Montero}, F. and {Altavilla}, G. and {{\'A}lvarez}, M.~A. and {Alves}, J. and {Anders}, F. and {Anderson}, R.~I. and {Anglada Varela}, E. and {Antoja}, T. and {Baines}, D. and {Baker}, S.~G. and {Balaguer-N{\'u}{\~n}ez}, L. and {Balbinot}, E. and {Balog}, Z. and {Barache}, C. and {Barbato}, D. and {Barros}, M. and {Barstow}, M.~A. and {Bartolom{\'e}}, S. and {Bassilana}, J. -L. and {Bauchet}, N. and {Becciani}, U. and {Bellazzini}, M. and {Berihuete}, A. and {Bernet}, M. and {Bertone}, S. and {Bianchi}, L. and {Binnenfeld}, A. and {Blanco-Cuaresma}, S. and {Blazere}, A. and {Boch}, T. and {Bombrun}, A. and {Bossini}, D. and {Bouquillon}, S. and {Bragaglia}, A. and {Bramante}, L. and {Breedt}, E. and {Bressan}, A. and {Brouillet}, N. and {Brugaletta}, E. and {Bucciarelli}, B. and {Burlacu}, A. and {Butkevich}, A.~G. and {Buzzi}, R. and {Caffau}, E. and {Cancelliere}, R. and {Cantat-Gaudin}, T. and {Carballo}, R. and {Carlucci}, T. and {Carnerero}, M.~I. and {Carrasco}, J.~M. and {Casamiquela}, L. and {Castellani}, M. and {Castro-Ginard}, A. and {Chaoul}, L. and {Charlot}, P. and {Chemin}, L. and {Chiaramida}, V. and {Chiavassa}, A. and {Chornay}, N. and {Comoretto}, G. and {Contursi}, G. and {Cooper}, W.~J. and {Cornez}, T. and {Cowell}, S. and {Crifo}, F. and {Cropper}, M. and {Crosta}, M. and {Crowley}, C. and {Dafonte}, C. and {Dapergolas}, A. and {David}, M. and {David}, P. and {de Laverny}, P. and {De Luise}, F. and {De March}, R. and {De Ridder}, J. and {de Souza}, R. and {de Torres}, A. and {del Peloso}, E.~F. and {del Pozo}, E. and {Delbo}, M. and {Delgado}, A. and {Delisle}, J. -B. and {Demouchy}, C. and {Dharmawardena}, T.~E. and {Di Matteo}, P. and {Diakite}, S. and {Diener}, C. and {Distefano}, E. and {Dolding}, C. and {Edvardsson}, B. and {Enke}, H. and {Fabre}, C. and {Fabrizio}, M. and {Faigler}, S. and {Fedorets}, G. and {Fernique}, P. and {Fienga}, A. and {Figueras}, F. and {Fournier}, Y. and {Fouron}, C. and {Fragkoudi}, F. and {Gai}, M. and {Garcia-Gutierrez}, A. and {Garcia-Reinaldos}, M. and {Garc{\'\i}a-Torres}, M. and {Garofalo}, A. and {Gavel}, A. and {Gavras}, P. and {Gerlach}, E. and {Geyer}, R. and {Giacobbe}, P. and {Gilmore}, G. and {Girona}, S. and {Giuffrida}, G. and {Gomel}, R. and {Gomez}, A. and {Gonz{\'a}lez-N{\'u}{\~n}ez}, J. and {Gonz{\'a}lez-Santamar{\'\i}a}, I. and {Gonz{\'a}lez-Vidal}, J.~J. and {Granvik}, M. and {Guillout}, P. and {Guiraud}, J. and {Guti{\'e}rrez-S{\'a}nchez}, R. and {Guy}, L.~P. and {Hatzidimitriou}, D. and {Hauser}, M. and {Haywood}, M. and {Helmer}, A. and {Helmi}, A. and {Sarmiento}, M.~H. and {Hidalgo}, S.~L. and {Hilger}, T. and {H{\l}adczuk}, N. and {Hobbs}, D. and {Holland}, G. and {Huckle}, H.~E. and {Jardine}, K. and {Jasniewicz}, G. and {Jean-Antoine Piccolo}, A. and {Jim{\'e}nez-Arranz}, {\'O}. and {Jorissen}, A. and {Juaristi Campillo}, J. and {Julbe}, F. and {Karbevska}, L. and {Kervella}, P. and {Khanna}, S. and {Kontizas}, M. and {Kordopatis}, G. and {Korn}, A.~J. and {K{\'o}sp{\'a}l}, {\'A}. and {Kostrzewa-Rutkowska}, Z. and {Kruszy{\'n}ska}, K. and {Kun}, M. and {Laizeau}, P. and {Lambert}, S. and {Lanza}, A.~F. and {Lasne}, Y. and {Le Campion}, J. -F. and {Lebreton}, Y. and {Lebzelter}, T. and {Leccia}, S. and {Leclerc}, N. and {Lecoeur-Taibi}, I. and {Liao}, S. and {Licata}, E.~L. and {Lindstr{\o}m}, H.~E.~P. and {Lister}, T.~A. and {Livanou}, E. and {Lobel}, A. and {Lorca}, A. and {Loup}, C. and {Madrero Pardo}, P. and {Magdaleno Romeo}, A. and {Managau}, S. and {Mann}, R.~G. and {Manteiga}, M. and {Marchant}, J.~M. and {Marconi}, M. and {Marcos}, J. and {Marcos Santos}, M.~M.~S. and {Mar{\'\i}n Pina}, D. and {Marinoni}, S. and {Marocco}, F. and {Marshall}, D.~J. and {Martin Polo}, L. and {Mart{\'\i}n-Fleitas}, J.~M. and {Marton}, G. and {Mary}, N. and {Masip}, A. and {Massari}, D. and {Mastrobuono-Battisti}, A. and {Mazeh}, T. and {McMillan}, P.~J. and {Messina}, S. and {Michalik}, D. and {Millar}, N.~R. and {Mints}, A. and {Molina}, D. and {Molinaro}, R. and {Moln{\'a}r}, L. and {Monari}, G. and {Mongui{\'o}}, M. and {Montegriffo}, P. and {Montero}, A. and {Mor}, R. and {Mora}, A. and {Morbidelli}, R. and {Morel}, T. and {Morris}, D. and {Muraveva}, T. and {Murphy}, C.~P. and {Musella}, I. and {Nagy}, Z. and {Noval}, L. and {Oca{\~n}a}, F. and {Ogden}, A. and {Ordenovic}, C. and {Osinde}, J.~O. and {Pagani}, C. and {Pagano}, I. and {Palaversa}, L. and {Palicio}, P.~A. and {Pallas-Quintela}, L. and {Panahi}, A. and {Payne-Wardenaar}, S. and {Pe{\~n}alosa Esteller}, X. and {Penttil{\"a}}, A. and {Pichon}, B. and {Piersimoni}, A.~M. and {Pineau}, F. -X. and {Plachy}, E. and {Plum}, G. and {Poggio}, E. and {Pr{\v{s}}a}, A. and {Pulone}, L. and {Racero}, E. and {Ragaini}, S. and {Rainer}, M. and {Raiteri}, C.~M. and {Rambaux}, N. and {Ramos}, P. and {Ramos-Lerate}, M. and {Re Fiorentin}, P. and {Regibo}, S. and {Richards}, P.~J. and {Rios Diaz}, C. and {Ripepi}, V. and {Riva}, A. and {Rix}, H. -W. and {Rixon}, G. and {Robichon}, N. and {Robin}, A.~C. and {Robin}, C. and {Roelens}, M. and {Rogues}, H.~R.~O. and {Rohrbasser}, L. and {Romero-G{\'o}mez}, M. and {Rowell}, N. and {Royer}, F. and {Ruz Mieres}, D. and {Rybicki}, K.~A. and {Sadowski}, G. and {S{\'a}ez N{\'u}{\~n}ez}, A. and {Sagrist{\`a} Sell{\'e}s}, A. and {Sahlmann}, J. and {Salguero}, E. and {Samaras}, N. and {Sanchez Gimenez}, V. and {Sanna}, N. and {Santove{\~n}a}, R. and {Sarasso}, M. and {Schultheis}, M. and {Sciacca}, E. and {Segol}, M. and {Segovia}, J.~C. and {S{\'e}gransan}, D. and {Semeux}, D. and {Shahaf}, S. and {Siddiqui}, H.~I. and {Siebert}, A. and {Siltala}, L. and {Silvelo}, A. and {Slezak}, E. and {Slezak}, I. and {Smart}, R.~L. and {Snaith}, O.~N. and {Solano}, E. and {Solitro}, F. and {Souami}, D. and {Souchay}, J. and {Spagna}, A. and {Spina}, L. and {Spoto}, F. and {Steele}, I.~A. and {Steidelm{\"u}ller}, H. and {Stephenson}, C.~A. and {S{\"u}veges}, M. and {Surdej}, J. and {Szabados}, L. and {Szegedi-Elek}, E. and {Taris}, F. and {Taylor}, M.~B. and {Teixeira}, R. and {Tolomei}, L. and {Tonello}, N. and {Torra}, F. and {Torra}, J. and {Torralba Elipe}, G. and {Trabucchi}, M. and {Tsounis}, A.~T. and {Turon}, C. and {Ulla}, A. and {Unger}, N. and {Vaillant}, M.~V. and {van Dillen}, E. and {van Reeven}, W. and {Vanel}, O. and {Vecchiato}, A. and {Viala}, Y. and {Vicente}, D. and {Voutsinas}, S. and {Weiler}, M. and {Wevers}, T. and {Wyrzykowski}, {\L}. and {Yoldas}, A. and {Yvard}, P. and {Zhao}, H. and {Zorec}, J. and {Zucker}, S. and {Zwitter}, T.},
        title = "{Gaia Data Release 3. Summary of the content and survey properties}",
      journal = {\aap},
         year = 2023,
        month = jun,
       volume = {674},
          eid = {A1},
        pages = {A1},
          doi = {10.1051/0004-6361/202243940},
archivePrefix = {arXiv},
       eprint = {2208.00211},
 primaryClass = {astro-ph.GA},
       adsurl = {https://ui.adsabs.harvard.edu/abs/2023A&A...674A...1G}
}

@BOOK{Kippenhahn1990,
       author = {{Kippenhahn}, Rudolf and {Weigert}, Alfred},
        title = "{Stellar Structure and Evolution}",
         year = 1990,
       adsurl = {https://ui.adsabs.harvard.edu/abs/1990sse..book.....K}
}

@ARTICLE{Goldberg2024,
       author = {{Goldberg}, Jared A. and {Joyce}, Meridith and {Moln{\'a}r}, L{\'a}szl{\'o}},
        title = "{A Buddy for Betelgeuse: Binarity as the Origin of the Long Secondary Period in {\ensuremath{\alpha}} Orionis}",
      journal = {\apj},
         year = 2024,
        month = dec,
       volume = {977},
       number = {1},
          eid = {35},
        pages = {35},
          doi = {10.3847/1538-4357/ad87f4},
archivePrefix = {arXiv},
       eprint = {2408.09089},
 primaryClass = {astro-ph.SR},
       adsurl = {https://ui.adsabs.harvard.edu/abs/2024ApJ...977...35G}
}

@ARTICLE{Molnar2023,
       author = {{Moln{\'a}r}, L{\'a}szl{\'o} and {Joyce}, Meridith and {Leung}, Shing-Chi},
        title = "{Comment on the Feasibility of Carbon Burning in Betelgeuse}",
      journal = {Research Notes of the American Astronomical Society},
         year = 2023,
        month = jun,
       volume = {7},
       number = {6},
          eid = {119},
        pages = {119},
          doi = {10.3847/2515-5172/acdb7a},
archivePrefix = {arXiv},
       eprint = {2306.05600},
 primaryClass = {astro-ph.SR},
       adsurl = {https://ui.adsabs.harvard.edu/abs/2023RNAAS...7..119M}
}

@ARTICLE{Moe2017,
       author = {{Moe}, Maxwell and {Di Stefano}, Rosanne},
        title = "{Mind Your Ps and Qs: The Interrelation between Period (P) and Mass-ratio (Q) Distributions of Binary Stars}",
      journal = {\apjs},
         year = 2017,
        month = jun,
       volume = {230},
       number = {2},
          eid = {15},
        pages = {15},
          doi = {10.3847/1538-4365/aa6fb6},
archivePrefix = {arXiv},
       eprint = {1606.05347},
 primaryClass = {astro-ph.SR},
       adsurl = {https://ui.adsabs.harvard.edu/abs/2017ApJS..230...15M}
}

@ARTICLE{deMink2014,
       author = {{de Mink}, S.~E. and {Sana}, H. and {Langer}, N. and {Izzard}, R.~G. and {Schneider}, F.~R.~N.},
        title = "{The Incidence of Stellar Mergers and Mass Gainers among Massive Stars}",
      journal = {\apj},
         year = 2014,
        month = feb,
       volume = {782},
       number = {1},
          eid = {7},
        pages = {7},
          doi = {10.1088/0004-637X/782/1/7},
archivePrefix = {arXiv},
       eprint = {1312.3650},
 primaryClass = {astro-ph.SR},
       adsurl = {https://ui.adsabs.harvard.edu/abs/2014ApJ...782....7D}
}

@ARTICLE{sana2012,
       author = {{Sana}, H. and {de Mink}, S.~E. and {de Koter}, A. and {Langer}, N. and {Evans}, C.~J. and {Gieles}, M. and {Gosset}, E. and {Izzard}, R.~G. and {Le Bouquin}, J.-B. and {Schneider}, F.~R.~N.},
        title = "{Binary Interaction Dominates the Evolution of Massive Stars}",
      journal = {Science},
         year = 2012,
        month = jul,
       volume = {337},
       number = {6093},
        pages = {444},
          doi = {10.1126/science.1223344},
archivePrefix = {arXiv},
       eprint = {1207.6397},
 primaryClass = {astro-ph.SR},
       adsurl = {https://ui.adsabs.harvard.edu/abs/2012Sci...337..444S}
}

@ARTICLE{Joyce2020,
       author = {{Joyce}, Meridith and {Leung}, Shing-Chi and {Moln{\'a}r}, L{\'a}szl{\'o} and {Ireland}, Michael and {Kobayashi}, Chiaki and {Nomoto}, Ken'ichi},
        title = "{Standing on the Shoulders of Giants: New Mass and Distance Estimates for Betelgeuse through Combined Evolutionary, Asteroseismic, and Hydrodynamic Simulations with MESA}",
      journal = {\apj},
         year = 2020,
        month = oct,
       volume = {902},
       number = {1},
          eid = {63},
        pages = {63},
          doi = {10.3847/1538-4357/abb8db},
archivePrefix = {arXiv},
       eprint = {2006.09837},
 primaryClass = {astro-ph.SR},
       adsurl = {https://ui.adsabs.harvard.edu/abs/2020ApJ...902...63J}
}

@ARTICLE{Mahy2022,
       author = {{Mahy}, L. and {Lanthermann}, C. and {Hutsem{\'e}kers}, D. and {Kluska}, J. and {Lobel}, A. and {Manick}, R. and {Miszalski}, B. and {Reggiani}, M. and {Sana}, H. and {Gosset}, E.},
        title = "{Multiplicity of Galactic luminous blue variable stars}",
      journal = {\aap},
         year = 2022,
        month = jan,
       volume = {657},
          eid = {A4},
        pages = {A4},
          doi = {10.1051/0004-6361/202040062},
archivePrefix = {arXiv},
       eprint = {2105.12380},
 primaryClass = {astro-ph.SR},
       adsurl = {https://ui.adsabs.harvard.edu/abs/2022A&A...657A...4M}
}

@ARTICLE{woosley2021,
       author = {{Woosley}, S.~E. and {Heger}, Alexander},
        title = "{The Pair-instability Mass Gap for Black Holes}",
      journal = {\apjl},
         year = 2021,
        month = may,
       volume = {912},
       number = {2},
          eid = {L31},
        pages = {L31},
          doi = {10.3847/2041-8213/abf2c4},
archivePrefix = {arXiv},
       eprint = {2103.07933},
 primaryClass = {astro-ph.SR},
       adsurl = {https://ui.adsabs.harvard.edu/abs/2021ApJ...912L..31W}
}

@ARTICLE{Nomoto1980,
       author = {{Nomoto}, K.},
        title = "{White dwarf models for type I supernovae and quiet supernovae, and presupernova evolution}",
      journal = {\ssr},
         year = 1980,
        month = nov,
       volume = {27},
       number = {3-4},
        pages = {563-570},
          doi = {10.1007/BF00168350},
       adsurl = {https://ui.adsabs.harvard.edu/abs/1980SSRv...27..563N}
}

@ARTICLE{Tauris2015,
       author = {{Tauris}, Thomas M. and {Langer}, Norbert and {Podsiadlowski}, Philipp},
        title = "{Ultra-stripped supernovae: progenitors and fate}",
      journal = {\mnras},
         year = 2015,
        month = aug,
       volume = {451},
       number = {2},
        pages = {2123-2144},
          doi = {10.1093/mnras/stv990},
archivePrefix = {arXiv},
       eprint = {1505.00270},
 primaryClass = {astro-ph.SR},
       adsurl = {https://ui.adsabs.harvard.edu/abs/2015MNRAS.451.2123T}
}

@ARTICLE{Timmes1996,
       author = {{Timmes}, F.~X. and {Woosley}, S.~E. and {Weaver}, Thomas A.},
        title = "{The Neutron Star and Black Hole Initial Mass Function}",
      journal = {\apj},
         year = 1996,
        month = feb,
       volume = {457},
        pages = {834},
          doi = {10.1086/176778},
archivePrefix = {arXiv},
       eprint = {astro-ph/9510136},
 primaryClass = {astro-ph},
       adsurl = {https://ui.adsabs.harvard.edu/abs/1996ApJ...457..834T}
}

@ARTICLE{Astroquery,
       author = {{Ginsburg}, Adam and {Sip{\H{o}}cz}, Brigitta M. and {Brasseur}, C.~E. and {Cowperthwaite}, Philip S. and {Craig}, Matthew W. and {Deil}, Christoph and {Guillochon}, James and {Guzman}, Giannina and {Liedtke}, Simon and {Lian Lim}, Pey and {Lockhart}, Kelly E. and {Mommert}, Michael and {Morris}, Brett M. and {Norman}, Henrik and {Parikh}, Madhura and {Persson}, Magnus V. and {Robitaille}, Thomas P. and {Segovia}, Juan-Carlos and {Singer}, Leo P. and {Tollerud}, Erik J. and {de Val-Borro}, Miguel and {Valtchanov}, Ivan and {Woillez}, Julien and {Astroquery Collaboration} and {a subset of astropy Collaboration}},
        title = "{astroquery: An Astronomical Web-querying Package in Python}",
      journal = {\aj},
         year = 2019,
        month = mar,
       volume = {157},
       number = {3},
          eid = {98},
        pages = {98},
          doi = {10.3847/1538-3881/aafc33},
archivePrefix = {arXiv},
       eprint = {1901.04520},
 primaryClass = {astro-ph.IM},
       adsurl = {https://ui.adsabs.harvard.edu/abs/2019AJ....157...98G}
}

@ARTICLE{BailerEDR3,
       author = {{Bailer-Jones}, C.~A.~L. and {Rybizki}, J. and {Fouesneau}, M. and {Demleitner}, M. and {Andrae}, R.},
        title = "{Estimating Distances from Parallaxes. V. Geometric and Photogeometric Distances to 1.47 Billion Stars in Gaia Early Data Release 3}",
      journal = {\aj},
         year = 2021,
        month = mar,
       volume = {161},
       number = {3},
          eid = {147},
        pages = {147},
          doi = {10.3847/1538-3881/abd806},
archivePrefix = {arXiv},
       eprint = {2012.05220},
 primaryClass = {astro-ph.SR},
       adsurl = {https://ui.adsabs.harvard.edu/abs/2021AJ....161..147B}
}

@ARTICLE{Maiz2021,
       author = {{Ma{\'\i}z Apell{\'a}niz}, J. and {Pantaleoni Gonz{\'a}lez}, M. and {Barb{\'a}}, R.~H.},
        title = "{Validation of the accuracy and precision of Gaia EDR3 parallaxes with globular clusters}",
      journal = {\aap},
         year = 2021,
        month = may,
       volume = {649},
          eid = {A13},
        pages = {A13},
          doi = {10.1051/0004-6361/202140418},
archivePrefix = {arXiv},
       eprint = {2101.10206},
 primaryClass = {astro-ph.IM},
       adsurl = {https://ui.adsabs.harvard.edu/abs/2021A&A...649A..13M}
}

@ARTICLE{Maiz2022,
       author = {{Ma{\'\i}z Apell{\'a}niz}, J.},
        title = "{An estimation of the Gaia EDR3 parallax bias from stellar clusters and Magellanic Clouds data}",
      journal = {\aap},
         year = 2022,
        month = jan,
       volume = {657},
          eid = {A130},
        pages = {A130},
          doi = {10.1051/0004-6361/202142365},
archivePrefix = {arXiv},
       eprint = {2110.01475},
 primaryClass = {astro-ph.IM},
       adsurl = {https://ui.adsabs.harvard.edu/abs/2022A&A...657A.130M}
}

@ARTICLE{Quintana2025,
       author = {{Quintana}, Alexis L. and {Wright}, Nicholas J. and {Mart{\'\i}nez Garc{\'\i}a}, Juan},
        title = "{A census of OB stars within 1 kpc and the star formation and core collapse supernova rates of the Milky Way}",
      journal = {\mnras},
         year = 2025,
        month = apr,
       volume = {538},
       number = {3},
        pages = {1367-1383},
          doi = {10.1093/mnras/staf083},
archivePrefix = {arXiv},
       eprint = {2503.08286},
 primaryClass = {astro-ph.SR},
       adsurl = {https://ui.adsabs.harvard.edu/abs/2025MNRAS.538.1367Q}
}

@ARTICLE{Maiz2018,
       author = {{Ma{\'\i}z Apell{\'a}niz}, J. and {Weiler}, M.},
        title = "{Reanalysis of the Gaia Data Release 2 photometric sensitivity curves using HST/STIS spectrophotometry}",
      journal = {\aap},
         year = 2018,
        month = nov,
       volume = {619},
          eid = {A180},
        pages = {A180},
          doi = {10.1051/0004-6361/201834051},
archivePrefix = {arXiv},
       eprint = {1808.02820},
 primaryClass = {astro-ph.IM},
       adsurl = {https://ui.adsabs.harvard.edu/abs/2018A&A...619A.180M}
}

@ARTICLE{Maiz2023,
       author = {{Ma{\'\i}z Apell{\'a}niz}, J. and {Holgado}, G. and {Pantaleoni Gonz{\'a}lez}, M. and {Caballero}, J.~A.},
        title = "{Stellar variability in Gaia DR3. I. Three-band photometric dispersions for 145 million sources}",
      journal = {\aap},
         year = 2023,
        month = sep,
       volume = {677},
          eid = {A137},
        pages = {A137},
          doi = {10.1051/0004-6361/202346759},
archivePrefix = {arXiv},
       eprint = {2304.14249},
 primaryClass = {astro-ph.SR},
       adsurl = {https://ui.adsabs.harvard.edu/abs/2023A&A...677A.137M}
}

@ARTICLE{Riello,
       author = {{Riello}, M. and {De Angeli}, F. and {Evans}, D.~W. and {Montegriffo}, P. and {Carrasco}, J.~M. and {Busso}, G. and {Palaversa}, L. and {Burgess}, P.~W. and {Diener}, C. and {Davidson}, M. and {Rowell}, N. and {Fabricius}, C. and {Jordi}, C. and {Bellazzini}, M. and {Pancino}, E. and {Harrison}, D.~L. and {Cacciari}, C. and {van Leeuwen}, F. and {Hambly}, N.~C. and {Hodgkin}, S.~T. and {Osborne}, P.~J. and {Altavilla}, G. and {Barstow}, M.~A. and {Brown}, A.~G.~A. and {Castellani}, M. and {Cowell}, S. and {De Luise}, F. and {Gilmore}, G. and {Giuffrida}, G. and {Hidalgo}, S. and {Holland}, G. and {Marinoni}, S. and {Pagani}, C. and {Piersimoni}, A.~M. and {Pulone}, L. and {Ragaini}, S. and {Rainer}, M. and {Richards}, P.~J. and {Sanna}, N. and {Walton}, N.~A. and {Weiler}, M. and {Yoldas}, A.},
        title = "{Gaia Early Data Release 3. Photometric content and validation}",
      journal = {\aap},
         year = 2021,
        month = may,
       volume = {649},
          eid = {A3},
        pages = {A3},
          doi = {10.1051/0004-6361/202039587},
archivePrefix = {arXiv},
       eprint = {2012.01916},
 primaryClass = {astro-ph.IM},
       adsurl = {https://ui.adsabs.harvard.edu/abs/2021A&A...649A...3R}
}

@INPROCEEDINGS{Topcat,
       author = {{Taylor}, M.~B.},
        title = "{TOPCAT \& STIL: Starlink Table/VOTable Processing Software}",
    booktitle = {Astronomical Data Analysis Software and Systems XIV},
         year = 2005,
       editor = {{Shopbell}, P. and {Britton}, M. and {Ebert}, R.},
       series = {Astronomical Society of the Pacific Conference Series},
       volume = {347},
        month = dec,
        pages = {29},
       adsurl = {https://ui.adsabs.harvard.edu/abs/2005ASPC..347...29T}
}

@ARTICLE{Astropy,
       author = {{Astropy Collaboration} and {Robitaille}, Thomas P. and {Tollerud}, Erik J. and {Greenfield}, Perry and {Droettboom}, Michael and {Bray}, Erik and {Aldcroft}, Tom and {Davis}, Matt and {Ginsburg}, Adam and {Price-Whelan}, Adrian M. and {Kerzendorf}, Wolfgang E. and {Conley}, Alexander and {Crighton}, Neil and {Barbary}, Kyle and {Muna}, Demitri and {Ferguson}, Henry and {Grollier}, Fr{\'e}d{\'e}ric and {Parikh}, Madhura M. and {Nair}, Prasanth H. and {Unther}, Hans M. and {Deil}, Christoph and {Woillez}, Julien and {Conseil}, Simon and {Kramer}, Roban and {Turner}, James E.~H. and {Singer}, Leo and {Fox}, Ryan and {Weaver}, Benjamin A. and {Zabalza}, Victor and {Edwards}, Zachary I. and {Azalee Bostroem}, K. and {Burke}, D.~J. and {Casey}, Andrew R. and {Crawford}, Steven M. and {Dencheva}, Nadia and {Ely}, Justin and {Jenness}, Tim and {Labrie}, Kathleen and {Lim}, Pey Lian and {Pierfederici}, Francesco and {Pontzen}, Andrew and {Ptak}, Andy and {Refsdal}, Brian and {Servillat}, Mathieu and {Streicher}, Ole},
        title = "{Astropy: A community Python package for astronomy}",
      journal = {\aap},
         year = 2013,
        month = oct,
       volume = {558},
          eid = {A33},
        pages = {A33},
          doi = {10.1051/0004-6361/201322068},
archivePrefix = {arXiv},
       eprint = {1307.6212},
 primaryClass = {astro-ph.IM},
       adsurl = {https://ui.adsabs.harvard.edu/abs/2013A&A...558A..33A}
}

@ARTICLE{Edenhofer2024,
       author = {{Edenhofer}, Gordian and {Zucker}, Catherine and {Frank}, Philipp and {Saydjari}, Andrew K. and {Speagle}, Joshua S. and {Finkbeiner}, Douglas and {En{\ss}lin}, Torsten A.},
        title = "{A parsec-scale Galactic 3D dust map out to 1.25 kpc from the Sun}",
      journal = {\aap},
         year = 2024,
        month = may,
       volume = {685},
          eid = {A82},
        pages = {A82},
          doi = {10.1051/0004-6361/202347628},
archivePrefix = {arXiv},
       eprint = {2308.01295},
 primaryClass = {astro-ph.GA},
       adsurl = {https://ui.adsabs.harvard.edu/abs/2024A&A...685A..82E}
}

@ARTICLE{Poggio2026,
       author = {{Poggio}, E. and {Drimmel}, R. and {Khanna}, S. and {Andrae}, R. and {Lattanzi}, M.~G.},
        title = "{The Flare and Spiral Structure of the Milky Way's Disk as Traced by Young Giant Stars}",
      journal = {\apj},
         year = 2026,
        month = oct,
       volume = {1010},
       number = {2},
          eid = {158},
        pages = {158},
          doi = {10.3847/1538-4357/aea55f},
       adsurl = {https://ui.adsabs.harvard.edu/abs/2026ApJ..1010..158P}
}

@ARTICLE{Fouesneau2023,
       author = {{Fouesneau}, M. and {Fr{\'e}mat}, Y. and {Andrae}, R. and {Korn}, A.~J. and {Soubiran}, C. and {Kordopatis}, G. and {Vallenari}, A. and {Heiter}, U. and {Creevey}, O.~L. and {Sarro}, L.~M. and {de Laverny}, P. and {Lanzafame}, A.~C. and {Lobel}, A. and {Sordo}, R. and {Rybizki}, J. and {Slezak}, I. and {{\'A}lvarez}, M.~A. and {Drimmel}, R. and {Garabato}, D. and {Delchambre}, L. and {Bailer-Jones}, C.~A.~L. and {Hatzidimitriou}, D. and {Lorca}, A. and {Le Fustec}, Y. and {Pailler}, F. and {Mary}, N. and {Robin}, C. and {Utrilla}, E. and {Abreu Aramburu}, A. and {Bakker}, J. and {Bellas-Velidis}, I. and {Bijaoui}, A. and {Blomme}, R. and {Bouret}, J. -C. and {Brouillet}, N. and {Brugaletta}, E. and {Burlacu}, A. and {Carballo}, R. and {Casamiquela}, L. and {Chaoul}, L. and {Chiavassa}, A. and {Contursi}, G. and {Cooper}, W.~J. and {Dafonte}, C. and {Demouchy}, C. and {Dharmawardena}, T.~E. and {Garc{\'\i}a-Lario}, P. and {Garc{\'\i}a-Torres}, M. and {Gomez}, A. and {Gonz{\'a}lez-Santamar{\'\i}a}, I. and {Jean-Antoine Piccolo}, A. and {Kontizas}, M. and {Lebreton}, Y. and {Licata}, E.~L. and {Lindstr{\o}m}, H.~E.~P. and {Livanou}, E. and {Magdaleno Romeo}, A. and {Manteiga}, M. and {Marocco}, F. and {Martayan}, C. and {Marshall}, D.~J. and {Nicolas}, C. and {Ordenovic}, C. and {Palicio}, P.~A. and {Pallas-Quintela}, L. and {Pichon}, B. and {Poggio}, E. and {Recio-Blanco}, A. and {Riclet}, F. and {Santove{\~n}a}, R. and {Schultheis}, M.~S. and {Segol}, M. and {Silvelo}, A. and {Smart}, R.~L. and {S{\"u}veges}, M. and {Th{\'e}venin}, F. and {Torralba Elipe}, G. and {Ulla}, A. and {van Dillen}, E. and {Zhao}, H. and {Zorec}, J.},
        title = "{Gaia Data Release 3. Apsis. II. Stellar parameters}",
      journal = {\aap},
         year = 2023,
        month = jun,
       volume = {674},
          eid = {A28},
        pages = {A28},
          doi = {10.1051/0004-6361/202243919},
archivePrefix = {arXiv},
       eprint = {2206.05992},
 primaryClass = {astro-ph.SR},
       adsurl = {https://ui.adsabs.harvard.edu/abs/2023A&A...674A..28F}
}

@ARTICLE{Mamajek,
       author = {{Pecaut}, Mark J. and {Mamajek}, Eric E.},
        title = "{Intrinsic Colors, Temperatures, and Bolometric Corrections of Pre-main-sequence Stars}",
      journal = {\apjs},
         year = "2013",
        month = "Sep",
       volume = {208},
       number = {1},
          eid = {9},
        pages = {9},
          doi = {10.1088/0067-0049/208/1/9},
archivePrefix = {arXiv},
       eprint = {1307.2657},
 primaryClass = {astro-ph.SR},
       adsurl = {https://ui.adsabs.harvard.edu/abs/2013ApJS..208....9P}
}

@ARTICLE{WangChen2019,
       author = {{Wang}, Shu and {Chen}, Xiaodian},
        title = "{The Optical to Mid-infrared Extinction Law Based on the APOGEE, Gaia DR2, Pan-STARRS1, SDSS, APASS, 2MASS, and WISE Surveys}",
      journal = {\apj},
         year = 2019,
        month = jun,
       volume = {877},
       number = {2},
          eid = {116},
        pages = {116},
          doi = {10.3847/1538-4357/ab1c61},
archivePrefix = {arXiv},
       eprint = {1904.04575},
 primaryClass = {astro-ph.GA},
       adsurl = {https://ui.adsabs.harvard.edu/abs/2019ApJ...877..116W}
}

@ARTICLE{Emcee,
       author = {{Foreman-Mackey}, Daniel and {Hogg}, David W. and {Lang}, Dustin and
         {Goodman}, Jonathan},
        title = "{emcee: The MCMC Hammer}",
      journal = {\pasp},
         year = "2013",
        month = "Mar",
       volume = {125},
       number = {925},
        pages = {306},
          doi = {10.1086/670067},
archivePrefix = {arXiv},
       eprint = {1202.3665},
 primaryClass = {astro-ph.IM},
       adsurl = {https://ui.adsabs.harvard.edu/abs/2013PASP..125..306F}
}

@ARTICLE{WEAVE,
       author = {{Jin}, Shoko and {Trager}, Scott C. and {Dalton}, Gavin B. and {Aguerri}, J. Alfonso L. and {Drew}, J.~E. and {Falc{\'o}n-Barroso}, Jes{\'u}s and {G{\"a}nsicke}, Boris T. and {Hill}, Vanessa and {Iovino}, Angela and {Pieri}, Matthew M. and {Poggianti}, Bianca M. and {Smith}, D.~J.~B. and {Vallenari}, Antonella and {Abrams}, Don Carlos and {Aguado}, David S. and {Antoja}, Teresa and {Arag{\'o}n-Salamanca}, Alfonso and {Ascasibar}, Yago and {Babusiaux}, Carine and {Balcells}, Marc and {Barrena}, R. and {Battaglia}, Giuseppina and {Belokurov}, Vasily and {Bensby}, Thomas and {Bonifacio}, Piercarlo and {Bragaglia}, Angela and {Carrasco}, Esperanza and {Carrera}, Ricardo and {Cornwell}, Daniel J. and {Dom{\'\i}nguez-Palmero}, Lilian and {Duncan}, Kenneth J. and {Famaey}, Benoit and {Fari{\~n}a}, Cecilia and {Gonzalez}, Oscar A. and {Guest}, Steve and {Hatch}, Nina A. and {Hess}, Kelley M. and {Hoskin}, Matthew J. and {Irwin}, Mike and {Knapen}, Johan H. and {Koposov}, Sergey E. and {Kuchner}, Ulrike and {Laigle}, Clotilde and {Lewis}, Jim and {Longhetti}, Marcella and {Lucatello}, Sara and {M{\'e}ndez-Abreu}, Jairo and {Mercurio}, Amata and {Molaeinezhad}, Alireza and {Mongui{\'o}}, Maria and {Morrison}, Sean and {Murphy}, David N.~A. and {Peralta de Arriba}, Luis and {P{\'e}rez}, Isabel and {P{\'e}rez-R{\`a}fols}, Ignasi and {Pic{\'o}}, Sergio and {Raddi}, Roberto and {Romero-G{\'o}mez}, Merc{\`e} and {Royer}, Fr{\'e}d{\'e}ric and {Siebert}, Arnaud and {Seabroke}, George M. and {Som}, Debopam and {Terrett}, David and {Thomas}, Guillaume and {Wesson}, Roger and {Worley}, C. Clare and {Alfaro}, Emilio J. and {Allende Prieto}, Carlos and {Alonso-Santiago}, Javier and {Amos}, Nicholas J. and {Ashley}, Richard P. and {Balaguer-N{\'u}{\~n}ez}, Lola and {Balbinot}, Eduardo and {Bellazzini}, Michele and {Benn}, Chris R. and {Berlanas}, Sara R. and {Bernard}, Edouard J. and {Best}, Philip and {Bettoni}, Daniela and {Bianco}, Andrea and {Bishop}, Georgia and {Blomqvist}, Michael and {Boeche}, Corrado and {Bolzonella}, Micol and {Bonoli}, Silvia and {Bosma}, Albert and {Britavskiy}, Nikolay and {Busarello}, Gianni and {Caffau}, Elisabetta and {Cantat-Gaudin}, Tristan and {Castro-Ginard}, Alfred and {Couto}, Guilherme and {Carbajo-Hijarrubia}, Juan and {Carter}, David and {Casamiquela}, Laia and {Conrado}, Ana M. and {Corcho-Caballero}, Pablo and {Costantin}, Luca and {Deason}, Alis and {de Burgos}, Abel and {De Grandi}, Sabrina and {Di Matteo}, Paola and {Dom{\'\i}nguez-G{\'o}mez}, Jes{\'u}s and {Dorda}, Ricardo and {Drake}, Alyssa and {Dutta}, Rajeshwari and {Erkal}, Denis and {Feltzing}, Sofia and {Ferr{\'e}-Mateu}, Anna and {Feuillet}, Diane and {Figueras}, Francesca and {Fossati}, Matteo and {Franciosini}, Elena and {Frasca}, Antonio and {Fumagalli}, Michele and {Gallazzi}, Anna and {Garc{\'\i}a-Benito}, Rub{\'e}n and {Gentile Fusillo}, Nicola and {Gebran}, Marwan and {Gilbert}, James and {Gledhill}, T.~M. and {Gonz{\'a}lez Delgado}, Rosa M. and {Greimel}, Robert and {Guarcello}, Mario Giuseppe and {Guerra}, Jose and {Gullieuszik}, Marco and {Haines}, Christopher P. and {Hardcastle}, Martin J. and {Harris}, Amy and {Haywood}, Misha and {Helmi}, Amina and {Hernandez}, Nauzet and {Herrero}, Artemio and {Hughes}, Sarah and {Ir{\v{s}}i{\v{c}}}, Vid and {Jablonka}, Pascale and {Jarvis}, Matt J. and {Jordi}, Carme and {Kondapally}, Rohit and {Kordopatis}, Georges and {Krogager}, Jens-Kristian and {La Barbera}, Francesco and {Lam}, Man I. and {Larsen}, S{\o}ren S. and {Lemasle}, Bertrand and {Lewis}, Ian J. and {Lhom{\'e}}, Emilie and {Lind}, Karin and {Lodi}, Marcello and {Longobardi}, Alessia and {Lonoce}, Ilaria and {Magrini}, Laura and {Ma{\'\i}z Apell{\'a}niz}, Jes{\'u}s and {Marchal}, Olivier and {Marco}, Amparo and {Martin}, Nicolas F. and {Matsuno}, Tadafumi and {Maurogordato}, Sophie and {Merluzzi}, Paola and {Miralda-Escud{\'e}}, Jordi and {Molinari}, Emilio and {Monari}, Giacomo and {Morelli}, Lorenzo and {Mottram}, Christopher J. and {Naylor}, Tim and {Negueruela}, Ignacio and {O{\~n}orbe}, Jose and {Pancino}, Elena and {Peirani}, S{\'e}bastien and {Peletier}, Reynier F. and {Pozzetti}, Lucia and {Rainer}, Monica and {Ramos}, Pau and {Read}, Shaun C. and {Rossi}, Elena Maria and {R{\"o}ttgering}, Huub J.~A. and {Rubi{\~n}o-Mart{\'\i}n}, Jose Alberto and {Sabater}, Jose and {San Juan}, Jos{\'e} and {Sanna}, Nicoletta and {Schallig}, Ellen and {Schiavon}, Ricardo P. and {Schultheis}, Mathias and {Serra}, Paolo and {Shimwell}, Timothy W. and {Sim{\'o}n-D{\'\i}az}, Sergio and {Smith}, Russell J. and {Sordo}, Rosanna and {Sorini}, Daniele and {Soubiran}, Caroline and {Starkenburg}, Else and {Steele}, Iain A. and {Stott}, John and {Stuik}, Remko and {Tolstoy}, Eline and {Tortora}, Crescenzo and {Tsantaki}, Maria and {Van der Swaelmen}, Mathieu and {van Weeren}, Reinout J. and {Vergani}, Daniela and {Verheijen}, Marc A.~W. and {Verro}, Kristiina and {Vink}, Jorick S. and {Vioque}, Miguel and {Walcher}, C. Jakob and {Walton}, Nicholas A. and {Wegg}, Christopher and {Weijmans}, Anne-Marie and {Williams}, Wendy L. and {Wilson}, Andrew J. and {Wright}, Nicholas J. and {Xylakis-Dornbusch}, Theodora and {Youakim}, Kris and {Zibetti}, Stefano and {Zurita}, Cristina},
        title = "{The wide-field, multiplexed, spectroscopic facility WEAVE: Survey design, overview, and simulated implementation}",
      journal = {\mnras},
         year = 2024,
        month = may,
       volume = {530},
       number = {3},
        pages = {2688-2730},
          doi = {10.1093/mnras/stad557},
archivePrefix = {arXiv},
       eprint = {2212.03981},
 primaryClass = {astro-ph.IM},
       adsurl = {https://ui.adsabs.harvard.edu/abs/2024MNRAS.530.2688J}
}

@BOOK{2MASS,
       author = {{Cutri}, R.~M. and {Skrutskie}, M.~F. and {van Dyk}, S. and {Beichman}, C.~A. and {Carpenter}, J.~M. and {Chester}, T. and {Cambresy}, L. and {Evans}, T. and {Fowler}, J. and {Gizis}, J. and {Howard}, E. and {Huchra}, J. and {Jarrett}, T. and {Kopan}, E.~L. and {Kirkpatrick}, J.~D. and {Light}, R.~M. and {Marsh}, K.~A. and {McCallon}, H. and {Schneider}, S. and {Stiening}, R. and {Sykes}, M. and {Weinberg}, M. and {Wheaton}, W.~A. and {Wheelock}, S. and {Zacarias}, N.},
        title = "{2MASS All Sky Catalog of point sources.}",
         year = 2003,
         publisher = {NASA/IPAC Infrared Science Archive},
       adsurl = {https://ui.adsabs.harvard.edu/abs/2003tmc..book.....C}
}

@ARTICLE{Marrese2017,
       author = {{Marrese}, P.~M. and {Marinoni}, S. and {Fabrizio}, M. and {Giuffrida}, G.},
        title = "{Gaia Data Release 1. Cross-match with external catalogues. Algorithm and results}",
      journal = {\aap},
         year = 2017,
        month = nov,
       volume = {607},
          eid = {A105},
        pages = {A105},
          doi = {10.1051/0004-6361/201730965},
archivePrefix = {arXiv},
       eprint = {1710.06739},
 primaryClass = {astro-ph.SR},
       adsurl = {https://ui.adsabs.harvard.edu/abs/2017A&A...607A.105M}
}

@ARTICLE{Marrese2019,
       author = {{Marrese}, P.~M. and {Marinoni}, S. and {Fabrizio}, M. and {Altavilla}, G.},
        title = "{Gaia Data Release 2. Cross-match with external catalogues: algorithms and results}",
      journal = {\aap},
         year = 2019,
        month = jan,
       volume = {621},
          eid = {A144},
        pages = {A144},
          doi = {10.1051/0004-6361/201834142},
archivePrefix = {arXiv},
       eprint = {1808.09151},
 primaryClass = {astro-ph.SR},
       adsurl = {https://ui.adsabs.harvard.edu/abs/2019A&A...621A.144M}
}

@ARTICLE{Drew,
       author = {{Drew}, Janet E. and {Greimel}, R. and {Irwin}, M.~J. and
         {Aungwerojwit}, A. and {Barlow}, M.~J. and {Corradi}, R.~L.~M. and
         {Drake}, J.~J. and {G{\"a}nsicke}, B.~T. and {Groot}, P. and
         {Hales}, A. and {Hopewell}, E.~C. and {Irwin}, J. and {Knigge}, C. and
         {Leisy}, P. and {Lennon}, D.~J. and {Mampaso}, A. and
         {Masheder}, M.~R.~W. and {Matsuura}, M. and {Morales-Rueda}, L. and
         {Morris}, R.~A.~H. and {Parker}, Q.~A. and {Phillipps}, S. and
         {Rodriguez-Gil}, P. and {Roelofs}, G. and {Skillen}, I. and
         {Sokoloski}, J.~L. and {Steeghs}, D. and {Unruh}, Y.~C. and
         {Viironen}, K. and {Vink}, J.~S. and {Walton}, N.~A. and {Witham}, A. and
         {Wright}, N. and {Zijlstra}, A.~A. and {Zurita}, A.},
        title = "{The INT Photometric H{\ensuremath{\alpha}} Survey of the Northern Galactic Plane (IPHAS)}",
      journal = {\mnras},
         year = "2005",
        month = "Sep",
       volume = {362},
       number = {3},
        pages = {753-776},
          doi = {10.1111/j.1365-2966.2005.09330.x},
archivePrefix = {arXiv},
       eprint = {astro-ph/0506726},
 primaryClass = {astro-ph},
       adsurl = {https://ui.adsabs.harvard.edu/abs/2005MNRAS.362..753D}
}

@ARTICLE{Mongui,
       author = {{Mongui{\'o}}, M. and {Greimel}, R. and {Drew}, J.~E. and {Barentsen}, G. and {Groot}, P.~J. and {Irwin}, M.~J. and {Casares}, J. and {G{\"a}nsicke}, B.~T. and {Carter}, P.~J. and {Corral-Santana}, J.~M. and {Gentile-Fusillo}, N.~P. and {Greiss}, S. and {van Haaften}, L.~M. and {Hollands}, M. and {Jones}, D. and {Kupfer}, T. and {Manser}, C.~J. and {Murphy}, D.~N.~A. and {McLeod}, A.~F. and {Oosting}, T. and {Parker}, Q.~A. and {Pyrzas}, S. and {Rodr{\'\i}guez-Gil}, P. and {van Roestel}, J. and {Scaringi}, S. and {Schellart}, P. and {Toloza}, O. and {Vaduvescu}, O. and {van Spaandonk}, L. and {Verbeek}, K. and {Wright}, N.~J. and {Eisl{\"o}ffel}, J. and {Fabregat}, J. and {Harris}, A. and {Morris}, R.~A.~H. and {Phillipps}, S. and {Raddi}, R. and {Sabin}, L. and {Unruh}, Y. and {Vink}, J.~S. and {Wesson}, R. and {Cardwell}, A. and {de Burgos}, A. and {Cochrane}, R.~K. and {Doostmohammadi}, S. and {Mocnik}, T. and {Stoev}, H. and {Su{\'a}rez-Andr{\'e}s}, L. and {Tudor}, V. and {Wilson}, T.~G. and {Zegmott}, T.~J.},
        title = "{IGAPS: the merged IPHAS and UVEX optical surveys of the northern Galactic plane}",
      journal = {\aap},
         year = 2020,
        month = jun,
       volume = {638},
          eid = {A18},
        pages = {A18},
          doi = {10.1051/0004-6361/201937333},
archivePrefix = {arXiv},
       eprint = {2002.05157},
 primaryClass = {astro-ph.IM},
       adsurl = {https://ui.adsabs.harvard.edu/abs/2020A&A...638A..18M}
}

@ARTICLE{VPHAS,
       author = {{Drew}, J.~E. and {Gonzalez-Solares}, E. and {Greimel}, R. and {Irwin}, M.~J. and {K{\"u}pc{\"u} Yoldas}, A. and {Lewis}, J. and {Barentsen}, G. and {Eisl{\"o}ffel}, J. and {Farnhill}, H.~J. and {Martin}, W.~E. and {Walsh}, J.~R. and {Walton}, N.~A. and {Mohr-Smith}, M. and {Raddi}, R. and {Sale}, S.~E. and {Wright}, N.~J. and {Groot}, P. and {Barlow}, M.~J. and {Corradi}, R.~L.~M. and {Drake}, J.~J. and {Fabregat}, J. and {Frew}, D.~J. and {G{\"a}nsicke}, B.~T. and {Knigge}, C. and {Mampaso}, A. and {Morris}, R.~A.~H. and {Naylor}, T. and {Parker}, Q.~A. and {Phillipps}, S. and {Ruhland}, C. and {Steeghs}, D. and {Unruh}, Y.~C. and {Vink}, J.~S. and {Wesson}, R. and {Zijlstra}, A.~A.},
        title = "{The VST Photometric H{\ensuremath{\alpha}} Survey of the Southern Galactic Plane and Bulge (VPHAS+)}",
      journal = {\mnras},
         year = 2014,
        month = may,
       volume = {440},
       number = {3},
        pages = {2036-2058},
          doi = {10.1093/mnras/stu394},
archivePrefix = {arXiv},
       eprint = {1402.7024},
 primaryClass = {astro-ph.GA},
       adsurl = {https://ui.adsabs.harvard.edu/abs/2014MNRAS.440.2036D}
}

@ARTICLE{Ekstrom,
       author = {{Ekstr{\"o}m}, S. and {Georgy}, C. and {Eggenberger}, P. and
         {Meynet}, G. and {Mowlavi}, N. and {Wyttenbach}, A. and {Granada}, A. and
         {Decressin}, T. and {Hirschi}, R. and {Frischknecht}, U. and
         {Charbonnel}, C. and {Maeder}, A.},
        title = "{Grids of stellar models with rotation. I. Models from 0.8 to 120 M$_{☉}$ at solar metallicity (Z = 0.014)}",
      journal = {\aap},
         year = "2012",
        month = "Jan",
       volume = {537},
          eid = {A146},
        pages = {A146},
          doi = {10.1051/0004-6361/201117751},
archivePrefix = {arXiv},
       eprint = {1110.5049},
 primaryClass = {astro-ph.SR},
       adsurl = {https://ui.adsabs.harvard.edu/abs/2012A&A...537A.146E}
}

@ARTICLE{Coelho,
       author = {{Coelho}, P.~R.~T.},
        title = "{A new library of theoretical stellar spectra with scaled-solar and {\ensuremath{\alpha}}-enhanced mixtures}",
      journal = {\mnras},
         year = "2014",
        month = "May",
       volume = {440},
       number = {2},
        pages = {1027-1043},
          doi = {10.1093/mnras/stu365},
archivePrefix = {arXiv},
       eprint = {1404.3243},
 primaryClass = {astro-ph.SR},
       adsurl = {https://ui.adsabs.harvard.edu/abs/2014MNRAS.440.1027C}
}

@INBOOK{Rauch,
       author = {{Rauch}, T. and {Deetjen}, J.~L.},
        title = "{Handling of Atomic Data}",
    booktitle = {Stellar Atmosphere Modeling, ASP Conference Proceedings, Vol. 288. Editors: Ivan Hubeny, Dimitri Mihalas, and Klaus Werner. San Francisco: Astronomical Society of the Pacific, ISBN: 1-58381-131-1, 2003, p.103},
         year = "2003",
       volume = {288},
       series = {Astronomical Society of the Pacific Conference Series},
        pages = {103},
     publisher = {Stellar Atmosphere Modeling, ASP Conference Proceedings, Vol. 288. in Tuebingen, Germany. Editors: Ivan Hubeny, Dimitri Mihalas, and Klaus Werner. San Francisco: Astronomical Society of the Pacific},
       adsurl = {https://ui.adsabs.harvard.edu/abs/2003ASPC..288..103R}
}

@INBOOK{Werner,
       author = {{Werner}, K. and {Deetjen}, J.~L. and {Dreizler}, S. and {Nagel}, T. and
         {Rauch}, T. and {Schuh}, S.~L.},
        title = "{Model Photospheres with Accelerated Lambda Iteration}",
    booktitle = {Stellar Atmosphere Modeling, ASP Conference Proceedings, Vol. 288. Abstracts from a conference held 8-12 April 2002 in Tuebingen, Germany. Editors: Ivan Hubeny, Dimitri Mihalas, and Klaus Werner. San Francisco: Astronomical Society of the Pacific, ISBN: 1-58381-131-1, 2003, p.31},
         year = "2003",
       volume = {288},
       publisher = {Stellar Atmosphere Modeling, ASP Conference Proceedings, Vol. 288. Editors: Ivan Hubeny, Dimitri Mihalas, and Klaus Werner. San Francisco: Astronomical Society of the Pacific},
       series = {Astronomical Society of the Pacific Conference Series},
        pages = {31},
       adsurl = {https://ui.adsabs.harvard.edu/abs/2003ASPC..288...31W}
}

@ARTICLE{Werner1999,
       author = {{Werner}, K. and {Dreizler}, S.},
        title = "{The classical stellar atmosphere problem.}",
      journal = {Journal of Computational and Applied Mathematics},
         year = 1999,
        month = sep,
       volume = {109},
       number = {1},
        pages = {65-93},
archivePrefix = {arXiv},
       eprint = {astro-ph/9906130},
 primaryClass = {astro-ph},
       adsurl = {https://ui.adsabs.harvard.edu/abs/1999JCoAM.109...65W}
}

@ARTICLE{Allard2012,
       author = {{Allard}, F. and {Homeier}, D. and {Freytag}, B.},
        title = "{Models of very-low-mass stars, brown dwarfs and exoplanets}",
      journal = {Philosophical Transactions of the Royal Society of London Series A},
         year = 2012,
        month = jun,
       volume = {370},
       number = {1968},
        pages = {2765-2777},
          doi = {10.1098/rsta.2011.0269},
archivePrefix = {arXiv},
       eprint = {1112.3591},
 primaryClass = {astro-ph.SR},
       adsurl = {https://ui.adsabs.harvard.edu/abs/2012RSPTA.370.2765A}
}

@ARTICLE{PantaleoniGonzalez2025,
       author = {{Pantaleoni Gonz{\'a}lez}, M. and {Ma{\'\i}z Apell{\'a}niz}, J. and {Barb{\'a}}, R.~H. and {Reed}, B. Cameron and {Berlanas}, S.~R. and {Parras Rico}, A. and {Bodaghee}, A.},
        title = "{The Alma catalogue of OB stars ─ III. A cross-match with Gaia DR3 and an extension based on new spectral classifications}",
      journal = {\mnras},
         year = 2025,
        month = oct,
       volume = {543},
       number = {1},
        pages = {63-82},
          doi = {10.1093/mnras/staf1409},
archivePrefix = {arXiv},
       eprint = {2508.14875},
 primaryClass = {astro-ph.SR},
       adsurl = {https://ui.adsabs.harvard.edu/abs/2025MNRAS.543...63P}
}

@ARTICLE{Reed2003,
       author = {{Reed}, B. Cameron},
        title = "{Catalog of Galactic OB Stars}",
      journal = {\aj},
         year = 2003,
        month = may,
       volume = {125},
       number = {5},
        pages = {2531-2533},
          doi = {10.1086/374771},
       adsurl = {https://ui.adsabs.harvard.edu/abs/2003AJ....125.2531R}
}

@ARTICLE{GOSC,
       author = {{Ma{\'\i}z-Apell{\'a}niz}, Jes{\'u}s and {Walborn}, Nolan R. and {Galu{\'e}}, H{\'e}ctor {\'A}. and {Wei}, Lisa H.},
        title = "{A Galactic O Star Catalog}",
      journal = {\apjs},
         year = 2004,
        month = mar,
       volume = {151},
       number = {1},
        pages = {103-148},
          doi = {10.1086/381380},
archivePrefix = {arXiv},
       eprint = {astro-ph/0311196},
 primaryClass = {astro-ph},
       adsurl = {https://ui.adsabs.harvard.edu/abs/2004ApJS..151..103M}
}

@ARTICLE{Zari2021,
       author = {{Zari}, E. and {Rix}, H. -W. and {Frankel}, N. and {Xiang}, M. and {Poggio}, E. and {Drimmel}, R. and {Tkachenko}, A.},
        title = "{Mapping luminous hot stars in the Galaxy}",
      journal = {\aap},
         year = 2021,
        month = jun,
       volume = {650},
          eid = {A112},
        pages = {A112},
          doi = {10.1051/0004-6361/202039726},
archivePrefix = {arXiv},
       eprint = {2102.08684},
 primaryClass = {astro-ph.GA},
       adsurl = {https://ui.adsabs.harvard.edu/abs/2021A&A...650A.112Z}
}

@ARTICLE{GaiaDR3GoldenSample,
       author = {{Gaia Collaboration} and {Creevey}, O.~L. and {Sarro}, L.~M. and {Lobel}, A. and {Pancino}, E. and {Andrae}, R. and {Smart}, R.~L. and {Clementini}, G. and {Heiter}, U. and {Korn}, A.~J. and {Fouesneau}, M. and {Fr{\'e}mat}, Y. and {De Angeli}, F. and {Vallenari}, A. and {Harrison}, D.~L. and {Th{\'e}venin}, F. and {Reyl{\'e}}, C. and {Sordo}, R. and {Garofalo}, A. and {Brown}, A.~G.~A. and {Eyer}, L. and {Prusti}, T. and {de Bruijne}, J.~H.~J. and {Arenou}, F. and {Babusiaux}, C. and {Biermann}, M. and {Ducourant}, C. and {Evans}, D.~W. and {Guerra}, R. and {Hutton}, A. and {Jordi}, C. and {Klioner}, S.~A. and {Lammers}, U.~L. and {Lindegren}, L. and {Luri}, X. and {Mignard}, F. and {Panem}, C. and {Pourbaix}, D. and {Randich}, S. and {Sartoretti}, P. and {Soubiran}, C. and {Tanga}, P. and {Walton}, N.~A. and {Bailer-Jones}, C.~A.~L. and {Bastian}, U. and {Drimmel}, R. and {Jansen}, F. and {Katz}, D. and {Lattanzi}, M.~G. and {van Leeuwen}, F. and {Bakker}, J. and {Cacciari}, C. and {Casta{\~n}eda}, J. and {Fabricius}, C. and {Galluccio}, L. and {Guerrier}, A. and {Masana}, E. and {Messineo}, R. and {Mowlavi}, N. and {Nicolas}, C. and {Nienartowicz}, K. and {Pailler}, F. and {Panuzzo}, P. and {Riclet}, F. and {Roux}, W. and {Seabroke}, G.~M. and {Gracia-Abril}, G. and {Portell}, J. and {Teyssier}, D. and {Altmann}, M. and {Audard}, M. and {Bellas-Velidis}, I. and {Benson}, K. and {Berthier}, J. and {Blomme}, R. and {Burgess}, P.~W. and {Busonero}, D. and {Busso}, G. and {C{\'a}novas}, H. and {Carry}, B. and {Cellino}, A. and {Cheek}, N. and {Damerdji}, Y. and {Davidson}, M. and {de Teodoro}, P. and {Nu{\~n}ez Campos}, M. and {Delchambre}, L. and {Dell'Oro}, A. and {Esquej}, P. and {Fern{\'a}ndez-Hern{\'a}ndez}, J. and {Fraile}, E. and {Garabato}, D. and {Garc{\'\i}a-Lario}, P. and {Gosset}, E. and {Haigron}, R. and {Halbwachs}, J. -L. and {Hambly}, N.~C. and {Hern{\'a}ndez}, J. and {Hestroffer}, D. and {Hodgkin}, S.~T. and {Holl}, B. and {Jan{\ss}en}, K. and {Jevardat de Fombelle}, G. and {Jordan}, S. and {Krone-Martins}, A. and {Lanzafame}, A.~C. and {L{\"o}ffler}, W. and {Marchal}, O. and {Marrese}, P.~M. and {Moitinho}, A. and {Muinonen}, K. and {Osborne}, P. and {Pauwels}, T. and {Recio-Blanco}, A. and {Riello}, M. and {Rimoldini}, L. and {Roegiers}, T. and {Rybizki}, J. and {Siopis}, C. and {Smith}, M. and {Sozzetti}, A. and {Utrilla}, E. and {van Leeuwen}, M. and {Abbas}, U. and {{\'A}brah{\'a}m}, P. and {Abreu Aramburu}, A. and {Aerts}, C. and {Aguado}, J.~J. and {Ajaj}, M. and {Aldea-Montero}, F. and {Altavilla}, G. and {{\'A}lvarez}, M.~A. and {Alves}, J. and {Anders}, F. and {Anderson}, R.~I. and {Anglada Varela}, E. and {Antoja}, T. and {Baines}, D. and {Baker}, S.~G. and {Balaguer-N{\'u}{\~n}ez}, L. and {Balbinot}, E. and {Balog}, Z. and {Barache}, C. and {Barbato}, D. and {Barros}, M. and {Barstow}, M.~A. and {Bartolom{\'e}}, S. and {Bassilana}, J. -L. and {Bauchet}, N. and {Becciani}, U. and {Bellazzini}, M. and {Berihuete}, A. and {Bernet}, M. and {Bertone}, S. and {Bianchi}, L. and {Binnenfeld}, A. and {Blanco-Cuaresma}, S. and {Boch}, T. and {Bombrun}, A. and {Bossini}, D. and {Bouquillon}, S. and {Bragaglia}, A. and {Bramante}, L. and {Breedt}, E. and {Bressan}, A. and {Brouillet}, N. and {Brugaletta}, E. and {Bucciarelli}, B. and {Burlacu}, A. and {Butkevich}, A.~G. and {Buzzi}, R. and {Caffau}, E. and {Cancelliere}, R. and {Cantat-Gaudin}, T. and {Carballo}, R. and {Carlucci}, T. and {Carnerero}, M.~I. and {Carrasco}, J.~M. and {Casamiquela}, L. and {Castellani}, M. and {Castro-Ginard}, A. and {Chaoul}, L. and {Charlot}, P. and {Chemin}, L. and {Chiaramida}, V. and {Chiavassa}, A. and {Chornay}, N. and {Comoretto}, G. and {Contursi}, G. and {Cooper}, W.~J. and {Cornez}, T. and {Cowell}, S. and {Crifo}, F. and {Cropper}, M. and {Crosta}, M. and {Crowley}, C. and {Dafonte}, C. and {Dapergolas}, A. and {David}, P. and {de Laverny}, P. and {De Luise}, F. and {De March}, R. and {De Ridder}, J. and {de Souza}, R. and {de Torres}, A. and {del Peloso}, E.~F. and {del Pozo}, E. and {Delbo}, M. and {Delgado}, A. and {Delisle}, J. -B. and {Demouchy}, C. and {Dharmawardena}, T.~E. and {Di Matteo}, P. and {Diakite}, S. and {Diener}, C. and {Distefano}, E. and {Dolding}, C. and {Enke}, H. and {Fabre}, C. and {Fabrizio}, M. and {Faigler}, S. and {Fedorets}, G. and {Fernique}, P. and {Figueras}, F. and {Fournier}, Y. and {Fouron}, C. and {Fragkoudi}, F. and {Gai}, M. and {Garcia-Gutierrez}, A. and {Garcia-Reinaldos}, M. and {Garc{\'\i}a-Torres}, M. and {Gavel}, A. and {Gavras}, P. and {Gerlach}, E. and {Geyer}, R. and {Giacobbe}, P. and {Gilmore}, G. and {Girona}, S. and {Giuffrida}, G. and {Gomel}, R. and {Gomez}, A. and {Gonz{\'a}lez-N{\'u}{\~n}ez}, J. and {Gonz{\'a}lez-Santamar{\'\i}a}, I. and {Gonz{\'a}lez-Vidal}, J.~J. and {Granvik}, M. and {Guillout}, P. and {Guiraud}, J. and {Guti{\'e}rrez-S{\'a}nchez}, R. and {Guy}, L.~P. and {Hatzidimitriou}, D. and {Hauser}, M. and {Haywood}, M. and {Helmer}, A. and {Helmi}, A. and {Hilger}, T. and {Sarmiento}, M.~H. and {Hidalgo}, S.~L. and {H{\l}adczuk}, N. and {Hobbs}, D. and {Holland}, G. and {Huckle}, H.~E. and {Jardine}, K. and {Jasniewicz}, G. and {Jean-Antoine Piccolo}, A. and {Jim{\'e}nez-Arranz}, {\'O}. and {Juaristi Campillo}, J. and {Julbe}, F. and {Karbevska}, L. and {Kervella}, P. and {Khanna}, S. and {Kordopatis}, G. and {K{\'o}sp{\'a}l}, {\'A}. and {Kostrzewa-Rutkowska}, Z. and {Kruszy{\'n}ska}, K. and {Kun}, M. and {Laizeau}, P. and {Lambert}, S. and {Lanza}, A.~F. and {Lasne}, Y. and {Le Campion}, J. -F. and {Lebreton}, Y. and {Lebzelter}, T. and {Leccia}, S. and {Leclerc}, N. and {Lecoeur-Taibi}, I. and {Liao}, S. and {Licata}, E.~L. and {Lindstr{\o}m}, H.~E.~P. and {Lister}, T.~A. and {Livanou}, E. and {Lorca}, A. and {Loup}, C. and {Madrero Pardo}, P. and {Magdaleno Romeo}, A. and {Managau}, S. and {Mann}, R.~G. and {Manteiga}, M. and {Marchant}, J.~M. and {Marconi}, M. and {Marcos}, J. and {Marcos Santos}, M.~M.~S. and {Mar{\'\i}n Pina}, D. and {Marinoni}, S. and {Marocco}, F. and {Marshall}, D.~J. and {Martin Polo}, L. and {Mart{\'\i}n-Fleitas}, J.~M. and {Marton}, G. and {Mary}, N. and {Masip}, A. and {Massari}, D. and {Mastrobuono-Battisti}, A. and {Mazeh}, T. and {McMillan}, P.~J. and {Messina}, S. and {Michalik}, D. and {Millar}, N.~R. and {Mints}, A. and {Molina}, D. and {Molinaro}, R. and {Moln{\'a}r}, L. and {Monari}, G. and {Mongui{\'o}}, M. and {Montegriffo}, P. and {Montero}, A. and {Mor}, R. and {Mora}, A. and {Morbidelli}, R. and {Morel}, T. and {Morris}, D. and {Muraveva}, T. and {Murphy}, C.~P. and {Musella}, I. and {Nagy}, Z. and {Noval}, L. and {Oca{\~n}a}, F. and {Ogden}, A. and {Ordenovic}, C. and {Osinde}, J.~O. and {Pagani}, C. and {Pagano}, I. and {Palaversa}, L. and {Palicio}, P.~A. and {Pallas-Quintela}, L. and {Panahi}, A. and {Payne-Wardenaar}, S. and {Pe{\~n}alosa Esteller}, X. and {Penttil{\"a}}, A. and {Pichon}, B. and {Piersimoni}, A.~M. and {Pineau}, F. -X. and {Plachy}, E. and {Plum}, G. and {Poggio}, E. and {Pr{\v{s}}a}, A. and {Pulone}, L. and {Racero}, E. and {Ragaini}, S. and {Rainer}, M. and {Raiteri}, C.~M. and {Ramos}, P. and {Ramos-Lerate}, M. and {Re Fiorentin}, P. and {Regibo}, S. and {Richards}, P.~J. and {Rios Diaz}, C. and {Ripepi}, V. and {Riva}, A. and {Rix}, H. -W. and {Rixon}, G. and {Robichon}, N. and {Robin}, A.~C. and {Robin}, C. and {Roelens}, M. and {Rogues}, H.~R.~O. and {Rohrbasser}, L. and {Romero-G{\'o}mez}, M. and {Rowell}, N. and {Royer}, F. and {Ruz Mieres}, D. and {Rybicki}, K.~A. and {Sadowski}, G. and {S{\'a}ez N{\'u}{\~n}ez}, A. and {Sagrist{\`a} Sell{\'e}s}, A. and {Sahlmann}, J. and {Salguero}, E. and {Samaras}, N. and {Sanchez Gimenez}, V. and {Sanna}, N. and {Santove{\~n}a}, R. and {Sarasso}, M. and {Schultheis}, M. and {Sciacca}, E. and {Segol}, M. and {Segovia}, J.~C. and {S{\'e}gransan}, D. and {Semeux}, D. and {Shahaf}, S. and {Siddiqui}, H.~I. and {Siebert}, A. and {Siltala}, L. and {Silvelo}, A. and {Slezak}, E. and {Slezak}, I. and {Snaith}, O.~N. and {Solano}, E. and {Solitro}, F. and {Souami}, D. and {Souchay}, J. and {Spagna}, A. and {Spina}, L. and {Spoto}, F. and {Steele}, I.~A. and {Steidelm{\"u}ller}, H. and {Stephenson}, C.~A. and {S{\"u}veges}, M. and {Surdej}, J. and {Szabados}, L. and {Szegedi-Elek}, E. and {Taris}, F. and {Taylor}, M.~B. and {Teixeira}, R. and {Tolomei}, L. and {Tonello}, N. and {Torra}, F. and {Torra}, J. and {Torralba Elipe}, G. and {Trabucchi}, M. and {Tsounis}, A.~T. and {Turon}, C. and {Ulla}, A. and {Unger}, N. and {Vaillant}, M.~V. and {van Dillen}, E. and {van Reeven}, W. and {Vanel}, O. and {Vecchiato}, A. and {Viala}, Y. and {Vicente}, D. and {Voutsinas}, S. and {Weiler}, M. and {Wevers}, T. and {Wyrzykowski}, {\L}. and {Yoldas}, A. and {Yvard}, P. and {Zhao}, H. and {Zorec}, J. and {Zucker}, S. and {Zwitter}, T.},
        title = "{Gaia Data Release 3. A golden sample of astrophysical parameters}",
      journal = {\aap},
         year = 2023,
        month = jun,
       volume = {674},
          eid = {A39},
        pages = {A39},
          doi = {10.1051/0004-6361/202243800},
archivePrefix = {arXiv},
       eprint = {2206.05870},
 primaryClass = {astro-ph.SR},
       adsurl = {https://ui.adsabs.harvard.edu/abs/2023A&A...674A..39G}
}

@ARTICLE{Chen2019,
       author = {{Chen}, B. -Q. and {Huang}, Y. and {Hou}, L. -G. and {Tian}, H. and {Li}, G. -X. and {Yuan}, H. -B. and {Wang}, H. -F. and {Wang}, C. and {Tian}, Z. -J. and {Liu}, X. -W.},
        title = "{The Galactic spiral structure as revealed by O- and early B-type stars}",
      journal = {\mnras},
         year = 2019,
        month = jul,
       volume = {487},
       number = {1},
        pages = {1400-1409},
          doi = {10.1093/mnras/stz1357},
archivePrefix = {arXiv},
       eprint = {1905.05542},
 primaryClass = {astro-ph.GA},
       adsurl = {https://ui.adsabs.harvard.edu/abs/2019MNRAS.487.1400C}
}

@ARTICLE{Khalatyan2024,
       author = {{Khalatyan}, A. and {Anders}, F. and {Chiappini}, C. and {Queiroz}, A.~B.~A. and {Nepal}, S. and {dal Ponte}, M. and {Jordi}, C. and {Guiglion}, G. and {Valentini}, M. and {Torralba Elipe}, G. and {Steinmetz}, M. and {Pantaleoni-Gonz{\'a}lez}, M. and {Malhotra}, S. and {Jim{\'e}nez-Arranz}, {\'O}. and {Enke}, H. and {Casamiquela}, L. and {Ard{\`e}vol}, J.},
        title = "{Transferring spectroscopic stellar labels to 217 million Gaia DR3 XP stars with SHBoost}",
      journal = {\aap},
         year = 2024,
        month = nov,
       volume = {691},
          eid = {A98},
        pages = {A98},
          doi = {10.1051/0004-6361/202451427},
archivePrefix = {arXiv},
       eprint = {2407.06963},
 primaryClass = {astro-ph.SR},
       adsurl = {https://ui.adsabs.harvard.edu/abs/2024A&A...691A..98K}
}

@ARTICLE{Poggio2021,
       author = {{Poggio}, E. and {Drimmel}, R. and {Cantat-Gaudin}, T. and {Ramos}, P. and {Ripepi}, V. and {Zari}, E. and {Andrae}, R. and {Blomme}, R. and {Chemin}, L. and {Clementini}, G. and {Figueras}, F. and {Fouesneau}, M. and {Fr{\'e}mat}, Y. and {Lobel}, A. and {Marshall}, D.~J. and {Muraveva}, T. and {Romero-G{\'o}mez}, M.},
        title = "{Galactic spiral structure revealed by Gaia EDR3}",
      journal = {\aap},
         year = 2021,
        month = jul,
       volume = {651},
          eid = {A104},
        pages = {A104},
          doi = {10.1051/0004-6361/202140687},
archivePrefix = {arXiv},
       eprint = {2103.01970},
 primaryClass = {astro-ph.GA},
       adsurl = {https://ui.adsabs.harvard.edu/abs/2021A&A...651A.104P}
}

@ARTICLE{Poggio2018,
       author = {{Poggio}, E. and {Drimmel}, R. and {Lattanzi}, M.~G. and {Smart}, R.~L. and {Spagna}, A. and {Andrae}, R. and {Bailer-Jones}, C.~A.~L. and {Fouesneau}, M. and {Antoja}, T. and {Babusiaux}, C. and {Evans}, D.~W. and {Figueras}, F. and {Katz}, D. and {Reyl{\'e}}, C. and {Robin}, A.~C. and {Romero-G{\'o}mez}, M. and {Seabroke}, G.~M.},
        title = "{The Galactic warp revealed by Gaia DR2 kinematics}",
      journal = {\mnras},
         year = 2018,
        month = nov,
       volume = {481},
       number = {1},
        pages = {L21-L25},
          doi = {10.1093/mnrasl/sly148},
archivePrefix = {arXiv},
       eprint = {1805.03171},
 primaryClass = {astro-ph.GA},
       adsurl = {https://ui.adsabs.harvard.edu/abs/2018MNRAS.481L..21P}
}

@ARTICLE{Garcia,
       author = {{Garc{\'\i}a P{\'e}rez}, Ana E. and {Allende Prieto}, Carlos and {Holtzman}, Jon A. and {Shetrone}, Matthew and {M{\'e}sz{\'a}ros}, Szabolcs and {Bizyaev}, Dmitry and {Carrera}, Ricardo and {Cunha}, Katia and {Garc{\'\i}a-Hern{\'a}ndez}, D.~A. and {Johnson}, Jennifer A. and {Majewski}, Steven R. and {Nidever}, David L. and {Schiavon}, Ricardo P. and {Shane}, Neville and {Smith}, Verne V. and {Sobeck}, Jennifer and {Troup}, Nicholas and {Zamora}, Olga and {Weinberg}, David H. and {Bovy}, Jo and {Eisenstein}, Daniel J. and {Feuillet}, Diane and {Frinchaboy}, Peter M. and {Hayden}, Michael R. and {Hearty}, Fred R. and {Nguyen}, Duy C. and {O'Connell}, Robert W. and {Pinsonneault}, Marc H. and {Wilson}, John C. and {Zasowski}, Gail},
        title = "{ASPCAP: The APOGEE Stellar Parameter and Chemical Abundances Pipeline}",
      journal = {\aj},
         year = 2016,
        month = jun,
       volume = {151},
       number = {6},
          eid = {144},
        pages = {144},
          doi = {10.3847/0004-6256/151/6/144},
archivePrefix = {arXiv},
       eprint = {1510.07635},
 primaryClass = {astro-ph.SR},
       adsurl = {https://ui.adsabs.harvard.edu/abs/2016AJ....151..144G}
}

@ARTICLE{Abdu,
       author = {{Abdurro'uf} and {Accetta}, Katherine and {Aerts}, Conny and {Silva Aguirre}, V{\'\i}ctor and {Ahumada}, Romina and {Ajgaonkar}, Nikhil and {Filiz Ak}, N. and {Alam}, Shadab and {Allende Prieto}, Carlos and {Almeida}, Andr{\'e}s and {Anders}, Friedrich and {Anderson}, Scott F. and {Andrews}, Brett H. and {Anguiano}, Borja and {Aquino-Ort{\'\i}z}, Erik and {Arag{\'o}n-Salamanca}, Alfonso and {Argudo-Fern{\'a}ndez}, Maria and {Ata}, Metin and {Aubert}, Marie and {Avila-Reese}, Vladimir and {Badenes}, Carles and {Barb{\'a}}, Rodolfo H. and {Barger}, Kat and {Barrera-Ballesteros}, Jorge K. and {Beaton}, Rachael L. and {Beers}, Timothy C. and {Belfiore}, Francesco and {Bender}, Chad F. and {Bernardi}, Mariangela and {Bershady}, Matthew A. and {Beutler}, Florian and {Bidin}, Christian Moni and {Bird}, Jonathan C. and {Bizyaev}, Dmitry and {Blanc}, Guillermo A. and {Blanton}, Michael R. and {Boardman}, Nicholas Fraser and {Bolton}, Adam S. and {Boquien}, M{\'e}d{\'e}ric and {Borissova}, Jura and {Bovy}, Jo and {Brandt}, W.~N. and {Brown}, Jordan and {Brownstein}, Joel R. and {Brusa}, Marcella and {Buchner}, Johannes and {Bundy}, Kevin and {Burchett}, Joseph N. and {Bureau}, Martin and {Burgasser}, Adam and {Cabang}, Tuesday K. and {Campbell}, Stephanie and {Cappellari}, Michele and {Carlberg}, Joleen K. and {Wanderley}, F{\'a}bio Carneiro and {Carrera}, Ricardo and {Cash}, Jennifer and {Chen}, Yan-Ping and {Chen}, Wei-Huai and {Cherinka}, Brian and {Chiappini}, Cristina and {Choi}, Peter Doohyun and {Chojnowski}, S. Drew and {Chung}, Haeun and {Clerc}, Nicolas and {Cohen}, Roger E. and {Comerford}, Julia M. and {Comparat}, Johan and {da Costa}, Luiz and {Covey}, Kevin and {Crane}, Jeffrey D. and {Cruz-Gonzalez}, Irene and {Culhane}, Connor and {Cunha}, Katia and {Dai}, Y. Sophia and {Damke}, Guillermo and {Darling}, Jeremy and {Davidson}, James W., Jr. and {Davies}, Roger and {Dawson}, Kyle and {De Lee}, Nathan and {Diamond-Stanic}, Aleksandar M. and {Cano-D{\'\i}az}, Mariana and {S{\'a}nchez}, Helena Dom{\'\i}nguez and {Donor}, John and {Duckworth}, Chris and {Dwelly}, Tom and {Eisenstein}, Daniel J. and {Elsworth}, Yvonne P. and {Emsellem}, Eric and {Eracleous}, Mike and {Escoffier}, Stephanie and {Fan}, Xiaohui and {Farr}, Emily and {Feng}, Shuai and {Fern{\'a}ndez-Trincado}, Jos{\'e} G. and {Feuillet}, Diane and {Filipp}, Andreas and {Fillingham}, Sean P. and {Frinchaboy}, Peter M. and {Fromenteau}, Sebastien and {Galbany}, Llu{\'\i}s and {Garc{\'\i}a}, Rafael A. and {Garc{\'\i}a-Hern{\'a}ndez}, D.~A. and {Ge}, Junqiang and {Geisler}, Doug and {Gelfand}, Joseph and {G{\'e}ron}, Tobias and {Gibson}, Benjamin J. and {Goddy}, Julian and {Godoy-Rivera}, Diego and {Grabowski}, Kathleen and {Green}, Paul J. and {Greener}, Michael and {Grier}, Catherine J. and {Griffith}, Emily and {Guo}, Hong and {Guy}, Julien and {Hadjara}, Massinissa and {Harding}, Paul and {Hasselquist}, Sten and {Hayes}, Christian R. and {Hearty}, Fred and {Hern{\'a}ndez}, Jes{\'u}s and {Hill}, Lewis and {Hogg}, David W. and {Holtzman}, Jon A. and {Horta}, Danny and {Hsieh}, Bau-Ching and {Hsu}, Chin-Hao and {Hsu}, Yun-Hsin and {Huber}, Daniel and {Huertas-Company}, Marc and {Hutchinson}, Brian and {Hwang}, Ho Seong and {Ibarra-Medel}, H{\'e}ctor J. and {Chitham}, Jacob Ider and {Ilha}, Gabriele S. and {Imig}, Julie and {Jaekle}, Will and {Jayasinghe}, Tharindu and {Ji}, Xihan and {Johnson}, Jennifer A. and {Jones}, Amy and {J{\"o}nsson}, Henrik and {Katkov}, Ivan and {Khalatyan}, Arman, Dr. and {Kinemuchi}, Karen and {Kisku}, Shobhit and {Knapen}, Johan H. and {Kneib}, Jean-Paul and {Kollmeier}, Juna A. and {Kong}, Miranda and {Kounkel}, Marina and {Kreckel}, Kathryn and {Krishnarao}, Dhanesh and {Lacerna}, Ivan and {Lane}, Richard R. and {Langgin}, Rachel and {Lavender}, Ramon and {Law}, David R. and {Lazarz}, Daniel and {Leung}, Henry W. and {Leung}, Ho-Hin and {Lewis}, Hannah M. and {Li}, Cheng and {Li}, Ran and {Lian}, Jianhui and {Liang}, Fu-Heng and {Lin}, Lihwai and {Lin}, Yen-Ting and {Lin}, Sicheng and {Lintott}, Chris and {Long}, Dan and {Longa-Pe{\~n}a}, Pen{\'e}lope and {L{\'o}pez-Cob{\'a}}, Carlos and {Lu}, Shengdong and {Lundgren}, Britt F. and {Luo}, Yuanze and {Mackereth}, J. Ted and {de la Macorra}, Axel and {Mahadevan}, Suvrath and {Majewski}, Steven R. and {Manchado}, Arturo and {Mandeville}, Travis and {Maraston}, Claudia and {Margalef-Bentabol}, Berta and {Masseron}, Thomas and {Masters}, Karen L. and {Mathur}, Savita and {McDermid}, Richard M. and {Mckay}, Myles and {Merloni}, Andrea and {Merrifield}, Michael and {Meszaros}, Szabolcs and {Miglio}, Andrea and {Di Mille}, Francesco and {Minniti}, Dante and {Minsley}, Rebecca and {Monachesi}, Antonela and {Moon}, Jeongin and {Mosser}, Benoit and {Mulchaey}, John and {Muna}, Demitri and {Mu{\~n}oz}, Ricardo R. and {Myers}, Adam D. and {Myers}, Natalie and {Nadathur}, Seshadri and {Nair}, Preethi and {Nandra}, Kirpal and {Neumann}, Justus and {Newman}, Jeffrey A. and {Nidever}, David L. and {Nikakhtar}, Farnik and {Nitschelm}, Christian and {O'Connell}, Julia E. and {Garma-Oehmichen}, Luis and {Luan Souza de Oliveira}, Gabriel and {Olney}, Richard and {Oravetz}, Daniel and {Ortigoza-Urdaneta}, Mario and {Osorio}, Yeisson and {Otter}, Justin and {Pace}, Zachary J. and {Padilla}, Nelson and {Pan}, Kaike and {Pan}, Hsi-An and {Parikh}, Taniya and {Parker}, James and {Peirani}, Sebastien and {Pe{\~n}a Ram{\'\i}rez}, Karla and {Penny}, Samantha and {Percival}, Will J. and {Perez-Fournon}, Ismael and {Pinsonneault}, Marc and {Poidevin}, Fr{\'e}d{\'e}rick and {Poovelil}, Vijith Jacob and {Price-Whelan}, Adrian M. and {B{\'a}rbara de Andrade Queiroz}, Anna and {Raddick}, M. Jordan and {Ray}, Amy and {Rembold}, Sandro Barboza and {Riddle}, Nicole and {Riffel}, Rogemar A. and {Riffel}, Rog{\'e}rio and {Rix}, Hans-Walter and {Robin}, Annie C. and {Rodr{\'\i}guez-Puebla}, Aldo and {Roman-Lopes}, Alexandre and {Rom{\'a}n-Z{\'u}{\~n}iga}, Carlos and {Rose}, Benjamin and {Ross}, Ashley J. and {Rossi}, Graziano and {Rubin}, Kate H.~R. and {Salvato}, Mara and {S{\'a}nchez}, Seb{\'a}stian F. and {S{\'a}nchez-Gallego}, Jos{\'e} R. and {Sanderson}, Robyn and {Santana Rojas}, Felipe Antonio and {Sarceno}, Edgar and {Sarmiento}, Regina and {Sayres}, Conor and {Sazonova}, Elizaveta and {Schaefer}, Adam L. and {Schiavon}, Ricardo and {Schlegel}, David J. and {Schneider}, Donald P. and {Schultheis}, Mathias and {Schwope}, Axel and {Serenelli}, Aldo and {Serna}, Javier and {Shao}, Zhengyi and {Shapiro}, Griffin and {Sharma}, Anubhav and {Shen}, Yue and {Shetrone}, Matthew and {Shu}, Yiping and {Simon}, Joshua D. and {Skrutskie}, M.~F. and {Smethurst}, Rebecca and {Smith}, Verne and {Sobeck}, Jennifer and {Spoo}, Taylor and {Sprague}, Dani and {Stark}, David V. and {Stassun}, Keivan G. and {Steinmetz}, Matthias and {Stello}, Dennis and {Stone-Martinez}, Alexander and {Storchi-Bergmann}, Thaisa and {Stringfellow}, Guy S. and {Stutz}, Amelia and {Su}, Yung-Chau and {Taghizadeh-Popp}, Manuchehr and {Talbot}, Michael S. and {Tayar}, Jamie and {Telles}, Eduardo and {Teske}, Johanna and {Thakar}, Ani and {Theissen}, Christopher and {Tkachenko}, Andrew and {Thomas}, Daniel and {Tojeiro}, Rita and {Hernandez Toledo}, Hector and {Troup}, Nicholas W. and {Trump}, Jonathan R. and {Trussler}, James and {Turner}, Jacqueline and {Tuttle}, Sarah and {Unda-Sanzana}, Eduardo and {V{\'a}zquez-Mata}, Jos{\'e} Antonio and {Valentini}, Marica and {Valenzuela}, Octavio and {Vargas-Gonz{\'a}lez}, Jaime and {Vargas-Maga{\~n}a}, Mariana and {Alfaro}, Pablo Vera and {Villanova}, Sandro and {Vincenzo}, Fiorenzo and {Wake}, David and {Warfield}, Jack T. and {Washington}, Jessica Diane and {Weaver}, Benjamin Alan and {Weijmans}, Anne-Marie and {Weinberg}, David H. and {Weiss}, Achim and {Westfall}, Kyle B. and {Wild}, Vivienne and {Wilde}, Matthew C. and {Wilson}, John C. and {Wilson}, Robert F. and {Wilson}, Mikayla and {Wolf}, Julien and {Wood-Vasey}, W.~M. and {Yan}, Renbin and {Zamora}, Olga and {Zasowski}, Gail and {Zhang}, Kai and {Zhao}, Cheng and {Zheng}, Zheng and {Zheng}, Zheng and {Zhu}, Kai},
        title = "{The Seventeenth Data Release of the Sloan Digital Sky Surveys: Complete Release of MaNGA, MaStar, and APOGEE-2 Data}",
      journal = {\apjs},
         year = 2022,
        month = apr,
       volume = {259},
       number = {2},
          eid = {35},
        pages = {35},
          doi = {10.3847/1538-4365/ac4414},
archivePrefix = {arXiv},
       eprint = {2112.02026},
 primaryClass = {astro-ph.GA},
       adsurl = {https://ui.adsabs.harvard.edu/abs/2022ApJS..259...35A}
}

@ARTICLE{Creevey2023,
       author = {{Creevey}, O.~L. and {Sordo}, R. and {Pailler}, F. and {Fr{\'e}mat}, Y. and {Heiter}, U. and {Th{\'e}venin}, F. and {Andrae}, R. and {Fouesneau}, M. and {Lobel}, A. and {Bailer-Jones}, C.~A.~L. and {Garabato}, D. and {Bellas-Velidis}, I. and {Brugaletta}, E. and {Lorca}, A. and {Ordenovic}, C. and {Palicio}, P.~A. and {Sarro}, L.~M. and {Delchambre}, L. and {Drimmel}, R. and {Rybizki}, J. and {Torralba Elipe}, G. and {Korn}, A.~J. and {Recio-Blanco}, A. and {Schultheis}, M.~S. and {De Angeli}, F. and {Montegriffo}, P. and {Abreu Aramburu}, A. and {Accart}, S. and {{\'A}lvarez}, M.~A. and {Bakker}, J. and {Brouillet}, N. and {Burlacu}, A. and {Carballo}, R. and {Casamiquela}, L. and {Chiavassa}, A. and {Contursi}, G. and {Cooper}, W.~J. and {Dafonte}, C. and {Dapergolas}, A. and {de Laverny}, P. and {Dharmawardena}, T.~E. and {Edvardsson}, B. and {Le Fustec}, Y. and {Garc{\'\i}a-Lario}, P. and {Garc{\'\i}a-Torres}, M. and {Gomez}, A. and {Gonz{\'a}lez-Santamar{\'\i}a}, I. and {Hatzidimitriou}, D. and {Jean-Antoine Piccolo}, A. and {Kontiza}, M. and {Kordopatis}, G. and {Lanzafame}, A.~C. and {Lebreton}, Y. and {Licata}, E.~L. and {Lindstr{\o}m}, H.~E.~P. and {Livanou}, E. and {Magdaleno Romeo}, A. and {Manteiga}, M. and {Marocco}, F. and {Marshall}, D.~J. and {Mary}, N. and {Nicolas}, C. and {Pallas-Quintela}, L. and {Panem}, C. and {Pichon}, B. and {Poggio}, E. and {Riclet}, F. and {Robin}, C. and {Santove{\~n}a}, R. and {Silvelo}, A. and {Slezak}, I. and {Smart}, R.~L. and {Soubiran}, C. and {S{\"u}veges}, M. and {Ulla}, A. and {Utrilla}, E. and {Vallenari}, A. and {Zhao}, H. and {Zorec}, J. and {Barrado}, D. and {Bijaoui}, A. and {Bouret}, J. -C. and {Blomme}, R. and {Brott}, I. and {Cassisi}, S. and {Kochukhov}, O. and {Martayan}, C. and {Shulyak}, D. and {Silvester}, J.},
        title = "{Gaia Data Release 3. Astrophysical parameters inference system (Apsis). I. Methods and content overview}",
      journal = {\aap},
         year = 2023,
        month = jun,
       volume = {674},
          eid = {A26},
        pages = {A26},
          doi = {10.1051/0004-6361/202243688},
archivePrefix = {arXiv},
       eprint = {2206.05864},
 primaryClass = {astro-ph.GA},
       adsurl = {https://ui.adsabs.harvard.edu/abs/2023A&A...674A..26C}
}

@ARTICLE{Xiang2022,
       author = {{Xiang}, Maosheng and {Rix}, Hans-Walter and {Ting}, Yuan-Sen and {Kudritzki}, Rolf-Peter and {Conroy}, Charlie and {Zari}, Eleonora and {Shi}, Jian-Rong and {Przybilla}, Norbert and {Ramirez-Tannus}, Maria and {Tkachenko}, Andrew and {Gebruers}, Sarah and {Liu}, Xiao-Wei},
        title = "{Stellar labels for hot stars from low-resolution spectra. I. The HotPayne method and results for 330 000 stars from LAMOST DR6}",
      journal = {\aap},
         year = 2022,
        month = jun,
       volume = {662},
          eid = {A66},
        pages = {A66},
          doi = {10.1051/0004-6361/202141570},
archivePrefix = {arXiv},
       eprint = {2108.02878},
 primaryClass = {astro-ph.SR},
       adsurl = {https://ui.adsabs.harvard.edu/abs/2022A&A...662A..66X}
}

@ARTICLE{Liu2019,
       author = {{Liu}, Zhicun and {Cui}, Wenyuan and {Liu}, Chao and {Huang}, Yang and {Zhao}, Gang and {Zhang}, Bo},
        title = "{A Catalog of OB Stars from LAMOST Spectroscopic Survey}",
      journal = {\apjs},
         year = 2019,
        month = apr,
       volume = {241},
       number = {2},
          eid = {32},
        pages = {32},
          doi = {10.3847/1538-4365/ab0a0d},
archivePrefix = {arXiv},
       eprint = {1902.07607},
 primaryClass = {astro-ph.SR},
       adsurl = {https://ui.adsabs.harvard.edu/abs/2019ApJS..241...32L}
}

@ARTICLE{Quintana2026,
       author = {{Quintana}, Alexis L. and {Wright}, Nicholas J. and {Kormann}, Lilly A. and {Alves}, Jo{\~a}o and {Katz}, David and {Casamiquela}, Laia and {Di Matteo}, Paola and {Haywood}, Misha and {Laporte}, Chervin},
        title = "{A new Gaia census of OB associations within 1 kpc}",
      journal = {\mnras},
         year = 2026,
        month = jun,
       volume = {549},
       number = {1},
          eid = {stag853},
        pages = {stag853},
          doi = {10.1093/mnras/stag853},
archivePrefix = {arXiv},
       eprint = {2512.05854},
 primaryClass = {astro-ph.GA},
       adsurl = {https://ui.adsabs.harvard.edu/abs/2026MNRAS.549ag853Q}
}

@ARTICLE{Berlanas2019,
       author = {{Berlanas}, S.~R. and {Wright}, N.~J. and {Herrero}, A. and {Drew}, J.~E. and {Lennon}, D.~J.},
        title = "{Disentangling the spatial substructure of Cygnus OB2 from Gaia DR2}",
      journal = {\mnras},
         year = 2019,
        month = apr,
       volume = {484},
       number = {2},
        pages = {1838-1842},
          doi = {10.1093/mnras/stz117},
archivePrefix = {arXiv},
       eprint = {1901.02959},
 primaryClass = {astro-ph.SR},
       adsurl = {https://ui.adsabs.harvard.edu/abs/2019MNRAS.484.1838B}
}

@ARTICLE{Morgan1953,
       author = {{Morgan}, W.~W. and {Whitford}, A.~E. and {Code}, A.~D.},
        title = "{Studies in Galactic Structure. I. a Preliminary Determination of the Space Distribution of the Blue Giants.}",
      journal = {\apj},
         year = 1953,
        month = sep,
       volume = {118},
        pages = {318},
          doi = {10.1086/145754},
       adsurl = {https://ui.adsabs.harvard.edu/abs/1953ApJ...118..318M}
}

@ARTICLE{PantaleoniGonzalez2021,
       author = {{Pantaleoni Gonz{\'a}lez}, M. and {Ma{\'\i}z Apell{\'a}niz}, J. and {Barb{\'a}}, R.~H. and {Reed}, B. Cameron},
        title = "{The Alma catalogue of OB stars - II. A cross-match with Gaia DR2 and an updated map of the solar neighbourhood}",
      journal = {\mnras},
         year = 2021,
        month = jun,
       volume = {504},
       number = {2},
        pages = {2968-2982},
          doi = {10.1093/mnras/stab688},
archivePrefix = {arXiv},
       eprint = {2103.02748},
 primaryClass = {astro-ph.SR},
       adsurl = {https://ui.adsabs.harvard.edu/abs/2021MNRAS.504.2968P}
}

@ARTICLE{HouHan2014,
       author = {{Hou}, L.~G. and {Han}, J.~L.},
        title = "{The observed spiral structure of the Milky Way}",
      journal = {\aap},
         year = 2014,
        month = sep,
       volume = {569},
          eid = {A125},
        pages = {A125},
          doi = {10.1051/0004-6361/201424039},
archivePrefix = {arXiv},
       eprint = {1407.7331},
 primaryClass = {astro-ph.GA},
       adsurl = {https://ui.adsabs.harvard.edu/abs/2014A&A...569A.125H}
}

@ARTICLE{Skowron2019,
       author = {{Skowron}, Dorota M. and {Skowron}, Jan and {Mr{\'o}z}, Przemek and {Udalski}, Andrzej and {Pietrukowicz}, Pawe{\l} and {Soszy{\'n}ski}, Igor and {Szyma{\'n}ski}, Micha{\l} K. and {Poleski}, Rados{\l}aw and {Koz{\l}owski}, Szymon and {Ulaczyk}, Krzysztof and {Rybicki}, Krzysztof and {Iwanek}, Patryk},
        title = "{A three-dimensional map of the Milky Way using classical Cepheid variable stars}",
      journal = {Science},
         year = 2019,
        month = aug,
       volume = {365},
       number = {6452},
        pages = {478-482},
          doi = {10.1126/science.aau3181},
archivePrefix = {arXiv},
       eprint = {1806.10653},
 primaryClass = {astro-ph.GA},
       adsurl = {https://ui.adsabs.harvard.edu/abs/2019Sci...365..478S}
}

@ARTICLE{Reid2019,
       author = {{Reid}, M.~J. and {Menten}, K.~M. and {Brunthaler}, A. and {Zheng}, X.~W. and {Dame}, T.~M. and {Xu}, Y. and {Li}, J. and {Sakai}, N. and {Wu}, Y. and {Immer}, K. and {Zhang}, B. and {Sanna}, A. and {Moscadelli}, L. and {Rygl}, K.~L.~J. and {Bartkiewicz}, A. and {Hu}, B. and {Quiroga-Nu{\~n}ez}, L.~H. and {van Langevelde}, H.~J.},
        title = "{Trigonometric Parallaxes of High-mass Star-forming Regions: Our View of the Milky Way}",
      journal = {\apj},
         year = 2019,
        month = nov,
       volume = {885},
       number = {2},
          eid = {131},
        pages = {131},
          doi = {10.3847/1538-4357/ab4a11},
archivePrefix = {arXiv},
       eprint = {1910.03357},
 primaryClass = {astro-ph.GA},
       adsurl = {https://ui.adsabs.harvard.edu/abs/2019ApJ...885..131R}
}

@BOOK{Sparke2000,
       author = {{Sparke}, Linda S. and {Gallagher}, John S., III},
        title = "{Galaxies in the universe : an introduction}",
         year = 2000,
       adsurl = {https://ui.adsabs.harvard.edu/abs/2000gaun.book.....S},
       publisher = {Cambridge University Press}
}

@ARTICLE{Russeil2003,
       author = {{Russeil}, D.},
        title = "{Star-forming complexes and the spiral structure of our Galaxy}",
      journal = {\aap},
         year = 2003,
        month = jan,
       volume = {397},
        pages = {133-146},
          doi = {10.1051/0004-6361:20021504},
       adsurl = {https://ui.adsabs.harvard.edu/abs/2003A&A...397..133R}
}

@ARTICLE{Churchwell2009,
       author = {{Churchwell}, Ed and {Babler}, Brian L. and {Meade}, Marilyn R. and {Whitney}, Barbara A. and {Benjamin}, Robert and {Indebetouw}, Remy and {Cyganowski}, Claudia and {Robitaille}, Thomas P. and {Povich}, Matthew and {Watson}, Christer and {Bracker}, Steve},
        title = "{The Spitzer/GLIMPSE Surveys: A New View of the Milky Way}",
      journal = {\pasp},
         year = 2009,
        month = mar,
       volume = {121},
       number = {877},
        pages = {213},
          doi = {10.1086/597811},
       adsurl = {https://ui.adsabs.harvard.edu/abs/2009PASP..121..213C}
}

@ARTICLE{MarcoNegueruela2016,
       author = {{Marco}, Amparo and {Negueruela}, Ignacio},
        title = "{Open clusters in Auriga OB2}",
      journal = {\mnras},
         year = 2016,
        month = jun,
       volume = {459},
       number = {1},
        pages = {880-901},
          doi = {10.1093/mnras/stw640},
archivePrefix = {arXiv},
       eprint = {1604.03881},
 primaryClass = {astro-ph.SR},
       adsurl = {https://ui.adsabs.harvard.edu/abs/2016MNRAS.459..880M}
}

@ARTICLE{QuintanaNegueruelaBerlanas2025,
       author = {{Quintana}, Alexis L. and {Negueruela}, Ignacio and {Berlanas}, Sara R.},
        title = "{Quantifying the scale of star formation across the Perseus spiral arm using young clusters around Cas OB5}",
      journal = {\aap},
         year = 2025,
        month = may,
       volume = {697},
          eid = {A47},
        pages = {A47},
          doi = {10.1051/0004-6361/202452615},
archivePrefix = {arXiv},
       eprint = {2504.01748},
 primaryClass = {astro-ph.GA},
       adsurl = {https://ui.adsabs.harvard.edu/abs/2025A&A...697A..47Q}
}

@ARTICLE{Quintana2023,
       author = {{Quintana}, Alexis L. and {Wright}, Nicholas J. and {Jeffries}, Robin D.},
        title = "{Mapping the distribution of OB stars and associations in Auriga}",
      journal = {\mnras},
         year = 2023,
        month = jun,
       volume = {522},
       number = {2},
        pages = {3124-3137},
          doi = {10.1093/mnras/stad1160},
archivePrefix = {arXiv},
       eprint = {2304.08370},
 primaryClass = {astro-ph.SR},
       adsurl = {https://ui.adsabs.harvard.edu/abs/2023MNRAS.522.3124Q}
}

@ARTICLE{Choi2014,
       author = {{Choi}, Y.~K. and {Hachisuka}, K. and {Reid}, M.~J. and {Xu}, Y. and {Brunthaler}, A. and {Menten}, K.~M. and {Dame}, T.~M.},
        title = "{Trigonometric Parallaxes of Star Forming Regions in the Perseus Spiral Arm}",
      journal = {\apj},
         year = 2014,
        month = aug,
       volume = {790},
       number = {2},
          eid = {99},
        pages = {99},
          doi = {10.1088/0004-637X/790/2/99},
archivePrefix = {arXiv},
       eprint = {1407.1609},
 primaryClass = {astro-ph.GA},
       adsurl = {https://ui.adsabs.harvard.edu/abs/2014ApJ...790...99C}
}

@ARTICLE{Vazquez2008,
       author = {{V{\'a}zquez}, Ruben A. and {May}, Jorge and {Carraro}, Giovanni and {Bronfman}, Leonardo and {Moitinho}, Andr{\'e} and {Baume}, Gustavo},
        title = "{Spiral Structure in the Outer Galactic Disk. I. The Third Galactic Quadrant}",
      journal = {\apj},
         year = 2008,
        month = jan,
       volume = {672},
       number = {2},
        pages = {930-939},
          doi = {10.1086/524003},
archivePrefix = {arXiv},
       eprint = {0709.3973},
 primaryClass = {astro-ph},
       adsurl = {https://ui.adsabs.harvard.edu/abs/2008ApJ...672..930V}
}

@ARTICLE{Ge2024,
       author = {{Ge}, Q.~A. and {Li}, J.~J. and {Hao}, C.~J. and {Lin}, Z.~H. and {Hou}, L.~G. and {Liu}, D.~J. and {Li}, Y.~J. and {Bian}, S.~B.},
        title = "{Evolution of the Local Spiral Structure Revealed by OB-type Stars in Gaia DR3}",
      journal = {\aj},
         year = 2024,
        month = jul,
       volume = {168},
       number = {1},
          eid = {25},
        pages = {25},
          doi = {10.3847/1538-3881/ad5201},
       adsurl = {https://ui.adsabs.harvard.edu/abs/2024AJ....168...25G}
}

@ARTICLE{Kerr2023,
       author = {{Kerr}, Ronan and {Kraus}, Adam L. and {Rizzuto}, Aaron C.},
        title = "{SPYGLASS. IV. New Stellar Survey of Recent Star Formation within 1 kpc}",
      journal = {\apj},
         year = 2023,
        month = sep,
       volume = {954},
       number = {2},
          eid = {134},
        pages = {134},
          doi = {10.3847/1538-4357/ace5b3},
archivePrefix = {arXiv},
       eprint = {2306.08150},
 primaryClass = {astro-ph.GA},
       adsurl = {https://ui.adsabs.harvard.edu/abs/2023ApJ...954..134K}
}

@ARTICLE{Uyaniker2001,
       author = {{Uyan{\i}ker}, B. and {F{\"u}rst}, E. and {Reich}, W. and {Aschenbach}, B. and {Wielebinski}, R.},
        title = "{The Cygnus superbubble revisited}",
      journal = {\aap},
     keywords = {ISM: CYGNUS SUPERBUBBLE, RADIO CONTINUUM: ISM, X-RAY: ISM, GALAXY: STRUCTURE\},
         year = 2001,
        month = may,
       volume = {371},
        pages = {675-697},
          doi = {10.1051/0004-6361:20010387},
       adsurl = {https://ui.adsabs.harvard.edu/abs/2001A&A...371..675U},
      adsnote = {Provided by the SAO/NASA Astrophysics Data System}
}

@ARTICLE{Wright2015,
       author = {{Wright}, Nicholas J. and {Drew}, Janet E. and {Mohr-Smith}, Michael},
        title = "{The massive star population of Cygnus OB2}",
      journal = {\mnras},
         year = 2015,
        month = may,
       volume = {449},
       number = {1},
        pages = {741-760},
          doi = {10.1093/mnras/stv323},
archivePrefix = {arXiv},
       eprint = {1502.05718},
 primaryClass = {astro-ph.SR},
       adsurl = {https://ui.adsabs.harvard.edu/abs/2015MNRAS.449..741W}
}

@ARTICLE{SanchezSanjuan2024,
       author = {{S{\'a}nchez-Sanju{\'a}n}, Sergio and {Hern{\'a}ndez}, Jes{\'u}s and {P{\'e}rez-Villegas}, {\'A}ngeles and {Rom{\'a}n-Z{\'u}{\~n}iga}, Carlos and {Aguilar}, Luis and {Ballesteros-Paredes}, Javier and {Bonilla-Barroso}, Andrea},
        title = "{Kinematic study of the Orion Complex: analysing the young stellar clusters from big and small structures}",
      journal = {\mnras},
         year = 2024,
        month = nov,
       volume = {534},
       number = {3},
        pages = {2566-2584},
          doi = {10.1093/mnras/stae2157},
archivePrefix = {arXiv},
       eprint = {2409.09206},
 primaryClass = {astro-ph.GA},
       adsurl = {https://ui.adsabs.harvard.edu/abs/2024MNRAS.534.2566S}
}

@ARTICLE{CantatGaudin2019,
       author = {{Cantat-Gaudin}, T. and others},
        title = "{A ring in a shell: the large-scale 6D structure of the Vela OB2 complex}",
      journal = {\aap},
         year = 2019,
        month = jan,
       volume = {621},
          eid = {A115},
        pages = {A115},
          doi = {10.1051/0004-6361/201834003},
archivePrefix = {arXiv},
       eprint = {1808.00573},
 primaryClass = {astro-ph.SR},
       adsurl = {https://ui.adsabs.harvard.edu/abs/2019A&A...621A.115C}
}

@ARTICLE{Chen2025,
       author = {{Chen}, Bingqiu and {Li}, Guangxing and {Yuan}, Haibo and {Xiang}, Maosheng and {Zhou}, Jixuan and {Chen}, Pinjian and {Krause}, Martin and {Coombs}, Ashley},
        title = "{A large, long-lived, slowly-expanding superbubble across the Perseus arm}",
      journal = {Nature Communications},
         year = 2025,
        month = nov,
       volume = {16},
       number = {1},
          eid = {10558},
        pages = {10558},
          doi = {10.1038/s41467-025-65591-5},
archivePrefix = {arXiv},
       eprint = {2512.21927},
 primaryClass = {astro-ph.GA},
       adsurl = {https://ui.adsabs.harvard.edu/abs/2025NatCo..1610558C}
}

@ARTICLE{HuntReffert2023,
       author = {{Hunt}, Emily L. and {Reffert}, Sabine},
        title = "{Improving the open cluster census. II. An all-sky cluster catalogue with Gaia DR3}",
      journal = {\aap},
         year = 2023,
        month = may,
       volume = {673},
          eid = {A114},
        pages = {A114},
          doi = {10.1051/0004-6361/202346285},
archivePrefix = {arXiv},
       eprint = {2303.13424},
 primaryClass = {astro-ph.GA},
       adsurl = {https://ui.adsabs.harvard.edu/abs/2023A&A...673A.114H}
}

@ARTICLE{HuntReffert2024,
       author = {{Hunt}, Emily L. and {Reffert}, Sabine},
        title = "{Improving the open cluster census. III. Using cluster masses, radii, and dynamics to create a cleaned open cluster catalogue}",
      journal = {\aap},
         year = 2024,
        month = jun,
       volume = {686},
          eid = {A42},
        pages = {A42},
          doi = {10.1051/0004-6361/202348662},
archivePrefix = {arXiv},
       eprint = {2403.05143},
 primaryClass = {astro-ph.GA},
       adsurl = {https://ui.adsabs.harvard.edu/abs/2024A&A...686A..42H}
}

@ARTICLE{HuntReffert2021,
       author = {{Hunt}, Emily L. and {Reffert}, Sabine},
        title = "{Improving the open cluster census. I. Comparison of clustering algorithms applied to Gaia DR2 data}",
      journal = {\aap},
         year = 2021,
        month = feb,
       volume = {646},
          eid = {A104},
        pages = {A104},
          doi = {10.1051/0004-6361/202039341},
archivePrefix = {arXiv},
       eprint = {2012.04267},
 primaryClass = {astro-ph.GA},
       adsurl = {https://ui.adsabs.harvard.edu/abs/2021A&A...646A.104H}
}

@ARTICLE{QuintanaHuntParul2025,
       author = {{Quintana}, Alexis L. and {Hunt}, Emily L. and {Parul}, Hanna},
        title = "{How many stars form in compact clusters in the local Milky Way?}",
      journal = {\aap},
         year = 2025,
        month = sep,
       volume = {701},
          eid = {L2},
        pages = {L2},
          doi = {10.1051/0004-6361/202556366},
archivePrefix = {arXiv},
       eprint = {2508.12788},
 primaryClass = {astro-ph.GA},
       adsurl = {https://ui.adsabs.harvard.edu/abs/2025A&A...701L...2Q}
}

@ARTICLE{Hunt2026,
       author = {{Hunt}, Emily L. and {Cantat-Gaudin}, Tristan and {Anders}, Friedrich and {Malhotra}, Sagar and {Spina}, Lorenzo and {Castro-Ginard}, Alfred and {Cavallo}, Lorenzo},
        title = "{The selection function of the Gaia DR3 open cluster census}",
      journal = {\aap},
         year = 2026,
        month = feb,
       volume = {706},
          eid = {A341},
        pages = {A341},
          doi = {10.1051/0004-6361/202557781},
archivePrefix = {arXiv},
       eprint = {2510.18343},
 primaryClass = {astro-ph.GA},
       adsurl = {https://ui.adsabs.harvard.edu/abs/2026A&A...706A.341H}
}

@ARTICLE{CantatGaudin2019_Perseus,
       author = {{Cantat-Gaudin}, T. and {Krone-Martins}, A. and {Sedaghat}, N. and {Farahi}, A. and {de Souza}, R.~S. and {Skalidis}, R. and {Malz}, A.~I. and {Mac{\^e}do}, S. and {Moews}, B. and {Jordi}, C. and {Moitinho}, A. and {Castro-Ginard}, A. and {Ishida}, E.~E.~O. and {Heneka}, C. and {Boucaud}, A. and {Trindade}, A.~M.~M.},
        title = "{Gaia DR2 unravels incompleteness of nearby cluster population: new open clusters in the direction of Perseus}",
      journal = {\aap},
         year = 2019,
        month = apr,
       volume = {624},
          eid = {A126},
        pages = {A126},
          doi = {10.1051/0004-6361/201834453},
archivePrefix = {arXiv},
       eprint = {1810.05494},
 primaryClass = {astro-ph.GA},
       adsurl = {https://ui.adsabs.harvard.edu/abs/2019A&A...624A.126C}
}

@ARTICLE{Wright2019,
       author = {{Wright}, Nicholas J. and {Jeffries}, R.~D. and {Jackson}, R.~J. and {Bayo}, A. and {Bonito}, R. and {Damiani}, F. and {Kalari}, V. and {Lanzafame}, A.~C. and {Pancino}, E. and {Parker}, R.~J. and {Prisinzano}, L. and {Randich}, S. and {Vink}, J.~S. and {Alfaro}, E.~J. and {Bergemann}, M. and {Franciosini}, E. and {Gilmore}, G. and {Gonneau}, A. and {Hourihane}, A. and {Jofr{\'e}}, P. and {Koposov}, S.~E. and {Lewis}, J. and {Magrini}, L. and {Micela}, G. and {Morbidelli}, L. and {Sacco}, G.~G. and {Worley}, C.~C. and {Zaggia}, S.},
        title = "{The Gaia-ESO Survey: asymmetric expansion of the Lagoon Nebula cluster NGC 6530 from GES and Gaia DR2}",
      journal = {\mnras},
         year = 2019,
        month = jun,
       volume = {486},
       number = {2},
        pages = {2477-2493},
          doi = {10.1093/mnras/stz870},
archivePrefix = {arXiv},
       eprint = {1903.12176},
 primaryClass = {astro-ph.SR},
       adsurl = {https://ui.adsabs.harvard.edu/abs/2019MNRAS.486.2477W}
}

@ARTICLE{Stoop2023,
       author = {{Stoop}, M. and {Kaper}, L. and {de Koter}, A. and {Guo}, D. and {Lamers}, H.~J.~G.~L.~M. and {Rieder}, S.},
        title = "{The early evolution of young massive clusters. The kinematic history of NGC 6611/M16}",
      journal = {\aap},
         year = 2023,
        month = feb,
       volume = {670},
          eid = {A108},
        pages = {A108},
          doi = {10.1051/0004-6361/202244511},
archivePrefix = {arXiv},
       eprint = {2207.08452},
 primaryClass = {astro-ph.GA},
       adsurl = {https://ui.adsabs.harvard.edu/abs/2023A&A...670A.108S}
}

@ARTICLE{Wright2020,
       author = {{Wright}, Nicholas J.},
        title = "{OB Associations and their origins}",
      journal = {\nar},
         year = 2020,
        month = nov,
       volume = {90},
          eid = {101549},
        pages = {101549},
          doi = {10.1016/j.newar.2020.101549},
archivePrefix = {arXiv},
       eprint = {2011.09483},
 primaryClass = {astro-ph.SR},
       adsurl = {https://ui.adsabs.harvard.edu/abs/2020NewAR..9001549W}
}

@ARTICLE{MaizApellaniz2020,
       author = {{Ma{\'\i}z Apell{\'a}niz}, J. and {Crespo Bellido}, P. and {Barb{\'a}}, R.~H. and {Fern{\'a}ndez Aranda}, R. and {Sota}, A.},
        title = "{The Villafranca catalog of Galactic OB groups. I. Systems with O2-O3.5 stars}",
      journal = {\aap},
         year = 2020,
        month = nov,
       volume = {643},
          eid = {A138},
        pages = {A138},
          doi = {10.1051/0004-6361/202038228},
archivePrefix = {arXiv},
       eprint = {2009.05773},
 primaryClass = {astro-ph.GA},
       adsurl = {https://ui.adsabs.harvard.edu/abs/2020A&A...643A.138M}
}

@ARTICLE{MaizApellaniz2022,
       author = {{Ma{\'\i}z Apell{\'a}niz}, J. and {Barb{\'a}}, R.~H. and {Fern{\'a}ndez Aranda}, R. and {Pantaleoni Gonz{\'a}lez}, M. and {Crespo Bellido}, P. and {Sota}, A. and {Alfaro}, E.~J.},
        title = "{The Villafranca catalog of Galactic OB groups. II. From Gaia DR2 to EDR3 and ten new systems with O stars}",
      journal = {\aap},
         year = 2022,
        month = jan,
       volume = {657},
          eid = {A131},
        pages = {A131},
          doi = {10.1051/0004-6361/202142364},
archivePrefix = {arXiv},
       eprint = {2110.01464},
 primaryClass = {astro-ph.GA},
       adsurl = {https://ui.adsabs.harvard.edu/abs/2022A&A...657A.131M}
}

@ARTICLE{Berlanas2023,
       author = {{Berlanas}, S.~R. and {Ma{\'\i}z Apell{\'a}niz}, J. and {Herrero}, A. and {Mahy}, L. and {Blomme}, R. and {Negueruela}, I. and {Dorda}, R. and {Comer{\'o}n}, F. and {Gosset}, E. and {Pantaleoni Gonz{\'a}lez}, M. and {Molina Lera}, J.~A. and {Sota}, A. and {Furst}, T. and {Alfaro}, E.~J. and {Bergemann}, M. and {Carraro}, G. and {Drew}, J.~E. and {Morbidelli}, L. and {Vink}, J.~S.},
        title = "{Gaia-ESO survey: Massive stars in the Carina Nebula. I. A new census of OB stars}",
      journal = {\aap},
         year = 2023,
        month = mar,
       volume = {671},
          eid = {A20},
        pages = {A20},
          doi = {10.1051/0004-6361/202245335},
archivePrefix = {arXiv},
       eprint = {2301.08310},
 primaryClass = {astro-ph.SR},
       adsurl = {https://ui.adsabs.harvard.edu/abs/2023A&A...671A..20B}
}

@ARTICLE{Berlanas2025,
       author = {{Berlanas}, S.~R. and {Mahy}, L. and {Herrero}, A. and {Ma{\'\i}z Apell{\'a}niz}, J. and {Blomme}, R. and {Comer{\'o}n}, F. and {Negueruela}, I. and {Molina Lera}, J.~A. and {Pantaleoni Gonz{\'a}lez}, M. and {Daflon}, S. and {Santos}, W. and {Kalari}, V.~M.},
        title = "{Gaia-ESO survey: Massive stars in the Carina Nebula: II. The spectroscopic analysis of the O-star population}",
      journal = {\aap},
         year = 2025,
        month = mar,
       volume = {695},
          eid = {A248},
        pages = {A248},
          doi = {10.1051/0004-6361/202453269},
archivePrefix = {arXiv},
       eprint = {2501.16508},
 primaryClass = {astro-ph.SR},
       adsurl = {https://ui.adsabs.harvard.edu/abs/2025A&A...695A.248B}
}

@ARTICLE{BobylevBajkova2016,
       author = {{Bobylev}, V.~V. and {Bajkova}, A.~T.},
        title = "{Estimating the vertical disk scale height using young galactic objects}",
      journal = {Baltic Astronomy},
         year = 2016,
        month = jan,
       volume = {25},
        pages = {261-266},
          doi = {10.1515/astro-2017-0128},
archivePrefix = {arXiv},
       eprint = {1609.01129},
 primaryClass = {astro-ph.GA},
       adsurl = {https://ui.adsabs.harvard.edu/abs/2016BaltA..25..261B}
}

@ARTICLE{Vergely2022,
       author = {{Vergely}, J.~L. and {Lallement}, R. and {Cox}, N.~L.~J.},
        title = "{Three-dimensional extinction maps: Inverting inter-calibrated extinction catalogues}",
      journal = {\aap},
         year = 2022,
        month = aug,
       volume = {664},
          eid = {A174},
        pages = {A174},
          doi = {10.1051/0004-6361/202243319},
archivePrefix = {arXiv},
       eprint = {2205.09087},
 primaryClass = {astro-ph.GA},
       adsurl = {https://ui.adsabs.harvard.edu/abs/2022A&A...664A.174V}
}

@ARTICLE{Wang2025,
       author = {{Wang}, Tao and {Yuan}, Haibo and {Chen}, Bingqiu and {Li}, Guangxing and {Huang}, Bowen and {Guo}, Helong and {Zhang}, Ruoyi},
        title = "{A Comprehensive All-sky Catalog of 3345 Molecular Clouds from Three-dimensional Dust Extinction}",
      journal = {\apjs},
         year = 2025,
        month = sep,
       volume = {280},
       number = {1},
          eid = {16},
        pages = {16},
          doi = {10.3847/1538-4365/aded89},
archivePrefix = {arXiv},
       eprint = {2509.07670},
 primaryClass = {astro-ph.GA},
       adsurl = {https://ui.adsabs.harvard.edu/abs/2025ApJS..280...16W}
}

@ARTICLE{Palicio2025,
       author = {{Palicio}, P.~A. and {Recio-Blanco}, A. and {Tepper-Garc{\'\i}a}, T. and {Poggio}, E. and {Peirani}, S. and {Dubois}, Y. and {McMillan}, P.~J. and {Bland-Hawthorn}, J. and {Kraljic}, K. and {Barbillon}, M.},
        title = "{Signatures of simulated spiral arms on radial actions}",
      journal = {\aap},
         year = 2025,
        month = mar,
       volume = {695},
          eid = {A193},
        pages = {A193},
          doi = {10.1051/0004-6361/202453611},
archivePrefix = {arXiv},
       eprint = {2412.17515},
 primaryClass = {astro-ph.GA},
       adsurl = {https://ui.adsabs.harvard.edu/abs/2025A&A...695A.193P}
}

@INPROCEEDINGS{MaizApellanizWeiler2025,
       author = {{Ma{\'\i}z Apell{\'a}niz}, J. and {Weiler}, M.},
        title = "{Differences Between Gaia DR2 and Gaia EDR3 Photometry: Demonstration, Consequences, and Applications}",
    booktitle = {Highlights of Spanish Astrophysics XII},
         year = 2025,
       editor = {{Manteiga}, M. and {Gonz{\'a}lez-Galindo}, F. and {Labiano-Ortega}, A. and {Mart{\'\i}nez-Gonz{\'a}lez}, M.~J. and {Rea}, N. and {Romero-G{\'o}mez}, M. and {Ulla-Miguel}, A. and {Yepes}, G. and {Rodr{\'\i}guez-L{\'o}pez}, C. and {G{\'o}mez-Garc{\'\i}a}, A. and {Dafonte}, C.},
        month = may,
        pages = {223},
          doi = {10.48550/arXiv.2407.21388},
archivePrefix = {arXiv},
       eprint = {2407.21388},
 primaryClass = {astro-ph.IM},
       adsurl = {https://ui.adsabs.harvard.edu/abs/2025hsa..conf..223M}
}

@BOOK{Schlesinger1930,
       author = {{Schlesinger}, Frank},
        title = "{Catalogue of Bright Stars}",
       publisher = {New Haven, Conn., The Tuttle, Morehouse \& Taylor company},
         year = 1930,
       adsurl = {https://ui.adsabs.harvard.edu/abs/1930cbs..book.....S}
}

@ARTICLE{Fabricius2021,
       author = {{Fabricius}, C. and {Luri}, X. and {Arenou}, F. and {Babusiaux}, C. and {Helmi}, A. and {Muraveva}, T. and {Reyl{\'e}}, C. and {Spoto}, F. and {Vallenari}, A. and {Antoja}, T. and {Balbinot}, E. and {Barache}, C. and {Bauchet}, N. and {Bragaglia}, A. and {Busonero}, D. and {Cantat-Gaudin}, T. and {Carrasco}, J.~M. and {Diakit{\'e}}, S. and {Fabrizio}, M. and {Figueras}, F. and {Garcia-Gutierrez}, A. and {Garofalo}, A. and {Jordi}, C. and {Kervella}, P. and {Khanna}, S. and {Leclerc}, N. and {Licata}, E. and {Lambert}, S. and {Marrese}, P.~M. and {Masip}, A. and {Ramos}, P. and {Robichon}, N. and {Robin}, A.~C. and {Romero-G{\'o}mez}, M. and {Rubele}, S. and {Weiler}, M.},
        title = "{Gaia Early Data Release 3. Catalogue validation}",
      journal = {\aap},
         year = 2021,
        month = may,
       volume = {649},
          eid = {A5},
        pages = {A5},
          doi = {10.1051/0004-6361/202039834},
archivePrefix = {arXiv},
       eprint = {2012.06242},
 primaryClass = {astro-ph.GA},
       adsurl = {https://ui.adsabs.harvard.edu/abs/2021A&A...649A...5F}
}

@BOOK{Hoffleit1991,
       author = {{Hoffleit}, Dorrit and {Jaschek}, Carlos},
        title = "{The Bright star catalogue}",
         year = 1991,
        publisher = "Yale University Observatory",
       adsurl = {https://ui.adsabs.harvard.edu/abs/1991bsc..book.....H}
}

@ARTICLE{Maiz2025,
       author = {{Ma{\'\i}z Apell{\'a}niz}, J.},
        title = "{Gaia and massive stars in 2025}",
      journal = {arXiv e-prints},
         year = 2025,
        month = oct,
          eid = {arXiv:2510.21385},
        pages = {arXiv:2510.21385},
          doi = {10.48550/arXiv.2510.21385},
archivePrefix = {arXiv},
       eprint = {2510.21385},
 primaryClass = {astro-ph.SR},
       adsurl = {https://ui.adsabs.harvard.edu/abs/2025arXiv251021385M}
}

@INPROCEEDINGS{Fremat2024,
       author = {{Fr{\'e}mat}, Y.},
        title = "{Astrophysical Parameters associated to 'Hot' stars in Gaia DR3}",
    booktitle = {EES2023: Proceedings of the Evry Schatzman School},
         year = 2024,
       editor = {{Babusiaux}, C. and {Reyl{\'e}}, C.},
        month = aug,
        pages = {59},
          doi = {10.48550/arXiv.2411.04887},
archivePrefix = {arXiv},
       eprint = {2411.04887},
 primaryClass = {astro-ph.SR},
       adsurl = {https://ui.adsabs.harvard.edu/abs/2024ees..conf...59F}
}

@ARTICLE{Ardevol2023,
       author = {{Ard{\`e}vol}, J. and {Mongui{\'o}}, M. and {Figueras}, F. and {Romero-G{\'o}mez}, M. and {Carrasco}, J.~M.},
        title = "{Exploring the structure and kinematics of the Milky Way through A stars}",
      journal = {\aap},
         year = 2023,
        month = oct,
       volume = {678},
          eid = {A111},
        pages = {A111},
          doi = {10.1051/0004-6361/202346925},
archivePrefix = {arXiv},
       eprint = {2308.01901},
 primaryClass = {astro-ph.GA},
       adsurl = {https://ui.adsabs.harvard.edu/abs/2023A&A...678A.111A}
}

@ARTICLE{Lindegren2021,
       author = {{Lindegren}, L. and {Klioner}, S.~A. and {Hern{\'a}ndez}, J. and {Bombrun}, A. and {Ramos-Lerate}, M. and {Steidelm{\"u}ller}, H. and {Bastian}, U. and {Biermann}, M. and {de Torres}, A. and {Gerlach}, E. and {Geyer}, R. and {Hilger}, T. and {Hobbs}, D. and {Lammers}, U. and {McMillan}, P.~J. and {Stephenson}, C.~A. and {Casta{\~n}eda}, J. and {Davidson}, M. and {Fabricius}, C. and {Gracia-Abril}, G. and {Portell}, J. and {Rowell}, N. and {Teyssier}, D. and {Torra}, F. and {Bartolom{\'e}}, S. and {Clotet}, M. and {Garralda}, N. and {Gonz{\'a}lez-Vidal}, J.~J. and {Torra}, J. and {Abbas}, U. and {Altmann}, M. and {Anglada Varela}, E. and {Balaguer-N{\'u}{\~n}ez}, L. and {Balog}, Z. and {Barache}, C. and {Becciani}, U. and {Bernet}, M. and {Bertone}, S. and {Bianchi}, L. and {Bouquillon}, S. and {Brown}, A.~G.~A. and {Bucciarelli}, B. and {Busonero}, D. and {Butkevich}, A.~G. and {Buzzi}, R. and {Cancelliere}, R. and {Carlucci}, T. and {Charlot}, P. and {Cioni}, M.-R.~L. and {Crosta}, M. and {Crowley}, C. and {del Peloso}, E.~F. and {del Pozo}, E. and {Drimmel}, R. and {Esquej}, P. and {Fienga}, A. and {Fraile}, E. and {Gai}, M. and {Garcia-Reinaldos}, M. and {Guerra}, R. and {Hambly}, N.~C. and {Hauser}, M. and {Jan{\ss}en}, K. and {Jordan}, S. and {Kostrzewa-Rutkowska}, Z. and {Lattanzi}, M.~G. and {Liao}, S. and {Licata}, E. and {Lister}, T.~A. and {L{\"o}ffler}, W. and {Marchant}, J.~M. and {Masip}, A. and {Mignard}, F. and {Mints}, A. and {Molina}, D. and {Mora}, A. and {Morbidelli}, R. and {Murphy}, C.~P. and {Pagani}, C. and {Panuzzo}, P. and {Pe{\~n}alosa Esteller}, X. and {Poggio}, E. and {Re Fiorentin}, P. and {Riva}, A. and {Sagrist{\`a} Sell{\'e}s}, A. and {Sanchez Gimenez}, V. and {Sarasso}, M. and {Sciacca}, E. and {Siddiqui}, H.~I. and {Smart}, R.~L. and {Souami}, D. and {Spagna}, A. and {Steele}, I.~A. and {Taris}, F. and {Utrilla}, E. and {van Reeven}, W. and {Vecchiato}, A.},
        title = "{Gaia Early Data Release 3. The astrometric solution}",
      journal = {\aap},
         year = 2021,
        month = may,
       volume = {649},
          eid = {A2},
        pages = {A2},
          doi = {10.1051/0004-6361/202039709},
archivePrefix = {arXiv},
       eprint = {2012.03380},
 primaryClass = {astro-ph.IM},
       adsurl = {https://ui.adsabs.harvard.edu/abs/2021A&A...649A...2L}
}

@ARTICLE{DucheneKraus2013,
       author = {{Duch{\^e}ne}, Gaspard and {Kraus}, Adam},
        title = "{Stellar Multiplicity}",
      journal = {\araa},
         year = 2013,
        month = aug,
       volume = {51},
       number = {1},
        pages = {269-310},
          doi = {10.1146/annurev-astro-081710-102602},
archivePrefix = {arXiv},
       eprint = {1303.3028},
 primaryClass = {astro-ph.SR},
       adsurl = {https://ui.adsabs.harvard.edu/abs/2013ARA&A..51..269D}
}

@ARTICLE{Hosek2020,
       author = {{Hosek}, Jr., Matthew W. and {Lu}, Jessica R. and {Lam}, Casey Y. and {Gautam}, Abhimat K. and {Lockhart}, Kelly E. and {Kim}, Dongwon and {Jia}, Siyao},
        title = "{SPISEA: A Python-based Simple Stellar Population Synthesis Code for Star Clusters}",
      journal = {\aj},
         year = 2020,
        month = sep,
       volume = {160},
       number = {3},
          eid = {143},
        pages = {143},
          doi = {10.3847/1538-3881/aba533},
archivePrefix = {arXiv},
       eprint = {2006.06691},
 primaryClass = {astro-ph.SR},
       adsurl = {https://ui.adsabs.harvard.edu/abs/2020AJ....160..143H}
}

@ARTICLE{Marigo2017,
       author = {{Marigo}, Paola and {Girardi}, L{\'e}o and {Bressan}, Alessandro and {Rosenfield}, Philip and {Aringer}, Bernhard and {Chen}, Yang and {Dussin}, Marco and {Nanni}, Ambra and {Pastorelli}, Giada and {Rodrigues}, Tha{\'\i}se S. and {Trabucchi}, Michele and {Bladh}, Sara and {Dalcanton}, Julianne and {Groenewegen}, Martin A.~T. and {Montalb{\'a}n}, Josefina and {Wood}, Peter R.},
        title = "{A New Generation of PARSEC-COLIBRI Stellar Isochrones Including the TP-AGB Phase}",
      journal = {\apj},
         year = 2017,
        month = jan,
       volume = {835},
       number = {1},
          eid = {77},
        pages = {77},
          doi = {10.3847/1538-4357/835/1/77},
archivePrefix = {arXiv},
       eprint = {1701.08510},
 primaryClass = {astro-ph.SR},
       adsurl = {https://ui.adsabs.harvard.edu/abs/2017ApJ...835...77M}
}

@INPROCEEDINGS{CastelliKurucz,
       author = {{Castelli}, F. and {Kurucz}, R.~L.},
        title = "{New Grids of ATLAS9 Model Atmospheres}",
    booktitle = {Modelling of Stellar Atmospheres},
         year = 2003,
       editor = {{Piskunov}, N. and {Weiss}, W.~W. and {Gray}, D.~F.},
       series = {IAU Symposium},
       volume = {210},
        month = jan,
        pages = {A20},
          doi = {10.48550/arXiv.astro-ph/0405087},
archivePrefix = {arXiv},
       eprint = {astro-ph/0405087},
 primaryClass = {astro-ph},
       adsurl = {https://ui.adsabs.harvard.edu/abs/2003IAUS..210P.A20C}
}

@ARTICLE{Husser2013,
       author = {{Husser}, T.-O. and {Wende-von Berg}, S. and {Dreizler}, S. and {Homeier}, D. and {Reiners}, A. and {Barman}, T. and {Hauschildt}, P.~H.},
        title = "{A new extensive library of PHOENIX stellar atmospheres and synthetic spectra}",
      journal = {\aap},
         year = 2013,
        month = may,
       volume = {553},
          eid = {A6},
        pages = {A6},
          doi = {10.1051/0004-6361/201219058},
archivePrefix = {arXiv},
       eprint = {1303.5632},
 primaryClass = {astro-ph.SR},
       adsurl = {https://ui.adsabs.harvard.edu/abs/2013A&A...553A...6H}
}

@ARTICLE{LadaLada2003,
       author = {{Lada}, Charles J. and {Lada}, Elizabeth A.},
        title = "{Embedded Clusters in Molecular Clouds}",
      journal = {\araa},
         year = 2003,
        month = jan,
       volume = {41},
        pages = {57-115},
          doi = {10.1146/annurev.astro.41.011802.094844},
archivePrefix = {arXiv},
       eprint = {astro-ph/0301540},
 primaryClass = {astro-ph},
       adsurl = {https://ui.adsabs.harvard.edu/abs/2003ARA&A..41...57L}
}

@ARTICLE{Almeida2025,
       author = {{Almeida}, Duarte and {Moitinho}, Andr{\'e} and {Moreira}, Sandro},
        title = "{Open cluster dissolution rate and the initial cluster mass function in the solar neighbourhood: Modelling the age and mass distributions of clusters observed by Gaia}",
      journal = {\aap},
         year = 2025,
        month = jan,
       volume = {693},
          eid = {A305},
        pages = {A305},
          doi = {10.1051/0004-6361/202451853},
archivePrefix = {arXiv},
       eprint = {2412.19204},
 primaryClass = {astro-ph.GA},
       adsurl = {https://ui.adsabs.harvard.edu/abs/2025A&A...693A.305A}
}

@ARTICLE{Bonatto2006,
       author = {{Bonatto}, C. and {Kerber}, L.~O. and {Bica}, E. and {Santiago}, B.~X.},
        title = "{Probing disk properties with open clusters}",
      journal = {\aap},
         year = 2006,
        month = jan,
       volume = {446},
       number = {1},
        pages = {121-135},
          doi = {10.1051/0004-6361:20053573},
archivePrefix = {arXiv},
       eprint = {astro-ph/0509804},
 primaryClass = {astro-ph},
       adsurl = {https://ui.adsabs.harvard.edu/abs/2006A&A...446..121B}
}

@ARTICLE{BlandHawthornGerhard2016,
       author = {{Bland-Hawthorn}, Joss and {Gerhard}, Ortwin},
        title = "{The Galaxy in Context: Structural, Kinematic, and Integrated Properties}",
      journal = {\araa},
         year = 2016,
        month = sep,
       volume = {54},
        pages = {529-596},
          doi = {10.1146/annurev-astro-081915-023441},
archivePrefix = {arXiv},
       eprint = {1602.07702},
 primaryClass = {astro-ph.GA},
       adsurl = {https://ui.adsabs.harvard.edu/abs/2016ARA&A..54..529B}
}

@ARTICLE{vanderKruit2011,
       author = {{van der Kruit}, P.~C. and {Freeman}, K.~C.},
        title = "{Galaxy Disks}",
      journal = {\araa},
         year = 2011,
        month = sep,
       volume = {49},
       number = {1},
        pages = {301-371},
          doi = {10.1146/annurev-astro-083109-153241},
archivePrefix = {arXiv},
       eprint = {1101.1771},
 primaryClass = {astro-ph.GA},
       adsurl = {https://ui.adsabs.harvard.edu/abs/2011ARA&A..49..301V}
}

@INPROCEEDINGS{Elmegreen2011,
       author = {{Elmegreen}, B.~G.},
        title = "{Star Formation in Spiral Arms}",
    booktitle = {EAS Publications Series},
         year = 2011,
       editor = {{Charbonnel}, Corinne and {Montmerle}, Thierry},
       series = {EAS Publications Series},
       volume = {51},
        month = nov,
    publisher = {EDP},
        pages = {19-30},
          doi = {10.1051/eas/1151002},
archivePrefix = {arXiv},
       eprint = {1101.3109},
 primaryClass = {astro-ph.GA},
       adsurl = {https://ui.adsabs.harvard.edu/abs/2011EAS....51...19E}
}

@ARTICLE{CastroGinard2021,
       author = {{Castro-Ginard}, A. and {McMillan}, P.~J. and {Luri}, X. and {Jordi}, C. and {Romero-G{\'o}mez}, M. and {Cantat-Gaudin}, T. and {Casamiquela}, L. and {Tarricq}, Y. and {Soubiran}, C. and {Anders}, F.},
        title = "{Milky Way spiral arms from open clusters in Gaia EDR3}",
      journal = {\aap},
         year = 2021,
        month = aug,
       volume = {652},
          eid = {A162},
        pages = {A162},
          doi = {10.1051/0004-6361/202039751},
archivePrefix = {arXiv},
       eprint = {2105.04590},
 primaryClass = {astro-ph.GA},
       adsurl = {https://ui.adsabs.harvard.edu/abs/2021A&A...652A.162C}
}

@ARTICLE{Rezeai2018,
       author = {{Rezaei Kh.}, Sara and {Bailer-Jones}, Coryn A.~L. and {Hogg}, David W. and {Schultheis}, Mathias},
        title = "{Detection of the Milky Way spiral arms in dust from 3D mapping}",
      journal = {\aap},
         year = 2018,
        month = oct,
       volume = {618},
          eid = {A168},
        pages = {A168},
          doi = {10.1051/0004-6361/201833284},
archivePrefix = {arXiv},
       eprint = {1808.00015},
 primaryClass = {astro-ph.GA},
       adsurl = {https://ui.adsabs.harvard.edu/abs/2018A&A...618A.168R}
}

@ARTICLE{Gaia,
       author = {{Gaia Collaboration} and {Prusti}, T. and {de Bruijne}, J.~H.~J. and {Brown}, A.~G.~A. and {Vallenari}, A. and {Babusiaux}, C. and {Bailer-Jones}, C.~A.~L. and {Bastian}, U. and {Biermann}, M. and {Evans}, D.~W. and {Eyer}, L. and {Jansen}, F. and {Jordi}, C. and {Klioner}, S.~A. and {Lammers}, U. and {Lindegren}, L. and {Luri}, X. and {Mignard}, F. and {Milligan}, D.~J. and {Panem}, C. and {Poinsignon}, V. and {Pourbaix}, D. and {Randich}, S. and {Sarri}, G. and {Sartoretti}, P. and {Siddiqui}, H.~I. and {Soubiran}, C. and {Valette}, V. and {van Leeuwen}, F. and {Walton}, N.~A. and {Aerts}, C. and {Arenou}, F. and {Cropper}, M. and {Drimmel}, R. and {H{\o}g}, E. and {Katz}, D. and {Lattanzi}, M.~G. and {O'Mullane}, W. and {Grebel}, E.~K. and {Holland}, A.~D. and {Huc}, C. and {Passot}, X. and {Bramante}, L. and {Cacciari}, C. and {Casta{\~n}eda}, J. and {Chaoul}, L. and {Cheek}, N. and {De Angeli}, F. and {Fabricius}, C. and {Guerra}, R. and {Hern{\'a}ndez}, J. and {Jean-Antoine-Piccolo}, A. and {Masana}, E. and {Messineo}, R. and {Mowlavi}, N. and {Nienartowicz}, K. and {Ord{\'o}{\~n}ez-Blanco}, D. and {Panuzzo}, P. and {Portell}, J. and {Richards}, P.~J. and {Riello}, M. and {Seabroke}, G.~M. and {Tanga}, P. and {Th{\'e}venin}, F. and {Torra}, J. and {Els}, S.~G. and {Gracia-Abril}, G. and {Comoretto}, G. and {Garcia-Reinaldos}, M. and {Lock}, T. and {Mercier}, E. and {Altmann}, M. and {Andrae}, R. and {Astraatmadja}, T.~L. and {Bellas-Velidis}, I. and {Benson}, K. and {Berthier}, J. and {Blomme}, R. and {Busso}, G. and {Carry}, B. and {Cellino}, A. and {Clementini}, G. and {Cowell}, S. and {Creevey}, O. and {Cuypers}, J. and {Davidson}, M. and {De Ridder}, J. and {de Torres}, A. and {Delchambre}, L. and {Dell'Oro}, A. and {Ducourant}, C. and {Fr{\'e}mat}, Y. and {Garc{\'\i}a-Torres}, M. and {Gosset}, E. and {Halbwachs}, J. -L. and {Hambly}, N.~C. and {Harrison}, D.~L. and {Hauser}, M. and {Hestroffer}, D. and {Hodgkin}, S.~T. and {Huckle}, H.~E. and {Hutton}, A. and {Jasniewicz}, G. and {Jordan}, S. and {Kontizas}, M. and {Korn}, A.~J. and {Lanzafame}, A.~C. and {Manteiga}, M. and {Moitinho}, A. and {Muinonen}, K. and {Osinde}, J. and {Pancino}, E. and {Pauwels}, T. and {Petit}, J. -M. and {Recio-Blanco}, A. and {Robin}, A.~C. and {Sarro}, L.~M. and {Siopis}, C. and {Smith}, M. and {Smith}, K.~W. and {Sozzetti}, A. and {Thuillot}, W. and {van Reeven}, W. and {Viala}, Y. and {Abbas}, U. and {Abreu Aramburu}, A. and {Accart}, S. and {Aguado}, J.~J. and {Allan}, P.~M. and {Allasia}, W. and {Altavilla}, G. and {{\'A}lvarez}, M.~A. and {Alves}, J. and {Anderson}, R.~I. and {Andrei}, A.~H. and {Anglada Varela}, E. and {Antiche}, E. and {Antoja}, T. and {Ant{\'o}n}, S. and {Arcay}, B. and {Atzei}, A. and {Ayache}, L. and {Bach}, N. and {Baker}, S.~G. and {Balaguer-N{\'u}{\~n}ez}, L. and {Barache}, C. and {Barata}, C. and {Barbier}, A. and {Barblan}, F. and {Baroni}, M. and {Barrado y Navascu{\'e}s}, D. and {Barros}, M. and {Barstow}, M.~A. and {Becciani}, U. and {Bellazzini}, M. and {Bellei}, G. and {Bello Garc{\'\i}a}, A. and {Belokurov}, V. and {Bendjoya}, P. and {Berihuete}, A. and {Bianchi}, L. and {Bienaym{\'e}}, O. and {Billebaud}, F. and {Blagorodnova}, N. and {Blanco-Cuaresma}, S. and {Boch}, T. and {Bombrun}, A. and {Borrachero}, R. and {Bouquillon}, S. and {Bourda}, G. and {Bouy}, H. and {Bragaglia}, A. and {Breddels}, M.~A. and {Brouillet}, N. and {Br{\"u}semeister}, T. and {Bucciarelli}, B. and {Budnik}, F. and {Burgess}, P. and {Burgon}, R. and {Burlacu}, A. and {Busonero}, D. and {Buzzi}, R. and {Caffau}, E. and {Cambras}, J. and {Campbell}, H. and {Cancelliere}, R. and {Cantat-Gaudin}, T. and {Carlucci}, T. and {Carrasco}, J.~M. and {Castellani}, M. and {Charlot}, P. and {Charnas}, J. and {Charvet}, P. and {Chassat}, F. and {Chiavassa}, A. and {Clotet}, M. and {Cocozza}, G. and {Collins}, R.~S. and {Collins}, P. and {Costigan}, G. and {Crifo}, F. and {Cross}, N.~J.~G. and {Crosta}, M. and {Crowley}, C. and {Dafonte}, C. and {Damerdji}, Y. and {Dapergolas}, A. and {David}, P. and {David}, M. and {De Cat}, P. and {de Felice}, F. and {de Laverny}, P. and {De Luise}, F. and {De March}, R. and {de Martino}, D. and {de Souza}, R. and {Debosscher}, J. and {del Pozo}, E. and {Delbo}, M. and {Delgado}, A. and {Delgado}, H.~E. and {di Marco}, F. and {Di Matteo}, P. and {Diakite}, S. and {Distefano}, E. and {Dolding}, C. and {Dos Anjos}, S. and {Drazinos}, P. and {Dur{\'a}n}, J. and {Dzigan}, Y. and {Ecale}, E. and {Edvardsson}, B. and {Enke}, H. and {Erdmann}, M. and {Escolar}, D. and {Espina}, M. and {Evans}, N.~W. and {Eynard Bontemps}, G. and {Fabre}, C. and {Fabrizio}, M. and {Faigler}, S. and {Falc{\~a}o}, A.~J. and {Farr{\`a}s Casas}, M. and {Faye}, F. and {Federici}, L. and {Fedorets}, G. and {Fern{\'a}ndez-Hern{\'a}ndez}, J. and {Fernique}, P. and {Fienga}, A. and {Figueras}, F. and {Filippi}, F. and {Findeisen}, K. and {Fonti}, A. and {Fouesneau}, M. and {Fraile}, E. and {Fraser}, M. and {Fuchs}, J. and {Furnell}, R. and {Gai}, M. and {Galleti}, S. and {Galluccio}, L. and {Garabato}, D. and {Garc{\'\i}a-Sedano}, F. and {Gar{\'e}}, P. and {Garofalo}, A. and {Garralda}, N. and {Gavras}, P. and {Gerssen}, J. and {Geyer}, R. and {Gilmore}, G. and {Girona}, S. and {Giuffrida}, G. and {Gomes}, M. and {Gonz{\'a}lez-Marcos}, A. and {Gonz{\'a}lez-N{\'u}{\~n}ez}, J. and {Gonz{\'a}lez-Vidal}, J.~J. and {Granvik}, M. and {Guerrier}, A. and {Guillout}, P. and {Guiraud}, J. and {G{\'u}rpide}, A. and {Guti{\'e}rrez-S{\'a}nchez}, R. and {Guy}, L.~P. and {Haigron}, R. and {Hatzidimitriou}, D. and {Haywood}, M. and {Heiter}, U. and {Helmi}, A. and {Hobbs}, D. and {Hofmann}, W. and {Holl}, B. and {Holland}, G. and {Hunt}, J.~A.~S. and {Hypki}, A. and {Icardi}, V. and {Irwin}, M. and {Jevardat de Fombelle}, G. and {Jofr{\'e}}, P. and {Jonker}, P.~G. and {Jorissen}, A. and {Julbe}, F. and {Karampelas}, A. and {Kochoska}, A. and {Kohley}, R. and {Kolenberg}, K. and {Kontizas}, E. and {Koposov}, S.~E. and {Kordopatis}, G. and {Koubsky}, P. and {Kowalczyk}, A. and {Krone-Martins}, A. and {Kudryashova}, M. and {Kull}, I. and {Bachchan}, R.~K. and {Lacoste-Seris}, F. and {Lanza}, A.~F. and {Lavigne}, J. -B. and {Le Poncin-Lafitte}, C. and {Lebreton}, Y. and {Lebzelter}, T. and {Leccia}, S. and {Leclerc}, N. and {Lecoeur-Taibi}, I. and {Lemaitre}, V. and {Lenhardt}, H. and {Leroux}, F. and {Liao}, S. and {Licata}, E. and {Lindstr{\o}m}, H.~E.~P. and {Lister}, T.~A. and {Livanou}, E. and {Lobel}, A. and {L{\"o}ffler}, W. and {L{\'o}pez}, M. and {Lopez-Lozano}, A. and {Lorenz}, D. and {Loureiro}, T. and {MacDonald}, I. and {Magalh{\~a}es Fernandes}, T. and {Managau}, S. and {Mann}, R.~G. and {Mantelet}, G. and {Marchal}, O. and {Marchant}, J.~M. and {Marconi}, M. and {Marie}, J. and {Marinoni}, S. and {Marrese}, P.~M. and {Marschalk{\'o}}, G. and {Marshall}, D.~J. and {Mart{\'\i}n-Fleitas}, J.~M. and {Martino}, M. and {Mary}, N. and {Matijevi{\v{c}}}, G. and {Mazeh}, T. and {McMillan}, P.~J. and {Messina}, S. and {Mestre}, A. and {Michalik}, D. and {Millar}, N.~R. and {Miranda}, B.~M.~H. and {Molina}, D. and {Molinaro}, R. and {Molinaro}, M. and {Moln{\'a}r}, L. and {Moniez}, M. and {Montegriffo}, P. and {Monteiro}, D. and {Mor}, R. and {Mora}, A. and {Morbidelli}, R. and {Morel}, T. and {Morgenthaler}, S. and {Morley}, T. and {Morris}, D. and {Mulone}, A.~F. and {Muraveva}, T. and {Musella}, I. and {Narbonne}, J. and {Nelemans}, G. and {Nicastro}, L. and {Noval}, L. and {Ord{\'e}novic}, C. and {Ordieres-Mer{\'e}}, J. and {Osborne}, P. and {Pagani}, C. and {Pagano}, I. and {Pailler}, F. and {Palacin}, H. and {Palaversa}, L. and {Parsons}, P. and {Paulsen}, T. and {Pecoraro}, M. and {Pedrosa}, R. and {Pentik{\"a}inen}, H. and {Pereira}, J. and {Pichon}, B. and {Piersimoni}, A.~M. and {Pineau}, F. -X. and {Plachy}, E. and {Plum}, G. and {Poujoulet}, E. and {Pr{\v{s}}a}, A. and {Pulone}, L. and {Ragaini}, S. and {Rago}, S. and {Rambaux}, N. and {Ramos-Lerate}, M. and {Ranalli}, P. and {Rauw}, G. and {Read}, A. and {Regibo}, S. and {Renk}, F. and {Reyl{\'e}}, C. and {Ribeiro}, R.~A. and {Rimoldini}, L. and {Ripepi}, V. and {Riva}, A. and {Rixon}, G. and {Roelens}, M. and {Romero-G{\'o}mez}, M. and {Rowell}, N. and {Royer}, F. and {Rudolph}, A. and {Ruiz-Dern}, L. and {Sadowski}, G. and {Sagrist{\`a} Sell{\'e}s}, T. and {Sahlmann}, J. and {Salgado}, J. and {Salguero}, E. and {Sarasso}, M. and {Savietto}, H. and {Schnorhk}, A. and {Schultheis}, M. and {Sciacca}, E. and {Segol}, M. and {Segovia}, J.~C. and {Segransan}, D. and {Serpell}, E. and {Shih}, I. -C. and {Smareglia}, R. and {Smart}, R.~L. and {Smith}, C. and {Solano}, E. and {Solitro}, F. and {Sordo}, R. and {Soria Nieto}, S. and {Souchay}, J. and {Spagna}, A. and {Spoto}, F. and {Stampa}, U. and {Steele}, I.~A. and {Steidelm{\"u}ller}, H. and {Stephenson}, C.~A. and {Stoev}, H. and {Suess}, F.~F. and {S{\"u}veges}, M. and {Surdej}, J. and {Szabados}, L. and {Szegedi-Elek}, E. and {Tapiador}, D. and {Taris}, F. and {Tauran}, G. and {Taylor}, M.~B. and {Teixeira}, R. and {Terrett}, D. and {Tingley}, B. and {Trager}, S.~C. and {Turon}, C. and {Ulla}, A. and {Utrilla}, E. and {Valentini}, G. and {van Elteren}, A. and {Van Hemelryck}, E. and {van Leeuwen}, M. and {Varadi}, M. and {Vecchiato}, A. and {Veljanoski}, J. and {Via}, T. and {Vicente}, D. and {Vogt}, S. and {Voss}, H. and {Votruba}, V. and {Voutsinas}, S. and {Walmsley}, G. and {Weiler}, M. and {Weingrill}, K. and {Werner}, D. and {Wevers}, T. and {Whitehead}, G. and {Wyrzykowski}, {\L}. and {Yoldas}, A. and {{\v{Z}}erjal}, M. and {Zucker}, S. and {Zurbach}, C. and {Zwitter}, T. and {Alecu}, A. and {Allen}, M. and {Allende Prieto}, C. and {Amorim}, A. and {Anglada-Escud{\'e}}, G. and {Arsenijevic}, V. and {Azaz}, S. and {Balm}, P. and {Beck}, M. and {Bernstein}, H. -H. and {Bigot}, L. and {Bijaoui}, A. and {Blasco}, C. and {Bonfigli}, M. and {Bono}, G. and {Boudreault}, S. and {Bressan}, A. and {Brown}, S. and {Brunet}, P. -M. and {Bunclark}, P. and {Buonanno}, R. and {Butkevich}, A.~G. and {Carret}, C. and {Carrion}, C. and {Chemin}, L. and {Ch{\'e}reau}, F. and {Corcione}, L. and {Darmigny}, E. and {de Boer}, K.~S. and {de Teodoro}, P. and {de Zeeuw}, P.~T. and {Delle Luche}, C. and {Domingues}, C.~D. and {Dubath}, P. and {Fodor}, F. and {Fr{\'e}zouls}, B. and {Fries}, A. and {Fustes}, D. and {Fyfe}, D. and {Gallardo}, E. and {Gallegos}, J. and {Gardiol}, D. and {Gebran}, M. and {Gomboc}, A. and {G{\'o}mez}, A. and {Grux}, E. and {Gueguen}, A. and {Heyrovsky}, A. and {Hoar}, J. and {Iannicola}, G. and {Isasi Parache}, Y. and {Janotto}, A. -M. and {Joliet}, E. and {Jonckheere}, A. and {Keil}, R. and {Kim}, D. -W. and {Klagyivik}, P. and {Klar}, J. and {Knude}, J. and {Kochukhov}, O. and {Kolka}, I. and {Kos}, J. and {Kutka}, A. and {Lainey}, V. and {LeBouquin}, D. and {Liu}, C. and {Loreggia}, D. and {Makarov}, V.~V. and {Marseille}, M.~G. and {Martayan}, C. and {Martinez-Rubi}, O. and {Massart}, B. and {Meynadier}, F. and {Mignot}, S. and {Munari}, U. and {Nguyen}, A. -T. and {Nordlander}, T. and {Ocvirk}, P. and {O'Flaherty}, K.~S. and {Olias Sanz}, A. and {Ortiz}, P. and {Osorio}, J. and {Oszkiewicz}, D. and {Ouzounis}, A. and {Palmer}, M. and {Park}, P. and {Pasquato}, E. and {Peltzer}, C. and {Peralta}, J. and {P{\'e}turaud}, F. and {Pieniluoma}, T. and {Pigozzi}, E. and {Poels}, J. and {Prat}, G. and {Prod'homme}, T. and {Raison}, F. and {Rebordao}, J.~M. and {Risquez}, D. and {Rocca-Volmerange}, B. and {Rosen}, S. and {Ruiz-Fuertes}, M.~I. and {Russo}, F. and {Sembay}, S. and {Serraller Vizcaino}, I. and {Short}, A. and {Siebert}, A. and {Silva}, H. and {Sinachopoulos}, D. and {Slezak}, E. and {Soffel}, M. and {Sosnowska}, D. and {Strai{\v{z}}ys}, V. and {ter Linden}, M. and {Terrell}, D. and {Theil}, S. and {Tiede}, C. and {Troisi}, L. and {Tsalmantza}, P. and {Tur}, D. and {Vaccari}, M. and {Vachier}, F. and {Valles}, P. and {Van Hamme}, W. and {Veltz}, L. and {Virtanen}, J. and {Wallut}, J. -M. and {Wichmann}, R. and {Wilkinson}, M.~I. and {Ziaeepour}, H. and {Zschocke}, S.},
        title = "{The Gaia mission}",
      journal = {\aap},
         year = 2016,
        month = nov,
       volume = {595},
          eid = {A1},
        pages = {A1},
          doi = {10.1051/0004-6361/20162927210.48550/arXiv.1609.04153},
archivePrefix = {arXiv},
       eprint = {1609.04153},
 primaryClass = {astro-ph.IM},
       adsurl = {https://ui.adsabs.harvard.edu/abs/2016A&A...595A...1G}
}

@INPROCEEDINGS{Zucker2023,
       author = {{Zucker}, C. and {Alves}, J. and {Goodman}, A. and {Meingast}, S. and {Galli}, P.},
        title = "{The Solar Neighborhood in the Age of Gaia}",
    booktitle = {Protostars and Planets VII},
         year = 2023,
       editor = {{Inutsuka}, S. and {Aikawa}, Y. and {Muto}, T. and {Tomida}, K. and {Tamura}, M.},
       series = {Astronomical Society of the Pacific Conference Series},
       volume = {534},
        month = jul,
        pages = {43},
          doi = {10.48550/arXiv.2212.00067},
archivePrefix = {arXiv},
       eprint = {2212.00067},
 primaryClass = {astro-ph.GA},
       adsurl = {https://ui.adsabs.harvard.edu/abs/2023ASPC..534...43Z}
}

@ARTICLE{Kormann2026,
       author = {{Kormann}, Lilly A. and {Alves}, Jo{\~a}o and {Pantaleoni Gonz{\'a}lez}, Michelangelo and {Swiggum}, Cameren and {En{\ss}lin}, Torsten A. and {Edenhofer}, Gordian},
        title = "{The superclouds of the local Milky Way}",
      journal = {\aap},
         year = 2026,
        month = feb,
       volume = {706},
          eid = {A161},
        pages = {A161},
          doi = {10.1051/0004-6361/202556469},
       adsurl = {https://ui.adsabs.harvard.edu/abs/2026A&A...706A.161K}
}

@ARTICLE{Nogueras2019,
       author = {{Nogueras-Lara}, F. and {Sch{\"o}del}, R. and {Najarro}, F. and {Gallego-Calvente}, A.~T. and {Gallego-Cano}, E. and {Shahzamanian}, B. and {Neumayer}, N.},
        title = "{Variability of the near-infrared extinction curve towards the Galactic centre}",
      journal = {\aap},
         year = 2019,
        month = oct,
       volume = {630},
          eid = {L3},
        pages = {L3},
          doi = {10.1051/0004-6361/201936322},
archivePrefix = {arXiv},
       eprint = {1909.02494},
 primaryClass = {astro-ph.SR},
       adsurl = {https://ui.adsabs.harvard.edu/abs/2019A&A...630L...3N}
}

@INPROCEEDINGS{Brown2019,
       author = {{Brown}, Anthony G.~A.},
        title = "{The Future of the Gaia Universe}",
    booktitle = {The Gaia Universe},
         year = 2019,
        month = apr,
          eid = {18},
        pages = {18},
          doi = {10.5281/zenodo.2637972},
       adsurl = {https://ui.adsabs.harvard.edu/abs/2019gaia.confE..18B}
}

@ARTICLE{Crowther2012,
       author = {{Crowther}, Paul},
        title = "{Birth, life and death of massive stars}",
      journal = {Astronomy and Geophysics},
         year = 2012,
        month = aug,
       volume = {53},
       number = {4},
        pages = {4.30-4.36},
          doi = {10.1111/j.1468-4004.2012.53430.x},
       adsurl = {https://ui.adsabs.harvard.edu/abs/2012A&G....53d..30C}
}

@PHDTHESIS{Payne1925,
       AUTHOR = {{Payne}, Cecilia Helena},
        TITLE = "{Stellar Atmospheres; a Contribution to the Observational Study of High Temperature in the Reversing Layers of Stars.}",
       school = {RADCLIFFE COLLEGE.},
         year = 1925,
        month = jan,
       adsurl = {https://ui.adsabs.harvard.edu/abs/1925PhDT.........1P}
}

@ARTICLE{BaumgardatKroupa2007,
       author = {{Baumgardt}, H. and {Kroupa}, P.},
        title = "{A comprehensive set of simulations studying the influence of gas expulsion on star cluster evolution}",
      journal = {\mnras},
         year = 2007,
        month = oct,
       volume = {380},
       number = {4},
        pages = {1589-1598},
          doi = {10.1111/j.1365-2966.2007.12209.x},
archivePrefix = {arXiv},
       eprint = {0707.1944},
 primaryClass = {astro-ph},
       adsurl = {https://ui.adsabs.harvard.edu/abs/2007MNRAS.380.1589B}
}

@ARTICLE{Krumholz2019,
       author = {{Krumholz}, Mark R. and {McKee}, Christopher F. and {Bland-Hawthorn}, Joss},
        title = "{Star Clusters Across Cosmic Time}",
      journal = {\araa},
         year = 2019,
        month = aug,
       volume = {57},
        pages = {227-303},
          doi = {10.1146/annurev-astro-091918-104430},
archivePrefix = {arXiv},
       eprint = {1812.01615},
 primaryClass = {astro-ph.GA},
       adsurl = {https://ui.adsabs.harvard.edu/abs/2019ARA&A..57..227K}
}

@ARTICLE{Jones2013,
       author = {{Jones}, S. and {Hirschi}, R. and {Nomoto}, K. and {Fischer}, T. and {Timmes}, F.~X. and {Herwig}, F. and {Paxton}, B. and {Toki}, H. and {Suzuki}, T. and {Mart{\'\i}nez-Pinedo}, G. and {Lam}, Y.~H. and {Bertolli}, M.~G.},
        title = "{Advanced Burning Stages and Fate of 8-10 M $_{{\ensuremath{\odot}}}$ Stars}",
      journal = {\apj},
         year = 2013,
        month = aug,
       volume = {772},
       number = {2},
          eid = {150},
        pages = {150},
          doi = {10.1088/0004-637X/772/2/150},
archivePrefix = {arXiv},
       eprint = {1306.2030},
 primaryClass = {astro-ph.SR},
       adsurl = {https://ui.adsabs.harvard.edu/abs/2013ApJ...772..150J}
}

@ARTICLE{DeRossi2010,
       author = {{de Rossi}, M.~E. and {Tissera}, P.~B. and {Pedrosa}, S.~E.},
        title = "{Impact of supernova feedback on the Tully-Fisher relation}",
      journal = {\aap},
         year = 2010,
        month = sep,
       volume = {519},
          eid = {A89},
        pages = {A89},
          doi = {10.1051/0004-6361/201014058},
archivePrefix = {arXiv},
       eprint = {1005.4960},
 primaryClass = {astro-ph.CO},
       adsurl = {https://ui.adsabs.harvard.edu/abs/2010A&A...519A..89D}
}

@BOOK{Ambartsumian1947,
       author = {{Ambartsumian}, V.~A.},
        title = "{The evolution of stars and astrophysics}",
         year = 1947,
         publisher = {Armenian SSR Academy of Sciences Press},
       adsurl = {https://ui.adsabs.harvard.edu/abs/1947esa..book.....A}
}

@ARTICLE{McKeeWilliams1997,
       author = {{McKee}, Christopher F. and {Williams}, Jonathan P.},
        title = "{The Luminosity Function of OB Associations in the Galaxy}",
      journal = {\apj},
         year = "1997",
        month = "Feb",
       volume = {476},
       number = {1},
        pages = {144-165},
          doi = {10.1086/303587},
       adsurl = {https://ui.adsabs.harvard.edu/abs/1997ApJ...476..144M}
}

@ARTICLE{4MOST,
       author = {{de Jong}, R.~S. and {Agertz}, O. and {Berbel}, A.~A. and {Aird}, J. and {Alexander}, D.~A. and {Amarsi}, A. and {Anders}, F. and {Andrae}, R. and {Ansarinejad}, B. and {Ansorge}, W. and {Antilogus}, P. and {Anwand-Heerwart}, H. and {Arentsen}, A. and {Arnadottir}, A. and {Asplund}, M. and {Auger}, M. and {Azais}, N. and {Baade}, D. and {Baker}, G. and {Baker}, S. and {Balbinot}, E. and {Baldry}, I.~K. and {Banerji}, M. and {Barden}, S. and {Barklem}, P. and {Barth{\'e}l{\'e}my-Mazot}, E. and {Battistini}, C. and {Bauer}, S. and {Bell}, C.~P.~M. and {Bellido-Tirado}, O. and {Bellstedt}, S. and {Belokurov}, V. and {Bensby}, T. and {Bergemann}, M. and {Bestenlehner}, J.~M. and {Bielby}, R. and {Bilicki}, M. and {Blake}, C. and {Bland-Hawthorn}, J. and {Boeche}, C. and {Boland}, W. and {Boller}, T. and {Bongard}, S. and {Bongiorno}, A. and {Bonifacio}, P. and {Boudon}, D. and {Brooks}, D. and {Brown}, M.~J.~I. and {Brown}, R. and {Br{\"u}ggen}, M. and {Brynnel}, J. and {Brzeski}, J. and {Buchert}, T. and {Buschkamp}, P. and {Caffau}, E. and {Caillier}, P. and {Carrick}, J. and {Casagrande}, L. and {Case}, S. and {Casey}, A. and {Cesarini}, I. and {Cescutti}, G. and {Chapuis}, D. and {Chiappini}, C. and {Childress}, M. and {Christlieb}, N. and {Church}, R. and {Cioni}, M. -R.~L. and {Cluver}, M. and {Colless}, M. and {Collett}, T. and {Comparat}, J. and {Cooper}, A. and {Couch}, W. and {Courbin}, F. and {Croom}, S. and {Croton}, D. and {Daguis{\'e}}, E. and {Dalton}, G. and {Davies}, L.~J.~M. and {Davis}, T. and {de Laverny}, P. and {Deason}, A. and {Dionies}, F. and {Disseau}, K. and {Doel}, P. and {D{\"o}scher}, D. and {Driver}, S.~P. and {Dwelly}, T. and {Eckert}, D. and {Edge}, A. and {Edvardsson}, B. and {Youssoufi}, D.~E. and {Elhaddad}, A. and {Enke}, H. and {Erfanianfar}, G. and {Farrell}, T. and {Fechner}, T. and {Feiz}, C. and {Feltzing}, S. and {Ferreras}, I. and {Feuerstein}, D. and {Feuillet}, D. and {Finoguenov}, A. and {Ford}, D. and {Fotopoulou}, S. and {Fouesneau}, M. and {Frenk}, C. and {Frey}, S. and {Gaessler}, W. and {Geier}, S. and {Gentile Fusillo}, N. and {Gerhard}, O. and {Giannantonio}, T. and {Giannone}, D. and {Gibson}, B. and {Gillingham}, P. and {Gonz{\'a}lez-Fern{\'a}ndez}, C. and {Gonzalez-Solares}, E. and {Gottloeber}, S. and {Gould}, A. and {Grebel}, E.~K. and {Gueguen}, A. and {Guiglion}, G. and {Haehnelt}, M. and {Hahn}, T. and {Hansen}, C.~J. and {Hartman}, H. and {Hauptner}, K. and {Hawkins}, K. and {Haynes}, D. and {Haynes}, R. and {Heiter}, U. and {Helmi}, A. and {Aguayo}, C.~H. and {Hewett}, P. and {Hinton}, S. and {Hobbs}, D. and {Hoenig}, S. and {Hofman}, D. and {Hook}, I. and {Hopgood}, J. and {Hopkins}, A. and {Hourihane}, A. and {Howes}, L. and {Howlett}, C. and {Huet}, T. and {Irwin}, M. and {Iwert}, O. and {Jablonka}, P. and {Jahn}, T. and {Jahnke}, K. and {Jarno}, A. and {Jin}, S. and {Jofre}, P. and {Johl}, D. and {Jones}, D. and {J{\"o}nsson}, H. and {Jordan}, C. and {Karovicova}, I. and {Khalatyan}, A. and {Kelz}, A. and {Kennicutt}, R. and {King}, D. and {Kitaura}, F. and {Klar}, J. and {Klauser}, U. and {Kneib}, J. -P. and {Koch}, A. and {Koposov}, S. and {Kordopatis}, G. and {Korn}, A. and {Kosmalski}, J. and {Kotak}, R. and {Kovalev}, M. and {Kreckel}, K. and {Kripak}, Y. and {Krumpe}, M. and {Kuijken}, K. and {Kunder}, A. and {Kushniruk}, I. and {Lam}, M.~I. and {Lamer}, G. and {Laurent}, F. and {Lawrence}, J. and {Lehmitz}, M. and {Lemasle}, B. and {Lewis}, J. and {Li}, B. and {Lidman}, C. and {Lind}, K. and {Liske}, J. and {Lizon}, J. -L. and {Loveday}, J. and {Ludwig}, H. -G. and {McDermid}, R.~M. and {Maguire}, K. and {Mainieri}, V. and {Mali}, S. and {Mandel}, H. and {Mandel}, K. and {Mannering}, L. and {Martell}, S. and {Martinez Delgado}, D. and {Matijevic}, G. and {McGregor}, H. and {McMahon}, R. and {McMillan}, P. and {Mena}, O. and {Merloni}, A. and {Meyer}, M.~J. and {Michel}, C. and {Micheva}, G. and {Migniau}, J. -E. and {Minchev}, I. and {Monari}, G. and {Muller}, R. and {Murphy}, D. and {Muthukrishna}, D. and {Nandra}, K. and {Navarro}, R. and {Ness}, M. and {Nichani}, V. and {Nichol}, R. and {Nicklas}, H. and {Niederhofer}, F. and {Norberg}, P. and {Obreschkow}, D. and {Oliver}, S. and {Owers}, M. and {Pai}, N. and {Pankratow}, S. and {Parkinson}, D. and {Paschke}, J. and {Paterson}, R. and {Pecontal}, A. and {Parry}, I. and {Phillips}, D. and {Pillepich}, A. and {Pinard}, L. and {Pirard}, J. and {Piskunov}, N. and {Plank}, V. and {Pl{\"u}schke}, D. and {Pons}, E. and {Popesso}, P. and {Power}, C. and {Pragt}, J. and {Pramskiy}, A. and {Pryer}, D. and {Quattri}, M. and {Queiroz}, A.~B. d. A. and {Quirrenbach}, A. and {Rahurkar}, S. and {Raichoor}, A. and {Ramstedt}, S. and {Rau}, A. and {Recio-Blanco}, A. and {Reiss}, R. and {Renaud}, F. and {Revaz}, Y. and {Rhode}, P. and {Richard}, J. and {Richter}, A.~D. and {Rix}, H. -W. and {Robotham}, A.~S.~G. and {Roelfsema}, R. and {Romaniello}, M. and {Rosario}, D. and {Rothmaier}, F. and {Roukema}, B. and {Ruchti}, G. and {Rupprecht}, G. and {Rybizki}, J. and {Ryde}, N. and {Saar}, A. and {Sadler}, E. and {Sahl{\'e}n}, M. and {Salvato}, M. and {Sassolas}, B. and {Saunders}, W. and {Saviauk}, A. and {Sbordone}, L. and {Schmidt}, T. and {Schnurr}, O. and {Scholz}, R. -D. and {Schwope}, A. and {Seifert}, W. and {Shanks}, T. and {Sheinis}, A. and {Sivov}, T. and {Sk{\'u}lad{\'o}ttir}, {\'A}. and {Smartt}, S. and {Smedley}, S. and {Smith}, G. and {Smith}, R. and {Sorce}, J. and {Spitler}, L. and {Starkenburg}, E. and {Steinmetz}, M. and {Stilz}, I. and {Storm}, J. and {Sullivan}, M. and {Sutherland}, W. and {Swann}, E. and {Tamone}, A. and {Taylor}, E.~N. and {Teillon}, J. and {Tempel}, E. and {ter Horst}, R. and {Thi}, W. -F. and {Tolstoy}, E. and {Trager}, S. and {Traven}, G. and {Tremblay}, P. -E. and {Tresse}, L. and {Valentini}, M. and {van de Weygaert}, R. and {van den Ancker}, M. and {Veljanoski}, J. and {Venkatesan}, S. and {Wagner}, L. and {Wagner}, K. and {Walcher}, C.~J. and {Waller}, L. and {Walton}, N. and {Wang}, L. and {Winkler}, R. and {Wisotzki}, L. and {Worley}, C.~C. and {Worseck}, G. and {Xiang}, M. and {Xu}, W. and {Yong}, D. and {Zhao}, C. and {Zheng}, J. and {Zscheyge}, F. and {Zucker}, D.},
        title = "{4MOST: Project overview and information for the First Call for Proposals}",
      journal = {The Messenger},
         year = 2019,
        month = mar,
       volume = {175},
        pages = {3-11},
          doi = {10.18727/0722-6691/5117},
archivePrefix = {arXiv},
       eprint = {1903.02464},
 primaryClass = {astro-ph.IM},
       adsurl = {https://ui.adsabs.harvard.edu/abs/2019Msngr.175....3D}
}

@ARTICLE{Brown2021,
       author = {{Brown}, Anthony G.~A.},
        title = "{Microarcsecond Astrometry: Science Highlights from Gaia}",
      journal = {\araa},
         year = 2021,
        month = sep,
       volume = {59},
        pages = {59-115},
          doi = {10.1146/annurev-astro-112320-035628},
archivePrefix = {arXiv},
       eprint = {2102.11712},
 primaryClass = {astro-ph.IM},
       adsurl = {https://ui.adsabs.harvard.edu/abs/2021ARA&A..59...59B}
}

@ARTICLE{GaiaDR3_AsymmetricalDisk,
       author = {{Gaia Collaboration} and {Drimmel}, R. and {Romero-G{\'o}mez}, M. and {Chemin}, L. and {Ramos}, P. and {Poggio}, E. and {Ripepi}, V. and {Andrae}, R. and {Blomme}, R. and {Cantat-Gaudin}, T. and {Castro-Ginard}, A. and {Clementini}, G. and {Figueras}, F. and {Fouesneau}, M. and {Fr{\'e}mat}, Y. and {Jardine}, K. and {Khanna}, S. and {Lobel}, A. and {Marshall}, D.~J. and {Muraveva}, T. and {Brown}, A.~G.~A. and {Vallenari}, A. and {Prusti}, T. and {de Bruijne}, J.~H.~J. and {Arenou}, F. and {Babusiaux}, C. and {Biermann}, M. and {Creevey}, O.~L. and {Ducourant}, C. and {Evans}, D.~W. and {Eyer}, L. and {Guerra}, R. and {Hutton}, A. and {Jordi}, C. and {Klioner}, S.~A. and {Lammers}, U.~L. and {Lindegren}, L. and {Luri}, X. and {Mignard}, F. and {Panem}, C. and {Pourbaix}, D. and {Randich}, S. and {Sartoretti}, P. and {Soubiran}, C. and {Tanga}, P. and {Walton}, N.~A. and {Bailer-Jones}, C.~A.~L. and {Bastian}, U. and {Jansen}, F. and {Katz}, D. and {Lattanzi}, M.~G. and {van Leeuwen}, F. and {Bakker}, J. and {Cacciari}, C. and {Casta{\~n}eda}, J. and {De Angeli}, F. and {Fabricius}, C. and {Galluccio}, L. and {Guerrier}, A. and {Heiter}, U. and {Masana}, E. and {Messineo}, R. and {Mowlavi}, N. and {Nicolas}, C. and {Nienartowicz}, K. and {Pailler}, F. and {Panuzzo}, P. and {Riclet}, F. and {Roux}, W. and {Seabroke}, G.~M. and {Sordo}, R. and {Th{\'e}venin}, F. and {Gracia-Abril}, G. and {Portell}, J. and {Teyssier}, D. and {Altmann}, M. and {Audard}, M. and {Bellas-Velidis}, I. and {Benson}, K. and {Berthier}, J. and {Burgess}, P.~W. and {Busonero}, D. and {Busso}, G. and {C{\'a}novas}, H. and {Carry}, B. and {Cellino}, A. and {Cheek}, N. and {Damerdji}, Y. and {Davidson}, M. and {de Teodoro}, P. and {Nu{\~n}ez Campos}, M. and {Delchambre}, L. and {Dell'Oro}, A. and {Esquej}, P. and {Fern{\'a}ndez-Hern{\'a}ndez}, J. and {Fraile}, E. and {Garabato}, D. and {Garc{\'\i}a-Lario}, P. and {Gosset}, E. and {Haigron}, R. and {Halbwachs}, J.-L. and {Hambly}, N.~C. and {Harrison}, D.~L. and {Hern{\'a}ndez}, J. and {Hestroffer}, D. and {Hodgkin}, S.~T. and {Holl}, B. and {Jan{\ss}en}, K. and {Jevardat de Fombelle}, G. and {Jordan}, S. and {Krone-Martins}, A. and {Lanzafame}, A.~C. and {L{\"o}ffler}, W. and {Marchal}, O. and {Marrese}, P.~M. and {Moitinho}, A. and {Muinonen}, K. and {Osborne}, P. and {Pancino}, E. and {Pauwels}, T. and {Recio-Blanco}, A. and {Reyl{\'e}}, C. and {Riello}, M. and {Rimoldini}, L. and {Roegiers}, T. and {Rybizki}, J. and {Sarro}, L.~M. and {Siopis}, C. and {Smith}, M. and {Sozzetti}, A. and {Utrilla}, E. and {van Leeuwen}, M. and {Abbas}, U. and {{\'A}brah{\'a}m}, P. and {Abreu Aramburu}, A. and {Aerts}, C. and {Aguado}, J.~J. and {Ajaj}, M. and {Aldea-Montero}, F. and {Altavilla}, G. and {{\'A}lvarez}, M.~A. and {Alves}, J. and {Anders}, F. and {Anderson}, R.~I. and {Anglada Varela}, E. and {Antoja}, T. and {Baines}, D. and {Baker}, S.~G. and {Balaguer-N{\'u}{\~n}ez}, L. and {Balbinot}, E. and {Balog}, Z. and {Barache}, C. and {Barbato}, D. and {Barros}, M. and {Barstow}, M.~A. and {Bartolom{\'e}}, S. and {Bassilana}, J.-L. and {Bauchet}, N. and {Becciani}, U. and {Bellazzini}, M. and {Berihuete}, A. and {Bernet}, M. and {Bertone}, S. and {Bianchi}, L. and {Binnenfeld}, A. and {Blanco-Cuaresma}, S. and {Boch}, T. and {Bombrun}, A. and {Bossini}, D. and {Bouquillon}, S. and {Bragaglia}, A. and {Bramante}, L. and {Breedt}, E. and {Bressan}, A. and {Brouillet}, N. and {Brugaletta}, E. and {Bucciarelli}, B. and {Burlacu}, A. and {Butkevich}, A.~G. and {Buzzi}, R. and {Caffau}, E. and {Cancelliere}, R. and {Carballo}, R. and {Carlucci}, T. and {Carnerero}, M.~I. and {Carrasco}, J.~M. and {Casamiquela}, L. and {Castellani}, M. and {Chaoul}, L. and {Charlot}, P. and {Chiaramida}, V. and {Chiavassa}, A. and {Chornay}, N. and {Comoretto}, G. and {Contursi}, G. and {Cooper}, W.~J. and {Cornez}, T. and {Cowell}, S. and {Crifo}, F. and {Cropper}, M.},
        title = "{Gaia Data Release 3. Mapping the asymmetric disc of the Milky Way}",
      journal = {\aap},
         year = 2023,
        month = jun,
       volume = {674},
          eid = {A37},
        pages = {A37},
          doi = {10.1051/0004-6361/202243797},
archivePrefix = {arXiv},
       eprint = {2206.06207},
 primaryClass = {astro-ph.GA},
       adsurl = {https://ui.adsabs.harvard.edu/abs/2023A&A...674A..37G}
}

@ARTICLE{Morgan1951,
       author = {{Morgan}, W.~W.},
        title = "{Application of the principle of natural groups to the classification of stellar spectra.}",
      journal = {Publications of Michigan Observatory},
         year = 1951,
        month = jan,
       volume = {10},
        pages = {33},
       adsurl = {https://ui.adsabs.harvard.edu/abs/1951POMic..10...33M}
}

@ARTICLE{Herrero2022,
       author = {{Herrero}, A. and {Berlanas}, S.~R. and {Gil de Paz}, A. and {Comer{\'o}n}, F. and {Puls}, J. and {Ram{\'\i}rez Alegr{\'\i}a}, S. and {Garcia}, M. and {Lennon}, D.~J. and {Najarro}, F. and {Sim{\'o}n-D{\'\i}az}, S. and {Urbaneja}, M.~A. and {Gallego}, J. and {Carrasco}, E. and {Iglesias}, J. and {Cedazo}, R. and {Garc{\'\i}a Vargas}, M.~L. and {Castillo-Morales}, {\'A}. and {Pascual}, S. and {Cardiel}, N. and {P{\'e}rez-Calpena}, A. and {G{\'o}mez-Alvarez}, P. and {Mart{\'\i}nez-Delgado}, I.},
        title = "{The nature of the Cygnus extreme B supergiant 2MASS J20395358+4222505}",
      journal = {\mnras},
         year = 2022,
        month = apr,
       volume = {511},
       number = {3},
        pages = {3113-3124},
          doi = {10.1093/mnras/stab3660},
archivePrefix = {arXiv},
       eprint = {2202.11806},
 primaryClass = {astro-ph.SR},
       adsurl = {https://ui.adsabs.harvard.edu/abs/2022MNRAS.511.3113H}
}

@ARTICLE{Levine2006,
       author = {{Levine}, E.~S. and {Blitz}, Leo and {Heiles}, Carl},
        title = "{The Spiral Structure of the Outer Milky Way in Hydrogen}",
      journal = {Science},
         year = 2006,
        month = jun,
       volume = {312},
       number = {5781},
        pages = {1773-1777},
          doi = {10.1126/science.1128455},
archivePrefix = {arXiv},
       eprint = {astro-ph/0605728},
 primaryClass = {astro-ph},
       adsurl = {https://ui.adsabs.harvard.edu/abs/2006Sci...312.1773L}
}

@ARTICLE{Leung2023,
       author = {{Leung}, Henry W. and {Bovy}, Jo and {Mackereth}, J. Ted and {Hunt}, Jason A.~S. and {Lane}, Richard R. and {Wilson}, John C.},
        title = "{A measurement of the distance to the Galactic centre using the kinematics of bar stars}",
      journal = {\mnras},
         year = 2023,
        month = feb,
       volume = {519},
       number = {1},
        pages = {948-960},
          doi = {10.1093/mnras/stac3529},
archivePrefix = {arXiv},
       eprint = {2204.12551},
 primaryClass = {astro-ph.GA},
       adsurl = {https://ui.adsabs.harvard.edu/abs/2023MNRAS.519..948L}
}

@ARTICLE{Willcox2026,
       author = {{Willcox}, R. and {Schneider}, F.~R.~N. and {Laplace}, E. and {Podsiadlowski}, P. and {Maltsev}, K. and {Mandel}, I. and {Marchant}, P. and {Sana}, H. and {Li}, T.~G.~F. and {Hertog}, T.},
        title = "{Good things always come in 3s: Trimodality in the binary black-hole chirp-mass distribution supports bimodal black-hole formation}",
      journal = {\aap},
         year = 2026,
        month = oct,
       volume = {714},
          eid = {A314},
        pages = {A314},
          doi = {10.1051/0004-6361/202557611},
archivePrefix = {arXiv},
       eprint = {2510.07573},
 primaryClass = {astro-ph.SR},
       adsurl = {https://scixplorer.org/abs/2026A&A...714A.314W}
}

@ARTICLE{Schneider2021,
       author = {{Schneider}, F.~R.~N. and {Podsiadlowski}, Ph. and {M{\"u}ller}, B.},
        title = "{Pre-supernova evolution, compact-object masses, and explosion properties of stripped binary stars}",
      journal = {\aap},
         year = 2021,
        month = jan,
       volume = {645},
          eid = {A5},
        pages = {A5},
          doi = {10.1051/0004-6361/202039219},
archivePrefix = {arXiv},
       eprint = {2008.08599},
 primaryClass = {astro-ph.SR},
       adsurl = {https://ui.adsabs.harvard.edu/abs/2021A&A...645A...5S}
}

@ARTICLE{Maltsev2025,
       author = {{Maltsev}, K. and {Schneider}, F.~R.~N. and {Mandel}, I. and {M{\"u}ller}, B. and {Heger}, A. and {R{\"o}pke}, F.~K. and {Laplace}, E.},
        title = "{Explodability criteria for the neutrino-driven supernova mechanism}",
      journal = {\aap},
         year = 2025,
        month = aug,
       volume = {700},
          eid = {A20},
        pages = {A20},
          doi = {10.1051/0004-6361/202554931},
archivePrefix = {arXiv},
       eprint = {2503.23856},
 primaryClass = {astro-ph.SR},
       adsurl = {https://ui.adsabs.harvard.edu/abs/2025A&A...700A..20M}
}

@ARTICLE{Nomoto1987,
       author = {{Nomoto}, Ken'ichi},
        title = "{Evolution of 8--10 M$_{sun}$ Stars toward Electron Capture Supernovae. II. Collapse of an O + NE + MG Core}",
      journal = {\apj},
         year = 1987,
        month = nov,
       volume = {322},
        pages = {206},
          doi = {10.1086/165716},
       adsurl = {https://ui.adsabs.harvard.edu/abs/1987ApJ...322..206N}
}

@ARTICLE{Burrows2025,
       author = {{Burrows}, Adam and {Wang}, Tianshu and {Vartanyan}, David},
        title = "{Channels of Stellar-mass Black Hole Formation}",
      journal = {\apj},
         year = 2025,
        month = jul,
       volume = {987},
       number = {2},
          eid = {164},
        pages = {164},
          doi = {10.3847/1538-4357/addd04},
archivePrefix = {arXiv},
       eprint = {2412.07831},
 primaryClass = {astro-ph.SR},
       adsurl = {https://ui.adsabs.harvard.edu/abs/2025ApJ...987..164B}
}

@ARTICLE{Eggenberger2026,
       author = {{Eggenberger Anderson}, Oliver and {O'Connor}, Evan and {Kovalenko}, Liubov and {Andresen}, Haakon and {Couch}, Sean M.},
        title = "{Black Hole Supernovae Outcomes Across a Wide Progenitor Range}",
      journal = {\apj},
         year = 2026,
        month = nov,
       volume = {1011},
       number = {1},
          eid = {64},
        pages = {64},
          doi = {10.3847/1538-4357/aeaa45},
archivePrefix = {arXiv},
       eprint = {2605.01405},
 primaryClass = {astro-ph.HE},
       adsurl = {https://scixplorer.org/abs/2026ApJ..1011...64A}
}

@ARTICLE{Burrows1988,
       author = {{Burrows}, Adam},
        title = "{Supernova Neutrinos}",
      journal = {\apj},
         year = 1988,
        month = nov,
       volume = {334},
        pages = {891},
          doi = {10.1086/166885},
       adsurl = {https://ui.adsabs.harvard.edu/abs/1988ApJ...334..891B}
}

@ARTICLE{Posdialowski2004,
       author = {{Podsiadlowski}, Ph. and {Langer}, N. and {Poelarends}, A.~J.~T. and {Rappaport}, S. and {Heger}, A. and {Pfahl}, E.},
        title = "{The Effects of Binary Evolution on the Dynamics of Core Collapse and Neutron Star Kicks}",
      journal = {\apj},
         year = 2004,
        month = sep,
       volume = {612},
       number = {2},
        pages = {1044-1051},
          doi = {10.1086/421713},
archivePrefix = {arXiv},
       eprint = {astro-ph/0309588},
 primaryClass = {astro-ph},
       adsurl = {https://ui.adsabs.harvard.edu/abs/2004ApJ...612.1044P}
}

@ARTICLE{Takahashi2013,
       author = {{Takahashi}, Koh and {Yoshida}, Takashi and {Umeda}, Hideyuki},
        title = "{Evolution of Progenitors for Electron Capture Supernovae}",
      journal = {\apj},
         year = 2013,
        month = jul,
       volume = {771},
       number = {1},
          eid = {28},
        pages = {28},
          doi = {10.1088/0004-637X/771/1/28},
archivePrefix = {arXiv},
       eprint = {1302.6402},
 primaryClass = {astro-ph.SR},
       adsurl = {https://ui.adsabs.harvard.edu/abs/2013ApJ...771...28T}
}

@ARTICLE{Temaj2024,
       author = {{Temaj}, D. and {Schneider}, F.~R.~N. and {Laplace}, E. and {Wei}, D. and {Podsiadlowski}, Ph.},
        title = "{Convective-core overshooting and the final fate of massive stars}",
      journal = {\aap},
         year = 2024,
        month = feb,
       volume = {682},
          eid = {A123},
        pages = {A123},
          doi = {10.1051/0004-6361/202347434},
archivePrefix = {arXiv},
       eprint = {2311.05701},
 primaryClass = {astro-ph.SR},
       adsurl = {https://ui.adsabs.harvard.edu/abs/2024A&A...682A.123T}
}

@ARTICLE{Firestone2014,
       author = {{Firestone}, R.~B.},
        title = "{Observation of 23 Supernovae That Exploded <300 pc from Earth during the past 300 kyr}",
      journal = {\apj},
         year = 2014,
        month = jul,
       volume = {789},
       number = {1},
          eid = {29},
        pages = {29},
          doi = {10.1088/0004-637X/789/1/29},
       adsurl = {https://ui.adsabs.harvard.edu/abs/2014ApJ...789...29F}
}

@ARTICLE{Zari2023,
       author = {{Zari}, Eleonora and {Frankel}, Neige and {Rix}, Hans-Walter},
        title = "{Did the Milky Way just light up? The recent star formation history of the Galactic disc}",
      journal = {\aap},
         year = 2023,
        month = jan,
       volume = {669},
          eid = {A10},
        pages = {A10},
          doi = {10.1051/0004-6361/202244194},
archivePrefix = {arXiv},
       eprint = {2206.02616},
 primaryClass = {astro-ph.GA},
       adsurl = {https://ui.adsabs.harvard.edu/abs/2023A&A...669A..10Z}
}

@ARTICLE{Boissier2003,
       author = {{Boissier}, S. and {Prantzos}, N. and {Boselli}, A. and {Gavazzi}, G.},
        title = "{The star formation rate in disc galaxies: thresholds and dependence on gas amount}",
      journal = {\mnras},
         year = 2003,
        month = dec,
       volume = {346},
       number = {4},
        pages = {1215-1230},
          doi = {10.1111/j.1365-2966.2003.07170.x},
archivePrefix = {arXiv},
       eprint = {astro-ph/0309372},
 primaryClass = {astro-ph},
       adsurl = {https://ui.adsabs.harvard.edu/abs/2003MNRAS.346.1215B}
}

@ARTICLE{Oemler2017,
       author = {{Oemler}, Jr., Augustus and {Abramson}, Louis E. and {Gladders}, Michael D. and {Dressler}, Alan and {Poggianti}, Bianca M. and {Vulcani}, Benedetta},
        title = "{The Star Formation Histories of Disk Galaxies: The Live, the Dead, and the Undead}",
      journal = {\apj},
         year = 2017,
        month = jul,
       volume = {844},
       number = {1},
          eid = {45},
        pages = {45},
          doi = {10.3847/1538-4357/aa789e},
archivePrefix = {arXiv},
       eprint = {1611.05932},
 primaryClass = {astro-ph.GA},
       adsurl = {https://ui.adsabs.harvard.edu/abs/2017ApJ...844...45O}
}

@ARTICLE{TaylorCordes1993,
       author = {{Taylor}, J.~H. and {Cordes}, J.~M.},
        title = "{Pulsar Distances and the Galactic Distribution of Free Electrons}",
      journal = {\apj},
         year = 1993,
        month = jul,
       volume = {411},
        pages = {674},
          doi = {10.1086/172870},
       adsurl = {https://ui.adsabs.harvard.edu/abs/1993ApJ...411..674T}
}

%\begin{thebibliography}{99}
%\end{thebibliography}

%%%%%%%%%%%%%%%%%%%%%%%%%%%%%%%%%%%%%%%%%%%%%%%%%%

%%%%%%%%%%%%%%%%% APPENDICES %%%%%%%%%%%%%%%%%%%%%

\appendix

\section{Quality of the SED fitting process}
\label{quality_SED_fitting}

The SED fitting process introduced in Section \ref{SEDfitter} relies on the fitting of 14 near-UV to near-IR photometric bands alongside the corrected \textit{Gaia} DR3 parallax. The photometric measurements on each band are only included if they pass the quality criteria mentioned in Section \ref{data}, with the exception of the \textit{Gaia} G-band which is used for the initial selection of candidate OB stars in Eq. \eqref{mgsec}. For every source, there is therefore a minimum measurement of one photometric band used combined with the corrected \textit{Gaia} DR3 parallax, although in practice, the median number of valid photometric bands for the 1,049,399 candidate OB stars is equal to 6 (often corresponding to the \textit{Gaia} DR3 and 2MASS bands, given that IGAPS and VPHAS+ data saturate easily for visually bright stars and only cover the Galactic plane).

The $\ln(f)$ parameter constitutes an indicator of the quality of the fit, and effectively replaces $\chi^2$ within the framework set by the \texttt{emcee} package. In principle, the smaller the value of $\ln(f)$, the better the fit. Out our sample of 1,049,399 candidate OB stars, 55 \% have been fitted with a median $\ln(f) < -6$ (excellent fit), another  \% with a median $\ln(f)$ between -6 and -4 (good fit)\footnote{Just like in \citetalias{Quintana2025}, we fitted up to 3000 supplementary burn-in iterations for sources initially fitted with $\ln(f) > -4$ that needed more time to converge.} and the remaining 13 \% with $\ln(f)$ > - 4 (mediocre to bad fits).

Nevertheless, the inclusion of more photometric bands does not systematically translate into a smaller value of $\ln(f)$. For example, the median and 1$\sigma$ values of $\ln(f)$ of sources with all three \textit{Gaia} DR3 photometric bands but without any valid measurement in either IGAPS or VPHAS+ are respectively equal to -7.3 and 0.8, whereas the equivalent values for sources with these same bands and at least one valid measurement in either IGAPS or VPHAS+ are equal to -4.7 and 1.5. This is because the MCMC fit is a multi-modal issue, which can notably be affected by:

\begin{itemize}
    \item \textit{Stellar variability}: The SED fitting method relies on the combination of photometric data from several surveys whose observations took place at different epochs. In the case of some variable stars, this potentially translates into similar bands across different surveys (e.g., $G_{\rm BP}$ in \textit{Gaia} and $g$ in IGAPS and VPHAS+) with a significant difference in apparent magnitude ($>$0.3 mag), resulting in a greater value of $\ln(f)$.
    \item \textit{Degeneracy between parallax and reddening}: in our method, $A_V$ is not a free parameter but rather constrained by the extinction map from \citet{Edenhofer2024}. There are some few instances where the reddening is not probably resolved by the extinction map, typically for high-extinction sources, such that their observed photometry could be correctly fitted but not their parallax.
    \item \textit{Intrinsically bright and highly-obscured stars}: these sources are characterized by large values of apparent colour (e.g., $G-K$) small values of absolute magnitude. They are therefore typically fitted with a high fractional age, and often correspond to supergiants.
\end{itemize}

An expanded discussion of such instances is provided in \citet{Quintana2023_PhDThesis}, albeit they originate from a previous version of the SED fitter. All these phenomena are absorbed into the $\ln(f)$ values and consequently taken into account. For these reasons, we have not rejected anomalous sources with poor SED fits as this could affect the completeness of our catalogue; quality control and sanity checks are rather provided in the paper through comparisons with the literature. The online version of our catalogue will however include the $\ln(f)$ values alongside an indication of which photometric band was used for the fitting of every source.

\section{Implementation and validation of the 3D extinction map}
\label{extinction_effects}

\subsection{Propagation of uncertainties}
\label{propagation}

Prior to the SED fitting process, we built extinction curves for the sightline of all the 1,049,399 candidate OB stars as follows. We queried the value of the \texttt{Edenhofer2024} mean extinction on the corresponding Galactic Coordinates and with a distance array with a step of 10 pc (roughly the distance discretization from \citealt{Edenhofer2024}), and assumed the uncertainties to be five times the standard deviation of the extinction\footnote{The actual deviation of extinction uncertainties described in Appendix D from \citet{Edenhofer2024} is more complex and too challenging to implement in our own work. Our simpler choice here stems from the validation catalogue within 2 kpc, where the average ratio between the standard deviation of the extinction and the extinction uncertainties $\sim$0.18.}. We adopted $A_V$ = 0 mag below 69 pc (the minimum distance that the dust map resolves), used the main map between 69 and 1250 pc and the reconstructed map between 1250 and 2000 pc. For sightlines whose maximum distance at which the extinction is resolved is below 2000 pc, we assumed this value of extinction constant up to 2000 pc. Exploiting both versions of the map introduces a potential discontinuity at $d = $ 1250 pc (that is visible on the right panel from Fig. \ref{Parameters_Candidate_OBstars}), although this is partially mitigated by the uncertainties. Due to the differences in quality cuts, there could be a bias towards lower extinction in the reconstructed version of the dust map (Edenhofer 2025, private communication).

The SED fitting process was only run after all the extinction curves were built, as they were used as inputs to fit the observed SED. As the stellar distance is a posterior parameter of the SED fitter, and given that $A_V$ is interpolated as a function of distance, the uncertainties on the extinction map is propagated into all the SED-fitted stellar physical parameters.

%\subsection{General validation of the reconstructed version of \texttt{Edenhofer2024}}
%\label{general_extinction_validation}
%To validate the exploitation of this version, we compare with the results that we would have obtained with the \texttt{Bayestar2019} map (used in previous versions of the SED fitter) We have randomly selected 1000 stars from our list of candidate OB stars with a median SED-fitted distance between 1.25 and 2 kpc, and with $\delta > -30\degr$ (corresponding to the coverage of the \texttt{Bayestar2019} dust map, see \citealt{Green2019}). Similar extinction curves were built, extracting uncertainties on the $A_V$ values from the percentiels given when querying \texttt{Bayestar2019} in the \texttt{dustmaps} package. Then we ran this alternative version of the SED fitter on these 1000 stars.}

%[...]

\subsection{Validation of the reconstructed version of \texttt{Edenhofer2024} in Cyg OB2}
\label{extinction_cygob2}

Our SED fitting method introduced in Section \ref{SEDfitter} interpolates $A_V$ as a function of distance through a constraint of the \texttt{Edenhofer2024} dust map, as detailed in Appendix \ref{propagation}, because keeping $A_V$ as a free parameter introduces some degeneracy to distinguish cool giants from hot stars (see \citealt{Quintana2023_PhDThesis}). Nevertheless, relying on an extinction map potentially engenders biases. Especially, compared with \citetalias{Quintana2025}, our SED fitting process exploits the reconstructed version of the 3D dust map from \citet{Edenhofer2024} between 1.25 and 2 kpc, whose precision is lower than for the main map below 1.25 kpc.

Because our objectives consist of mapping out the massive star-forming regions in the local Milky Way, we want to verify if the reconstructed map can properly resolve them. We then turn to Cyg OB2, as it is one of the most extinguished regions of the volume covered by our census (e.g. \citealt{Wright2015}), suited to test the robustness and reliability of \texttt{Edenhofer2024} into the most extreme conditions. Cyg OB2 was the subject of a dedicated target study for a previous version of the SED fitter \citep{QuintanaWright2021}. We have thus crossmatched our catalogue of SED-fitted OB stars as well as the Cygnus OB association members from \citet{QuintanaWright2021} with the catalogue of massive OB stars in Cyg OB2 from \citet{Wright2015} and the subset of O-type stars from \citet{Berlanas2020}, both of which include spectroscopic measurements of A$_V$ for individual Cyg OB2 members.

\begin{figure*}
    \centering
    \includegraphics[scale = 0.3]{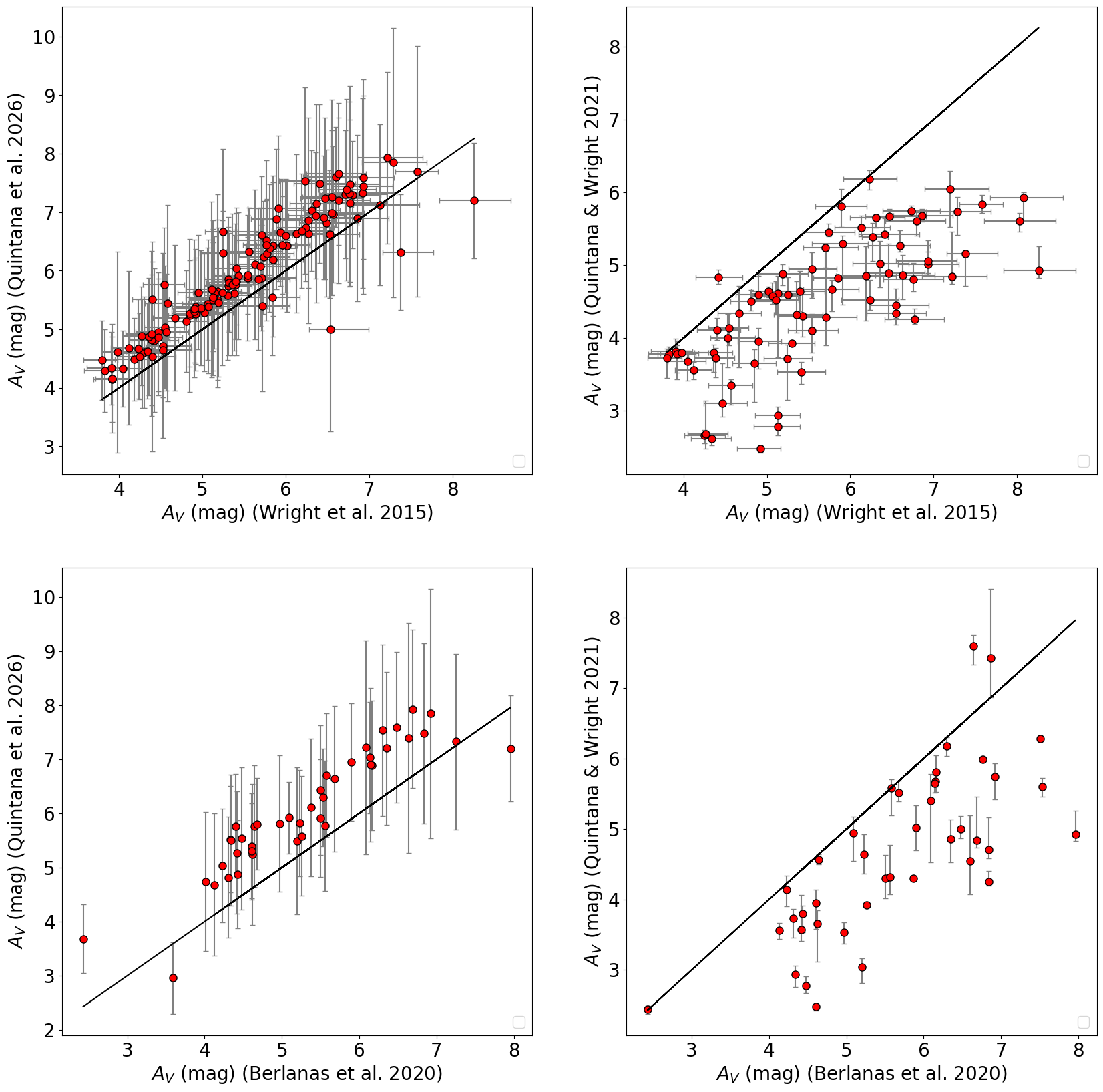}
    \caption{Comparison between the A$_V$ derived in our work (left panels) and from \citet{QuintanaWright2021} (right panels) with the corresponding Cyg OB2 members from \citet{Wright2015} (upper panels) and \citet{Berlanas2020} (lower panels).}
    \label{Comp_Extinction_CygOB2}
\end{figure*}

In Fig. \ref{Comp_Extinction_CygOB2} is displayed the comparison between these measurements. Immediately apparent is the net improvement between the reddening values estimated in \citet{QuintanaWright2021} and the present study. The A$_V$ values from \citet{QuintanaWright2021} are nearly consistently under-estimated compared with spectroscopy. They were interpolated from the \texttt{Bayestar2019} 3D dust map from \citet{Green2019}, which were found to be under-estimated by a factor of $\sim$22 \% in \citet{Delchambre2023}, a correction applied in the improved version of the SED fitter in \citet{Quintana2023}.

By contrast, our A$_V$ values from this study (interpolated from \texttt{Edenhofer2024}) align much better with spectroscopy: although they are slightly over-estimated, they agree within the error bars. The comparison from Fig. \ref{Comp_Extinction_CygOB2} thereby validates the usage of the reconstructed version of the \texttt{Edenhofer2024} dust map for the latest version of the SED fitter. 

\section{Comparison of our SED-fitted distances with geometric distances}
\label{comp_dist}

\begin{figure}
    \centering
    \includegraphics[scale=0.3]{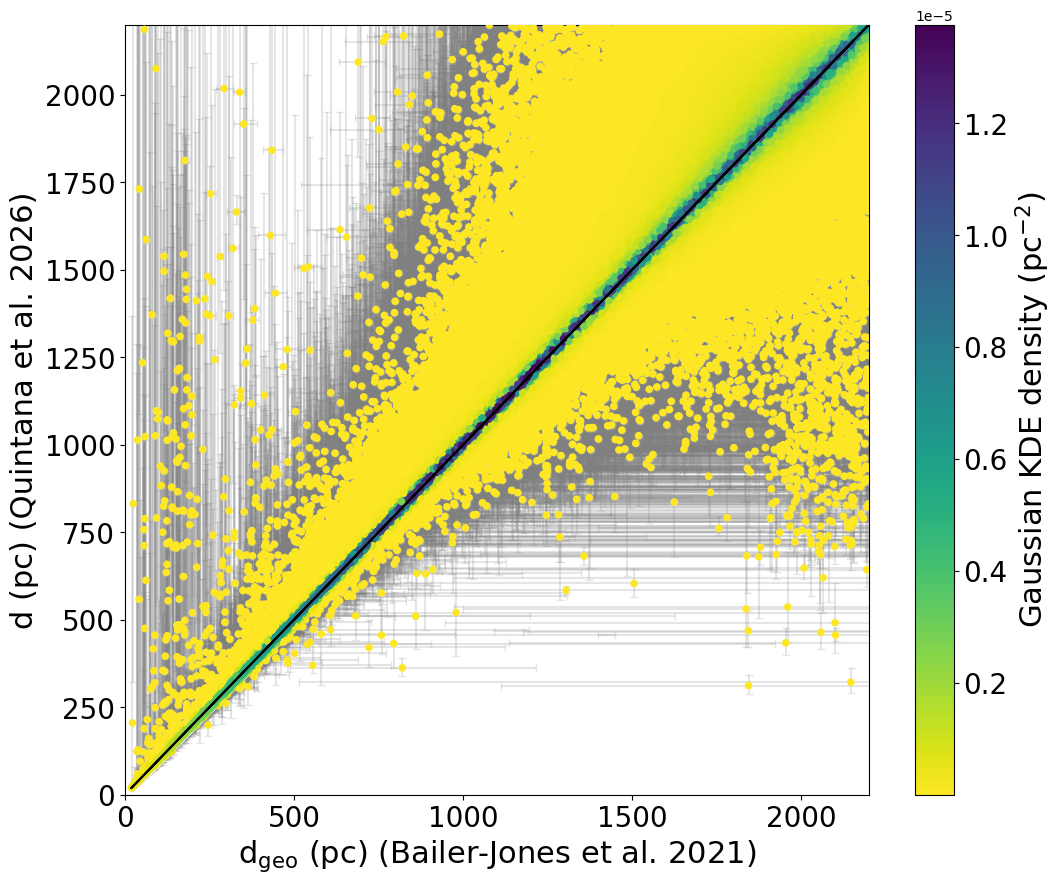}
    \caption{Comparison between our SED-fitted distances (ordinate) with the geometric distances from \citet{BailerEDR3} for our 945,361 candidate OB-type stars within 2.2 kpc, colour-coded by Gaussian KDE density. The black line corresponds to the 1:1 line.}
    \label{Comp_Dist}
\end{figure}

To further assess the reliability of our SED fitting method, we have compared our line-of-sight distances with the geometric distances from \citet{BailerEDR3} in Fig. \ref{Comp_Dist} that were used to make the initial selection of candidate OB-type stars of the sample.

Within our area of interest (i.e., within 2.2 kpc from the Sun), the overwhelming majority of the stars sit on the 1:1 line, consistent with other studies showing that the deviations with the estimates from \citet{BailerEDR3} only start to become significant at larger distances for OB(A) stars (see Fig. 9 from \citetalias{Zari2021} and Fig. 3 from \citetalias{PantaleoniGonzalez2025}). There is still some deviations in Fig. \ref{Comp_Dist} that we attribute to intrinsic differences in the adopted method (as our distance inference is also based on the observed photometry from multiple surveys), as we have used the same priors as in \citet{BailerJones2015}.

\section{Comparison with other catalogues of OB(A) stars}
\label{comparison_obstars}

We have compiled several \textit{Gaia}-based catalogues of OB(A) stars from the recent literature in order to further verify the validity and completeness of our catalogue. For each catalogue, we have performed the crossmatch with our 105,971 OB stars within 2 kpc using the source IDs, directly for those compiled with \textit{Gaia} DR3, while for catalogues based on \textit{Gaia} DR2 data, we exploited the \texttt{dr2\_neighborhood} table in the \textit{Gaia} archive. We used the distances provided in these catalogues to perform the cuts within 2 kpc if available, otherwise we used the geometric distances from \citet{BailerEDR3}.

\begin{table*}
	\centering
	\caption{Comparison between our 105,971 OB stars within 2 kpc with external catalogues of OB(A) stars. Type refers to the nature of the external catalogue irrespective of the stars ultimately selected for crossmatching. N corresponds to the total number of stars in the external catalogue, N$_{\rm C}$ to the total number of selected stars in the external catalogue (as described in Section \ref{comparison_obstars}) within $\sqrt{X^2+Y^2} <$ 2 kpc, N$_{\rm O}$  to our number of OB stars which successfully overlap with the external catalogue,  $R_F = \frac{N_{\rm O}}{N_{\rm C}}$ to the recovery fraction}.
    \label{CompCatalogues}
	\renewcommand{\arraystretch}{1.3} 
	\begin{tabular}{ccccccccc} 
		\hline
		Catalogue & Type & Data used & N & N$_{\rm C}$ & N$_{\rm O}$ & $R_F$ (\%)   \\
		\hline
        \citetalias{Chen2019} & O- and early B-type stars & \textit{Gaia} DR2, VPHAS+ DR2 and spectroscopic catalogues & 14,880 & 4611 & 3436 & 75 \\
        \citetalias{PantaleoniGonzalez2025} & Massive stars & \textit{Gaia} DR3 and spectroscopic catalogues & 20,808 & 8235 & 6246 & 76 \\
        \citetalias{Zari2021} & Luminous OBA stars & \textit{Gaia} DR3 and 2MASS & 417,535 & 103,945 & 37,756 & 36  \\
        \citetalias{Khalatyan2024} & B-type stars & \textit{Gaia} DR3, 2MASS and AllWISE & 454,510 & 116,625 & 55,838 & 48 \\
        \citetalias{Poggio2021} & Upper main-sequence stars & \textit{Gaia} DR3 and 2MASS & 606,996 & 162,791 & 43,864 & 27 \\ 
        \citetalias{GaiaDR3GoldenSample} & OBA stars & \textit{Gaia} DR3 & 3,023,388 & 206,225 & 73,281 & 36 \\
		\hline
	\end{tabular}
\end{table*}

The results of the crossmatching process between our catalogue and theirs are displayed in Table \ref{CompCatalogues}. In addition, Fig. \ref{OBmap_others} shows the normalised surface density map across the X-Y plane of these external catalogues, similarly to ours in Fig. \ref{OBmap}, in order to compare our maps with those obtained with their datasets. The following subsections are then focused on the comparison with individual catalogues.

\begin{figure*}
    \centering
    \includegraphics[scale =0.17]{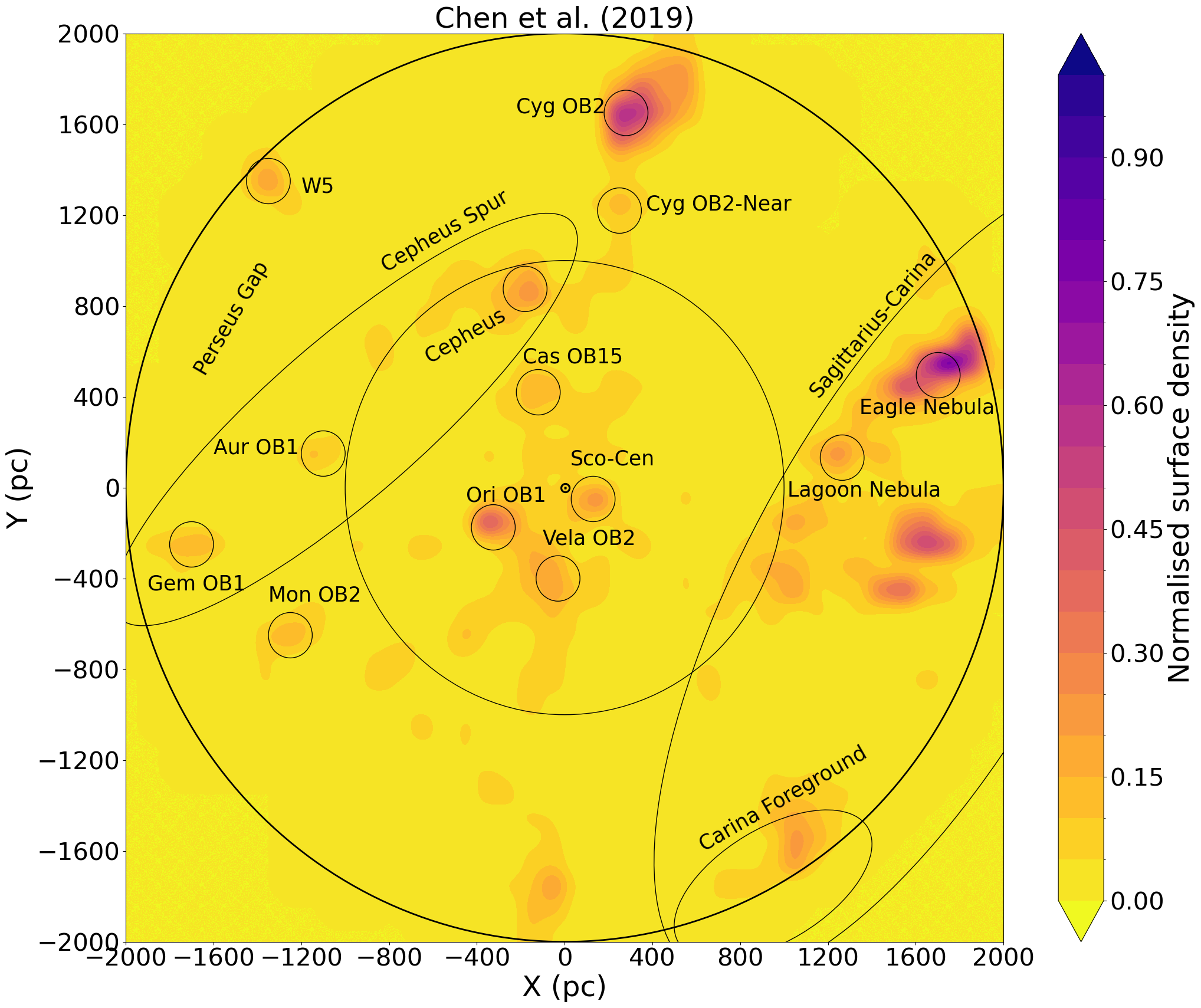}
    \includegraphics[scale =0.17]{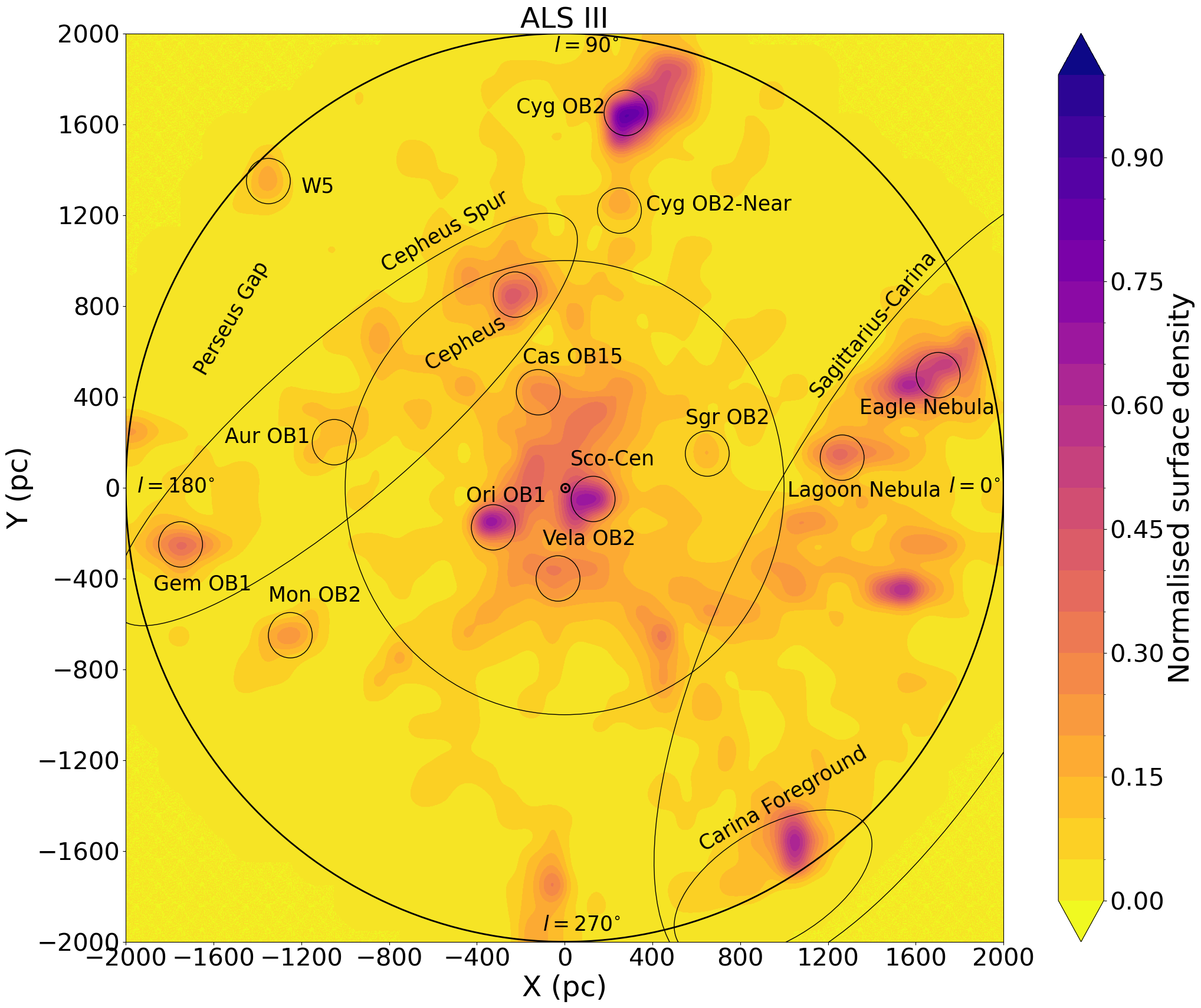}
    \includegraphics[scale =0.17]{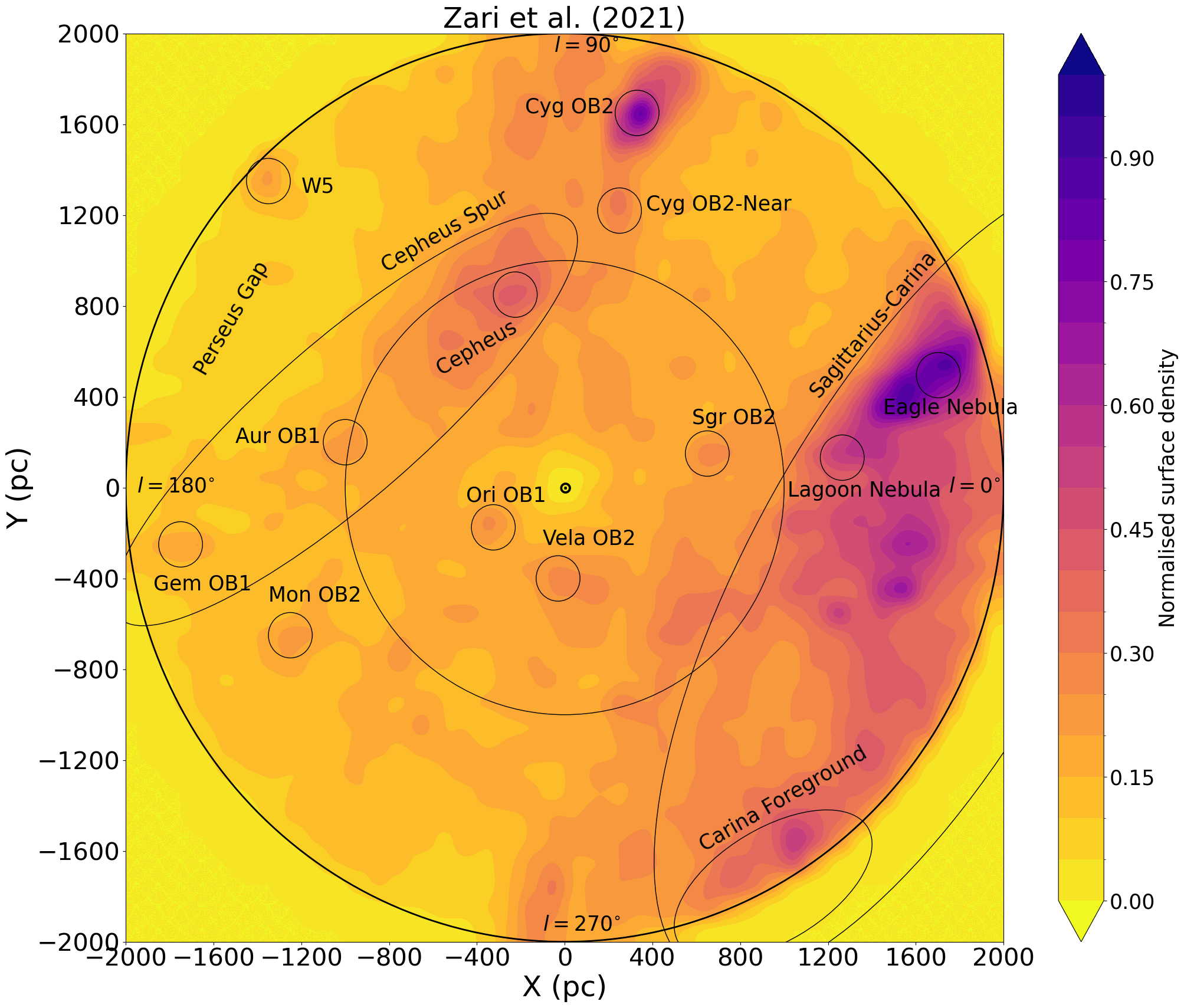}
    \includegraphics[scale =0.17]{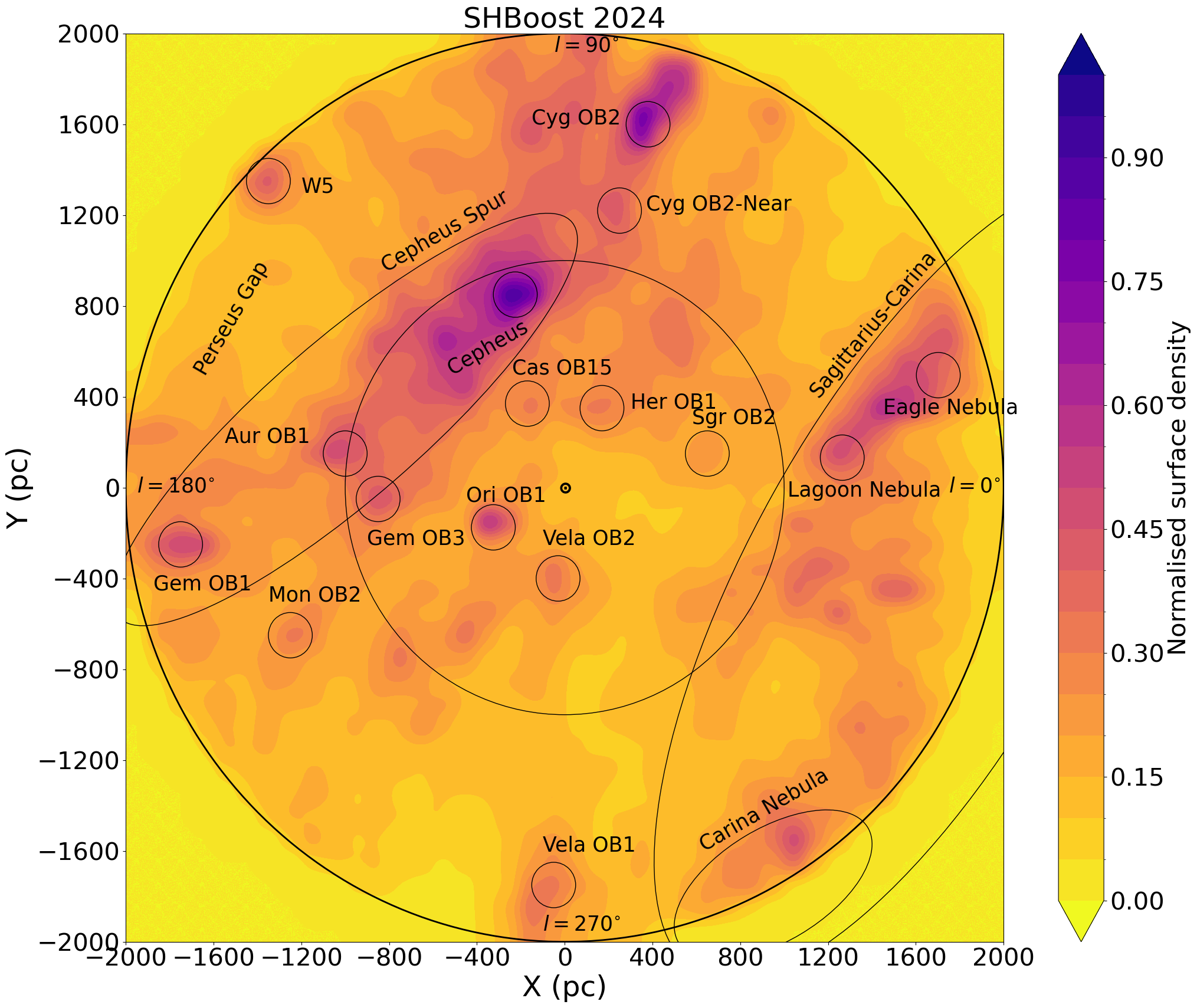}
    \includegraphics[scale =0.17]{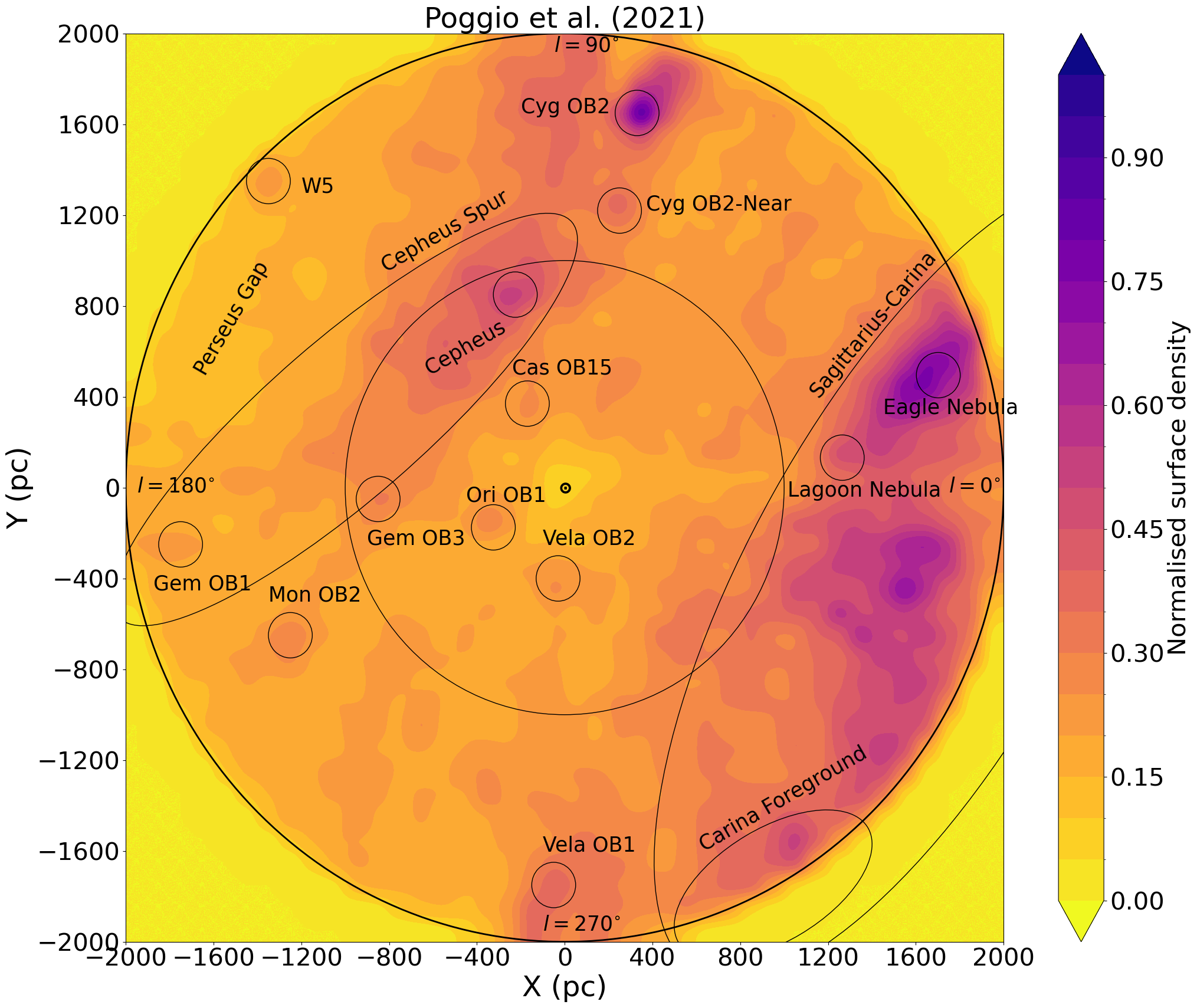}
     \includegraphics[scale =0.17]{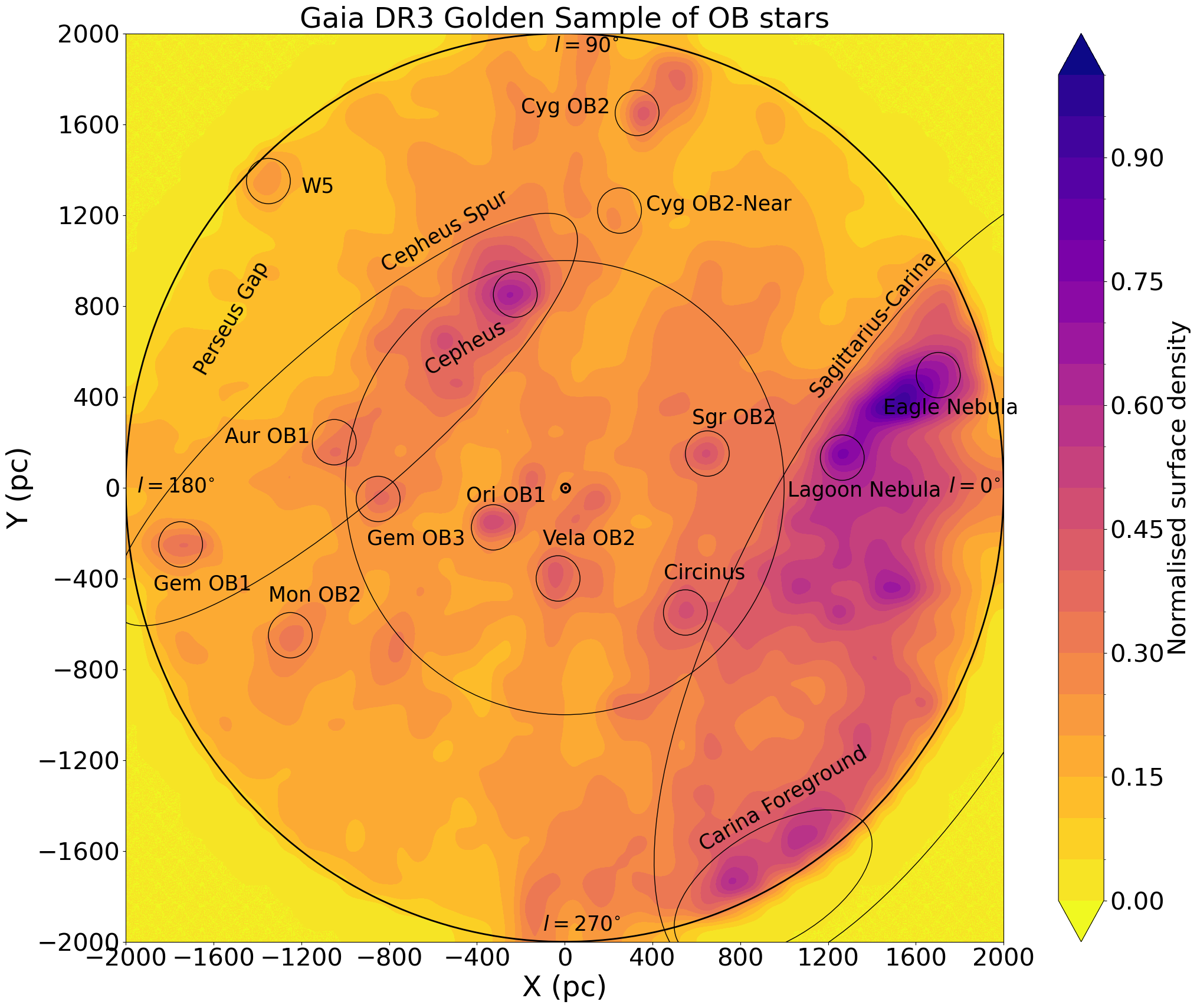}
    \caption{Same as Fig. \ref{OBmap} but for the selected external catalogues of OB(A) stars listed in Table \ref{CompCatalogues}. For all these catalogues, the number of stars displayed is equal to N$_{\rm C}$ in Table \ref{CompCatalogues}.}
    \label{OBmap_others}
\end{figure*}

\subsection{Chen et al. (2019)}
 \citet[][hereafter C19]{Chen2019} used \textit{Gaia} DR2 to produce a catalogue of O- and early B-type stars based on VPHAS+ DR2 photometry as well as spectroscopic data from the literature, resulting in a catalogue of 14,880 OB stars, including 4611 within 2 kpc.

 Our catalogue includes 3436 stars out of these 4611 stars, corresponding to a recovery fraction $R_F$ of $\sim$75 \% (to be compared with $R_F\sim$80 \% in \citetalias{Quintana2025} with the equivalent subset from \citetalias{Chen2019} within 1 kpc). This difference probably stems from differences in distance estimation (as \citetalias{Chen2019} is based on \textit{Gaia} DR2 data): as shown in Fig. \ref{OBmap_others}, W5 and Vela OB1 are located within 2 kpc in \citetalias{Chen2019}, contrary to us. The 102,535 remaining OB stars from our catalogue are, as expected, overwhelmingly late B-type stars (their median $\log(T_{\rm eff})$ is equal to 4.05 compared with the median $\log(T_{\rm eff})$ of 4.28 of our 3436 stars in common with \citetalias{Chen2019}).

\subsection{ALS III}

The third version of the Alma Luminous Star catalogue from \citet[][hereafter PG25]{PantaleoniGonzalez2025} is the latest update of the original ALS catalogue from \citet{Reed2003}, combining the information from \textit{Gaia} DR3 photometry and astrometry and spectroscopic surveys such as the Galactic O-type Star Catalogue \citep[GOSC, ][]{GOSC}. The complete catalogue includes 20,808 stars, classified between several subsets, from which we have selected  `M' (likely massive stars), `I' (intermediate-mass stars, as they sit between the 10,000 K and 20,000 K extinction tracks, see Fig. 2 from \citetalias{PantaleoniGonzalez2025}), as well as those unmatched with \textit{Gaia} (`U'), with bad astrometry (`A') and bad colours (`C), as our selection process was different than theirs. We have removed their sources without distance measurement (with a AFlag of `N') to obtain a catalogue of 8235 stars within 2 kpc to compare with.

Our catalogue includes 6246 stars out of these 8235 stars, hence corresponding to $R_F$ = 76 \%, a significant improvement over the equivalent value of $\sim$51 \% for our census of OB stars within 1 kpc from \citetalias{Quintana2025} compared with the ALS II catalogue from \citet{PantaleoniGonzalez2021}. We attribute this to improvements to both of our catalogues. As expected, our overlap corresponds again to the upper-left part of the HR diagram, as the 6246 stars in common have been fitted with a median SED-fitted $\log(T_{\rm eff})$ of 4.20, compared with the median $\log(T_{\rm eff})$ of 4.04 for the 99,725 remaining stars from our catalogue, which were thereby excluded from ALS III due to being late B-type stars. Conversely, most of the 1989 remaining  stars from this ALS III subset of 8235 stars are likely cooler and evolved objects.

The reliability of ALS III is illustrated in Fig. \ref{OBmap_others}, since the most prominent OB associations and massive star-forming regions are recovered (e.g. Sco-Cen, Ori OB1 and Cyg OB2), but it also displays considerably more structures than the similar catalogue from \citetalias{Chen2019}.

\subsection{Zari et al. (2021)}

\citet[][hereafter Z21]{Zari2021} combined \textit{Gaia} DR3 and 2MASS photometry to identify luminous ($M_{Ks} < 0$ mag) OBA stars in the Milky Way. In doing so they compiled a catalogue of 417,535 stars, including 103,945 within 2 kpc.

 Our catalogue includes 37,756 stars out of these 103,945 stars, corresponding to $R_F = 36$ \%. This is a similar value as in \citetalias{Quintana2025}, implying the contrasts between both catalogues also stem from the differences in our selection process. Our 68,395 remaining stars have a median $\log(T_{\rm eff})$ of 4.03 (compared with the median $\log(T_{\rm eff})$ of 4.11 for our 37,756 stars in common), suggesting that they are dominated by late B-type stars that were too faint to be included in \citetalias{Zari2021}. Likewise, 50,529 out of the 66,369 remaining stars from \citetalias{Zari2021} (out of their 103,945 selected stars) successfully overlap with our 1,049,399 candidate OB stars, implying that they were not included in our final catalogue of 105,971 SED-fitted OB stars within 2 kpc because they were colder (A-type) stars, or are located beyond the edges of our map. Fig. \ref{OBmap_others} reveals that we recover similar large-scale features as \citetalias{Zari2021}, particularly the Perseus Gap and the Sagittarius-Carina arm.

\subsection{SHBoost 2024}

 \citet[][hereafter K24]{Khalatyan2024}, as introduced in Section \ref{spectro}, produced an all-sky list of 454,510 B-type stars based on their SHBoost 2024 catalogue, selecting those with $\log(T_{\rm eff}) \in$ [4.0,4.5] and $\log g < 6$. To these cuts, together with $\sqrt{X^2+Y^2} <$ 2 kpc (based on their measurements), we obtain a list of 116,625 B-type stars within 2 kpc. 

 Our catalogue includes 55,838 stars out of these 116,625 stars, corresponding to $R_F = 48 \%$. Again most of the differences can be attributed to the selection processes and differences in temperature estimations, as out of their 61,267 remaining stars, 47,392 stars are in our initialcatalogue of 1,049,399 candidate OB stars and have been rejected because they were slightly too cool (median $\log(T_{\rm eff})$ of 3.96). This is directly reflected in the temperature comparison of Fig. \ref{CompTeff}. Compared with other OB(A) maps, Fig. \ref{OBmap_others} shows that the Cepheus Spur is more pronounced in \citet{Khalatyan2024} at the cost of the Sagittarius-Carina arm.

 \subsection{Poggio et al. (2021)}

 \citet[][hereafter P21]{Poggio2021} is a catalogue of 606,996 young upper-main sequence stars (referred as UMS in their study), initially defined in \citet{Poggio2018}, and subsequently updated with \textit{Gaia} DR3 and 2MASS data. It includes 162,791 stars within 2 kpc, with a strong overlap with the catalogue from \citetalias{Zari2021}.

 Our catalogue includes 43,864 stars out of these 162,791 stars, corresponding to $R_F = $27 \%. Out of the 62,107 stars only present in our catalogue, 1488 have $G > 15.5$ mag (and have therefore been excluded from \citetalias{Poggio2021}), while 51,637 have $BP-RP > 0$ mag. Conversely, 80,670 out of the 118,927 remaining stars from the subset of 162,791 stars from \citetalias{Poggio2021} were successfully crossmatched in our catalogue of 1,049,399 candidate OB stars: just like for \citetalias{Zari2021}, they are absent from our catalogue of 105,971 SED-fitted OB stars within 2 kpc due to being too cold (A-type stars) or too distant. 
 
 The similarities between the maps from \citetalias{Zari2021} and \citetalias{Poggio2021} are apparent in Fig. \ref{OBmap_others}, especially the dearth of stars around the Sun that can be attributed to the exclusion of sources with 2MASS photometry. However, the Perseus Gap is less pronounced in the map from \citetalias{Poggio2021}, whereas the Sagittarius-Carina arm is more emphasized on the fourth quadrant (bottom right), as this catalogue was more tailored to trace the Galactic spiral arm structure. 

 \subsection{Gaia Golden Sample of OBA stars}

\citet[][hereafter G23]{GaiaDR3GoldenSample} created a clean catalogue of 3,023,388 OBA stars based on the \textit{Gaia} DR3 Apsis modules, defined as stars with $T_{\rm eff} >$ 7500 K in either GSP-Phot or ESP-HS. Following their convention, we have selected the stars with $T_{\rm eff} >$ 10,000 K, and combined with the geometric distances from \citet{BailerEDR3} to select a subsample of 206,225 O- and B-type stars within 2 kpc.

Our catalogue includes 73,281 stars out of these 206,225 stars, corresponding to $R_F = 36$ \%, a similar fraction compared with the equivalent comparison in \citetalias{Quintana2025} within 1 kpc. Once more, this difference can be mostly explained by our distinct method of temperature estimation, as 94,197 out of the 132,944 remaining stars from the subset of \citetalias{GaiaDR3GoldenSample} overlap successfully with our catalogue of 1,049,399 candidate OB stars. Likewise, 7567 out of the remaining 32,690 OB stars within 2 kpc from our catalogue overlap with the general golden sample of OBA stars from \citetalias{GaiaDR3GoldenSample}, implying that a significant fraction of our OB stars could have been classified as A-type stars in their catalogue. 

In \citetalias{Quintana2025} we showed that the golden sample of OBA stars from \citetalias{GaiaDR3GoldenSample} suffered from a higher false positive rate when compared with the spectral types from SIMBAD, and illustrated this by contrasting catalogues in a \textit{Gaia} CMD (Fig. B1 from \citetalias{Quintana2025}). We argue that a similar selection bias could be present here, as the distribution from Fig. \ref{OBmap_others} shows that \citetalias{GaiaDR3GoldenSample} is noticeably noisier than the other maps of OB(A) stars: notably, it is the catalogue where Cyg OB2 stands out the least.

\section{Adopted technique for mapping overdensities of OB stars}
\label{technique_overdensities}

To complement the normalised surface density map from the left panel of Fig. \ref{OBmap} where we followed the approach from \citetalias{Zari2021} and \citetalias{Quintana2025}, we adopt the method from \citetalias{Poggio2021} that allows us to clearly separate the overdensities from the underdensities of OB stars in our map. 

In this framework, the stellar overdensity $\Delta_{\Sigma}$ is defined as:

\begin{equation}
\Delta_{\Sigma} = \frac{\Sigma (X,Y) - <\Sigma (X,Y)>}{<\Sigma (X,Y)>}
\end{equation}

\noindent where $\Sigma (X,Y)$ stands for the local density at the Galactic Cartesian coordinate $(X,Y)$ whilst $<\Sigma (X,Y)>$ is the mean density. Both quantities are computed as in \citetalias{Poggio2021}, using a bivariate kernel estimator from , such that:

\begin{equation}
\label{density}
\small{\Sigma(X,Y)} = \frac{c}{N \, h²_{\rm local}} \sum_{n=i}^{N} \bigg[K \, \bigg(\frac{X- x_i}{h_{\rm local}}\bigg) \, K \, \bigg(\frac{Y- y_i}{h_{\rm local}}\bigg)\bigg]
\end{equation}

\noindent where $N$ is the total number of OB stars in our catalogue, $(x_i,y_i)$ correspond to the $X$ and $Y$ coordinates of each OB star, h is the local bandwidth and the Epanechnikov kernel function is defined as (and same for the Y-coordinate):

%\begin{equation}
\[
K \, \bigg(\frac{X- x_i}{h_{\rm local}}\bigg) = 
\begin{cases}
  \frac{3}{4} \bigg(1-\bigg(\frac{X-x_i}{h_{\rm local}}\bigg)^2\bigg) &  \text{if $|(X-x_i)/h_{\rm local}| < 1$} \\
  0 & \text{otherwise}
\end{cases}
\]
%\end{equation}

In Eq. \ref{density}, compared with \citetalias{Poggio2021}, we have introduced a normalisation factor $c$ because we are covering a limited volume and thereby applying a cut in distance. This factor is set between 0 and 1, and corresponds to the fraction of a uniformly randomized samples of 100,000 points in the X-Y plane between $x_i - h_{\rm local}$ and $x_i + h_{\rm local}$ (and equivalent for the $Y$ direction) that are contained within the 2 kpc circle, allowing us to cancel the border effects akin to the approach from \citet{Palicio2025}.

$<\Sigma (X,Y)>$ is also calculated using Eq. \eqref{density}, but with a distinct value of the mean density bandwidth (labelled $h_{\rm mean}$), such that its normalisation factor $c$ is defined with randomized coordinates between $x_i - h_{\rm mean}$ and $x_i + h_{\rm mean}$.

For this work, we have adopted a value of $h_{\rm local} = 100$ pc (to better reflect the scale of OB associations and star-forming complexes) and $h_{\rm mean} = 500$ pc (smaller than the value from \citetalias{Poggio2021} as our map covers a smaller volume than theirs). 

\section{Comparison with other spiral arm models}
\label{spiral_arm_models}

In Section \ref{galactic_structure}, we compared the spatial distribution of our OB stars with the spiral arm model from \citet{Reid2019} and the contours shaped by the \textit{Gaia} OB stars from \citet{GaiaDR3_AsymmetricalDisk}. We stress there that there is currently no clear consensus on the large-scale spiral structure of the Milky Way, and several models are available in the literature.

Here we will therefore compare our spatial distribution of OB stars with other models of the spiral arms. To that end, we take advantage of the \texttt{SpiralMap} module from Python \citep{SpiralMap}, from which we choose the following models:

\begin{enumerate}
    \item \texttt{Drimmel\_Ceph\_2024} is the model from \citet{Drimmel2025}, based on a sample of 2857 dynamically young Cepheids.
    \item \texttt{Hou\_Han\_II\_2014} is the model from \citet{HouHan2014}, that leveraged 900 masers alongside 1300 GMCs and 2500 H{\sc ii} regions.
    \item \texttt{Taylor\_Cordes\_1992} is the model from \citet{TaylorCordes1993} that also exploited known H{\sc ii} regions to map out the spiral arm structure of the Milky Way.
    \item \texttt{Vallee\_1995} is the model from \citet{Vallee1995} that combines magnetic field data with interstellar dust \& gas as well as stars.
\end{enumerate}

\begin{figure*}
    \centering
    \includegraphics[scale = 0.17]{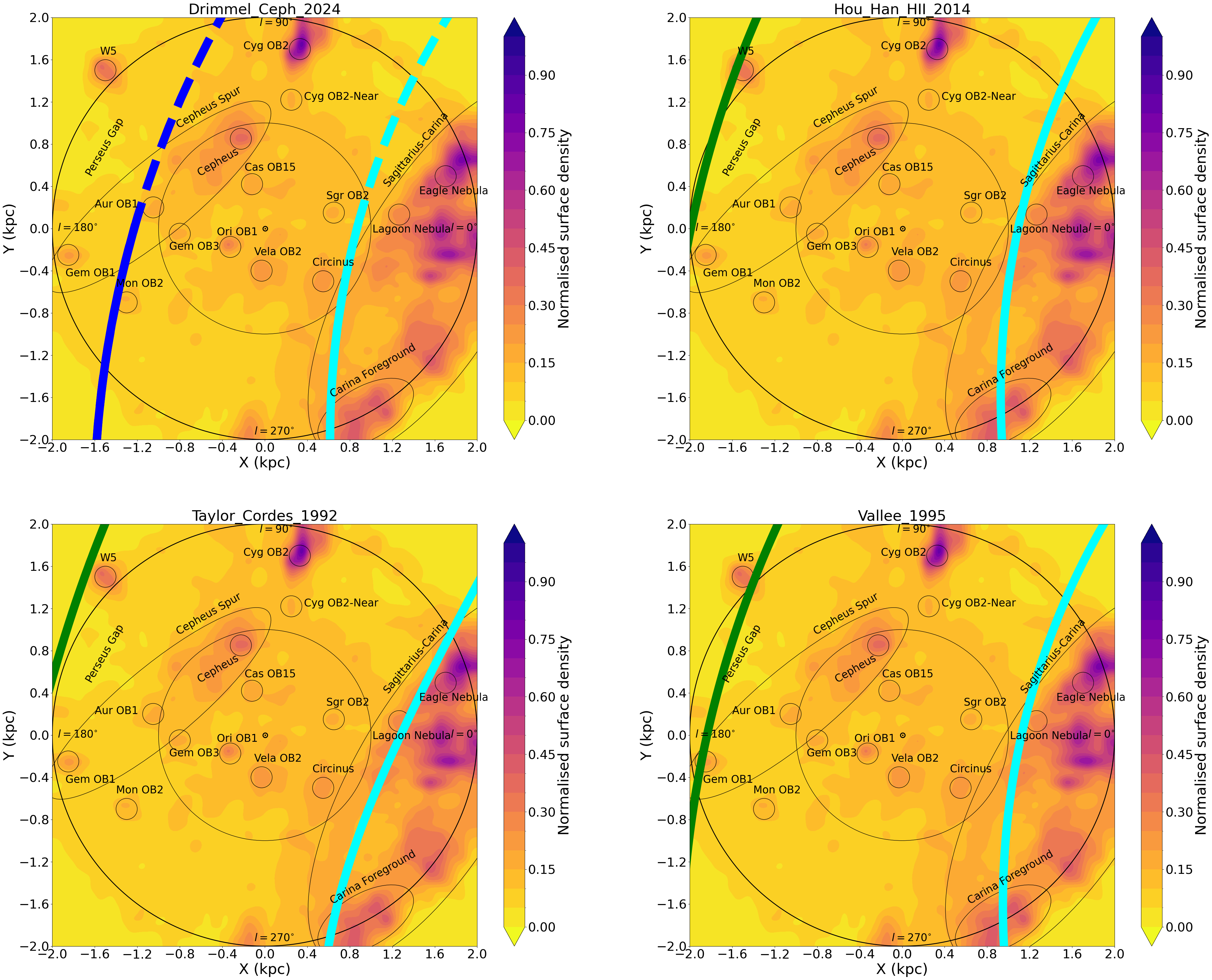}
    \caption{Same as Fig. \ref{OBmap} but with the spiral arm models from \citet{Drimmel2025} (upper left panel), \citet{HouHan2014} (upper right panel), \citet{TaylorCordes1993} (lower left panel) and \citet{Vallee1995} (lower right panel). We have adopted the same colour convention as for the left panel Fig. \ref{OBmap_SpiralArms} for identical spiral arms, with the Perseus, Orion-Cygnus and Sagittarius-Carina arms displayed in green, blue and cyan, respectively. For each model, we again adopted a distance of the Sun to the Galactic Centre of $R_0 = 8.23$ kpc from \citet{Leung2023}.}
    \label{OBmap_SpiralArms_Others}
\end{figure*}

All these models have been plotted on top of our normalised surface density map of OB stars in Fig. \ref{OBmap_SpiralArms_Others}. Effectively, they present a complementary view of the structure of the local Milky Way, with individual spiral arms characterized by different pitch angles. While \texttt{Taylor\_Cordes\_1992}, \texttt{Vallee\_1995} and \texttt{Drimmel\_Ceph\_2024} all have four arms, \texttt{Hou\_Han\_II\_2014} is characterized by a number of six arms.

Out of these four models, only \texttt{Drimmel\_Ceph\_2024} includes the local (Orion-Cygnus) arm, but is also the model deviating the most from our overdensities of OB stars. Its fit of the Perseus arm is beyond the coverage of our sample, whereas the fit of its Orion-Cygnus arm does not cross Cyg OB2, Cepheus and Ori OB1 contrary to the model from \citet{Reid2019} (see left panel from Fig. \ref{OBmap_SpiralArms}). 

The Cepheids catalog from \texttt{Drimmel\_Ceph\_2024} has a better coverage for Y $<$ 0 pc, where our catalogue of OB stars contains fewer stars. This is therefore difficult to compare (especially for the Orion-Cygnus arm). On the other hand, the lower part of the Sagittarius-Carina arm seems in reasonable agreement; however,the upper parts of the plot (Y $>$0 pc) seem to deviate most from our structures. \texttt{Drimmel\_Ceph\_2024} is defined for Y $<$ 0 pc (hence the dashed lines for positive Y values), hence the disagreement with the extrapolation from \texttt{Drimmel\_Ceph\_2024} on such small scales (as Cepheids are mapped out at large very distances).

\section{Observed OB-type star ccSN explosion and direct-collapse BH progenitor candidates in the Hertzsprung-Russel diagram}

\label{imminent}

In Section \ref{waiting_times}, we identified OB primary star candidates for imminent ccSN explosions (within the next 1 Myr) or direct-collapse BH formation events (within the next 0.5 Myr) in the local field, and showed their spatial distribution in Fig. \ref{XY_Progenitors}.

\begin{figure}
    \centering
    \includegraphics[scale=0.32]{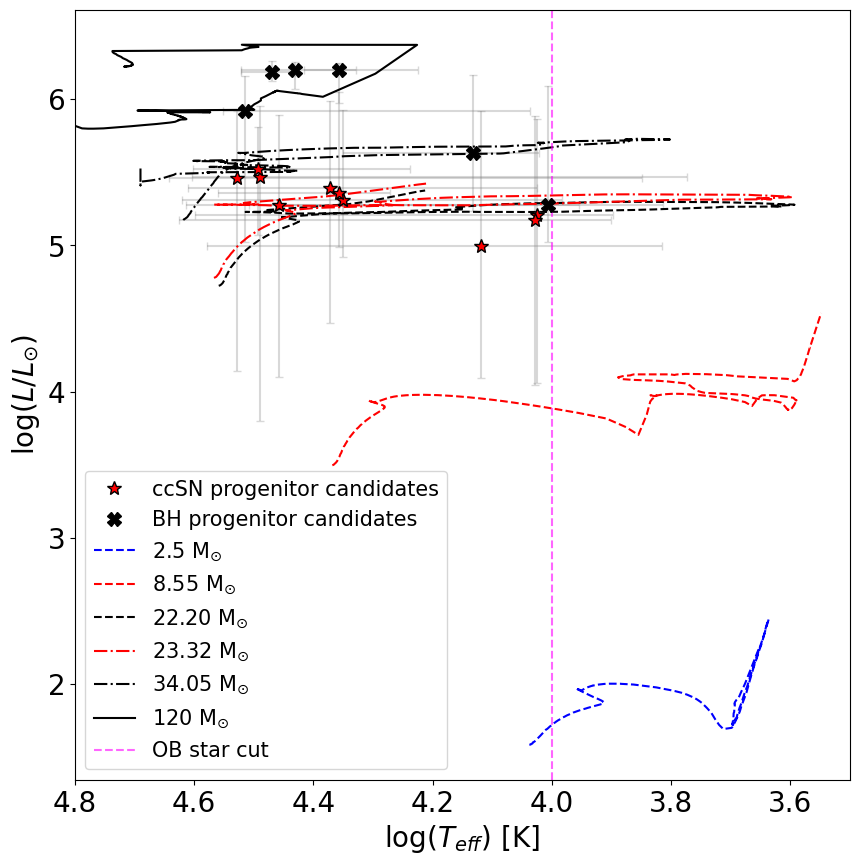}
    \caption{HR diagram of the candidate successful (as blue stars) and failed ccSN (as green crosses) progenitors that we predict to explode/collapse within the next 1 or 0.5 Myr, respectively, based on their SED-fitted parameters (see in Table \ref{Next_explosions}). For comparison, we also display the stellar evolutionary models from \citetalias{Ekstrom} at $M_\mathrm{ZAMS}/M_\odot \in \{ 8.55, 22.20, 23.34, 34.05\}$ demarcating the thresholds between successful and failed ccSN outcomes (see Table \ref{MassiveStars_Fates}). We also indicate our separation criterion between OB and non-OB stars (the latter not being part of our analysis) as well as the models at $M_\mathrm{ZAMS}$ =
     2.5 M$_{\odot}$ (threshold between A-type and B-type stars according to \citealt{Mamajek}) and the most massive $M_\mathrm{ZAMS}$ =120 M$_{\odot}$ from \citetalias{Ekstrom} for comparison.}
    \label{HR_Diagram_Fates}
\end{figure}

Fig. \ref{HR_Diagram_Fates} then displays their distribution across the HR diagram, plotted alongside the stellar evolutionary models from \citetalias{Ekstrom} at the  ZAMS masses thresholds marking the transitions between explosive and non-explosive compact remnant formation adopted in Table \ref{MassiveStars_Fates}. This figure highlights that aside from the most luminous direct-collapse BH progenitors ($\log(L/L_{\odot}) > 5.6$), the error bars from our astro-photometric fit are too large to faithfully predict the final fates of our ccSN explosion and direct-collapse BH progenitors of our catalogue, justifying the approach of a synthetic modelling that we will adopt in Paper II.

% Don't change these lines
\bsp	% typesetting comment
\label{lastpage}
\end{document}